\documentclass[useAMS,usenatbib,usegraphicx,usedcolumn]{mn2e}

\usepackage{multirow}

\newcommand{\mg}{Mg\,{\scriptsize I}}
\newcommand{\mgII}{Mg\,{\scriptsize II}}
\newcommand{\caI}{Ca\,{\scriptsize I}}
\newcommand{\caII}{Ca\,{\scriptsize II}}
\newcommand{\multi}{{\scriptsize MULTI}}
\newcommand{\marcs}{{\scriptsize MARCS}}
\newcommand{\nist}{{\scriptsize NIST}}
\newcommand{\vald}{{\scriptsize VALD}}

\title[A grid of NLTE corrections for Mg and Ca for evolved stars]{A grid of NLTE corrections for magnesium and calcium in late-type giant and supergiant stars: application to {\it Gaia}}
\author[T. Merle, F. Th\'evenin, B. Pichon, L. Bigot]{T. Merle\thanks{E-mail:
thibault.merle@oca.eu (TM)}, F. Th\'evenin, B. Pichon, L. Bigot\\
Universit\'e de Nice Sophia-antipolis; CNRS; Observatoire de la C\^ote d'Azur, Laboratoire Cassiop\'ee\\
B.P. 4229, 06304 Nice Cedex 4, France\\
}

\begin{document}

\date{Accepted ?. Received ?; in original form ?}

\pagerange{\pageref{firstpage}--\pageref{lastpage}} \pubyear{2011}

\maketitle

\label{firstpage}

\begin{abstract}
We investigate NLTE effects for magnesium and calcium in the atmospheres of late-type giant and supergiant stars. 
The aim of this paper is to provide a grid of NLTE/LTE equivalent width ratios $W/W^*$ of Mg and Ca lines for the following range of stellar parameters: $T_{\rm {eff}} \in [3500, 5250]$~K, $\log\ g \in [0.5, 2.0]$~dex and [Fe/H$] \in [-4.0, 0.5]$~dex. 
We use realistic model atoms with the best physics available and taking into account the fine structure.  
The Mg and Ca lines of interest are in optical and near IR ranges.
A special interest concerns the lines in the {\it Gaia} spectrograph (RVS) wavelength domain [8470, 8740]~\AA.
The NLTE corrections are provided as function of stellar parameters in an electronic table as well as in a polynomial form for the {\it Gaia}/RVS lines.
\end{abstract}


\section{Introduction}

Individual stars of different ages carry information that allow astronomers to detail the sequence of events which built up the Galaxy. 
Calcium (Ca) and magnesium (Mg) are key elements for the understanding of these events since these $\alpha$-elements are mainly produced by supernovae of type II. 
Therefore, they are both important tracers to derive trends of chemical evolution of the Galaxy.
These two elements are easily observed in spectra of late-type stars through strong or weak lines. 
Among them, there are the \mg\ b green triplet, the \caI\ red triplet and the \caII\ IR triplet (CaT) lines which are measurable in the most extreme metal-poor stars \citep{Christlieb02} even at moderate spectral resolutions.

The lines of these $\alpha$-elements provide diagnostic tools to probe a wide range of astrophysical mechanisms in stellar populations.
The study of $\alpha$-element synthesis in halo metal-poor stars constrains the rate of SNe II \citep{Nakamura99}.     
The Mg element is a reliable reference of the early evolutionary time scale of the Galaxy (e.g. \citealt{Cayrel04}, \citealt{Gehren06} and \citealt{Andrievsky10}).
Moreover, the \caI\ triplet at $\lambda \lambda$ 6102, 6122 and 6162, can be used as a gravity indicator in detailed analyses of dwarf and subgiant stars \citep{Cayrel96}.
Two Ca lines of different ionization stages can also be used to constraint surface gravity in extremely metal-poor stars (\caI\ 4226~\AA\ and \caII\ 8498~\AA) as shown by \citet{Mashonkina07}. 
Moreover, the line strengths of CaT lines can be used as metallicity indicator for stellar population in distant galaxies \citep{Id97} or to search for the most extreme metal-poor stars in the vicinity of nearby dwarf spheroidal galaxies \citep{Starkenburg10}.
The CaT lines are also used to probe stellar activity by the measure of the central line depression as done by \citet{Andretta05} and \citet{Busa07}. 

It is well known that the assumption of Local Thermodynamic Equilibrium (LTE) is the main source of uncertainties (after the quality of the atomic data). 
In his review, Asplund (2005) summarized the need to take Non-LTE (NLTE) effects into account to get reliable and precise chemical abundances. 
Such effects are crucial in the atmospheres of metal-poor stars due to several mechanisms: the overionization in the presence of a weak opacity, the photon pumping and the photon suction. 
However, since the NLTE computations are still time consuming as soon as we deal with realistic  model atoms,  most of the studies are still done under the LTE assumption for large surveys (e.g., \citealt{Edvardsson93} for late-type dwarf stars, \citealt{Liu10} and \citealt{DeSilva11} for late-type giant and subgiant stars).

Several studies of NLTE effects on \mg\ lines were performed for stars of various metallicities \citep{Mashonkina96, Zhao98, Zhao00, Gratton99, Shimanskaya00, Idiart00, Mishenina04, Andrievsky10}. 
All studies showed an increase of NLTE effects with decreasing metallicity which imply corrections up to $+0.2$~dex. 
The first attempt to build a realistic model atom of \mg\ was performed by \citet{Carlsson92} to explain the \mg\ solar emission lines near to 12 $\mu$m. 
They demonstrated the photospheric nature of these lines by using the photon suction mechanism which required a model atom with Rydberg states to ensure the collisional-dominated coupling with the ionization stage.   
The most complete Mg model atom was built by \citet{Przybilla01} with 88 energy levels for \mg\ and 37 levels for \mgII.  However, they did not take into account the fine structure of the triplet system. 
The first \mg\ model atom that includes the fine structure was built by \citet{Idiart00} who applied it to the determination of the Mg abundance for more than two hundred late-type stars. 
We can mention an update of the \mg\ model atom of \citet{Carlsson92} done by \citet{Sundqvist08} in order to reproduce IR emission lines in K giants, taking into account the electronic collisions between states with the principal quantum number $n \le 15$. 

NLTE effects on Ca were investigated by \citet{Jorgensen92} who gave a relation between equivalent widths of CaT lines as functions of the surface gravities and the effective temperatures. 
As for the \mg, \citet{Idiart00} included the fine structure in their model atom of \caI.  
They combined these results to the NLTE Fe abundance predictions \citep{Thevenin99} and performed an NLTE analysis of the same set of late-type dwarf and subgiant stars.
They concluded that the [Ca/Fe] remained mostly unchanged but some parallel structures appeared in the NLTE [Ca/Fe] vs [Fe/H] diagram.
The most complete Ca atom was done by \citet{Mashonkina07} who built a model atom through two ionization stages with 63 energy levels for \caI\ and 37 levels for \caII, including fine structure only for the first states. 
They used the most recent and accurate atomic data from the Opacity Project \citep{Seaton94}. 
They concluded that NLTE abundance corrections can be as large as 0.5 dex (for the \caI\ 4226 \AA\ resonance line) in extremely metal-poor stars. 
Using these NLTE corrections, \citet{Asplund09} found good agreement of solar \caI\ and \caII\ abundances.

In the view of the ESA {\it Gaia} mission (\citealt{Wilkinson05}), it is important to investigate such NLTE abundance corrections in order to distinguish different stellar populations. 
The onboard spectrograph, named RVS (Radial Velocity Spectrometer, \citealt{Katz04}) which was designed primarily to get the radial velocities can also be used in principle to extract individual chemical abundances.   
The spectral window considered was selected in the near-infrared [8470-8740]~\AA\ because of the presence of the CaT lines which will be visible even in the most metal-poor stars and at a considered low resolution ($R~\sim$~11500 at best).
The counterpart ground-based observations is the RAVE survey \citep{Steinmetz06} which will provide about 1 million of spectra. 
These two surveys will analyze a representative sample of stars throughout the Galaxy, allowing an unprecedented kinematic and chemical study of the stellar populations.
The huge amount of stars will prevent us to analyze them one by one. 
It is necessary to have precalculated correction tables that can be used for automatic stellar parameter determination and this is the aim of this work. 


The structure of the paper is as follows: in Sect.~2, we  present the model atoms of \mg, \caI\ and \caII.
In Sect.~3, we test these models by comparing NLTE synthetic lines with several solar (Sect.~3.1) and Arcturus (Sect.~3.2) lines of \mg\ and \caII\ in the {\it Gaia}/RVS region.
Then, in Sect.~4.1, we present a grid of NLTE corrections of equivalent widths of spectral lines, selected in Sect.~4.2, useful for the {\it Gaia} mission and the RAVE survey.
The behaviours of \mg, \caI\ and \caII\ NLTE lines, the comparisons with previous works, the estimate of uncertainties on atomic data, and the polynomial fits of the {\it Gaia}/RVS lines are discussed in Sect.~4.   

\section{\mg, \caI\ and \caII\ model atoms}

The construction of a model atom consists in collecting and computing atomic data which are: 
\begin{itemize}
\item the excited states of the atom characterized by their energy levels and their statistical weights; 
\item the radiative transitions characterized by the oscillator strength  values $f$ for the bound--bound transitions, and by the photoionization cross-sections for the bound--free ones; 
\item the collisional transitions with electrons and neutral hydrogen characterized by the collisional strength $\Omega$; 
\item the elastic radiations and collisions with neutral hydrogen characterized by the radiative ($\gamma_{{\rm rad}}$) and the collisional ($\gamma_{{\rm coll}}$) damping factors.
\end{itemize}

\subsection{Construction of model atoms}\label{section21}
{\it Energy levels.} 
The energy levels come from the \nist\footnote{available at: http://physics.nist.gov/asd3} \citep{Ralchenko08} and the TopBase\footnote{available at: http://cdsweb.u-strasbg.fr/topbase/topbase.html} \citep{Cunto92} on line atomic databases. 
When theoretical levels are present in the TopBase but absent in the \nist, we add them in the model atoms by shifting them with respect to the \nist\ ionization energies. 
We consider the fine structure of energy levels as long as they exist in \nist\ database. 
If some fine levels are missing for the highest levels in atomic databases, we compute them as specified in the next sub-sections. 

\begin{figure}
\includegraphics[width=\linewidth]{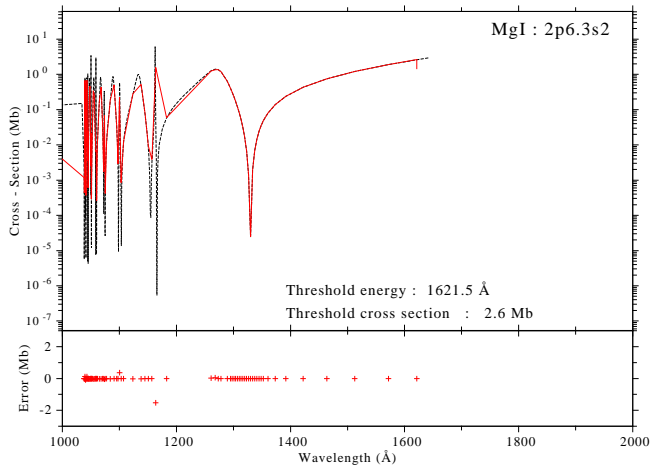}
\caption{Photoionization cross-section as a function of the energy of the incident photon for the $3s^{2}\ ^{1}S$ ground state  of \mg.
We use a slide window to smooth data after the resonance peak and undersample them. 
Dashed line stands for data from TopBase, red full line stands for smoothed and undersampled data used in our atomic model.}
\label{Phot_MgI}
\end{figure}

{\it Radiative transitions.} 
The oscillator strengths of bound-bound transitions considered here are from \vald\footnote{available at: http://vald.astro.univie.ac.at}\citep{Kupka00}, Kurucz\footnote{available at:\\ http://www.cfa.harvard.edu/amp/ampdata/kurucz23/sekur.html} \citep{Kurucz95} and \nist\ on line databases. We take the \vald\ log $gf$ values as default for the whole transitions of an atom and choose the most accurate one among \vald, \nist, \cite{Hirata95}, and Kurucz for the selected lines. 
For the bound-free transitions, we use the theoretical tables of photoionization cross-sections provided by the TopBase. 
We shift the TopBase energy threshold of photoionization to the corresponding \nist\ values to keep consistency with the measured \nist\ energy levels.
In order to save computational time, we undersampled the photoionization cross-section tables taking care to keep large resonance peaks near the photoionization threshold but smoothing faint peaks far from the threshold using a slide window (see Fig.~\ref{Phot_MgI}). 
We assume that the photoionization cross-sections are the same for the fine structure of a level.
Some photoionization data are missing for high levels, therefore we calculated photoionization threshold using hydrogenic cross-section formulation of \citet{Karzas1961},
allowing us to compute Gaunt's factor as  functions of the quantum numbers $n$ and $l$.
This approach leads to a better estimation than the standard formulation of \citet{Menzel35} and \citet{Mihalas78}.

{\it Collisional transitions.} 
The inelastic collisions between the considered atoms with the other particles of the medium tend to set the atom in the Boltzmann Statistical Equilibrium (SE).
A careful treatment of these collisional transitions is therefore important for the SE that defines the level populations.
Tables of quantum calculations of the collisional cross-sections with electrons are implemented when available.
Otherwise, we use the modified Impact Parameter Method (IPM) of \citet{Seaton62a} for allowed bound--bound collisions with electrons, which uses values of oscillator strengths corresponding to radiative ones, and depends on the electronic density and temperature structures of the model atmosphere. 
\citet{Seaton62a} emphasized that the collisional rates are more reliable when $(13.6/ \Delta E) f$ is large, where $\Delta E$ is the energy of the transition in eV. 
Then, for intercombination lines with low $f$ values, the collisional rate can be under-estimated.
For the forbidden transitions, when no quantum computations are available, it is generally assumed that the collisional strength $\Omega_{ij}$ does not depend on the energy of the colliding electrons. 
We choose $\Omega_{ij} = 1$ (as in \citealt{Allen73} and \citealt{Zhao98}) since quantum calculations give this order of magnitude for some forbidden optical lines (e.g. \citealt{Burgess95}).  
For forbidden transitions between fine levels, we choose $\Omega_{ij} = 10$ in order to couple them by collisions. 
Collisional ionization with electrons are computed following \citet{Seaton62b} in \citet{Mihalas78} where the collisional ionization rates are taken to be proportional to the photoionization cross-sections. 
\begin{figure*}
\includegraphics[width=0.8\linewidth]{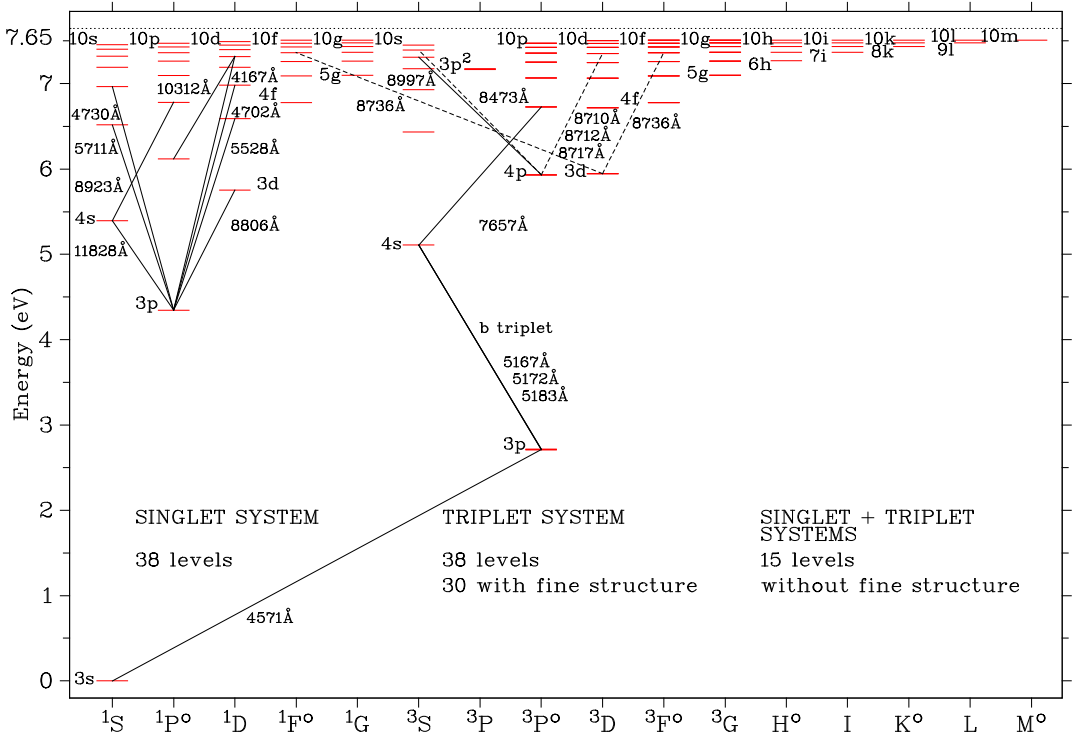}
\caption{\mg\ Grotrian diagram. The electronic configuration of the ground state is [Ne]$3s^2$.
Two multiplicity
systems are present due to spin combination of the incomplete sub-level $n$ = 3.
All levels with fine structure
are complete until $n$ = 10 and $l$ = 4 (but not visible at this scale);
fine levels and multiplicity systems were merged for 5 $\leq\ l \leq\ $9.
Only one doubly-excited state $3p^2$ is present in triplet system since in the singlet system the value 
of the corresponding energy level is greater than the ionization stage of \mg.
Horizontally dotted line stands for the ionization stage. 
Transitions represented by solid lines are those for which we deliver NLTE corrections; transitions represented by dashed lines are those in the {\it Gaia}/RVS wavelength range.}
\label{MgI_terms}
\end{figure*}

 \begin{figure*}
\includegraphics[width=0.8\linewidth]{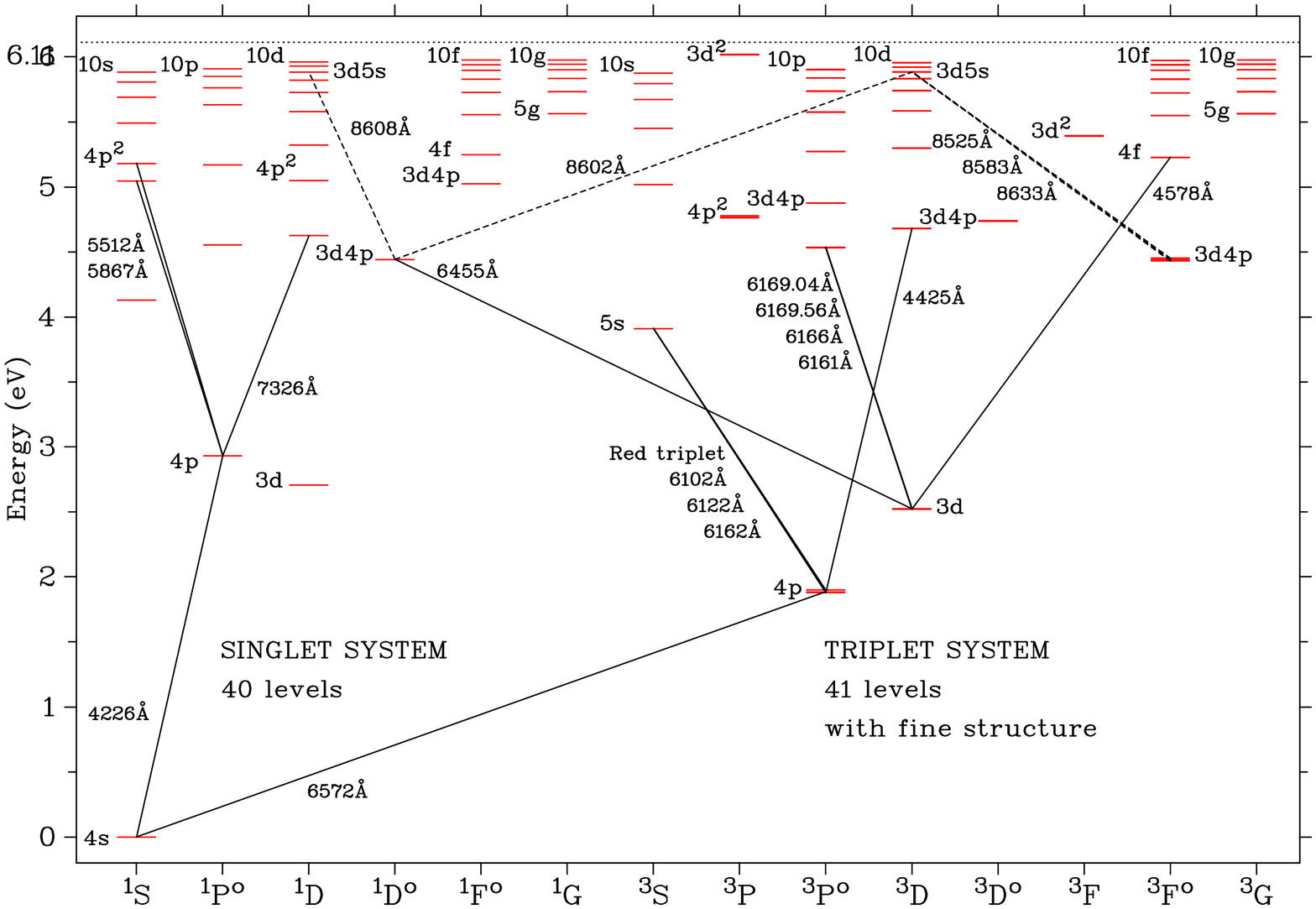}
\caption{\caI\ Grotrian diagram. The electronic configuration of the ground state is [Ar]$4s^{2}$.
All levels with fine structure are complete until $n$ = 10 and $l$ = 4.
Twelve doubly excited terms are present: $4p^{2}\ ^{1}SD\ ^{3}P$, $3d^{2}\ ^{3}PF$ (the terms for the singlet system $^{1}SDG$ are missing but thought to be above the
ionization threshold), $3d4p ^{1}DF\ ^{3}PDF^{o}$ (the terms for the singlet system $^{1}P^{o}$ are missing) and $3d5s\ ^{1}D\ ^{3}D$. 
Horizontally dotted line stands for the ionization level. 
Transitions represented by solid lines are those for which we deliver NLTE corrections; transitions represented by dashed lines are those in the {\it Gaia}/RVS wavelength range.}
\label{CaI_terms}
\end{figure*}

It is known today that the inelastic collisional cross-sections with neutral hydrogen atoms are largely over-estimated by the Drawin's formula \citep{Drawin68} when applied to dwarf stars (see \citealt{Caccin93}, \citealt{Barklem03}, \citealt{Barklem11}). 
This over-estimation tends to force LTE by exaggerating the collisional rates with respect to the radiative ones.
For giant and supergiant stars that we consider in this paper, we assume, {\it a priori}, that the low gas density in the atmospheres of theses stars make the H-collisions inefficient.  
Note that for the \mg\ and \caI\ atoms efforts are being made to compute with quantum mechanics the exact cross-sections for inelastic collisions with neutral hydrogen \citep{guitou2010}. 
These results will be implemented in a dedicated forthcoming paper. 

{\it Damping parameters.} 
The radiative damping is calculated using the Einstein coefficients of spontaneous de-excitation of each level present in the model atom:
\begin{equation}
\gamma_{{\rm rad}} = \sum_{k<j} A_{jk} + \sum_{k<i} A_{ik},
\end{equation} 
where $i$ stands for the lower level and $j$ for the upper level.  
For the collisional damping $\gamma_{{\rm coll}}$, we only take into account elastic collisions with neutral hydrogen, namely the Van der Waals' damping.
Stark broadening is ignored in the present calculations, therefore: $\gamma_{{\rm coll}} = \gamma_{{\rm VdW}}$.
We use quantum calculations of hydrogen elastic collisions with neutral species from \cite*{Anstee95, Barklem97, Barklem98'}, hereafter named ABO theory. 
The great improvement of quantum mechanical calculation compared with the traditional Uns\"{o}ld recipe \citep{Unsoeld} is to lead to line widths that agree well with the ones observed without the need of an adjustable factor $F_H$ (e.g. \citealt{Mihalas78, Thevenin89}). 
As used later in Table~\ref{Lines} and \ref{Lines_MgI_Gaia}, we translate results from the ABO theory in terms of an enhancement factor $F_H$ which is the ratio between ABO broadening an classical Uns\"{o}ld broadening.
We notice that for transitions including terms higher than $l \ge 4$ no broadening parameters are available from ABO theory. 
For these lines, we use an averaged value of this enhancement factor $F_H = 2$. 
We tested several values in the range of 1 to 4 and found no significant effect on the resulting NLTE over LTE equivalent width ratios.

\begin{figure}
\includegraphics[width=\linewidth]{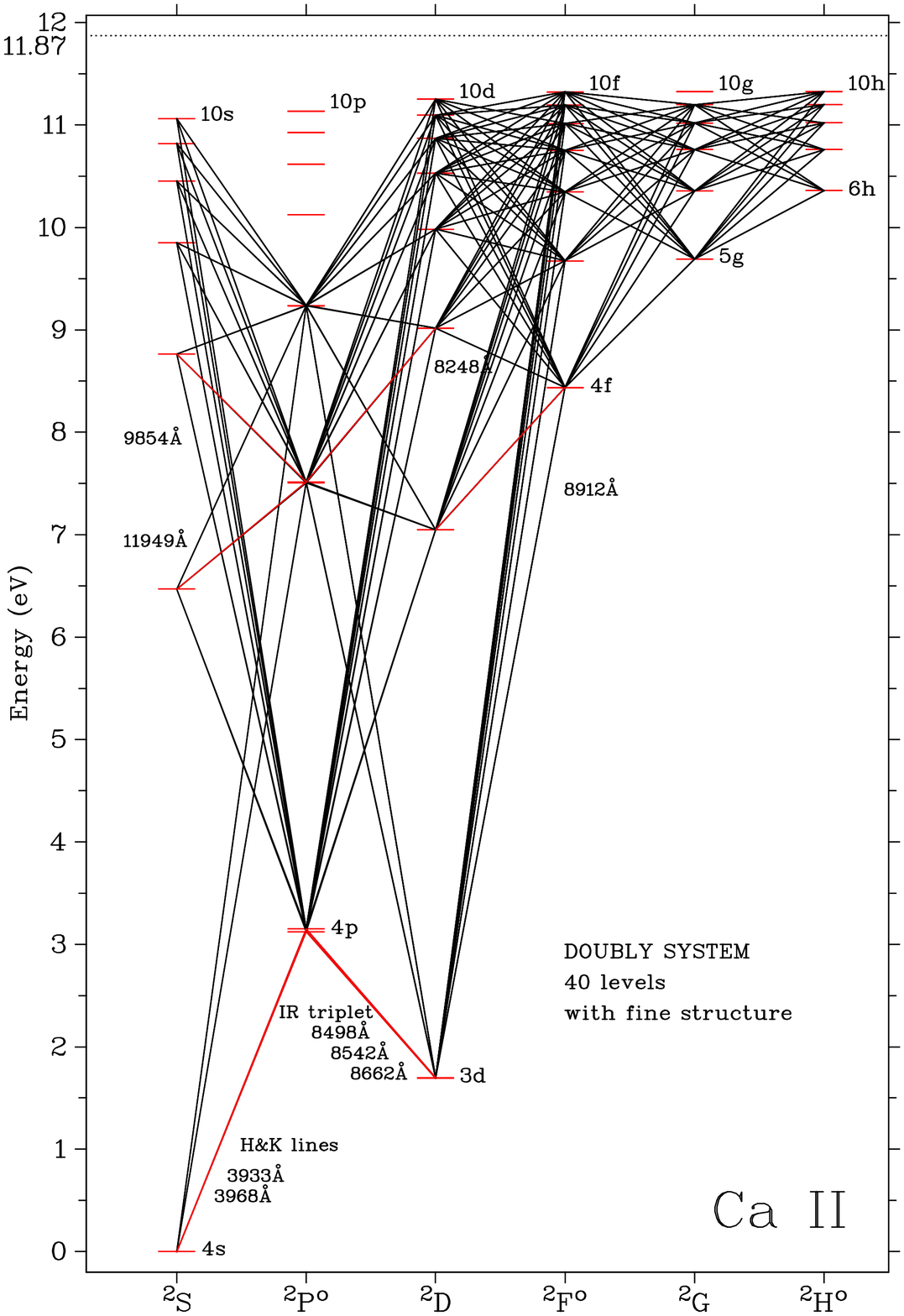}
\caption{\caII\ Grotrian diagram. First ionized calcium has a doubly multiplicity system. 
In this case, we have represented all the radiative bound--bound transitions used in the model atom with full lines. Transition represented by red lines are those for which we will deliver NLTE corrections. 
Horizontally dotted line represents the ionization stage.}
\label{CaII_terms}
\end{figure}

\subsection{Magnesium model atoms}
We developed two models for the neutral Mg atom:
\begin{itemize}
\item one for the lines outside the {\it Gaia}/RVS wavelength range (model A) with the detailed fine structure of the energy levels;
\item one dedicated to the multiplets in the {\it Gaia}/RVS wavelength range (model B) which need a special treatment to account for the blended multiplets rather than the separated components.
\end{itemize}

\subsubsection{\mg\ model atom A}
Since the neutral Mg is an alkaline earth metal, its electronic configuration leads to a singlet and a triplet systems. 
We consider all levels until $n = 10$ and $l = 9$ with fine structure for all levels with $l \leq 4$. 
For levels with $l > 4$, the multiplicity systems are merged (for the $H$, $I$, $K$, $L$ and $M$ terms) and no fine structure is considered.
For completeness and because some high energy levels are missing in the databases, we used the formulation of polarization theory given by \citet{Chang83} to include them. 
We take into account 38 spectral terms in the singlet system and 38 in the triplet system plus 15 merged terms (singlet + triplet systems without fine structure).
In total, 149 energy levels + ionization stage (at 7.646 eV) are considered for the SE computation (see the Grotrian diagram in Fig.~\ref{MgI_terms}).

1102 radiative bound-bound transitions have been taken into account: 127 allowed in the singlet system, 674 in the triplet system and 301 intercombination transitions
(semi-allowed transitions between the two systems). 
We used hydrogenic approximation \citep{Green57} to compute several log $gf$ missing in \vald.
For transitions between levels with and without explicit fine structure, we did two assumptions to estimate individual oscillator strengths: firstly, as the transitions are produced between Rydberg states, the wavelength shifts between components are small enough to be neglected; and secondly the oscillator strength of each component of one multiplet is given in the LS coupling.
The TopBase photoionization cross-sections come from \citet{Butler93}.
11026 electronic collisions between all levels and sub-levels are included using the IPM approximation when corresponding radiative transitions exist.
The elastic collisional broadening from the ABO theory is interpolated for 87 transitions ($s-p$, $p-d$ and $d-f$) in singlet and triplet systems, the Uns\"{o}ld recipe is used with $F_H = 2$ for the remaining transitions. 

\subsubsection{\mg\ model atom B for the {\it Gaia}/RVS lines}

In order to reproduce \mg\ lines in {\it Gaia}/RVS wavelength range, we construct a second \mg\ model atom which enables \multi\ to compute these lines as blended multiplets rather than separate components. 
These lines appear in the triplet system. 
We merge fine levels $E_i$ under interest taking average energy level $E_{\rm{level}}$ with respect to their statistical weights $g_i$: 
\begin{equation}
E_{\rm{level}} = \frac {\sum_{i=1}^{3} g_{i}E_{i}^{\rm{fine\ level}}}{\sum_{i=1}^{3} g_{i}},
\end{equation}
as illustrated in Fig.~\ref{Gro_MgI_RVS}. 
Averaged energy of each considered level and its fine structure are represented in Fig.~\ref{Gro_MgI_RVS}. 
Then only averaged levels of $4p\ ^{3}P^{o}$, $3d\ ^{3}D$, $7d\ ^{3}D$ and $7f\ ^{3}F^{o}$ are considered in this model atom B.  
The radiative transfer for a multiplet is then done under assuming that the same source function is used for the components of this multiplet \citep{Mihalas78}. 
Moreover we take the oscillator strength of the multiplet  as a linear combination of the weighted $gf$ of each component (in LS coupling). 
Actually, we can neglect the wavelength deviation between components and multiplet (relative deviation smaller than 1\% for the \mg\ lines):
\begin{equation}
g_{i}f_{ij}^{\rm{mult}} = \frac {1}{\lambda_{ij}} \sum_{u}\sum_{l} g_{l}\lambda_{ul}f_{lu} \simeq \sum_{u}\sum_{l} g_{l}f_{lu}
\end{equation}
with respect to the line selection rules and with $\lambda_{ij}$ (named Ritz wavelength) the average wavelength of the multiplet, $g_{i}=\sum_{l}g_{l}$, $l$ for lower fine levels of term $i$, and $u$ for upper fine levels of term $j$. 
We note that merging these fine levels implies shifts in wavelengths for the other lines. 
Then this atom will be exclusively used for the study of the lines in the {\it Gaia}/RVS wavelength range, except for the singlet line at 8473~\AA\ for which we use the \mg\ model atom~A.

\subsection{Calcium model atoms}
The neutral and first ionized Ca model atoms are built in the same way as the \mg\ model atom A, but we consider them separately. 
We build separate model atoms of \caI\ and \caII, but for comparison with \citet{Mashonkina07} we merge the two separate atoms of \caI\ and \caII\ to have a model atom of Ca through two ionization stages.

\subsubsection{\caI\ model atom}
Neutral Ca is also an alkaline earth metal with two multiplicity systems: singlet and triplet systems. 
However, more doubly excited energy levels exist which produce a more complex model atom than the \mg\ one. 
All energy levels until $n = 10$ and $l = 4$ are considered with fine structure and we take into account all doubly excited levels under ionization stage of \caI. 
This leads to consider 151 levels + ionization stage (at 6.113 eV). 
A Grotrian diagram is shown in Fig.~\ref{CaI_terms}. 
Twelve doubly excited levels are presented for $4p^2$, $3d^2$, $3d4p$ and $3d5s$ electronic configurations (see the legend of Fig.~\ref{CaI_terms} for more details).
The TopBase photoionization cross-sections come from \citet{Saraph}.
All transitions present in \vald\ are considered representing 2120 transitions: 191 in the singlet system, 1161 in the triplet system and 768 between these two systems. 
For elastic collisions with neutral hydrogen, the ABO theory provides broadening for 107 lines.

11628 electronic collisions are involved in the \caI\ model atom but only 11 collisional cross-sections with electrons for resonance transitions are available from quantum computations \citep{Samson01}. 
These authors used the Variable Phase Method (VPM) for \caI\ lines shown in Table~\ref{Lines}.  
For inelastic collisions with neutral hydrogen, we can refer to \citet{Mashonkina07} who show detailed tests of the importance of these inelastic collisions.
They conclude that for many transition lines these collisions are small for \caI\ and negligible for \caII\ to reproduce the line profiles.

\subsubsection{\caII\ model atom}
The \caII\ atom is like an alkali element, so it presents only one system of levels with a doubly multiplicity system (see Fig~\ref{CaII_terms}). 
We consider levels until $n$ = 10 and $l = 5$ for \caII. 
Fine structure is used, leading to 74 levels for \caII. 
422 radiative transitions are considered from the \vald\ database. 
Photoionization cross-sections from the TopBase \citep{Saraph} do not present any peaks. 
In the ABO theory, the line broadening due to elastic collisions with hydrogen is only known for the H, K and CaT lines \citep{Barklem98}.
2775 inelastic collisions with electrons are involved in the \caII\ model atom.
Quantum computations are available from \citet{Burgess95}, using the Coulomb Distorted Wave approximation (CDW), for 21 transitions between mean levels: 10 allowed and 11 forbidden as reported in Table~\ref{Lines}. 
For components between fine levels, we apply the same ratio from the oscillator strengths $gf$ to the collisional strength $\Omega$.
    
\subsubsection{Ca {\scriptsize I/II} model atom}\label{CaI/II}
In order to see if NLTE effects act on the ionization equilibrium, we build a Ca {\scriptsize I/II} model atom by merging the \caI\ and the \caII\ ones. 
Several energy levels of \caI\ are photoionized on excited levels of \caII.
It is the case for the doubly excited levels $3d4p ^{1}DF\ ^{3}PDF^{o}$ which photoionize on to $3d ^{2}D$ of \caII\ and for the $4p^{2}\ ^{1}SD\ ^{3}P$ which photoionize on to $4p ^{2}P^o$ of \caII. 
This model atom is used to compare results with those of \citet{Mashonkina07} in Sect.~\ref{Ca_section}.     
     
\section{Test on two benchmark stars: the Sun and Arcturus}

The new model atoms of these $\alpha$-elements are first tested on two well known benchmark stars: a main sequence star (the Sun) and a giant (Arcturus).
We compute flux profiles for several lines in the {\it Gaia}/RVS wavelength range using the NLTE radiative transfer code \multi\ \citep{Carlsson86}.
This code directly provides correction ratios of NLTE ($W$) versus LTE ($W^*$) equivalent widths (EW), hereafter named NLTE/LTE EW ratios or $W/W^*$.
 
We define the abundance of an element (El) relative to hydrogen by $A_{{\rm El}}$ using notation\footnote{$A_{{\rm El}}=\log_{10}\left( N_{{\rm El}}/ N_{{\rm H}}\right)+ 12$, $N$ the number density of a given element.} from \citet{Grevesse07}. 
Stellar abundances are given with respect to the solar ones on a logarithmic scale ${\rm [El/H]}$ as usually done\footnote{${\rm [El/H]} = A_{{\rm El}}^{\star} - A_{{\rm El}}^{\odot}$.}.
The abundances of $\alpha$-elements are given with respect to iron by [$\alpha$/Fe]. 

We also define NLTE abundance correction as:
\begin{equation}
\Delta {\rm [El/H]} = A^{{\rm NLTE}}_{{\rm El}} - A^{{\rm LTE}}_{{\rm El}},
\end{equation} 
where $A_{{\rm El}}^{{\rm LTE}}$ is the LTE abundance of the element El, and $A^{{\rm NLTE}}_{{\rm El}}$ the NLTE abundance.

\begin{figure}
\begin{center}
\includegraphics[width=\linewidth]{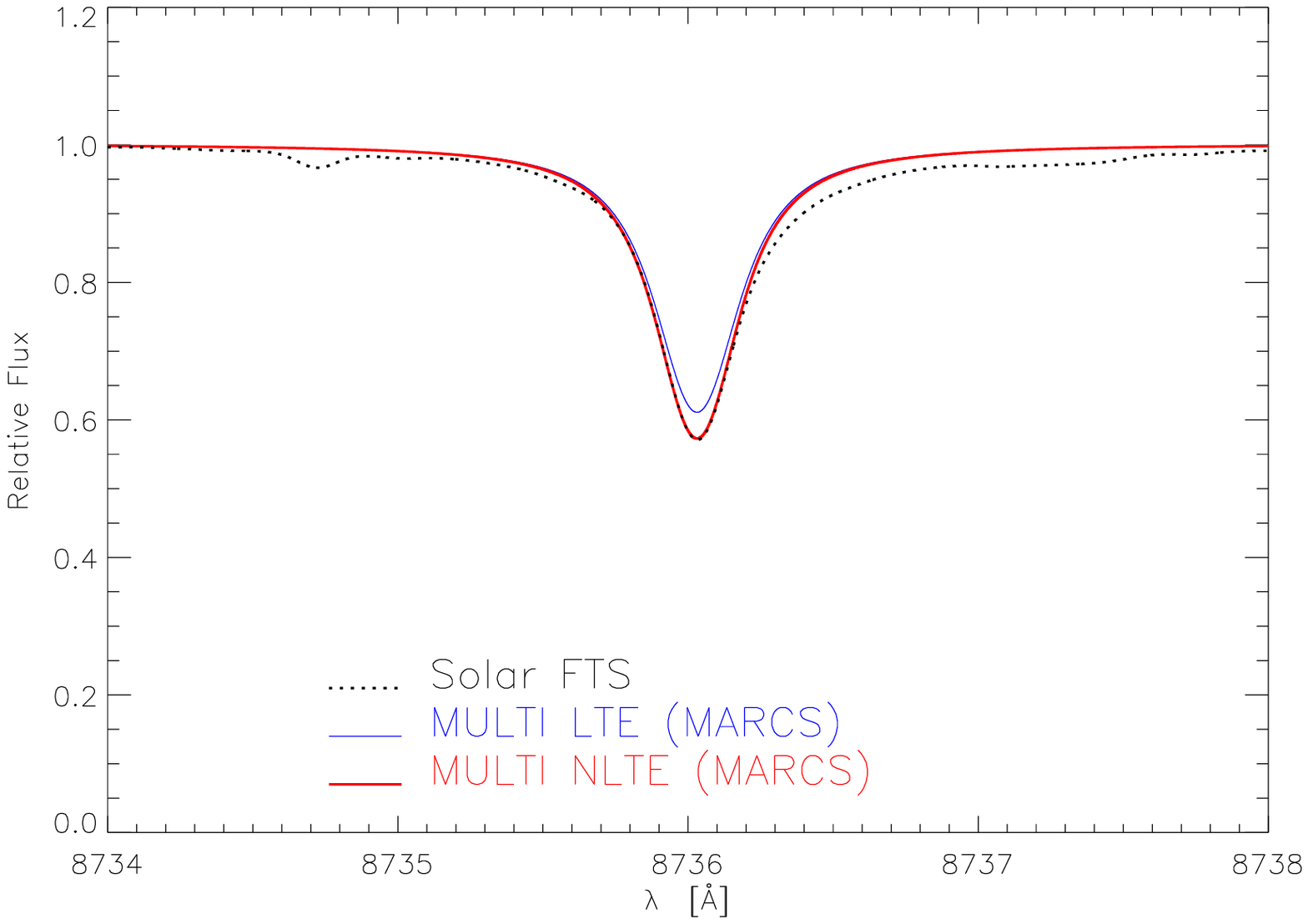}
\includegraphics[width=\linewidth]{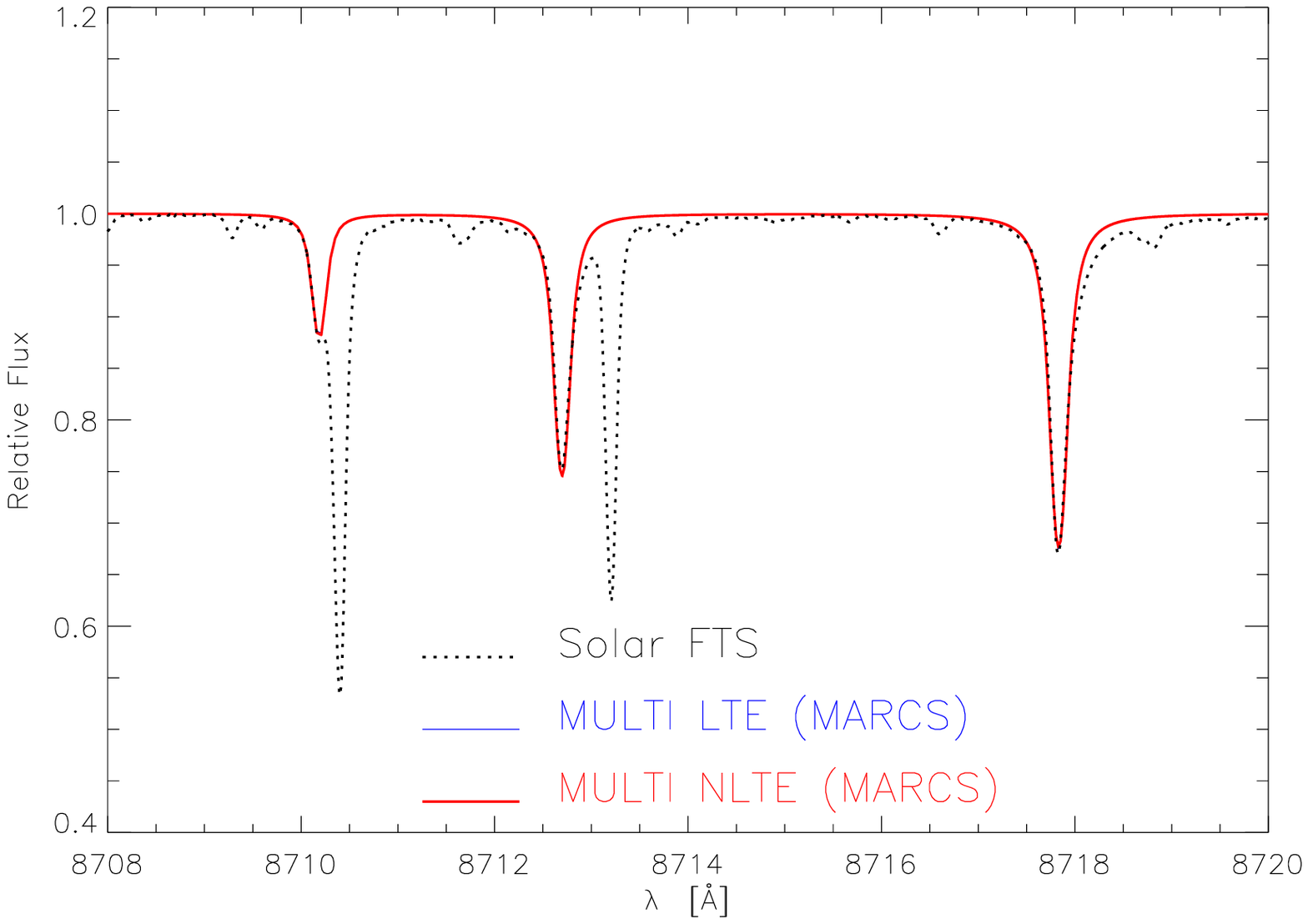}	
\includegraphics[width=\linewidth]{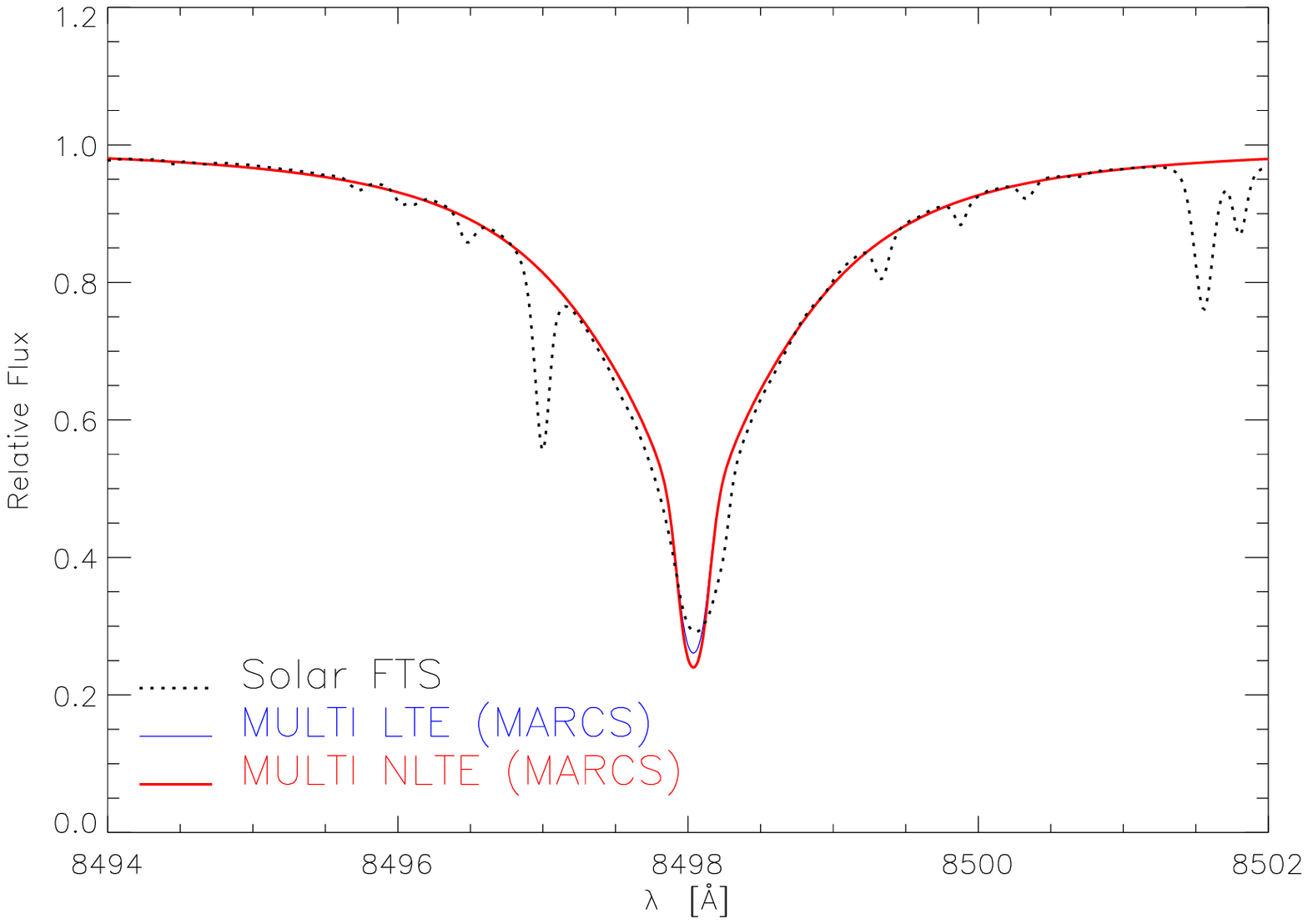}
\end{center}
\caption{Solar line fits in the {\it Gaia}/RVS wavelength range, using a solar plane--parallel \marcs\ model. 
The blue lines stand for LTE synthesis and the red lines for NLTE synthesis. 
Dashed lines stand for observations.
Top panel: \mg\ 8736 \AA\ multiplet. 
Middle panel: \mg\ triplet at $\lambda\lambda$ 8710, 8712 and 8717. 
Bottom panel: \caII\ 8498 \AA\ line.}
\label{Solar_lines}
\end{figure}
	
\begin{figure}
\begin{center}
\includegraphics[width=\linewidth]{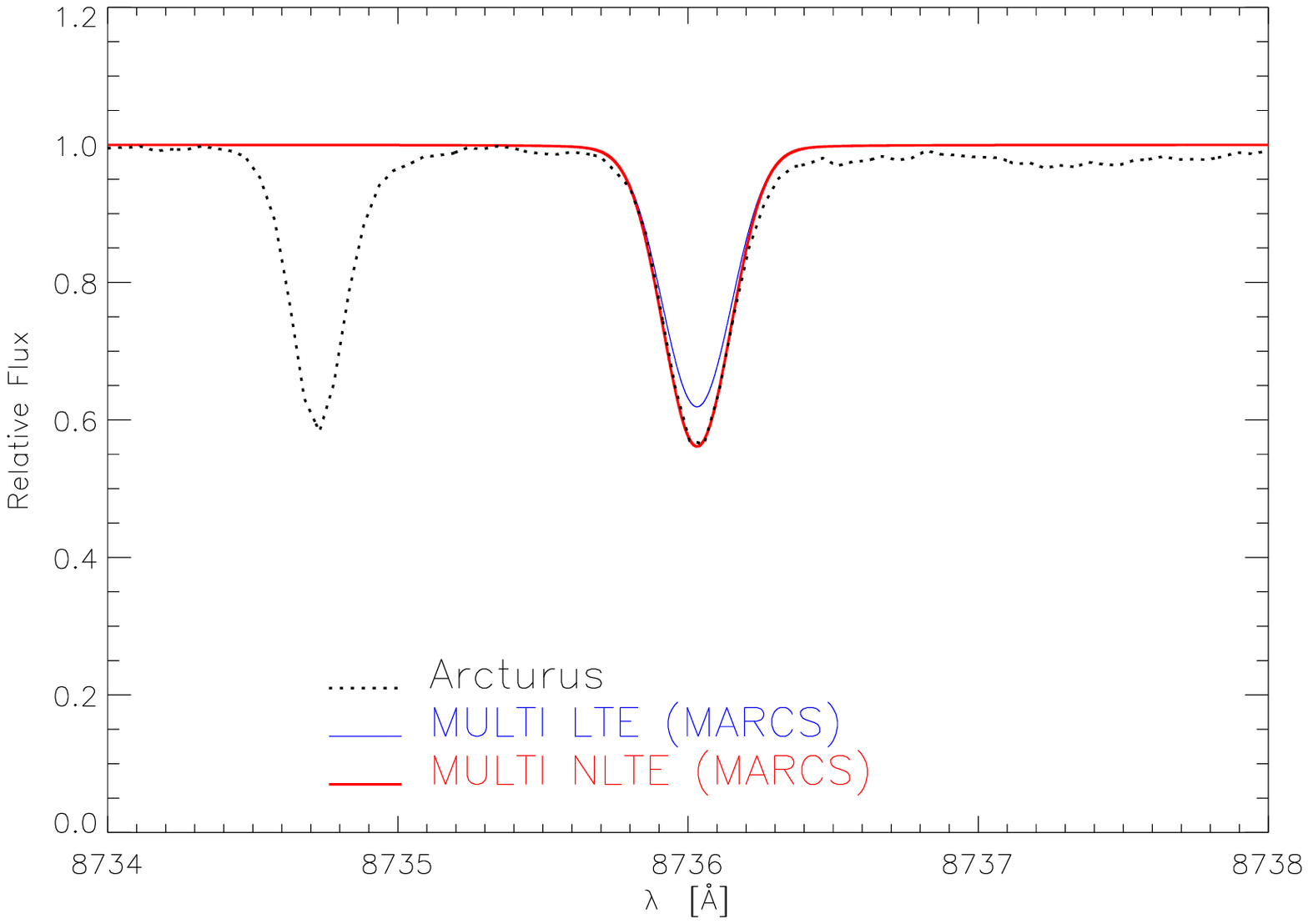}
\includegraphics[viewport=23 -21 527 341,scale=0.505]{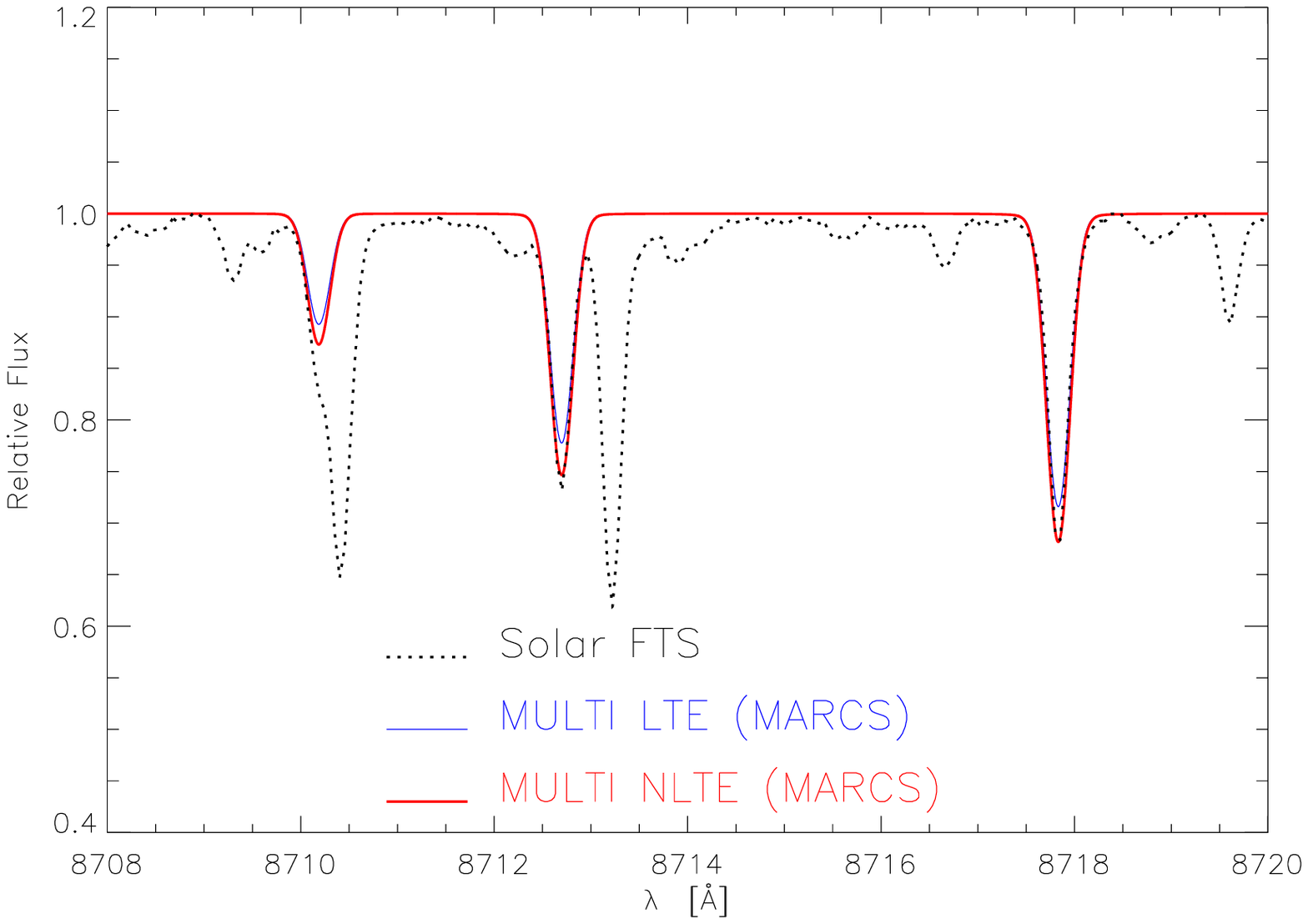}
\includegraphics[viewport=23 -51 527 311,scale=0.505]{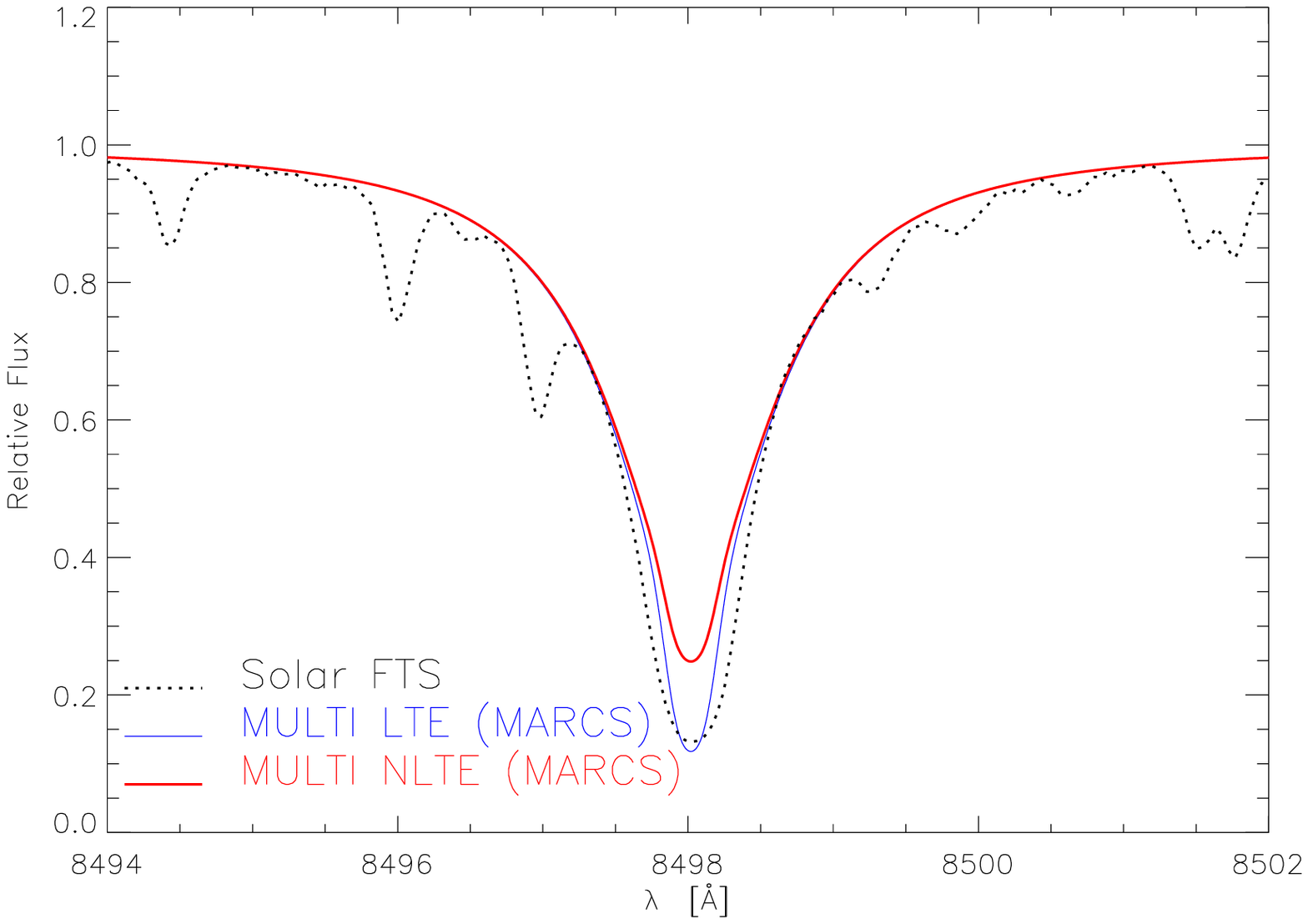}
\end{center}
\caption{Arcturus line fits in the {\it Gaia}/RVS wavelength range, using a spherical interpolated \marcs\ model atmosphere. 
The blue lines stand for LTE synthesis and the red lines for NLTE synthesis. 
Dotted lines stand for observations. 
Top panel: \mg\ 8736 \AA\ multiplet.
Middle panel: \mg\ triplet at $\lambda\lambda$ 8710, 8712 and 8717.
Note that NLTE effects are more pronounced than the Sun (Fig.~\ref{Solar_lines}).  
Bottom panel: \caII\ 8498 \AA\ line. 
Departure from LTE appears in the core.}
\label{Arcturus_lines}
\end{figure}

For very weak lines (e.g. \mg\ 8473~\AA) or for very strong lines (e.g. CaT), one can express $W/W^*$ as a function of the abundance ratio by using the curve of growth theory:
\begin{equation}
\frac{W}{W^*} =  \left( \frac{A_{{\rm El}}^{\rm NLTE}}{A_{{\rm El}}^{{\rm LTE}}} \right)^{-\beta},
\end{equation}  
where $\beta$ is a coefficient which depends on the strength of the line. If $\log (W/\lambda ) \le -4.8$, the line is considered weak and $\log (W/ \lambda )$ has a linear dependence on the abundance ($\beta = 1$); if  $\log (W/ \lambda ) \ge -4.2$ the line is considered strong and $\log (W/ \lambda )$ has a square root dependence on the abundance ($\beta = 1/2$). We deduce for these two extreme cases the NLTE abundance correction as:
\begin{equation}
 \Delta{\rm [El/H]} = - \frac{1}{\beta} \log \left( W/W^* \right).
 \label{1}
\end{equation}
We apply this relation to deduce NLTE abundance corrections $\Delta[$EL/H] from $W/W^*$ when we compare weak and strong lines with the available literature as done in Table~\ref{Comp_Zhao}, \ref{Comp_Shimanskaya} and \ref{Comp_Mashonkina}. 
Note that between these two extreme cases, $\beta$ is lower than 1/2. 
This can imply strong variations of $\Delta$[El/H] corresponding to the plateau of the curve of growth (e.g. \citealt{Mihalas78}, pp 319, 320 and 325).
We emphasize that Eq.~\ref{1} is an approximation that gives an estimate of the NLTE abundance correction. 
The right way to determine the NLTE abundance correction should be to vary the NLTE abundance of El until the NLTE EW fits the observed EW.

In the following sections, we test our model atoms on the Sun for which we know in advance that NLTE effects are small \citep{Asplund05} but that can be used to estimate the missing enhancement factor $F_H$ for some {\it Gaia} lines (outside of the ABO theory range).
We also test our model atoms on Arcturus, a well known red giant, using high resolution and high $S/N$ spectrum. 

\subsection{NLTE synthetic line profile fits of solar lines}

\begin{table*}
 \caption{Solar NLTE abundance corrections for \mg\ lines of astrophysical interest with ($\Delta$[Mg/H]$_H$, $S_H = 1$) and without ($\Delta$[Mg/H], $S_H = 0$) inelastic collisions with neutral hydrogen according to a scaled Drawin's formula. We applied Eq.(\ref{1}) to compute $\Delta$[Mg/H] and $\Delta$[Mg/H]$_H$.}
 \begin{tabular}{rrrrrrrlrr}
                 &            &         &         &           &\multicolumn{2}{c}{This work}&&\multicolumn{2}{c}{\citet{Zhao98} } \\  
                 \cline{6-7} \cline{9-10} 
 $\lambda$ [\AA] &$W$ [m\AA]&$W_{H}$ [m\AA]&$W/W^*$ [m\AA] &$W/W^*_H$ &$\Delta$[Mg/H] &$\Delta$[Mg/H]$_H$ &&$\Delta$[Mg/H] &$\Delta$[Mg/H]$_H$ \\
\hline
 4571.095 &   83 &   85 & 0.935 & 0.954 & 0.06 &  0.02 && 0.07 &  0.03 \\
 5167.320 &  675 &  706 & 0.954 & 0.998 & 0.04 &  0.00 && 0.04 &  0.01 \\
 5172.683 & 1139 & 1195 & 0.951 & 0.998 & 0.04 &  0.00 && 0.04 &  0.01 \\
 5183.603 & 1430 & 1502 & 0.949 & 0.998 & 0.05 &  0.00 && 0.04 &  0.01 \\
 5528.403 &  333 &  342 & 0.973 & 1.001 & 0.02 &$ 0.00$&& 0.06 &  0.00 \\
 8806.751 &  465 &  499 & 0.934 & 1.001 & 0.06 &$ 0.00$&& 0.09 &$-0.03$\\
 8923.563 &   43 &   47 & 0.929 & 1.001 & 0.03 &  0.00 && 0.11 &  0.02 \\
 11828.18 &  755 &  771 & 0.981 & 1.001 & 0.02 &$ 0.00$&& 0.05 &  0.02 \\ 
 \hline
 \end{tabular}
 \label{Comp_Zhao}
\end{table*}

We compare our NLTE disk integrated line profiles with the observed solar spectrum obtained by the Fourier Transform Spectrograph, here-after FTS (Brault \& Neckel 1987)\footnote{available at: ftp://ftp.hs.uni-hamburg.de/pub/outgoing/FTS-Atlas/}. 
This is particularly well adapted for the present work because of its high signal-to-noise ratio ($S/N > 500$) and its large resolving power ($R\sim 400000$). 
We use a 1D, static, and plane--parallel solar model atmosphere from \marcs\ grid (Gustafsson et al. 2008) with $T_{\rm{eff}}=5777$~K and $\log g = 4.44$. 
The solar abundances of Ca and Mg are from \citet{Grevesse07} corresponding to $A_{{\rm Mg}}^{\odot} = 7.53$ and $A_{{\rm Ca}}^{\odot} = 6.31$. 
The flux profiles are convolved with a gaussian profile to account for the effects of macroturbulence velocity ($\sim$~2~km~s$^{-1}$).   
As the projected rotational velocity is weak for the Sun, we neglect it.    

The NLTE effects on lines in the visible wavelength range for both \mg\ and \caI\ are small in the late type solar-like stars (e.g. comparing the NLTE abundances from \citealt{Idiart00} with the LTE ones from \citealt{Thevenin98}). 
However, NLTE effects become non-negligible for IR lines (see the conclusion of \citealt{Zhao98}). 
As an example, we compare NLTE and LTE profiles for \mg\ lines in the RVS wavelength range. 
In the top panel of Fig.~\ref{Solar_lines}, we show the \mg\ 8736~\AA\ multiplet computed with the model B.
This line is fitted in NLTE with a value of $F_H = 3$ and $A^{{\rm NLTE}}_{{\rm Mg}} = 7.63$. We find a ratio of $W/W^* =1.10$ and a NLTE EW of $204$~m\AA. 
The LTE fit gives a value of $A^{{\rm LTE}}_{{\rm Mg}} = 7.70$. 
Therefore, we have $\Delta[$Mg/H$]=-0.07$~dex of NLTE abundance correction for this line on the Sun.
The middle panel of Fig.~\ref{Solar_lines} presents the \mg\ triplet at $\lambda\lambda$ 8710, 8712 and 8717 computed assuming the same source function since we have merged lower and upper fine levels into one lower and one upper in model~B. 
Our best fit is obtained with $F_H = 2.5$, $A^{{\rm NLTE}}_{{\rm Mg}} = 7.63$ and we found that NLTE effects are negligible for this multiplet since $W/W^*= 1.02$.  

We compare our NLTE results for the Sun with \citet{Zhao98} who used Kurucz's model atmosphere with the {\scriptsize DETAIL} code \citep{Butler85} to perform NLTE computations. 
They studied the impact of the inelastic collision with hydrogen on the NLTE results, using Drawin's formula and empirical scaling laws.
We compute $\Delta$[Mg/H] using Eq.~(\ref{1}) for several lines in common with their study. 
For the comparison, we add collisions with neutral hydrogen, using Drawin's formula with a scaling factor $S_H = 1$. 
Results are presented in Table~\ref{Comp_Zhao} for solar lines. 
For the \mg~b triplet we obtain a very good agreement with \citet{Zhao98}. 
For other transitions, our NLTE corrections are lower than theirs. 
The effect of inelastic collisions with neutral hydrogen, through Drawin's formula, eliminates completely the NLTE effects except for the intercombination resonance line at 4571~\AA\ for which we have $\Delta$[Mg/H]$_H = 0.02$ (0.03 in \citealt{Zhao98}). 
In the Sun, NLTE corrections for these lines are all less than 0.06~dex. 
If we do not take into account Drawin's formula, all the NLTE corrections are positive. 
We have already mentioned in Sect.~\ref{section21} that Drawin's formula is not adapted for the inelastic collisions with neutral hydrogen.  
    
We also consider NLTE effects on the well studied CaT.
We note that the central depression of these lines are deeper in NLTE compared to LTE ones as seen in the bottom panel of Fig.~\ref{Solar_lines}.
The discrepancies with the observations are due to the fact that the cores of these lines form in the chromosphere which is not included in the \marcs\ model atmospheres.  
Since these strong lines are dominated by the wings that are formed in LTE, the equivalent widths are quite insensitive to NLTE in solar conditions ($W/W^* \sim 1.00$).
Note that the small asymmetry of the line core may be due to an isotopic shift as shown by \citet{Kurucz05} and \citet{Mashonkina07} but a contribution of 3D effect is also not to be excluded.

\subsection{NLTE synthetic line profile fits of Arcturus}
We compute NLTE line profiles using an interpolated \marcs\ model atmosphere (spherical geometry, $T_{\rm{eff}}=4300$~K, $\log g=1.5$, [Fe/H$]=-0.5$, [$\alpha$/Fe$]=+0.2$, $\xi=2$~km~s$^{-1}$, $M=1\ M_{\odot}$) obtained with the interpolation code of T.~Masseron\footnote{available at: http://marcs.astro.uu.se/software.php}. 
The spectrum of Arcturus comes from the Narval instrument at the Bernard Lyot Telescope\footnote{available at: http://magics.bagn.obs-mip.fr} ($R \sim 65000$) with hight signal-to-noise ratio ($S/N > 500$). 

The computation of Arcturus line profiles reveals that NLTE effects are larger than in the Sun in particular for the {\it Gaia}/RVS lines as shown in Fig.~\ref{Arcturus_lines}. 
When we fit the \mg\ 8736 \AA\ line with $A^{{\rm NLTE}}_{{\rm Mg}}=6.93$, we find an equivalent width of 128 m\AA\ and a ratio of $W/W^* = 1.30$. 
The LTE fit of the same line gives a value of $A^{{\rm LTE}}_{{\rm Mg}} = 7.23$. 
Therefore, we have $\Delta$[Mg/H$]=-0.30$ dex of NLTE abundance correction for this line in Arcturus. 
For the \mg\ triplet at $\lambda\lambda$ 8710, 8712 and 8717, the NLTE effects are smaller than for 8736 \AA\ line but larger than in the Sun. 
We find $W/W^* = 1.10$.

Concerning the \caII\ IR 8498 \AA\ line, departure from LTE always affects the line core. 
We are not able to reproduce satisfactorily the observed line core, neither in LTE or NLTE as shown in bottom panel of Fig.~\ref{Arcturus_lines}. 
Contrary to the solar case, the central depression of these lines are deeper in LTE compared to NLTE ones.
The best value of Ca abundance fitting the wings of this line in NLTE is $A^{{\rm NLTE}}_{{\rm Ca}} = 5.81$. 
The equivalent width ratio is $W/W^* = 0.95$. 
As in the Sun, the discrepancy in the line core is probably due to the presence of a chromosphere in such giant stars and in particular in Arcturus for which many studies have been devoted (e.g. \citealt{Ayres82}).

\section{NLTE effects versus atmospheric parameters}
We compute an NLTE correction table for a wide range of stellar parameters ($T_{\rm{eff}}$, $\log g$, [Fe/H]) of cool giant and supergiant stars with metallicities from $+0.5$ down to $-4$ dex.
The selected lines are those of astrophysical importance in the optical and near IR wavelength ranges, especially for those in the  domain of the {\it Gaia}/RVS.
Then, we describe the general trends of NLTE effects for the selected \mg, \caI\ and \caII\ lines of astrophysical interest (Table~\ref{Lines} and \ref{Lines_MgI_Gaia}) as functions of stellar parameters. 
The most important results deduced from our computed $W/W^*$ are an enhancement of NLTE effects with decreasing metallicity, a non monotonic variation with the effective temperature, and a significative dependence on the surface gravity for several lines. We also compare our results with previous works.

\subsection{The $W/W^*$ grid for giant and supergiant stars} 
We compute NLTE/LTE EW ratios for $T_{\rm{eff}}$, $\log g$ and [Fe/H] covering most of the late-type giant and supergiant stars. 
The effective temperature $T_{{\rm eff}}$ range is 3500(200)3900~K and 4000(250)5250~K, the surface gravity $\log g$ range is 0.5(0.5)2.0~dex, and the metallicity [Fe/H] range is $-4.00(0.50)-1.00(0.25)+0.50$~dex.
We use spherical, 1D, and static model atmospheres from the \marcs\ site\footnote{available at: http://marcs.astro.uu.se} \citep{Gus08} with chemical standard composition from \citet{Grevesse07}.  
In the standard composition of \marcs\ models, [$\alpha$/Fe] depends on [Fe/H] as:
\begin{equation}
\begin{array}{rcrccl}
       & {\rm [Fe/H]}\ \ge &  0 & \Rightarrow & \left[\alpha/{\rm Fe}\right] = & 0 \\
-1  <  & {\rm [Fe/H]}\   < &  0 & \Rightarrow & \left[\alpha/{\rm Fe}\right] = & -0.4\ {\rm [Fe/H]} \\
       & {\rm [Fe/H]}\ \le & -1 & \Rightarrow & \left[\alpha/{\rm Fe}\right] = & +0.4 

\end{array}
\label{alpha}
\end{equation}
This means that the Ca abundance used for NLTE computations follows the $\left[\alpha/{\rm Fe}\right]$ value of the model atmosphere. 
The \marcs\ models with $M = 1\ M_{\odot}$ and constant microturbulent velocity fields $\xi = 2$~km~s$^{-1}$ are adopted. 
We interpolated \marcs\ model atmospheres with the Masseron code for [Fe/H$]=-3.50$ and for few models missing in the \marcs\ database.

For the NLTE computations, we need to solve coherently the coupled equations of the SE and the radiative transfer which use the local approximate lambda operator implemented in the 1D, plane--parallel, and time independent code \multi, version 2.2 \citep{Carlsson86}. 
Continuous opacities come from Uppsala package \citep{Gus73}. 
We take into account 45000 lines for line opacity computations.
Opacities from molecular lines are not included.
We do not take into account the collisions with neutral hydrogen ($S_H=0$).
Starting population numbers are taken at LTE and are used to be compared with the NLTE results. 
With very large model atoms, the convergence of the Newton-Raphson scheme can be very slow and the number of iterations for some extremely metal-poor models can be very high. 
For the \mg\ and \caII\ model atoms, the minimal relative precision for the population number at any layer of the atmosphere is of $10^{-4}$ whereas it decreases to $10^{-3}$ for \caI\ for which the model atom is more complex.

\begin{table}
 \caption{Part of the results for the \mg\ 8736 \AA\ and \caII\ 8498 \AA\ lines. The full version of the results is available in the online version of the journal.
 NLTE/LTE EW ratios $W/W^*$ are given for $T_{{\rm eff}} = 5000$~K, $\log g = 1$ and 2, and metallicities if $W > 1$~m\AA. 
 We do not take into account collisions with neutral hydrogen ($S_H=0$). 
 The NLTE abundance correction $\Delta$[El/H] is given when the Eq.(\ref{1}) is applicable.}
 \begin{tabular}{p{0.8cm}p{0.5cm}p{0.8cm}p{0.8cm}p{0.8cm}p{0.9cm}p{0.9cm}}
\hline
&&&&\multicolumn{3}{c}{\mg\ 8736.012 \AA}\\
 \multirow{-2}*{$T_{\rm eff}$[K]} & \multirow{-2}*{$\log g$} & \multirow{-2}*{[Fe/H]} &\multirow{-2}*{[$\alpha$/Fe]}& $W$[m\AA] & $W/W^*$ & $\Delta$[Mg/H]\\
 \hline\\
$5000$&$1$&$+0.50$&0.0&$225$&$1.179$&\\ 
         &&$+0.25$&0.0&$205$&$1.226$&\\ 
         &&$+0.00$&0.0&$177$&$1.268$&\\ 
         &&$-0.25$&0.1&$158$&$1.299$&\\ 
         &&$-0.50$&0.2&$135$&$1.302$&$-0.11$\\ 
         &&$-0.75$&0.3&$111$&$1.272$&$-0.10$\\ 
         &&$-1.00$&0.4&$~89$&$1.210$&$-0.08$\\ 
         &&$-1.50$&0.4&$~33$&$0.927$&$+0.03$\\ 
         &&$-2.00$&0.4&$~~9$&$0.667$&$+0.18$\\ 
      &$2$&$+0.50$&0.0&$211$&$1.170$&\\
         &&$+0.25$&0.0&$192$&$1.196$&\\                  
         &&$+0.00$&0.0&$167$&$1.214$&\\ 
         &&$-0.25$&0.1&$151$&$1.230$&\\ 
         &&$-0.50$&0.2&$131$&$1.236$&$-0.09$\\ 
         &&$-0.75$&0.3&$110$&$1.218$&$-0.09$\\ 
         &&$-1.00$&0.4&$~88$&$1.170$&$-0.07$\\ 
         &&$-1.50$&0.4&$~34$&$0.919$&$+0.04$\\ 
         &&$-2.00$&0.4&$~10$&$0.683$&$+0.17$\\ 
\\
\hline
&&&& \multicolumn{3}{c}{\caII\ 8498.018 \AA}\\
 \multirow{-2}*{$T_{\rm eff}$[K]} & \multirow{-2}*{$\log g$} & \multirow{-2}*{[Fe/H]} &\multirow{-2}*{[$\alpha$/Fe]}& $W$[m\AA] & $W/W^*$ & $\Delta$[Mg/H]\\
 \hline \\
5000	&1	&	$+0.50$	&0.0&	3857	&	0.969	& $+0.03$\\
	&		&	$+0.00$	&0.0&	3086	&	0.973	& $+0.02$\\
	&		&	$-0.50$	&0.2&	2550	&	0.983	& $+0.01$\\
	&		&	$-1.00$	&0.4&	2036	&	0.993	& $+0.01$\\
	&		&	$-2.00$	&0.4&	863	&	1.037	&$-0.03$ \\
	&		&	$-3.00$	&0.4&	396	&	1.154	& \\
	&		&	$-4.00$	&0.4&	264	&	1.352	& \\
	& 2	&	$+0.50$	&0.0&	2255	&	0.965	& $+0.03$\\
	&		&	$+0.00$	&0.0&	1909	&	0.972	& $+0.02$\\
	&		&	$-0.50$	&0.2&	1715	&	0.984	& $+0.01$\\
	&		&	$-1.00$	&0.4&	1446	&	0.998	& $+0.00$\\
	&		&	$-2.00$	&0.4&	737	&	1.069	& $-0.06$\\
	&		&	$-3.00$	&0.4&	370	&	1.233	&\\
	&		&	$-4.00$	&0.4&	241	&	1.435	&\\
\\
\hline
 \end{tabular}
 \label{NLTEtable}
\end{table}

We emphasise that we did 1D NLTE plane--parallel radiative transfer in 1D theoretical spherical model atmospheres which may appear not consistent. However, as shown by \citet{Heiter06}, it is better to use spherical model atmosphere with plane--parallel radiative transfer rather than plane--parallel model atmosphere with plane--parallel radiative transfer for giant and supergiant stars. 

\begin{table*}
\caption{\mg, \caI\ and \caII\ line selection for NLTE corrections. The wavelengths $\lambda$ are given in the air. 
The oscillator strengths $gf$ and the radiative damping parameters $\gamma_{{\rm rad}}$ are given in a decimal logarithmic scale. 
The enhancement factor $F_H$ is defined as the ratio of the collisional damping parameter computed from ABO theory divided by the classical Uns\"{o}ld's formula (a default value of 2 is applied when lines are outside the range of ABO theory). 
The radiative damping parameters are calculated using the $\log gf$ values taken from \vald, except when specified. 
\mg\ line selection in the {\it Gaia}/RVS wavelength range is detailed in Table~\ref{Lines_MgI_Gaia}. 
The precision of log $gf$ are given with letters, following the \nist\ notation: A ($<3\%$), B+ ($<7\%$), B ($<10\%$), C+ ($<18\%$), C ($<25\%$), D ($< 50\%$), and E ($>50\%$). Inelastic collisions with electrons are specified in the last column (Coll.) where IPM stands for Impact Parameter Method from \citet{Seaton62a}, CDW stands for Coulomb Distorted Wave approximation from \citet{Burgess95} for \caII, and VPM stands for Variable Phase Method from \citet{Samson01} for \caI\ resonance lines.}
\label{Lines}

\begin{tabular}{rllrrllrrrlrrc}
&\multicolumn{2}{l}{Lower level}&&&\multicolumn{2}{l}{Upper level}&&&&&&&\\
$\lambda$ [\AA] &\multicolumn{2}{l}{Configuration}	& $g_i$ & $E_i$ [eV]	& \multicolumn{2}{l}{Configuration} & $g_j$	& $E_j$ [eV] & $\log gf$ & Prec. & log $\gamma_{{\rm rad}}$ & $F_{H}$ & Coll. \\
\hline
  \\
\mg \\
\\
4167.271&$3p	  $&$^{1}P^{o}$&3&4.345802&$ 7d$&$^{1}D	 $&5&7.320153&$-0.745^{a}$&C+&8.71&$     2.00\ \ $& IPM\\
4571.095&$4s^{2}$&$^{1}S	  $&1&0.000000&$ 3p$&$^{3}P^{o}$&3&2.711592&$-5.688^{b}$&C+&2.34&$1.34^{\alpha}$&   IPM\\
4702.990&$3p	  $&$^{1}P^{o}$&3&4.345802&$ 5d$&$^{1}D	 $&5&6.981349&$-0.440^{a}$&B+&8.72&$2.77^{\alpha}$& IPM\\
4730.028&$3p	  $&$^{1}P^{o}$&3&4.345802&$ 6s$&$^{1}S	 $&1&6.966284&$-2.400^{b}$&B+&8.71&$	    2.00\ \ $& IPM\\
5167.320&$3p     $&$^{3}P^{o}$&1&2.709105&$ 4s$&$^{3}S	 $&3&5.107827&$-0.931^{*}$&B &7.98&$2.37^{\alpha}$& IPM\\
5172.683&$3p	  $&$^{3}P^{o}$&3&2.711592&$ 4s$&$^{3}S	 $&3&5.107827&$-0.450^{*}$&B &7.98&$2.37^{\alpha}$& IPM\\
5183.603&$3p     $&$^{3}P^{o}$&5&2.716640&$ 4s$&$^{3}S	 $&3&5.107827&$-0.239^{*}$&B &7.98&$2.37^{\alpha}$& IPM\\
5528.403&$3p	  $&$^{1}P^{o}$&3&4.345802&$ 4d$&$^{1}D	 $&5&6.587855&$-0.490^{b}$&B+&7.38&$2.11^{\alpha}$& IPM\\
5711.086&$3p	  $&$^{1}P^{o}$&3&4.345802&$ 5s$&$^{1}S	 $&1&6.516139&$-1.720^{b}$&B+&8.71&$		 2.00\ \ $& IPM\\
7657.599&$4s     $&$^{3}S    $&3&5.107827&$ 5p$&$^{3}P^{o}$&5&6.726480&$-1.268^{a}$&C &8.00&$     2.00\ \ $& IPM\\
8473.688&$4p     $&$^{3}P^{o}$&5&5.932787&$ 9s$&$^{3}S    $&3&7.395551&$-2.020^{c}$&  & 7.20&$2.00\ \ $& IPM\\     
8806.751&$3p	  $&$^{1}P^{o}$&3&4.345803&$ 3d$&$^{1}D	 $&5&5.753246&$-0.134^{a}$&A &8.72&$1.39^{\alpha}$& IPM\\
8923.563&$4s	  $&$^{1}S	  $&1&5.932370&$ 5p$&$^{1}P^{o}$&3&7.354590&$-1.678^{a}$&B &7.81&$		 2.00\ \ $& IPM\\
8997.147&$4p     $&$^{3}P^{o}$&5&5.932787&$8s$&$^{3}S     $&3&7.310447&$-1.770\ \,$& & 7.24&$2.00\ \ $& IPM\\
10312.52&$4p	  $&$^{1}P^{o}$&3&5.945915&$ 7d$&$^{1}D  	 $&5&7.364755&$-1.730^{a}$&C+&8.19&$		 2.00\ \ $& IPM\\
11828.18&$3p	  $&$^{1}P^{o}$&3&4.345802&$ 4s$&$^{1}S    $&1&5.753246&$-0.333^{a}$&A &8.73&$2.91^{\alpha}$& IPM\\
 \\
\caI \\
 \\
4226.727 &$4s^{2}$ &$^{1}S$    &1&0.000000&$4p$    &$^{1}P^{o}$&3&2.932512&$ 0.244\ \,$&B+&8.34&$2.09^{\alpha}$& VPM\\
4425.435 &$4p$	  &$^{3}P^{o}$&1&1.879340&$4d$	 &$^{3}D$	 &3&4.680180&$-0.358\ \,$&C &7.94&$1.78^{\alpha}$& IPM\\
4578.549 &$3d$	  &$^{3}D$	  &3&2.521263&$4f$    &$^{3}F^{o}$&5&5.228440&$-0.697\ \,$&B+&7.36&$		 2.00\ \ $& IPM\\
5512.980 &$4p$	  &$^{1}P^{o}$&3&2.932512&$4p^{2}$&$^{1}S$	 &1&5.180837&$-0.447^{b}$&B+&8.48&$	    2.00\ \ $& IPM\\
5867.559 &$4p$	  &$^{1}P^{o}$&3&2.932512&$6s$	 &$^{1}S$	 &1&5.044971&$-1.570\ \,$&B &8.35&$		 2.00\ \ $& IPM\\
6102.719 &$4p$	  &$^{3}P^{o}$&1&1.879340&$5s$	 &$^{3}S$	 &3&3.910399&$-0.850*$   &B &7.87&$2.92^{\beta} $& IPM\\
6122.213 &$4p$     &$^{3}P^{o}$&3&1.885807&$5s$	 &$^{3}S$	 &3&3.910399&$-0.380*$   &B &7.87&$2.22^{\beta} $& IPM\\
6161.294 &$3d$     &$^{3}D$	  &5&2.522986&$5p$	 &$^{3}P^{o}$&5&4.534736&$-1.266\ \,$&B+&7.30&$2.11^{\alpha}$& IPM\\
6162.170 &$4p$	  &$^{3}P^{o}$&5&1.898935&$5s$	 &$^{3}S$	 &3&3.910399&$-0.170*$   &B &7.87&$2.44^{\beta} $& IPM\\
6166.438 &$3d$	  &$^{3}D$	  &3&2.521263&$5p$	 &$^{3}P^{o}$&1&4.531335&$-1.142\ \,$&B+&7.28&$2.11^{\alpha}$& IPM\\
6169.037 &$3d$	  &$^{3}D$	  &5&2.522986&$5p$    &$^{3}P^{o}$&3&4.532211&$-0.797\ \,$&B+&7.30&$2.12^{\alpha}$& IPM\\
6169.562 &$3d$	  &$^{3}D$    &7&2.525682&$5p$    &$^{3}P^{o}$&5&4.534736&$-0.478\ \,$&B+&7.30&$2.11^{\alpha}$& IPM\\
6455.592 &$3d$	  &$^{3}D$    &5&2.522986&$3d4p$  &$^{1}D^{o}$&5&4.443025&$-1.290^{b}$&B+&7.67&$		 2.00\ \ $& IPM\\
6572.777 &$4s^{2}$ &$^{1}S$    &1&0.000000&$4p$	 &$^{3}P^{o}$&3&1.885807&$-4.296^{b}$&D+&3.47&$1.57^{\alpha}$& VPM\\
7326.142 &$4p$	  &$^{1}P^{o}$ &3&2.932512&$4d$	  &$^{1}D$	 &5&4.624398&$-0.208\ \,$&B &8.37&$1.72^{\alpha}$& IPM\\
8525.707 &$3d4p$   &$^{3}F^{o}$ &5&4.430011&$3d5s$  &$^{3}D$   &3&5.883850&$-1.202^{c}$&  &7.70&$	    2.00\ \ $& IPM\\
8583.318 &$3d4p$   &$^{3}F^{o}$ &7&4.440954&$3d5s$  &$^{3}D$	 &5&5.885035&$-0.986^{c}$&  &7.79&$		 2.00\ \ $& IPM\\
8633.929 &$3d4p$   &$^{3}F^{o}$ &9&4.450647&$3d5s$  &$^{3}D$	 &7&5.886263&$-0.810^{c}$&  &7.77&$     2.00\ \ $& IPM\\
 \\
\caII \\
 \\
3933.663	&$4s$&$^{2}P	 $	&2&0.000000&$4p$&$^{2}P^{o}$&4&3.150984&$ 0.104^{b}$&A+&8.21&$1.42^{\alpha}$&CDW \\
3968.468	&$4s$&$^{2}P	 $	&2&0.000000&$4p$&$^{2}P^{o}$&2&3.123349&$-0.210^{b}$&A+&8.15&$1.43^{\alpha}$&CDW \\
8248.791	&$5p$&$^{2}P^{o}$	&4&7.514840&$5d$&$^{2}D	   $&6&9.017485&$ 0.556\ \,$&C &8.29&$     2.00\ \ $&IPM \\
8498.018	&$3d$&$^{2}D	 $	&4&1.692408&$4p$&$^{2}P^{o}$&4&3.150984&$-1.356^{d}$& &8.21&$2.16^{\alpha}$&CDW \\
8542.086	&$3d$&$^{2}D  	 $	&6&1.699932&$4p$&$^{2}P^{o}$&4&3.150984&$-0.405^{d}$& &8.21&$2.16^{\alpha}$&CDW \\
8662.135	&$3d$&$^{2}D	 $	&4&1.692408&$4p$&$^{2}P^{o}$&2&3.123349&$-0.668^{d}$& &8.15&$2.16^{\alpha}$&CDW \\
8912.059 &$4d$&$^{2}D	 $	&4&7.047168&$4f$&$^{2}F^{o}$&6&8.437981&$ 0.575^{c}$&  &8.85&$	    2.00\ \ $&CDW \\
9854.752 &$5p$&$^{2}P^{o}$	&2&7.505137&$6s$&$^{2}S	   $&2&8.762907&$-0.260^{a}$&D &8.26&$	    2.00\ \ $&IPM \\
11949.74	&$5s$&$^{2}S	 $	&2&6.467874&$5p$&$^{2}P^{o}$&2&7.505137&$-0.010\ \,$&D &8.43&$	    2.00\ \ $&CDW \\
\\
\hline 
\end{tabular}
\\
\begin{flushleft}
$^{a}$ \nist. $^{b}$ \citet{Hirata95}. $^{c}$ \citet{Kurucz95}. $^{d}$ \citet{Melendez07}. * \citet{Aldenius07} for \mg\ and \citet{Aldenius09} for \caI. \\  
$^{\alpha}$ ABO theory. $^{\beta}$ \citet{Cayrel96}. 
\end{flushleft}
\end{table*}

\begin{table*}
\caption{Details of the blended multiplets of \mg\ in the {\it Gaia}/RVS wavelengths.
The $\log gf$ are from \vald\ except when specified. 
The wavelengths in bold and between $\left<.\right>$ stand respectively for observed lines and Ritz wavelengths (theoretical ones). 
The notations are the same as in Table~\ref{Lines}.}
\label{Lines_MgI_Gaia}
\begin{tabular}{rllrrllrrrlrrr}
&\multicolumn{2}{l}{Lower level}&&&\multicolumn{2}{l}{Upper level}&&&&&&&\\
$\lambda$ [\AA] &\multicolumn{2}{l}{Configuration}	& $g_i$ & $E_i$ [eV]	& \multicolumn{2}{l}{Configuration} & $g_j$	& $E_j$ [eV] & $\log gf$ & Prec. & log $\gamma_{r}$ & $F_{H}$& Coll. \\
\hline \\
$\left<8715.255\right>$    &$4p$&$^{3}P^{o}$& 9&5.932370&$7d$&$^{3}D	   $&15&7.354590&               &  &	 &$	2.50$& IPM\\
\textbf{8710.169}&$4p$&$^{3}P^{o}$& 1&5.931542&$7d$&$^{3}D	   $&	3&7.354592&$	-1.565^{a}$&D &7.38&$	2.50$& IPM\\
\textbf{8712.678}&$4p$&$^{3}P^{o}$& 3&5.931951&$7d$&$^{3}D	   $&	8&7.354591&$-1.088\ \,   $&D &7.39&$	2.50$& IPM\\
8712.670         &$  $&$		     $& 3&5.931951&$	 $&$		      $& 3&7.354592&$	-1.690^{a}$&D &7.39&$	    $& \\
8712.684         &$  $&$		     $& 3&5.931951&$	 $&$	   	   $&	5&7.354590&$	-1.213^{a}$&D+&7.32&$	 $& \\
\textbf{8717.812}&$4p$&$^{3}P^{o}$& 5&5.932787&$7d$&$^{3}D	   $&15&7.354590&$-0.866\ \,   $&E &7.39&$	2.50$& IPM\\
8717.797         &$  $&$		     $& 5&5.932787&$	 $&$      	   $&	3&7.354592&$	-2.866^{a}$&E &7.38&$		$&	\\
8717.811         &$  $&$		     $& 5&5.932787&$  $&$		      $&	5&7.354590&$	-1.690^{a}$&D &7.32&$		$&	\\
8717.819         &$  $&$		     $& 5&5.932787&$  $&$   		   $&	7&7.354589&$	-0.941^{a}$&D+&7.39&$		$&	\\
\\
\textbf{8736.012}&$3d$&$^{3}D    $&15&5.945915&$7f$&$^{\ }F^{o} $&28&7.364755&$-0.351\ \,$&E &8.21&$	3.00$& IPM\\
	             &$  $&$		     $&12&5.945914&$	 $&$^{1}F^{o}	$&	7&7.364755&		        &  &  	 &$		$&\\
8736.000         &$  $&$		     $& 5&5.945913&$	 $&$		      $&	7&7.364755&$-3.210^{c}$&  &	 &$		$&\\
8736.014         &$  $&$		     $& 7&5.945915&$	 $&$		      $&	7&7.364755&$-1.788^{a}$&E &	 &$		$&\\
	             &$  $&$	        $&15&5.945915&$  $&$^{3}F^{o}	$&21&7.364755&		        &  &	 &$		$&\\
8736.000         &$  $&$	        $& 5&5.945913&$	 $&$        	$&	5&7.364755&$-1.786^{a}$&E &	 &$		$&\\
8736.000         &$  $&$		     $& 5&5.945913&$	 $&$      		$&	7&7.364755&$-0.890^{b}$&  &	 &$		$&\\
8736.014         &$  $&$	        $& 7&5.945915&$	 $&$		      $&	5&7.364755&$-3.337^{a}$&E &	 &$		$&\\
8736.014         &$  $&$		     $& 7&5.945915&$	 $&$    		   $&	7&7.364755&$-2.170^{b}$&  &	 &$		$&\\
8736.014         &$  $&$		     $& 7&5.945915&$	 $&$      		$&	9&7.364755&$-0.725^{a}$&C &	 &$		$&\\
8736.024         &$  $&$		     $& 3&5.945917&$  $&$	      	$&	5&7.364755&$-1.056^{a}$&D+&	 &$		$&\\
\hline
\end{tabular}
\begin{flushleft}
$^{a}$ \nist. $^{b}$ \citet{Hirata95}. $^{c}$ \citet{Kurucz95}.
\end{flushleft}
\end{table*}

Moreover the $W/W^*$ do not take into account the possible contribution of the stellar chromospheres for lines such as the \mg\ b triplet, the \caII\ H, K and the CaT. All the \marcs\ models are theoretical, 1D and LTE models without any model of chromosphere, i.e. without any rising of the temperature profile in the upper layers. 
This absence of model chromosphere can explain why we are not able to reproduce the line core of CaT in Arcturus (Fig.~\ref{Arcturus_lines}). 
For an extended discussion about the treatment of chromospheric contribution on a peculiar red giant star ($ \beta$~Cet), using this work, see \citet{Berio11}.

The NLTE results for the lines in Tables~\ref{Lines} and \ref{Lines_MgI_Gaia} are available in electronic forms. 
Illustrations of $W/W^*$ for a selection of these lines as function of [Fe/H], for several $T_{{\rm eff}}$ and parametrized in $\log g$ are shown in Appendix~A and B. 
Global NLTE results are summarized in Tables~\ref{Mg_WWS}, \ref{CaI_WWS} and \ref{CaII_WWS} for \mg, \caI\ and \caII\ lines, respectively.
We show in Table~\ref{NLTEtable}, an example of the NLTE results expressed in terms of $W/W^*$ for the \mg\ 8736~\AA\ line and the \caII\ IR 8498~\AA\ line as a function of stellar parameters. 
We also give the NLTE EW in order to know if the Eq.~(\ref{1}) is applicable to directly deduce the NLTE abundance correction $\Delta[$El/H]. 

\subsection{Line selection}
The lines are mostly selected as unblended and strong enough to be measurable in metal-poor stars.
Spectroscopic and micro-physics details on \mg, \caI\ and \caII\ selected lines are presented in Table~\ref{Lines}.
 
Wavelengths are given in air using the IAU standard dispersion law for conversion between vacuum and air 
(see \citealt{Morton91}).
For \mg, we mainly selected lines of the singlet system except for the \mg\ b triplet lines at $\lambda \lambda$ 5167, 5172, 5183, at 7657 and 8997 \AA\ lines, and the lines belonging to the {\it Gaia}/RVS domain. 
The important intercombination resonance line at 4571~\AA\ is also selected. This line has been recently studied \citep{Langangen09} and has been proposed to be used as temperature diagnostic in stellar atmospheres.
For \caI, we mainly selected lines coming from singly excited levels except for the 5512~\AA\ line and for the weak triplet in RVS because for lines coming from doubly excited levels we have less atomic data. 
Most of these optical lines are blended in their wings at solar metallicity but are less blended with decreasing metallicity. 
For the optical red triplet at $\lambda \lambda$ 6102, 6122, 6162, we prefer to use the enhancement factors $F_H$ from \citet{Cayrel96} who provide resolved $F_H$ for each transition of the triplet ($F_H = 2.92$, 2.22 and 2.44 respectively) rather than ABO theory which gives one unique value of $F_H=2.55$. 
Lines selected for \caII\ are the strong H \& K lines, the IR triplet and few other unblended IR lines. 

\begin{table*}
\caption{\mg\ $W/W^*$ values for $\log g = 1.5$. Values outside [0.9, 1.1] range are boldfaced. Blanks mean that $W < 1$~m\AA.}
\begin{tabular}{lccccrcccrccc}
&[Fe/H]&\multicolumn{3}{c}{$-3$}&&\multicolumn{3}{c}{$-2$}&&\multicolumn{3}{c}{0}\\
\cline{3-5} \cline{7-9} \cline{11-13}
$\lambda$[\AA]&$T_{{\rm eff}}$ [K]&3500&4250&5250&&3500&4250&5250&&3500&4250&5250\\
\hline\hline
4571.095&&0.90&\textbf{0.34}&\textbf{1.46}&&0.98&\textbf{0.69}&\textbf{0.27}&&0.99&0.97&\textbf{0.73}\\
\hline\hline
4167.271&&0.95&\textbf{0.63}&\textbf{0.57}&&1.01&0.96&\textbf{0.85}&&0.97&0.94&0.97\\
4702.990&&\textbf{1.18}&\textbf{0.82}&\textbf{0.65}&&1.08&1.03&0.95&&0.97&0.93&0.99\\
5528.403&&\textbf{1.37}&0.96&\textbf{0.71}&&\textbf{1.19}&\textbf{1.11}&1.04&&0.97&0.92&0.97\\
\hline
8806.751&&\textbf{1.39}&\textbf{1.15}&0.91&&\textbf{1.22}&\textbf{1.18}&\textbf{1.24}&&0.93&\textbf{0.86}&0.95\\
\hline \hline
5167.320&&0.95&\textbf{0.87}&\textbf{0.87}&&0.91&0.85&0.99&&0.93&\textbf{0.84}&\textbf{0.85} \\
5172.683&&0.94&\textbf{0.82}&0.95&&0.90&\textbf{0.83}&0.99&&0.93&\textbf{0.83}&\textbf{0.82} \\
5183.603&&0.94&\textbf{0.79}&0.97&&0.90&\textbf{0.82}&0.97&&0.93&\textbf{0.83}&\textbf{0.81} \\  
\hline \hline
4730.028&&\textbf{1.39}&\textbf{0.79}&    &&1.08&0.94&\textbf{0.77}&&0.98&0.92&\textbf{0.89}\\
5711.086&&\textbf{1.48}&0.86&    &&\textbf{1.22}&1.04&\textbf{0.83}&&1.01&0.94&0.94\\
\hline
11828.18&&\textbf{1.47}&\textbf{1.23}&0.96&&\textbf{1.26}&\textbf{1.24}&\textbf{1.32}&&0.98&\textbf{0.87}&0.99\\
\hline\hline
8923.563&&\textbf{1.30}&    &    &&\textbf{1.20}&    &\textbf{0.68}&&0.96&\textbf{0.89}&\textbf{0.86}\\
7657.599$^{\dag}$&&\textbf{1.75}&\textbf{0.85}&    &&\textbf{1.48}&\textbf{1.28}&\textbf{0.89}&&\textbf{1.12}&1.05&0.98\\
\hline\hline
10312.52&&    &    &    &&    &    &    &&0.97&0.95&1.01\\
\hline\hline
8997.147&&    &    &    &&\textbf{1.19}&1.09&    &&1.02&1.00&1.08\\
8473.688&&    &    &    &&    &    &    &&1.00&0.97&1.03\\
\hline\hline
$\left< 8715.255\right>^{\dag}$&&    &    &    &&    &\textbf{0.18}&    &&1.01&1.10&\textbf{1.22}\\
\hline\hline
8736.012&&\textbf{1.45}&\textbf{0.82}&    &&1.08&1.03&\textbf{0.56}&&\textbf{1.36}&\textbf{1.28}&\textbf{1.23}\\
\hline\hline
\end{tabular}
\begin{flushleft}
$^{\dag}$ Note that this line is sensitive to the surface gravity.
\end{flushleft}
\label{Mg_WWS}
\end{table*}
 
Lines in the {\it Gaia}/RVS are dominated by the CaT. 
There are also 5 \mg\ lines: a very weak line at 8473 \AA, a weak triplet at $\lambda\lambda$ 8710, 8712, 8717, and a stronger line at 8736 \AA. 
Details on the components of these lines are shown in Table~\ref{Lines_MgI_Gaia}.
A partial Grotrian diagram in Fig.~\ref{Gro_MgI_RVS} also details the fine structure of the levels involved in these multiplets. 
Note that only the mean level of the $7f\ ^3F^o$ is available on the \nist\ database.
The weakest 8710 and 8712~\AA\ components of the triplet are also blended on their red wings by iron lines as shown in Fig.~\ref{Solar_lines} and \ref{Arcturus_lines}. 
This multiplet at $\lambda\lambda$ 8710, 8712, 8717 is represented, using the \mg\ B model, with a Ritz wavelength at 8715 \AA\ for which we deliver NLTE/LTE EW ratio by assuming the same NLTE behaviour for the three components. 
For \caI, 5 transitions exist in this domain, between doubled excited levels $3d4p$ and $3d5s$, as shown in Fig.~\ref{CaI_terms}. 
These lines are very weak at solar metallicity and only $\lambda\lambda$ 8525, 8583 and 8633 are visible in the solar spectrum.
Moreover, in the spectrum of Arcturus, these three lines are a little bit strengthened but blended by CN lines. 
Thus, we decide to provide NLTE corrections only for these three lines. 
We notice that for the $\lambda \lambda$ 8498, 8542, 8662 IR triplet (CaT), we used theoretical $\log gf$ values ($-1.356$, $-0.405$, $-0.668$ respectively)  from \citet{Melendez07} which are in a very good agreement with values derived from 3D hydrodynamical line fits on the Sun ($-1.309$, $-0.410$, $-0.683$) from \citet{Bigot08}.

\subsection{The \mg\ lines}

\begin{figure}
 \includegraphics[width=\linewidth]{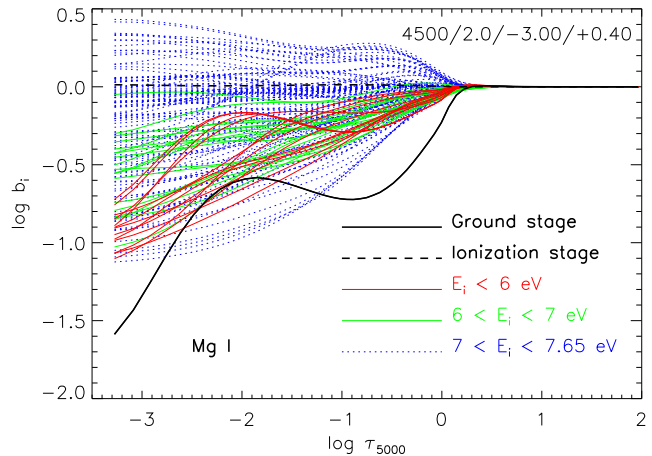}
 \caption{\mg\ departure coefficient as a function of optical depth in the continuum at 5000 \AA\ for a metal-poor giant stellar model. We separate levels as function of their energy $E_i$ with respect to the ground state as shown in the legend.}
 \label{MgI_pop}
\end{figure}

In order to see how the \mg\ is affected by the NLTE, the departure coefficients $b_i = n_i/n_i^*$ of the \mg\ model atom are plotted in Fig.~\ref{MgI_pop}, where $n_i$ and $n_i^*$ stands for NLTE and LTE level populations respectively, as a function of the standard optical depth $\tau_{5000}$ for a metal-poor giant model with atmospheric parameters of $T_{{\rm eff}}=4500$ K, $\log g = 2$, [Fe/H]~$= -3$ and [$\alpha$/Fe]~$= +0.4$. 
The ground state is very depopulated regarding the LTE case. 
This is due to an increase of the UV radiative field produced by a decrease of the metallicity. 
Moreover, we show that there is an overpopulation of a large part of higher levels explained by a collisional dominated recombination.
The population in the ionization stage is in LTE.

The \mg\ $\lambda\lambda$ 4167, 4702, 5528 and 8806 lines are transitions between the $3p\ ^1P^o$ level and $^1D$ terms.
Therefore, their $W/W^*$ have similar behaviours except for the 8806~\AA\ line which has the most pronounced deviation from LTE (see Table~\ref{Mg_WWS} and Fig.~\ref{MgI_lines}).
At lower metallicities there is a strong variation with $T_{\rm eff}$ and $W/W^*$ can vary between 0.4 and 1.9, which corresponds for weak lines to NLTE abundance corrections between $+0.4$ and $-0.3$ respectively. 
Their $W/W^*$ are not sensitive to the surface gravity except for the line at 8806 \AA\ which can have 0.1~dex of difference in abundance correction between $\log g = 1$ and $\log g = 2$ at 4500~K and at solar metallicity.
 
\begin{table*}
\caption{Comparison of \mg\ abundance corrections with \citet{Shimanskaya00} for five lines as functions of atmospheric parameters. We produce NLTE corrections also taking into account the hydrogen collisions by Drawin's formula. The Mg abundances of our results follow the enhancement of $\alpha$-elements of the model atmospheres. Blanks are used when Eq.(\ref{1}) is not applicable for our NLTE corrections or when $W$ is below 1 m\AA, and when NLTE corrections are not provided by \citet{Shimanskaya00}.} 
\begin{tabular}{cllrrrcrrr}
&$T_{{\rm eff}}$ [K]&&\multicolumn{3}{c}{4500}&&\multicolumn{3}{c}{5000}\\
&$\log g$  &    &\multicolumn{3}{c}{1}&&\multicolumn{3}{c}{2}\\
\cline{4-6} \cline{8-10}
&$$[Fe/H] & &0& $-2$&$-3$&&0& $-2$&$-3$\\
\hline
$\lambda$ [\AA]&NLTE correction&$S_H$\\
\hline\\
         &$\Delta$[Mg/H]      &0  &       &       &  0.36 &&       &  0.11 &  0.48 \\
4571.095 &$\Delta$[Mg/H]      &0.1&       &       &  0.24 &&       &  0.06 &  0.26 \\
         &$\Delta$[Mg/H]$^{\dag}$&0.1&  0.03 &$-0.01$&  0.14 &&  0.08 &  0.08 &       \\
\hline
         &$\Delta$[Mg/H]      &0  &  0.05 &       &  0.09 &&       &       &  0.13 \\
4702.990 &$\Delta$[Mg/H]      &0.1&$-0.01$&       &  0.00 &&       &       &  0.00 \\
         &$\Delta$[Mg/H]$^{\dag}$&0.1&  0.01 &$-0.11$&  0.02 &&  0.00 &  0.14 &  0.14 \\
\hline
         &$\Delta$[Mg/H]      &0  &  0.20 &  0.25 &       &&  0.18 &       &       \\
5183.603 &$\Delta$[Mg/H]      &0.1&  0.10 &  0.05 &       &&  0.10 &$-0.01$&       \\
         &$\Delta$[Mg/H]$^{\dag}$&0.1&       &  0.02 &$-0.29$&&       &  0.00 &$-0.14$\\
\hline
         &$\Delta$[Mg/H]      &0  &       &$-0.01$&  0.04 &&       &  0.02 &  0.10 \\
5711.086 &$\Delta$[Mg/H]      &0.1&       &  0.00 &  0.00 &&       &  0.00 &  0.00 \\
         &$\Delta$[Mg/H]$^{\dag}$&0.1&$-0.03$&$-0.03$&       &&$-0.03$&  0.12 &       \\
\hline         
         &$\Delta$[Mg/H]      &0  &$-0.11$&  0.09 &       &&$-0.09$&  0.21 &       \\
8736.012  &$\Delta$[Mg/H]      &0.1&$-0.01$&  0.00 &       &&  0.00 &  0.00 &       \\
         &$\Delta$[Mg/H]$^{\dag}$&0.1&$-0.05$&$-0.05$&       &&$-0.05$&       &       \\
         \\
         \hline
\end{tabular}
 \begin{flushleft}
  $^{\dag}$ Shimanskaya (private communication)
 \end{flushleft}
 \label{Comp_Shimanskaya}
\end{table*} 
   
The \mg\ $\lambda\lambda$ 4730, 5711 and 11828 lines also come from the $3p\ ^1P^o$ level and reach $^1S$ terms.
The $W/W^*$ of \mg\ 4730 and 5711~\AA\ lines have similar trends ($0.9\le W/W^*\le1.1$ for $-1\le[$Fe/H$]\le+0.5$, stronger NLTE variations otherwise) while the $W/W^*$ of the 11828~\AA\ line is very similar to the $W/W^*$ of the 8806~\AA\ line.

The \mg\ 8923~\AA\ line is not very sensitive to the NLTE effects except for models with $T_{{\rm eff}} > 4500$~K and $-2\le [$Fe/H$]\le -1$ for which we can have $\Delta$[Mg/H$]=+0.15$~dex of abundance correction for the hottest models. 

The weak \mg\ 10312~\AA\ line is not sensitive to NLTE effects since abundance correction is less than $\pm 0.05$~dex.

The 7657~\AA\ line is the counterpart of the 8923~\AA\ line in the triplet system. 
This line is in the saturated part of the curve of growth at solar metallicity and becomes weak for [Fe/H$] \le -1$. 
There is a systematic negative NLTE abundance correction for models with $T_{{\rm eff}} \le 4000$~K that can reach $-0.3$~dex.

The components of the \mg~b triplet follow the same NLTE trends.
When these components are strong with $T_{{\rm eff}} \ge 4000$~K and with $-1\le[$Fe/H$]\le+0.5$, the NLTE abundance corrections are positive and vary between $+0.1$ and $+0.3$~dex for the most metal-rich models.
For models with $T_{{\rm eff}} \ge 4500$~K and with $-4\le[$Fe/H$]\le-3$, strong variations exist for $W/W^*$ that can produce NLTE abundance correction as large as $+0.3$~dex at [Fe/H$] = -4$ in the linear regime.
The sensitivity of the surface gravity to the NLTE effect is maximum for the intermediate $T_{{\rm eff}}$ and for the most metal-rich models, and can reach $+0.1$~dex for models between $\log g =1$ and 2.

We compare our results with those of \citet{Shimanskaya00} for which grids of stellar parameters overlap. They provided us NLTE corrections for several lines (Shimanskaya, private communication).
They take into account inelastic collisions with hydrogen through the use of Drawin's formula with a scaling factor of $S_H = 0.1$.
Therefore, we added hydrogen inelastic collisions with the same scaling factor for six model atmospheres ($T_{{\rm eff}} = 4500$ and  $\log g = 1$) and ($T_{{\rm eff}} = 5000$~K and  $\log g = 2$) with [Fe/H] = 0, $-2$, and $-3$.
Results for five lines ($\lambda\lambda$ 4571, 4702, 5183, 5711 and 8736) are presented in Table~\ref{Comp_Shimanskaya}.
We formulate three remarks.
Firstly, the use of Drawin's formula reduces strongly NLTE effects, even with a small scaling factor. 
Secondly, our NLTE effects are smaller than \citet{Shimanskaya00} when we use Drawin's formula.
Thirdly, one should be cautious with this comparison because we used different model atoms, model atmospheres, atomic data and codes.
We noticed that we have a strong abundance correction for the intercombination resonance 4571 \AA\ line that can reach 0.5 dex even if we treat electronic collision of this line not using the IPM based on the low $\log gf$ value ($-5.688$), but using a collisional strength $\Omega_{ij}$ of 1.
Results from \citet{Shimanskaya00} give lower NLTE abundance correction for this line, maybe because they used isotopic shifts.
The 5711 \AA\ line is only affected by NLTE effects at lower metallicities.
The subordinate 5183 \AA\ line of the \mg~b triplet is strongly affected by NLTE effects even at solar metallicity with (0.1~dex) and without (0.2~dex) the use of Drawin's formula.  

\begin{figure}
\includegraphics[width=\linewidth]{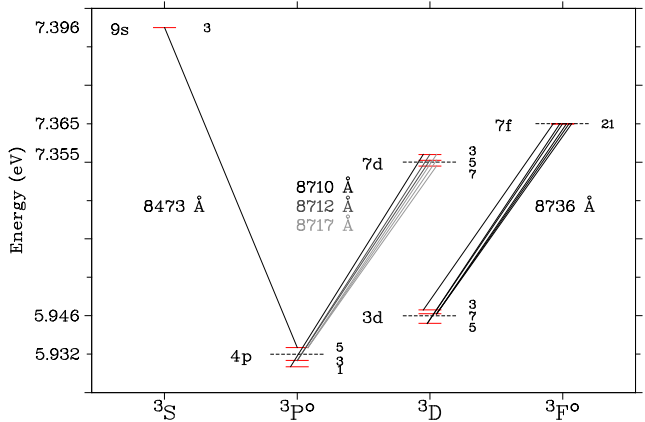}
\caption{Partial Grotrian diagram for \mg\ lines involved in the wavelength domain of {\it Gaia}/RVS. 
Mean levels are represented by dashed lines with their \nist\ values. 
Statistical weights are represented on the right side of the fine levels.}
\label{Gro_MgI_RVS}
\end{figure} 
  
\vspace{0.5cm}
{\it The {\it Gaia} lines} 
\vspace{0.5cm}

\begin{figure}
\includegraphics[width=\linewidth]{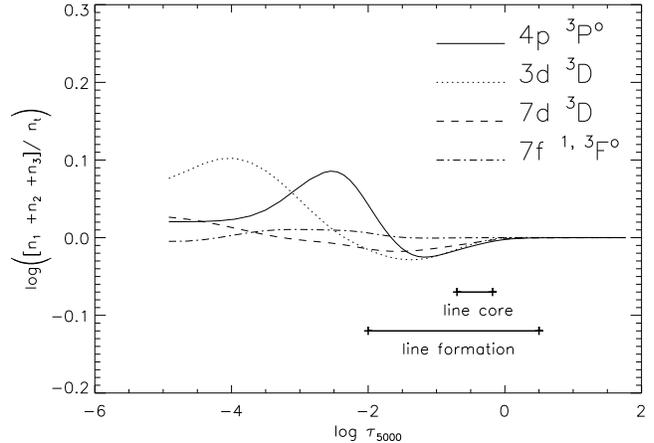}
\caption{Ratio between sum of detailed level populations ($n_{1}+n_{2}+n_{3})$ with average level population ($n_{t}$) in logarithmic scale as a function of standard optical depth. For the $7f$ configuration, we merge the terms of the 2 multiplicities. We have represented domain formations of lines in {\it Gaia}/RVS wavelengths for a solar \marcs\ model.}
\label{Nregroupe}
\end{figure}

Five \mg\ lines from the triplet system belong to the wavelength range of {\it Gaia}/RVS. 
These lines are multiplets with poorly known oscillator strengths.
The best accuracy for a component of the \mg~8736~\AA\ line is C (25\%) while all the other log~$gf$ have an accuracy of D or E, in the \nist's definition. 
In order to deliver NLTE effects for these multiplets, we use the model atom B with merged levels, except for the 8473~\AA\ line. 
To illustrate the fact that the SE is not disturbed when introducing the merged levels, we compare the level populations of the \mg\ model atoms A and B (so the number of levels drops from 150 at 141 in the SE). 
We compare the sum of populations of fine levels of the \mg\ model atom A with the population of the merged levels of the \mg\ model atom B. 
The logarithm of the formed ratio is plotted for each level considered in Fig.~\ref{Nregroupe}.
The difference between populations seems to be important, but when we integrate over the depth-scale, the relative error is smaller than 0.1\% for $4p\ ^{3}P^{o}$ and smaller than 0.05\% for the three higher levels.
We conclude that the merging of fine levels  for 4 average levels does not disturb the entire SE of the \mg\ atom. 

The $W/W^*$ for the \mg\ {\it Gaia}/RVS lines are plotted for several $T_{{\rm eff}}$  as a function of [Fe/H] for each $\log\, g$ in Fig.~\ref{Gaia_lines1}.
The 8473~\AA\ line is the least affected by NLTE effects but vanishes as soon as the metallicity becomes less than $-1$~dex.
The 8715~\AA\ Ritz line (representing a fictitious line with the $\lambda\lambda$ 8710, 8712 and 8717 triplet lines) suffers from large NLTE effects and it is the most sensitive to surface gravity. The NLTE abundance correction can vary of 0.25~dex for $T_{{\rm eff}}=3500$~K, [Fe/H$] = -1$ between $\log g =1$ and 2.
The 8736~\AA\ line, which is the most visible \mg\ line in this range, disappears under [Fe/H]~=~$-2$. 
This $W/W^*$ is larger than one for a large range of stellar parameters and can reach 1.4 at solar metallicity. 

In our knowledge, there is only NLTE results for the \mg~8736~\AA\ line in the literature by \citet{Shimanskaya00}.
A comparison for six model atmospheres are presented in Table~\ref{Comp_Shimanskaya}.
For models at solar metallicity, we are in 0.05 dex agreement.
But for the model at $T_{{\rm eff}} = 4500$, $\log g = 1$ and [Fe/H]$ = -2$, we find a positive NLTE abundance correction (for $S_H = 0$). 
This positive NLTE abundance correction increases with effective temperature and surface gravity whereas this NLTE abundance correction is negative and constant at $-0.05$~dex for Shimanskaya (private communication). 
This can be due to the difference in the Mg abundance chosen at this metallicity ([Mg/Fe$]=+0.4$ for us versus $+0.00$ for them) and to the value of $S_H$.

\begin{table*}
\caption{\caI\ $W/W^*$ values for $\log g = 1.5$. Values outside [0.9, 1.1] range are boldfaced. Blanks mean that $W < 1$~m\AA.}
\begin{tabular}{lccccrcccrccc}
$\lambda$[\AA]&[Fe/H]&\multicolumn{3}{c}{$-3$}&&\multicolumn{3}{c}{$-2$}&&\multicolumn{3}{c}{0}\\
\cline{3-5} \cline{7-9} \cline{11-13}
&$T_{{\rm eff}}$ [K]&3500&4250&5250&&3500&4250&5250&&3500&4250&5250\\
\hline\hline
 4226.727$^{\dag}$& &\textbf{0.54}&\textbf{0.59}&0.93& &\textbf{0.64}&\textbf{0.66}&\textbf{0.86}& &\textbf{0.84}&\textbf{0.58}&\textbf{0.81}\\
\hline
 6572.777$^{\dag}$& &\textbf{0.58}&\textbf{0.35}&    & &\textbf{0.85}&\textbf{0.67}&\textbf{0.51}& &0.93&\textbf{0.81}&\textbf{0.75} \\
 \hline\hline
 5512.980& &1.02&\textbf{0.61}&\textbf{0.52}& &1.04&0.92&\textbf{0.76}& &0.95&0.97&1.04 \\
 5867.559& &\textbf{1.28}&    &    & &1.04&0.92&\textbf{0.76}& &0.99&\textbf{0.87}&0.97 \\
 7326.142& &\textbf{1.13}&\textbf{0.71}&\textbf{0.51}& &1.10&1.08&\textbf{0.85}& &0.94&0.97&1.08 \\
 \hline
 4425.435& &\textbf{0.83}&\textbf{0.59}&\textbf{0.52}& &\textbf{0.85}&0.90&\textbf{0.82}& &\textbf{0.81}&\textbf{0.85}&0.96 \\
 \hline\hline
 6102.719& &0.96&\textbf{0.64}&\textbf{0.56}& &\textbf{0.88}&0.99&\textbf{0.86}& &\textbf{0.77}&\textbf{0.83}&0.96 \\
 6122.213& &0.96&\textbf{0.77}&\textbf{0.60}& &\textbf{0.86}&1.00&0.97& &\textbf{0.75}&\textbf{0.80}&0.98 \\
 6162.170& &0.92&\textbf{0.82}&\textbf{0.66}& &\textbf{0.84}&0.97&1.03& &\textbf{0.74}&\textbf{0.76}&0.98 \\
 \hline\hline
 6161.294& &\textbf{0.57}&\textbf{0.40}&    & &\textbf{0.76}&\textbf{0.60}&\textbf{0.58}& &\textbf{0.87}&\textbf{0.77}&\textbf{0.83} \\
 6166.438& &\textbf{0.55}&\textbf{0.40}&    & &\textbf{0.73}&\textbf{0.60}&\textbf{0.57}& &\textbf{0.87}&\textbf{0.78}&\textbf{0.85} \\
 6169.037& &\textbf{0.55}&\textbf{0.39}&\textbf{0.39}& &\textbf{0.79}&\textbf{0.66}&\textbf{0.57}& &\textbf{0.86}&\textbf{0.80}&\textbf{0.88} \\
 6169.562& &\textbf{0.59}&\textbf{0.42}&\textbf{0.46}& &\textbf{0.83}&\textbf{0.75}&\textbf{0.65}& &\textbf{0.84}&\textbf{0.83}&0.90 \\
 \hline\hline
 4578.549& &\textbf{0.49}&\textbf{0.36}&\textbf{0.46}& &\textbf{0.77}&\textbf{0.65}&\textbf{0.59}& &\textbf{0.88}&\textbf{0.87}&0.91 \\
 \hline
 6455.592& &\textbf{0.56}&\textbf{0.40}&    & &\textbf{0.77}&\textbf{0.62}&\textbf{0.62}& &0.92&\textbf{0.82}&0.91 \\ 
\hline\hline
 8525.707& &    &    &    & &    &    &    & &\textbf{0.73}&\textbf{0.60}&\textbf{0.62} \\
 8583.318& &    &    &    & &    &    &    & &\textbf{0.81}&\textbf{0.69}&\textbf{0.73} \\
 8633.929& &    &    &    & &    &    &    & &\textbf{0.88}&\textbf{0.76}&\textbf{0.86} \\
\hline\hline
\end{tabular}
\begin{flushleft}
$^{\dag}$  Note that this line is sensitive to the surface gravity.
\end{flushleft}
\label{CaI_WWS}
\end{table*}
 
\subsection{The \caI\ and \caII\ lines}
 
\subsubsection{The \caI\ lines}\label{Ca_section}

\begin{figure}
 \includegraphics[width=\linewidth]{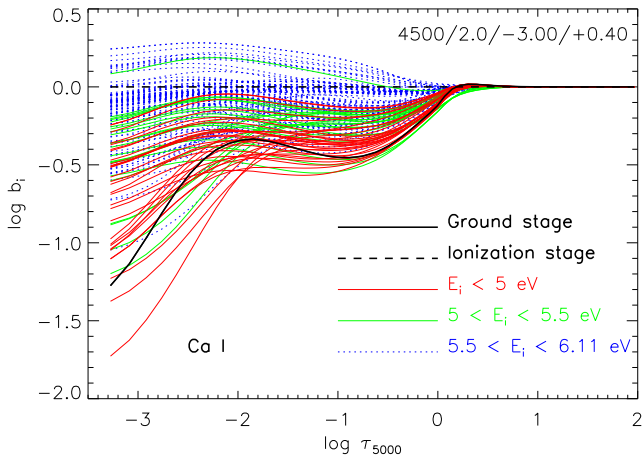}
 \caption{\caI\ departure coefficient as a function of optical depth in the continuum at 5000 \AA\ for a metal-poor giant stellar model ($T_{{\rm eff}} = 4500$~K, $\log g=2.0$, [Fe/H$]=-3.0$ and [$\alpha$/Fe$]=+0.4$). We separate levels  as functions of their energy $E_i$ in respect to the ground state as shown in the legend.}
 \label{CaI_pop}
\end{figure}

We show the departure coefficients $b_i$ for the \caI\ model atom in Fig.~\ref{CaI_pop}, with the same model atmosphere used for the \mg\ model atom in Fig.\ref{MgI_pop}, i.e. $T_{{\rm eff}}=4500$, $\log g = 2$ and [Fe/H$] = -3$. The ground stage of \caI\ is also depopulated by overionization but in a lesser extent than for \mg. In general, the \caI\ model is less affected by NLTE effects than the \mg\ model for the atmospheric parameters used in this paper. Two mechanisms appear: on the one hand, over-photoionization depopulates lower levels and on the other hand, collisional recombination and photon suction overpopulate levels close to the ionization stage.  

The \caI\ 4226~\AA\ resonance line shows that its $W/W^*$ is strongly dependent of the three atmospheric parameters (see Table~\ref{CaI_WWS} and Fig.~\ref{CaI_lines}). 
For all the models, the NLTE abundance corrections are positive with $\Delta[$Ca/H$]\ge 0.03$. 
This line is very sensitive to the surface gravity except for $T_{{\rm eff}}\ge 5000$~K and [Fe/H$]\le -2$.
The NLTE abundance correction can reach $+0.2$~dex between $\log g = 1$ and 2 for the most metal-rich models and $+0.5$~dex for the coolest and most metal-poor models. 

The \caI\ 6572~\AA\ intercombination and resonance line is also sensitive to the surface gravity but in a lesser extent. 
This line is weaker than the 4226~\AA\ one and shows that $W/W^*$ depends on the $\log g$ only for models with $T_{\rm eff} \ge 4000$~K.   

The \caI\ $\lambda\lambda$ 5512, 5867 and 7326 lines share the same lower level $4p\ ^1P^o$.
For these lines, $0.5\le W/W^* \le 1.2$ except for the coolest and most metal-poor model. 
We can adopt that they are formed under LTE condition in the range of $-1.5\le[$Fe/H$]\le 0$. 
NLTE abundance correction can be as large as $+0.3$~dex for [Fe/H$]\le -1.5$.
For $T_{{\rm eff}} < 4000$~K and [Fe/H$]< -3$, the 5512 and 7326~\AA\ lines become sensitive to the surface gravity. 

\begin{figure}
\includegraphics[width=\linewidth]{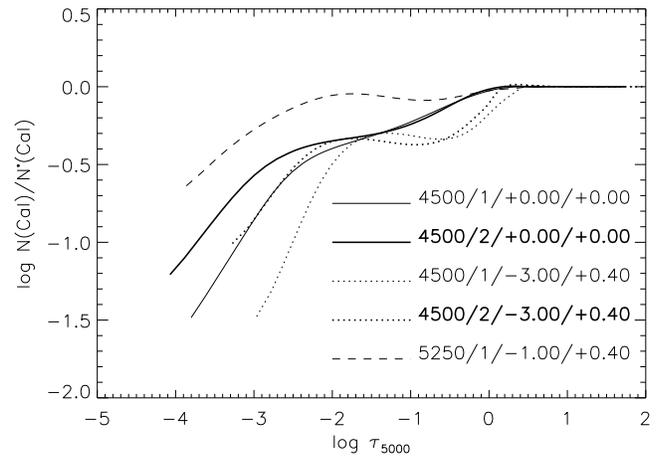}
\caption{Global \caI\ departure coefficients for different model atmospheres of $T_{{\rm eff}} = 4500$~K, $\log g = 1$ (thin lines), $\log g = 2$ (thick lines), and [Fe/H]~$ = 0$ (solid line), [Fe/H] $ = -3$ (dotted line); and of $T_{{\rm eff}} = 5250$~K, $\log g = 1$ and [Fe/H]$ =-1$ (dashed line). The fourth parameter on the plot is the $\alpha$-enhancement of the model atmosphere.}
\label{CaI_effects}
\end{figure}

For the \caI\ red triplet $\lambda\lambda$ 6102, 6122 and 6162 subordinate lines, the NLTE abundance correction are positive for almost all the models.
Large deviations from LTE exist for $T_{{\rm eff}}<4500$~K with $-1\le[$Fe/H$]\le +0.5$, and for $T_{{\rm eff}}>4500$~K with $-4\le[$Fe/H$]\le -2.5$.
They are not sensitive to the surface gravity.

The \caI\ lines at $\lambda\lambda$ 6161, 6166, 6169.0 and 6169.6 belong to the same multiplet.
Their $W/W^*$ behaviours are very similar and always less than one, that produces positive NLTE correction.
The NLTE abundance corrections can reach $+0.4$~dex for [Fe/H$]\le -3$.
Theses lines are sensitive to the surface gravity for models with $T_{{\rm eff}}<4000$~K and [Fe/H$]<-3$.

The line at 4578~\AA\ is also from the same lower term than the previous multiplet. 
The $W/W^*$ is very similar at low metallicities.
We can note that this line is formed in LTE for models with $-1\le[$Fe/H$]\le 0$ ($W/W^*$between 0.9 and 1).

The \caI\ 6455~\AA\ intercombination line also shares the same lower term than the previous multiplet and the 4578 \AA\ line. The $W/W^*$ is also similar and always lower than one.
We notice that the lower deviation from LTE comes near to [Fe/H$]=-1$. 

The \caI\ {\it Gaia}/RVS lines at 8525, 8583 and 8633~\AA\ are strongly affected by NLTE effects at solar metallicity (see Table~\ref{CaI_WWS}).
These lines are weak whatever the stellar parameters (NLTE EW $\sim$ 50, 70 and 90~m\AA\ at maximum respectively).
The stronger NLTE effects appear for 8525 and 8583~\AA\ lines and can reach $\Delta$[Ca/H$]= +0.4$~dex. 
For $T_{{\rm eff}} > 4250$~K (respectively $T_{{\rm eff}}<4250$~K), NLTE effects increase (respectively decrease) with decreasing metallicity as shown in Fig.~\ref{Gaia_lines1}. 

\begin{table*}
\caption{\caII\ $W/W^*$ values for $\log g = 1.5$. Values outside [0.9, 1.1] range are boldfaced. Blanks mean that $W < 1$~m\AA.}
\begin{tabular}{lccccrcccrccc}
&[Fe/H]&\multicolumn{3}{c}{$-3$}&&\multicolumn{3}{c}{$-2$}&&\multicolumn{3}{c}{0}\\
\cline{3-5} \cline{7-9} \cline{11-13}
$\lambda$[\AA]&$T_{{\rm eff}}$ [K]&3500&4250&5250&&3500&4250&5250&&3500&4250&5250\\
\hline\hline
3933.663& &0.97&0.95&0.98& &0.96&0.98&0.99& &0.99&0.99&0.99 \\
3968.468& &0.96&0.95&0.97& &0.97&0.98&0.99& &0.99&0.99&0.99 \\
\hline\hline
8498.018$^{\dag}$& &\textbf{1.13}&1.06&\textbf{1.23}& &1.02&1.00&1.07& &0.95&0.94&0.98 \\
8542.086$^{\dag}$& &\textbf{1.13}&1.02&\textbf{1.11}& &1.01&1.00&1.04& &0.95&0.95&0.98 \\
8662.135$^{\dag}$& &\textbf{1.11}&1.02&\textbf{1.14}& &1.00&1.00&1.05& &0.95&0.95&0.98 \\
\hline\hline
8248.791$^{\ddag}$& &    &    &\textbf{1.12}& &    &\textbf{1.17}&\textbf{1.24}& &\textbf{1.13}&\textbf{1.23}&\textbf{1.26} \\
9854.752$^{\ddag}$& &    &    &    & &    &    &\textbf{1.24}& &\textbf{1.12}&\textbf{1.22}&\textbf{1.27} \\
\hline\hline
8912.059& &    &\textbf{1.91}&\textbf{1.54}& &    &\textbf{1.55}&\textbf{1.64}& &\textbf{1.27}&\textbf{1.37}&\textbf{1.38} \\
\hline\hline
11949.74$^{\dag}$& &    &\textbf{1.69}&\textbf{1.57}& &\textbf{1.26}&\textbf{1.51}&\textbf{1.80}& &\textbf{1.27}&\textbf{1.43}&\textbf{1.51} \\
\end{tabular}
\begin{flushleft}
$^{\dag}$  Note that this line is sensitive to the surface gravity.
$^{\ddag}$ Note that this line is sensitive to the surface gravity for the hottest models.
\end{flushleft}
\label{CaII_WWS}
\end{table*}

We compare our NLTE results with those of \citet{Drake91} and found opposite effects concerning the variation of overionization with surface gravity. 
As emphasized in the review of \citet{Asplund05}, Drake unexpectedly found that the over ionization of \caI\ decreases with surface gravity which is explained by an opacity effect that did not convince \citet{Asplund05}. 
 For giants stars, we show in Fig.~\ref{CaI_effects} the opposite effects of overionization, comparing with fig.~6a by \cite{Drake91}. 
 We plot overall \caI\ departure coefficient for model atmospheres of $T_{{\rm eff}} = 4500$~K, $\log g = 1$ and $2$, and [Fe/H]~$ = 0$ and $-3$. 
 We see that for a given metallicity, the overall \caI\ departure coefficient tends to be closer to one with the increase of the surface gravity. 
This is the opposite of the two models in fig.~6a of Drake ($T_{{\rm eff}}=4500$~K, $\log g = 1$ and 0, and [Fe/H$]=0$) where the overionization of \caI\ increases when decreasing the surface gravity.  

Comparison with the work of \citet{Mashonkina07} is done even if the grids of stellar parameters do not overlap. 
They provided NLTE abundance corrections for a grid of dwarfs and subgiants.
We compare results for the Sun and for a spherical model at $T_{{\rm eff}} = 5000$, $\log g =3$, [Fe/H$]=-2$ and [$\alpha$/Fe$]=+0.4$. 
In order to be the most consistent for the comparison, we use our \caI/{\scriptsize II} model atom described in Sect.~\ref{CaI/II}. 
Results are shown in Table~\ref{Comp_Mashonkina}. We found a better agreement with these authors for the metal-poor star rather than for the Sun. 
Our NLTE abundance corrections for the solar \caI\ lines are mainly lower and in opposite sign compared to them. For the metal-poor star, agreement is very satisfactory except for the 4425 \AA\ line for which we have a halved abundance correction. 
These agreements are found using different geometries for the model atmospheres (plane--parallel Kurucz models for \citet{Mashonkina07} against spherical \marcs\ models for us) and different NLTE codes, but give confidence for the NLTE computations done for the grid of late-type giant and supergiant stars.         

\begin{table*}
 \caption{Comparison of \caI\ abundance corrections with \citet{Mashonkina07} for the Sun and for a metal poor sub-giant star without hydrogenic collisions ($S_H=0$). $W$ represents our NLTE computed EW and $W/W^*$ represents our NLTE/LTE EW ratios. When Eq.~(\ref{1}) is not applicable, blanks are used in $\Delta$[Ca/H] column. Blank in $\Delta$[Ca/H]$^{\dag}$ column denotes a NLTE EW less than 5 m\AA.}
 \begin{tabular}{crcrrlrcrr}
 & \multicolumn{4}{c}{Sun}&& \multicolumn{4}{c}{$T_{{\rm eff}}=5000,\ \log g = 3,\ [$Fe/H$]=-2$}\\
 \cline{2-5} \cline{7-10}
 $\lambda$ [\AA] & $W$ [m\AA]& $W/W^*$ & $\Delta$[Ca/H] & $\Delta$[Ca/H]$^{\dag}$ && $W$ [m\AA]& $W/W^*$ & $\Delta$[Ca/H] & $\Delta$[Ca/H]$^{\dag}$ \\
 \hline
 &&&&&\\
 \caI &&&&&\\
 &&&&&\\
 4226.727&1318& 0.988&  0.01 &0.07   &&543& 0.766& 0.23  & 0.24   \\
 4425.435& 174& 1.005&       &0.04   && 84& 0.830& 0.08  & 0.17   \\
 5512.980&  96& 1.056&$-0.01$&0.03   && 20& 0.766& 0.12  & 0.10   \\
 5867.559&  21& 1.018&$-0.01$&0.06   &&  2& 0.765& 0.12  & 0.08   \\
 6162.170& 260& 1.037&       &0.01   &&128& 0.990&       & 0.00   \\
 6166.438&  65& 1.013&$-0.01$&0.07   &&  9& 0.557& 0.25  & 0.30   \\
 &&&&&\\ 
 \caII &&&&&\\
 &&&&&\\
 8248.7907&  68& 1.184&$-0.15$&$-0.11$&&  5& 1.153&$-0.06$&        \\
 8498.0180&1244& 1.004&  0.00 &$-0.02$&&731& 1.087&$-0.07$&$-0.08$ \\
 \\
 \hline 
 \end{tabular}
 \\
 \begin{flushleft}
 $^{\dag}$ \citet{Mashonkina07}\\
 \end{flushleft}
 \label{Comp_Mashonkina}
\end{table*}

\subsubsection{The \caII\ lines}
A part of the results for \caII\ lines are shown in Table~\ref{CaII_WWS} and in Fig.~\ref{CaII_lines}.
For the strong resonance H \& K lines at $\lambda\lambda$ 3968 and 3933 respectively, $W/W^*$ is less but close than one except for temperatures below 3900~K.
Hence, the NLTE abundance correction does not exceed 0.1~dex between [Fe/H$]=+0.5$ and $-3$. 
It can vary between $-0.1$ and $0.2$~dex at the lowest metallicity of $-4$~dex.
We noticed a light sensitivity to the surface gravity for all effective temperatures.
For [Fe/H$]\le -2.5$ that can reach 0.1~dex of amplitude between $\log g = 1$ and $\log g = 2$ at 3500~K (see Table~\ref{CaII_WWS} and Fig.~\ref{CaII_lines}). 

The lines at $\lambda\lambda$ 8248 and 9854 come from the same $5p\ ^2P^o$ term but from different fine levels and have $W/W^*>1.1$.
The deviation from LTE increases with increasing effective temperature and can reach $-0.1$~dex of abundance correction.
These weak lines are sensitive to the change of the surface gravity for $T_{{\rm eff}} \ge 4500$~K.
These variations can reach 0.05~dex of amplitude between $\log g = 0.5$ and $\log g = 2$.

The NLTE abundance correction for the 8912~\AA\ line can vary between $-0.15$~dex at solar and subsolar metallicities and $-0.3$~dex at [Fe/H$] = -3.5$ and $T_{{\rm eff}}= 4500$~K.

The \caII\ IR 11949~\AA\ line shows an NLTE sensitivity to the surface gravity (as seen in Fig.~\ref{CaII_lines}) that increases with the effective temperature. 
$W/W^*>1.2$ for all of the atmospheric parameters.

\begin{figure}
\begin{center}
\includegraphics[width=\linewidth]{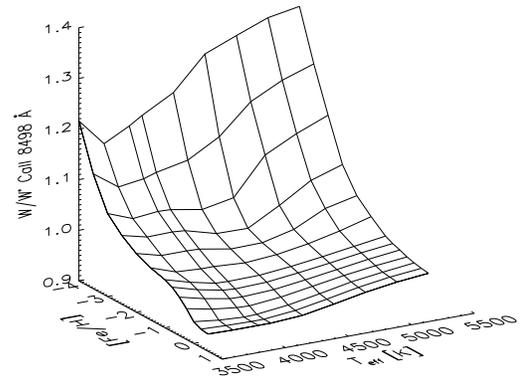}
\newline
\caption{EW ratio of \caII\ IR 8498~\AA\ line as a function of effective temperature $T_{{\rm eff}}$ and metallicity [Fe/H], for a given surface gravity $\log g = 1.5$.}
\label{surf}
\end{center}
\end{figure}

\begin{figure}
\includegraphics[width=\linewidth]{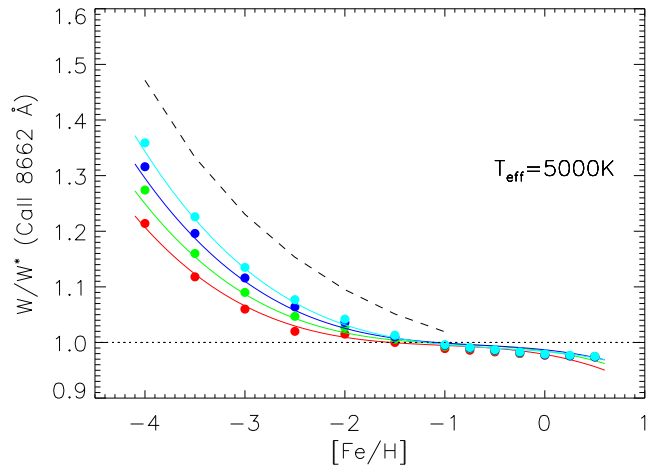}
\caption{NLTE/LTE EW ratio of \caII\ IR 8662 \AA\ line as function of metallicity [Fe/H] and $\log g$, for a given effective temperature $T_{{\rm eff}} = 5000$~K. 
Each colour of the $W/W^*$ represents a surface gravity (red, green, blue and cyan for $\log g = 0.5$, 1.0, 1.5 and 2.0~dex respectively).
Dots represent computed ratios while full lines represents our polynomial fit of the NLTE/LTE EW ratios. The dashed curve stands for the $W/W^*(8662\ {\rm \AA})$ polynomial fit computed by \citet{Starkenburg10}.}  
\label{caII}
\end{figure}

\begin{figure}
\includegraphics[width=\linewidth]{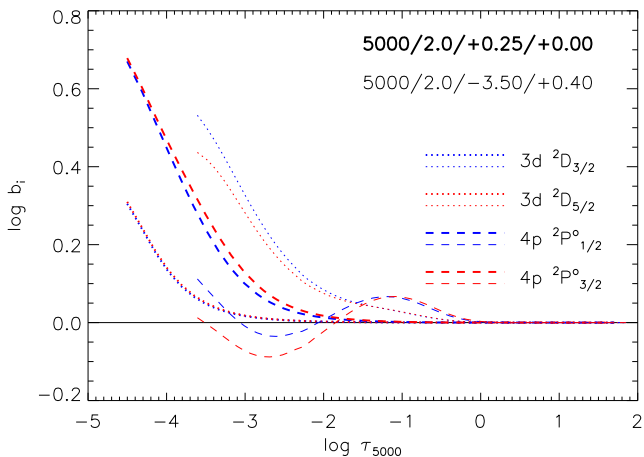}
\caption{Departure coefficients for fine levels implied in the CaT lines for comparison between a metal rich model atmosphere (bold curves) and an extremely metal poor model (thin curves). The two model atmospheres have $T_{{\rm eff}} = 5000$~K and $\log g = 2$, but different global metallicities ([Fe/H]$ = +0.25$ and $-3.5$) and different $\alpha$-enhancement factors ([$\alpha$/Fe]$ = 0$ and $+0.4$ respectively).}
\label{CaII_8498_pop}
\end{figure}

The NLTE/LTE EW of each component of the CaT follows the same trend. 
An example of the $W/W^*$ trend of the CaT is shown for the 8498 \AA\ line, which has the most pronounced NLTE effects among the triplet, in Fig.~\ref{surf} for a surface gravity of $\log g = 1.5$. 
NLTE effects on the CaT are weak at solar and sub-solar metallicities for all the effective temperatures. 
For lower metallicities, strong deviations from LTE appear. 
As shown in Fig.~\ref{surf}, \ref{caII} and \ref{Gaia_lines2}, the $W/W^*$ trends of the CaT as a function of the metallicity can be separated in two regimes: 
\begin{itemize}
\item the $W/W^*$ is a linear function of the metallicity [Fe/H] when $W/W^* \le 1$;
\item the $W/W^*$ increases strongly with decreasing metallicity when $W/W^*>1$ and can reach $1.45$.
\end{itemize}  
Roughly speaking, the CaT lines are formed in LTE ($W/W^*=1\pm 0.05$ which corresponds to $\Delta[$Ca/H$]=0.00 \pm 0.04$~dex) if $-2\le[$Fe/H$]\le0$ for $T_{{\rm eff}}$ and $\log g$ in our grid.
The CaT lines cannot be considered in LTE when [Fe/H$] < -2$, and when [Fe/H$]>0$ for $T_{{\rm eff}} < 4000$~K.  
The NLTE effects decrease with effective temperature and pass through a minimum at $\sim$ 4000~K  to increase again at lower $T_{{\rm eff}}$, as shown in Fig.~\ref{surf}. 
This effect may be due to the increase of molecular continuous opacities in \marcs\ models when the effective temperature decreases.

The behaviour of the $W/W^*$ with varying [Fe/H] of the CaT can be explained with the variations of the departure coefficients presented in Fig.~\ref{CaII_8498_pop}.
 As the stimulated emission can be neglected for optical and near IR lines, the line source function relative to the Planck function $S_\nu^l/B_\nu$ follows the departure coefficient ratio $b_j/b_i$ ($i$ and $j$ are the lower and upper levels respectively). 
The behaviour of the $W/W^*$ depends on the values of $b_i$ relative to $b_j$ and on the deviation between them. 
In the metal-rich model (in Fig.~\ref{CaII_8498_pop}), the levels are over-populated due to overionization of \caI\ and $b_i < b_j$ that implies $S_\nu^l > B_\nu$, and the emergent intensity is strengthened in the line compared to LTE intensity. 
Thus the EW in NLTE is lower than the EW in LTE. This explains why $W/W^* \le 1$ for models with $-1\le$[Fe/H]$\le +0.5$ in Fig.~\ref{caII}. 
For the metal-poor model in Fig.~\ref{CaII_8498_pop}, the mechanism is the opposite due to the change in the relative values of $b_i$ and $b_j$. 
For metallicities lower than $-1$, $b_i > b_j$ if $\tau_{5000} \le -1.7$ and then $S_\nu^l < B_\nu$. 
This implies a reduction of the emergent intensity in the line and then a larger value of EW in NLTE relative to the LTE. 
This explains why $W/W^* > 1$ for models with [Fe/H]$\le -1$ in Fig.~\ref{caII}. 
The large effects at lower metallicity is due to a larger extent of $\log |b_j/b_i|$ compared to solar and sub-solar metallicities.

The fine structure of the levels follows the same trends but level population differences increase with decreasing $\log \tau_{5000}$. 
The \caII\ IR 8498 \AA\ line has the greater NLTE effects relative to the two other lines because it has the largest amplitude on the deviation between the lower coefficient $b_i$ ($3d\ ^2D_{3/2}$, blue dotted line) and the upper coefficient $b_j$ ($4p\ ^2P^o_{3/2}$, red dashed line) as seen in Fig.~\ref{CaII_8498_pop}. 
The CaT lines are dominated by their wings even at [Fe/H]~$=-3$ as shown in fig.~1 of \citet{Starkenburg10}. These wings are formed in the deep photosphere in LTE conditions. Therefore the EW are weakly influenced by NLTE effects at solar metallicity and moderate for most metal-poor stars for which the correction can become very large (20-30\% at [Fe/H]$=-3$). 
These results are in quite good agreement with those of previous investigations with the same stellar parameters (e.g. \citealt{Jorgensen92}, \citealt{Andretta05}). 

We note that even if NLTE line formation has almost no influence on Ca abundance determination using the CaT at solar or sub-solar metallicity, it has an impact on the detection of stellar activity. 
Indeed, the central depression of the CaT is often used as an indicator of activity, (e.g. \citealt{Andretta05}, \citealt{Busa07}). 
The detail modelling of the line core formed in NLTE is therefore crucial.  

\begin{table}
\caption{Comparison of our $W/W^*_J$ with those of \citet{Jorgensen92} for the same stellar parameters. The NLTE/LTE EW ratio is defined as $W/W^*_J = W(8542{\rm \AA}+ 8662{\rm \AA})/ W^*(8542{\rm \AA}+ 8662{\rm \AA})$.}
\begin{center}
\begin{tabular}{ccrcc}
\\
$T_{{\rm eff}}$ [K] & log $g$ & [Fe/H] & $W/W^*_J$  &  $W/W^*_J\dag$\\
\hline
     &   &$ 0.00$&0.93 &0.95 \\
     & 1 &$-0.50$&0.95 &0.94 \\
4000 &   &$-1.00$&0.97 &0.95 \\
     &   &$ 0.00$&0.96 &0.97 \\
     & 2 &$-0.50$&0.97 &0.98 \\
     &   &$-1.00$&0.98 &0.99 \\
     \hline
     &   &$ 0.00$&0.98 &0.99 \\
     & 1 &$-0.50$&0.99 &0.99 \\
5000 &   &$-1.00$&0.99 &0.98 \\
     &   &$ 0.00$&0.98 &0.99 \\
     & 2 &$-0.50$&0.99 &0.99 \\
     &   &$-1.00$&1.00 &0.99 \\
\hline
\end{tabular}
\end{center}
$\dag$ \citet{Jorgensen92}
\label{tab_J}
\end{table}

Comparing our results with those of \citet{Jorgensen92}, we find that they are in very good agreement. They combine the two strongest lines of the CaT, noted here $W/W^*_J$ (8542 and 8662 \AA) and compute NLTE effects for a large set of gravities (from $\log g = 0.0$ to 4.0) but for a small metallicity range (from $-1.0$ to $0.2$). 
We give the comparison between our result for the combined lines ratio $W/W^*_J$ in Table~\ref{tab_J}. 
Deviation in NLTE effect with these authors are less than 2\%.  
The NLTE EW increases with increasing metallicity ($-1 <$ [Fe/H]$ < +0.25$) but our EW are larger by a factor of 2 with those of J\o rgensen et al. ones at [Fe/H] $= -1$. 
This may come from the fact that their model atmospheres are not $\alpha$-enhanced at lower metallicities. 
Note also that they do not have the photoionization cross-sections from the TopBase and ABO theory. 
Other authors computed NLTE effects for the CaT but for late-type dwarf stars (e.g. \citealt{Andretta05}). 
They found values of $W/W^*$ always larger than 1 reaching 1.35 at lower metallicity ([Fe/H]$ = -2$).    
 
We also compared the NLTE effects on CaT computed by \citet{Starkenburg10} and found a clear discrepancy at low metallicity. 
They provided a 2D polynomial fit of $W^*/W = f(T_{{\rm eff}}, {\rm [Fe/H]})$ for the lines at 8542~\AA\ and 8662~\AA, considering that these ratios are insensitive to the surface gravity ($1\le \log g \le 2$). 
We plotted in Fig.~\ref{caII} and \ref{CaII_lines} their polynomial fit for the 8662 \AA\ line. 
The discrepancies between our $W/W^*$ and theirs, at a given $T_{{\rm eff}}$, increase with decreasing metallicity and with decreasing surface gravity. 
At [Fe/H] $ = -4$, the deviation is about $10\%$ for $\log g = 2$ and about $18\%$ for $\log g = 1$. 
These differences may come from the fact that they used a different geometry (plane--parallel) in their study. 
We checked that the use of the scaling factor $S_H = 1$ for the collisions with neutral hydrogen cannot compensate this difference.
Note that they not clearly indicate the value of Ca abundance that they used for the computations of the $W/W^*$.

Since \citet{Starkenburg10} used the model atom of \citet{Mashonkina07}, we investigate the effects of the different atomic physics injected in the two model atoms of Ca~{\scriptsize I/II}.
We consider a stellar model ($T_{{\rm eff}} = 5000$~K, $\log g = 3$, [Fe/H] $= - 2$), which is outside of our grid,  but used in \citet{Mashonkina07} to compare with. 
Our predictions for the two \caII\ lines (8248 and 8498~\AA) in this particular case are shown in Table~\ref{Comp_Mashonkina}. 
We note a difference of 0.04 dex for the weak line at 8248 \AA\ for the Sun and a very good agreement (better than 0.01 dex) for the \caII\ IR 8498~\AA\ line for the metal-poor model atmosphere. 
Therefore, the discrepancy with \citet{Starkenburg10} does not come from the model atoms.
It must be investigated in more details since the calibration between the CaT and [Fe/H], as defined in their paper, may be affected for RGB metal-poor stars.  

Interesting opposite NLTE trends appear between \caI\ and \caII\ lines. 
In general, $W/W^*$ for \caI\ lines are smaller than 1.0, whereas $W/W^*$ for \caII\ lines are larger than 1.0, whatever the stellar parameters.
We note that the NLTE effects are anti-correlated between the \caI\ $\lambda\lambda$ 6102, 6122, 6162 triplet lines and the CaT lines in metal-poor model atmospheres. 
This may have a large impact on the Ca abundance analysis in metal-poor red giants; such remark has already been mentioned in the {\it Gaia} context \citep{Recio05}. 

\subsection{Uncertainty estimations}

\begin{table*}
 \caption{Fit coefficients of NLTE/LTE EW ratio $[W/W^*]_{{\rm fit}}$ for lines in the {\it Gaia}/RVS wavelength range. 
 The fits are valid for $3500 \le T_{{\rm eff}} \le 5500$~K, $0.5 \le \log g \le 2$ and [Fe/H$]_{\min} \le[$Fe/H$]\le +0.5$. 
 The Mg and Ca abundances follow the $\alpha$-enhancement given in Eq.~(\ref{alpha}).
 The blanks denote coefficients related to discarded terms.
 $\epsilon_{\max}$  and  $\epsilon_{{\rm mean}}$  are given (in \%) as the maximum and the mean absolute difference between $[W/W^*]_{{\rm fit}}$ and $W/W^*$.}
 \begin{tabular}{p{0.5cm}rrrcrrrcrrr}
             & \multicolumn{3}{c}{\mg} && \multicolumn{3}{c}{\caI} && \multicolumn{3}{c}{\caII}\\
\cline{2-4} \cline{6-8} \cline{10-12}
  $\lambda$ [\AA] & 8473.688 & $\left<8715.255\right>$ & 8736.012 && 8525.707&8583.318&8633.929 && 8498.018 & 8542.086 & 8662.135 \\
  \hline
{\scriptsize [Fe/H]$_{\rm min}$} &$-1.0$&$-1.5$&$-2.0$&& $-1.0$&$-1.0$&$-1.0$ &&$-4.0$&$-4.0$&$-4.0$\\
\hline
$a_0$    &   1.03E$+0$&   1.22E$+0$&   1.31E$+0$&&   5.95E$-1$&   7.05E$-1$&   7.97E$-1$&&   9.88E$-1$&   9.72E$-1$&   9.73E$-1$\\
$a_1$    &   3.32E$-2$&   1.43E$-1$&$-6.84$E$-2$&&$-2.81$E$-2$&            &   6.00E$-2$&&   3.68E$-2$&   3.71E$-2$&   3.82E$-2$\\
$a_2$    &$-1.79$E$-2$&$-9.26$E$-2$&$-2.36$E$-2$&&$-1.67$E$-2$&            &            &&   9.15E$-3$&   3.05E$-3$&   2.21E$-3$\\
$a_3$    &$-1.48$E$-1$&$-3.90$E$-2$&   1.61E$-1$&&   7.11E$-2$&   2.76E$-2$&            &&$-8.44$E$-2$&$-2.74$E$-2$&$-3.18$E$-2$\\
$a_4$    &   1.50E$-2$&$-1.04$E$-1$&$-6.34$E$-2$&&   4.74E$-2$&   5.13E$-2$&   6.63E$-2$&&   4.22E$-2$&   4.66E$-2$&   4.60E$-2$\\
$a_5$    &            &   6.79E$-2$&            &&   2.41E$-2$&   2.12E$-2$&   2.02E$-2$&&            &            &            \\
$a_6$    &            &$-2.76$E$-2$&            &&   1.22E$-2$&            &            &&            &   6.45E$-3$&            \\
$a_7$    &   2.24E$-2$&$-1.86$E$-2$&   8.05E$-2$&&   2.09E$-1$&   1.84E$-1$&   1.32E$-1$&&$-3.87$E$-2$&   5.33E$-3$&$-1.67$E$-2$\\
$a_8$    &   4.39E$-2$&   9.46E$-2$&$-2.16$E$-2$&&   5.01E$-2$&   3.87E$-2$&   2.81E$-2$&&$-3.81$E$-2$&$-2.93$E$-2$&$-2.91$E$-2$\\
$a_9$    &            &$-2.95$E$-1$&$-2.42$E$-1$&&$-2.49$E$-2$&$-3.69$E$-2$&$-3.44$E$-2$&&   1.12E$-1$&   7.77E$-2$&   9.38E$-2$\\
$a_{10}$ &$-2.31$E$-2$&            &   3.44E$-2$&&$-9.98$E$-2$&$-1.02$E$-1$&$-1.20$E$-1$&&$-2.54$E$-2$&$-2.43$E$-2$&$-2.32$E$-2$\\
$a_{11}$ &$-6.92$E$-3$&$-2.49$E$-2$&$-2.50$E$-2$&&   1.61E$-2$&            &            &&   1.15E$-2$&   1.42E$-2$&   1.34E$-2$\\
$a_{12}$ &            &            &            &&            &            &            &&            &$-1.08$E$-2$&$-9.03$E$-3$\\
$a_{13}$ &            &            &            &&            &            &            &&            &            &            \\
$a_{14}$ &   4.43E$-2$&   1.84E$-1$&   1.20E$-1$&&            &            &            &&$-3.22$E$-2$&$-5.32$E$-2$&$-5.47$E$-2$\\
$a_{15}$ &$-1.23$E$-2$&$-6.67$E$-2$&            &&            &            &            &&            &            &            \\
$a_{16}$ &            &   2.82E$-2$&            &&            &            &            &&            &$-5.81$E$-3$&            \\
$a_{17}$ &$-2.26$E$-2$&$-7.15$E$-2$&$-1.18$E$-1$&&   1.04E$-2$&   1.14E$-2$&   1.53E$-2$&&   4.17E$-2$&            &   2.07E$-2$\\
$a_{18}$ &   1.90E$-2$&            &   5.08E$-2$&&   1.06E$-2$&   6.51E$-3$&            &&   2.93E$-2$&   2.95E$-2$&   3.21E$-2$\\
$a_{19}$ &   3.75E$-2$&   1.72E$-1$&$-5.34$E$-2$&&   6.04E$-2$&   6.25E$-2$&   4.85E$-2$&&$-7.47$E$-2$&$-7.74$E$-2$&$-8.14$E$-2$\\
  \hline
  $\epsilon_{\max}$      & 3.4 & 7.0 & 9.3 && 4.7 & 4.5 & 4.6 && 6.0 & 7.2 & 8.3 \\
  $\epsilon_{{\rm mean}}$& 0.7 & 1.6 & 2.2 && 1.4 & 1.4 & 1.2 && 0.9 & 1.0 & 1.0 \\  
  $\epsilon_{\max}^\dag$       &  &  &  &&  &  &  && 2.3 & 4.6 & 4.2 \\  
  $\epsilon_{{\rm mean}}^\dag$ &  &  &  &&  &  &  && 0.6 & 0.8 & 0.8 \\  
  \hline
 \end{tabular}
 \begin{flushleft}
  $^\dag$ if [Fe/H$]\ge -2$ 
 \end{flushleft}
 \label{fit}
\end{table*}

Uncertainties of $W/W^*$ rely on the atomic parameters and on the assumptions of the stellar atmosphere modelling  (viz 1D, static, NLTE radiative plane parallel transfer with \multi\ code using 1D, static, LTE, spherical \marcs\ model atmospheres).
We computed $W/W^*$ varying atomic parameters in order to estimate uncertainties of the input atomic physics.
We selected four model atmospheres ($T_{{\rm eff}} = 5000$~K, $\log~g= 2$, [Fe/H$]=0, -2$; and $T_{{\rm eff}} = 4500$~K, $\log~g= 1$, [Fe/H$]=0, -2$ and three lines (one resonance, one subordinate, and one allowed lines) for each model atom ($\lambda\lambda$~4571, 5183, 8806 for \mg, $\lambda\lambda$~6162, 6166, 6572 for \caI, and $\lambda\lambda$~3933, 8248, 8498 for \caII).
The \mg~4571~\AA\ and the \caI~6572~\AA\ lines are the equivalent intercombination resonance lines for \mg\ and \caI\ model atoms.
For each line, the uncertainties of three atomic parameters were tested: the oscillator strengths, the photoionization cross-sections and the effective collisional strengths with electrons.
We emphasize that we performed these uncertainty estimations over a hundred of runs (3 among 45 lines over 4 among 453 model atmospheres) to give a rough idea of the impact of the uncertainties on atomic data.

We tested for each model atoms the change of the oscillator strengths of the three lines by $\pm 50\%$. 
The changes on the $W/W^*$ for the three lines are less than 8\%.
For instance, with the model at $T_{{\rm eff}} = 5000$~K, $\log~g= 2$, [Fe/H$]=-2$, we obtained $W/W^*($\mg~8806~\AA$) = 1.35^{+0.02}_{-0.06}$, $W/W^*($\caI~6572~\AA$) = 0.53^{+0.01}_{-0.02}$ and $W/W^*($\caII~8498~\AA$) = 1.07^{+0.03}_{-0.02}$.
We remind that the accuracy of the $\log gf$ used and given in Table~\ref{Lines} and \ref{Lines_MgI_Gaia} varies between $2\%$ and $50\%$.
Therefore, these uncertainties can be seen as an upper limit.
We noticed that the $W/W^*$ are more affected by changes of oscillator strength values for the metal-poor model atmospheres.

Varying the amplitudes of the photoionization cross-sections does not affect the lines of model atoms in the same way.
We changed by a factor $\pm 2$ the amplitude of the TopBase photoionization cross-sections of the lower levels of the lines considered. 
The \mg\ model atom is the most affected with variations on the $W/W^*$ that can reach $+8\%$ (for instance $W/W^*($\mg~5183~\AA$) = 0.79^{+0.03}_{-0.03}$ for the model with $T_{{\rm eff}} = 4500$~K, $\log~g= 1$, [Fe/H$]=0$).
Calcium is less affected, with $1\%$ of variations of $W/W^*$ for the \caI\ and less than $1\%$ for the \caII\ model atoms.   
As emphasized by \cite{Mashonkina07}, the changes of the photoionization cross-sections mainly affect the minority species (\caI\ in the range of stellar parameters considered here).
However, we found that the absolute corrections on the $W/W^*$ increase with decreasing metallicity, due to variations on the amplitude of photoionization cross-sections.

Finally, we tested the variations of the amplitude of the collisional strength with electrons. 
Using the IPM approximation, the change by a factor of $\pm 2$ of the effective collisional strength slightly affects the $W/W^*$ results (less than $\leq 2\%$).  

\subsection{Polynomial fit of the $W/W^*$ for the {\it Gaia} lines}

Due to the importance of the CaT and the \mg\ 8736~\AA\ for the {\it Gaia}/RVS mission, we provide a fit of the $W/W^*$ presented in the previous section as a multivariable polynomial for the five \mg\ lines, the three \caI\ lines and the CaT lines.  
The expression  of the fit is up to the third order in stellar parameters:

\begin{equation}
\left[\frac{W}{W^*}\right]_{{\rm fit}} \left(x,\ y,\ z\right) = 
\end{equation}
\[
\begin{array}{rlllllll}
 &   a_0         &+& a_1\ x      &+& a_2\ y       &+& a_3\ z       \\
+& a_4\ x^2      &+& a_5\ xy     &+& a_6\ y^2     &+& a_7\ xz      \\
+& a_8\ yz       &+& a_9\ z^2    &+& a_{10}\ x^3  &+& a_{11}\ x^2 y\\
+& a_{12}\ x y^2 &+& a_{13}\ y^3 &+& a_{14}\ x^2z &+& a_{15}\ xyz  \\
+& a_{16}\ y^2z  &+& a_{17}\ xz^2&+& a_{18}\ yz^2 &+& a_{19}\ z^3  \\
\end{array}
\]
with:  
\begin{equation}
\begin{array}{rcl}
x &=& (T_{{\rm eff}}-4375)/875\\
y &=& (\log g-1.25)/0.75\\
z &=&([{\rm Fe/H}]-[{\rm Fe/H}]_{{\rm c}})/(0.5\Delta[{\rm Fe/H}])\\
\end{array}
\end{equation}
the reduced and centred variables that lead to a variation in the $[-1, 1]$ range for our ranges of atmospheric parameters. 
$\Delta$[Fe/H] represents the metallicity range and [Fe/H]$_{\rm c}$ is the median metallicity, which depends on the line considered. 
As [Fe/H$]_{\max}=+0.50$ in all cases, the median metallicity and the metallicity range are expressed by:
\[
[{\rm Fe/H}]_{{\rm c}} = \frac{1}{2}[{\rm Fe/H}]_{\min} + 0.25
\]
\[
\Delta[{\rm Fe/H}] = [{\rm Fe/H}]_{\max} - [{\rm Fe/H}]_{\min}
\]
where [Fe/H$]_{\min}$ is specified in the Table~\ref{fit}.
As seen in Figs.~\ref{Gaia_lines1} and \ref{Gaia_lines2}, the NLTE/LTE EW ratios have different behaviours in stellar parameters for each line considered.
A second order formula \citep{Andretta05} or modified second order \citep{Starkenburg10} is not enough to fit the dependence, especially for [Fe/H] at low metallicity ($\leq -3$).
In order to find the coefficients $a_m$ ($m=0,\ 19$), we use the LSQ package of \citet{Miller92} and we rejected $W/W^*$ values for which $W< 1$~m\AA.
We present the results in Table~\ref{fit} for the \mg, \caI\ and \caII\ {\it Gaia}/RVS lines.
We noticed that some terms may be discarded without modifying the accuracy of the fit: the related coefficients are represented by blanks in Table~\ref{fit}.
For instance, we see that for \mg\ and \caI\ lines, the fits are insensitive to the $xy^2$ and $y^3$ terms; for the CaT, the fits are insensitive to the $xy$, $y^3$ and $xyz$ terms.

Examples of the fits are shown in Fig. \ref{caII} for the \caII\ IR 8662 \AA\ and in Figs.~\ref{Gaia_lines1} and \ref{Gaia_lines2} for the other {\it Gaia}/RVS lines. For the {\it Gaia}/RVS \mg\ and \caI\ lines we restrict the range of metallicities as indicated in Table~\ref{fit} since below the lower metallicities the line is too weak ($W \leq 1$ m\AA) to be considered. 
These fits can be used to estimate the NLTE/LTE EW ratios with an accuracy better than 10\%, in the range of the stellar parameters considered. 
The largest deviations of the fits appear for the most metal-poor model atmospheres for the \mg\ 8736~\AA\ and the CaT lines.
Hence, when requested accuracy is greater than $\epsilon_{{\rm mean}}$ (given in \% in Table~\ref{fit}), our fits can be used with confidence.
Otherwise, it is advised to directly use $W/W^*$ in the electronic tables.

\section{Conclusion}

We have performed NLTE computations for the \mg, \caI\ and \caII\ model atoms in late-type giant and supergiant stars.
We provide NLTE/LTE EW ratios for a grid of 453 \marcs\ model atmospheres ($3500\le T_{{\rm eff}}\le 5250$~K, $0.5\le \log g \le 2.0$~dex and $-4.00\le [$Fe/H$] \le+0.50$~dex) in electronic forms\footnote{available on the Vizier/CDS database}. 
The model atoms are based on the assumption that we do not take into account inelastic collisions with neutral hydrogen since realistic quantum mechanical collisional cross-sections are still unavailable.
We used a formulation for electronic collisions that underestimate the collisional rate. 
Such underestimation induced an upper limit on the NTLE/LTE EW ratios especially for the \mg\ lines in the {\it Gaia}/RVS wavelength range.
The use of fine structure in the model atoms do not affect strongly NLTE results since the $W/W^*$ of the components of a multiplet are very similar but permit a consistent representation of the physics. 
 This work will be extended to late-type main sequence stars when collisional cross-sections with hydrogen are available.

The main conclusions for the lines outside the {\it Gaia}/RVS wavelength range are as follows.
For the \mg\ and \caI\ lines, the assumption of LTE underestimates the Mg and the Ca abundances and gives mainly positive NLTE abundance correction.
For the \caII\ lines, the assumption of LTE overestimates the Ca abundance and gives mainly negative NLTE abundance corrections.
However, most of the \mg\ lines show $W/W^* > 1$ that can reach 2 for the coolest models with an increase of the sensitivity to the surface gravities.
The \mg\ b and \caI\ red triplets show $W/W^* < 1$ that can reach 0.5 for the most metal-poor and the hottest model atmospheres.
The NLTE effects for the \mg~4571~\AA\ and the \caI~6572~\AA\ intercombination and resonance lines are very sensitive to the metallicity and to the surface gravity.    
The NLTE effects on \caII\ H\&K lines are negligible except at [Fe/H$] = -4$.

For the {\it Gaia}/RVS lines, NLTE computations give the following trends.
The very weak 8473~\AA\ line is mainly formed in LTE whereas the \mg\ 8736~\AA\ and the \mg\ $\lambda\lambda$ 8710, 8712 and 8717 triplet lines are mainly formed in NLTE with a strong sensitivity to the surface gravity for the triplet.
The weak \caI\ triplet lines are mainly formed in NLTE but vanish as soon as [Fe/H$] < 1$. 
The famous CaT lines are mainly formed in LTE if [Fe/H$] \ge -2$.
The NLTE effects increase with a decrease of the metallicity and with an increase of the surface gravity.
We show that the $W/W^*$(CaT) can increase by 20\% for $\log g$ varying from 0.5 to 2.0.
For convenience, we provide a polynomial fit computed for the {\it Gaia}/RVS lines: the \mg\ $\lambda\lambda$ 8473, $\left<8715\right>$ triplet, 8736 multiplet, the \caI\ $\lambda\lambda$ 8525, 8583 and 8633, and the CaT lines. 
The polynomial can be extensively used by the automatic tools dedicated to the analysis and extract chemical abundances for millions of stars in the context of the large surveys as, for example, {\it Gaia} and RAVE.

\section*{Acknowledgments}
TM is granted by OCA and R\'egion PACA, and supported by Thal\`es Alenia Space.
This work was supported by the "Action Sp\'ecifique Gaia". 
Part of computations were performed with the "Mesocentre SIGAMM" machine, hosted by the Observatoire de la C\^ote d'Azur.
The authors acknowledge the role of the SAM collaboration (http://www.anst.uu.se/ulhei450/GaiaSAM) in stimulating this research
through regular workshops. 
TM also thanks J.~Tully and M.~Carlsson for helpful discussions.

\appendix

\section{$W/W^*$ for the lines outside  of the {\it Gaia}/RVS wavelength range}
We present the results of the NLTE computations as the evolution of the NLTE/LTE EW ratios $W/W^*$ as a function of the metallicity [Fe/H]. 
Each row represents a selected line.
Each panel of a row represents the $W/W^*$ for a given effective temperature (3500, 4250 and 5250~K).
Each colour of the $W/W^*$ in a panel represents a surface gravity (red, green, blue and cyan for $\log g = 0.5$, 1.0, 1.5 and 2.0~dex respectively).
The variations of $W/W^*$ are not plotted when $W<1$~m\AA.
We notice that \marcs\ model atmospheres are missing for $\log g = 0.5$, $T_{{\rm eff}} \in [3500, 4500]$, and [Fe/H$]=-4$; and for $\log g = 0.5$, $T_{{\rm eff}} \in [3500, 3900]$,  and [Fe/H$]=-3.00$. 
The dotted line stands for no deviation from LTE.
The dashed lines stand for a variation of $\pm 5\%$ of the $W/W^*$ (corresponding to a variation of $\pm 0.04$~dex of the $\Delta$[El/H]) in case of strong regime of the curve of growth theory. 
The solid lines stand for a variation of $\pm 10\%$ of the $W/W^*$ (corresponding to a variation of $\pm 0.04$~dex of the $\Delta$[El/H]) in case of weak regime of the curve of growth theory.
 
\begin{figure*}
\hbox{
\includegraphics[width=5.45cm]{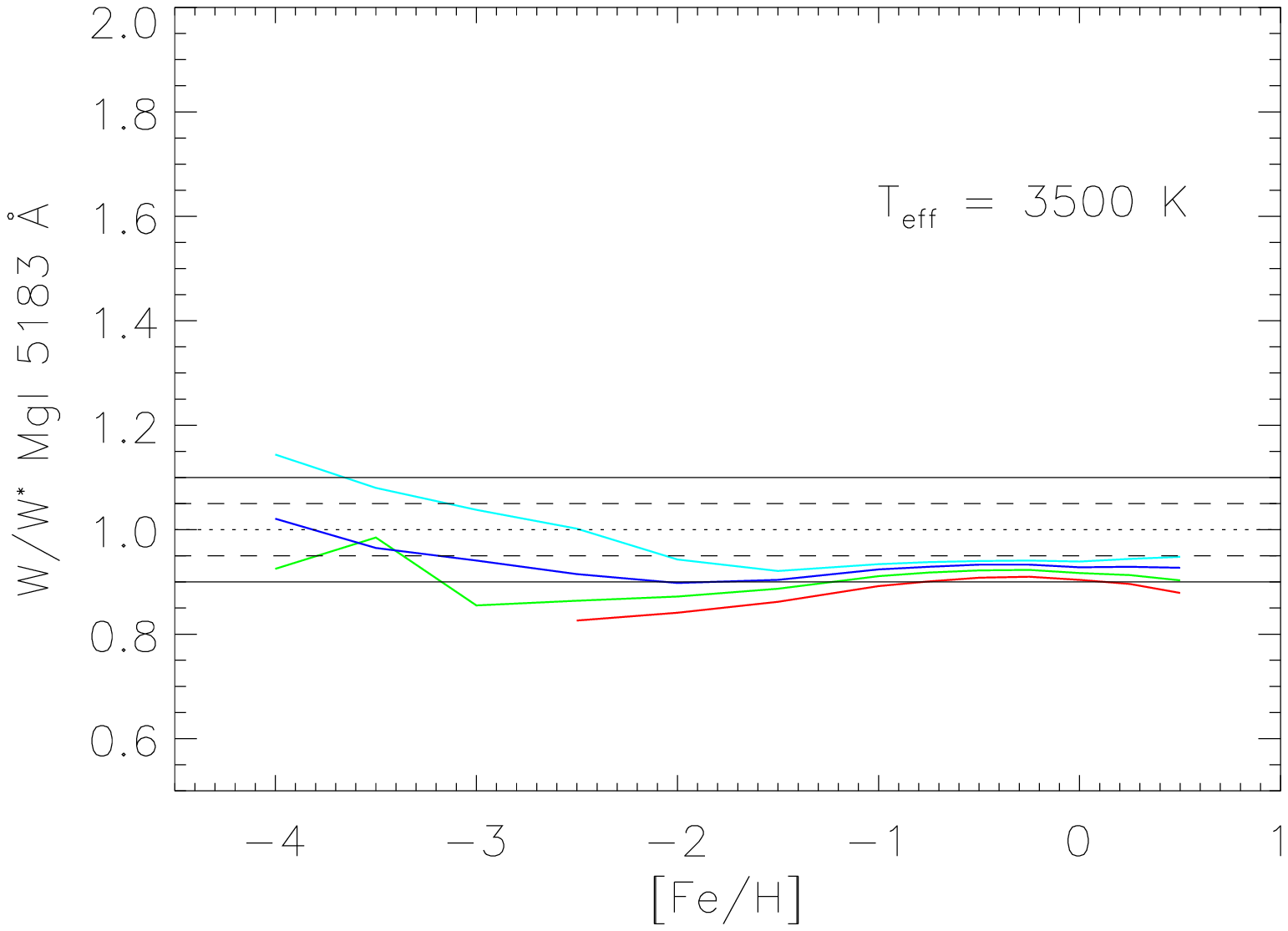}
\includegraphics[width=5.45cm]{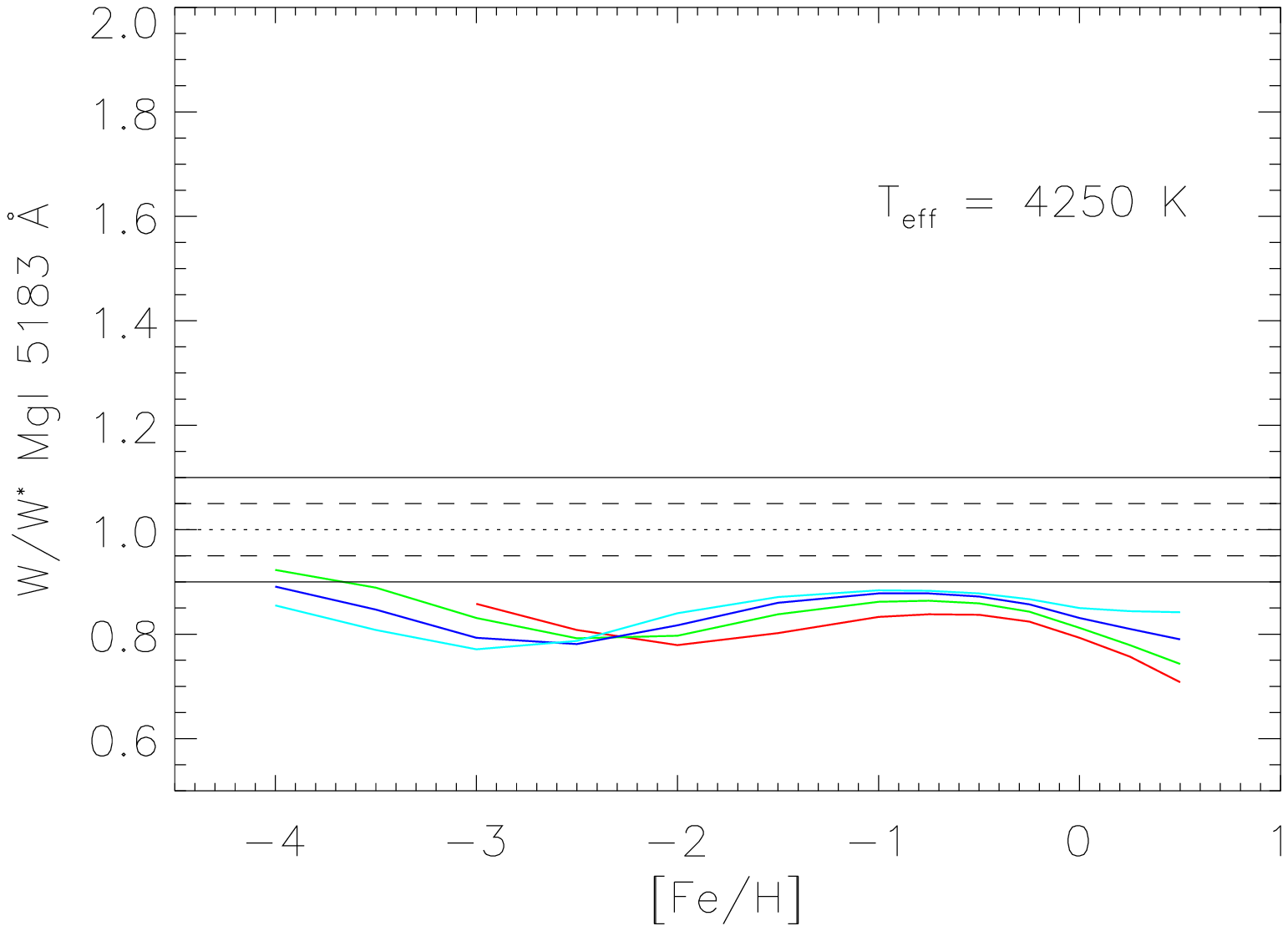}
\includegraphics[width=5.45cm]{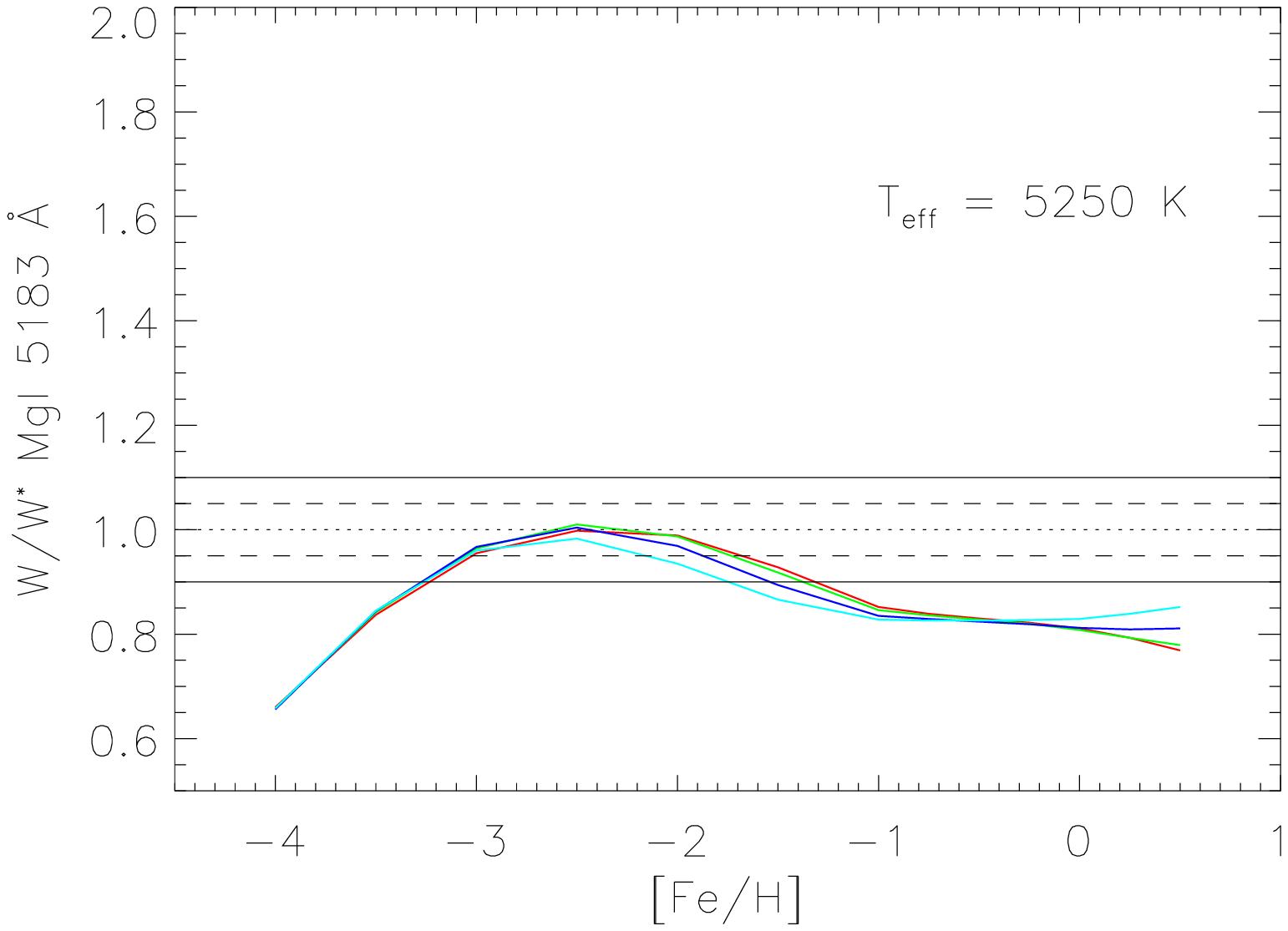}
}
\hbox{
\includegraphics[width=5.45cm]{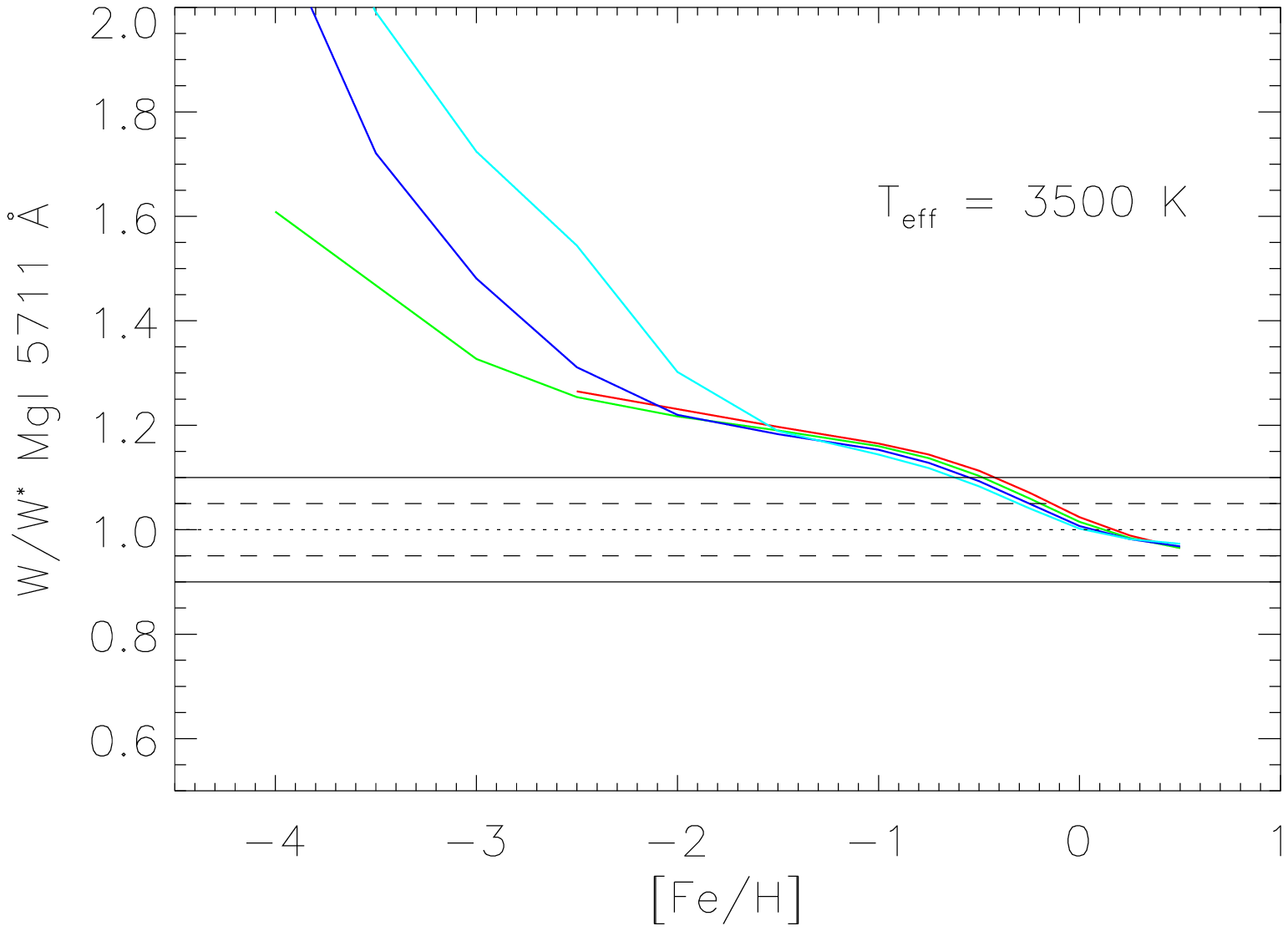}
\includegraphics[width=5.45cm]{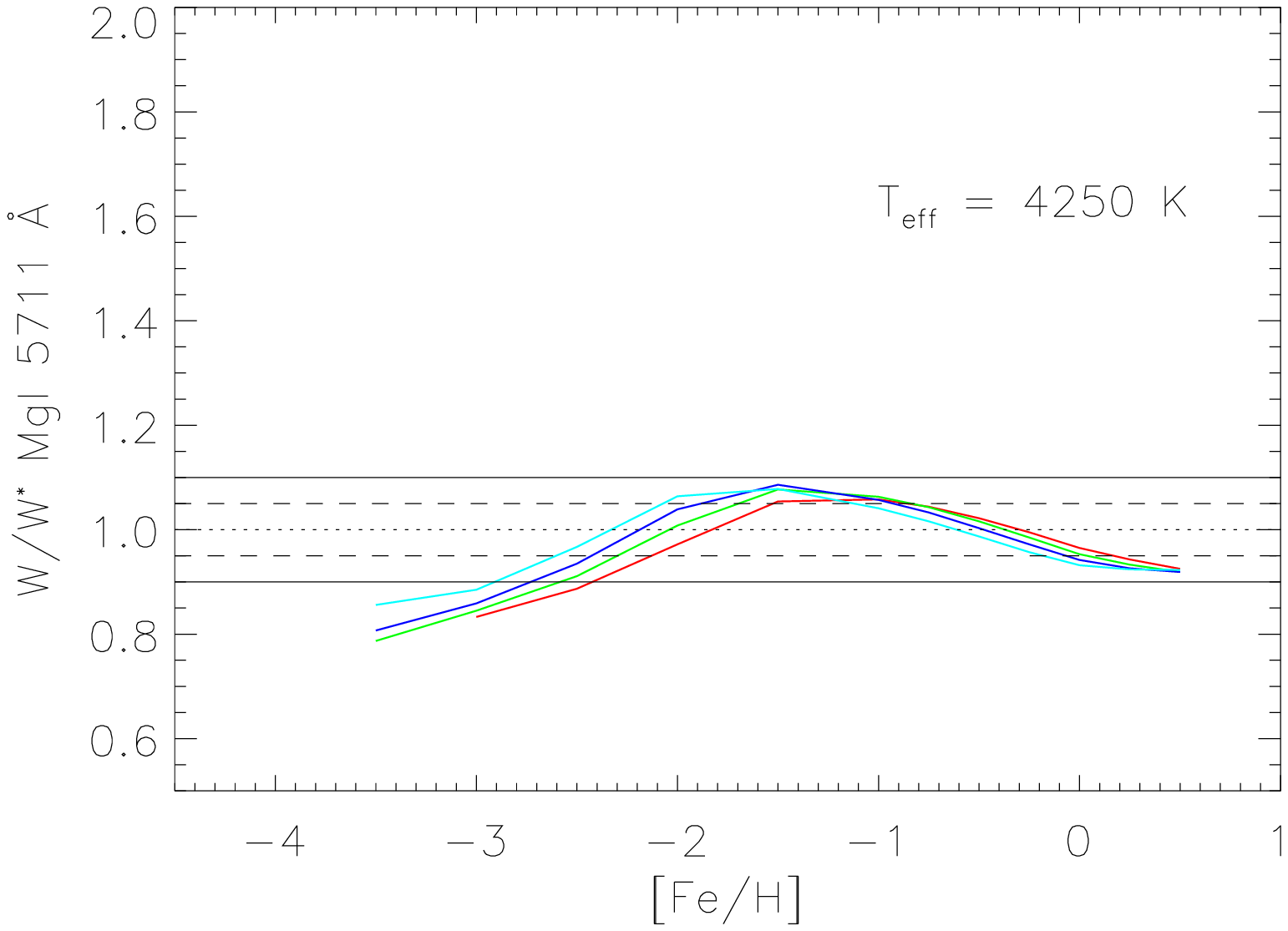}
\includegraphics[width=5.45cm]{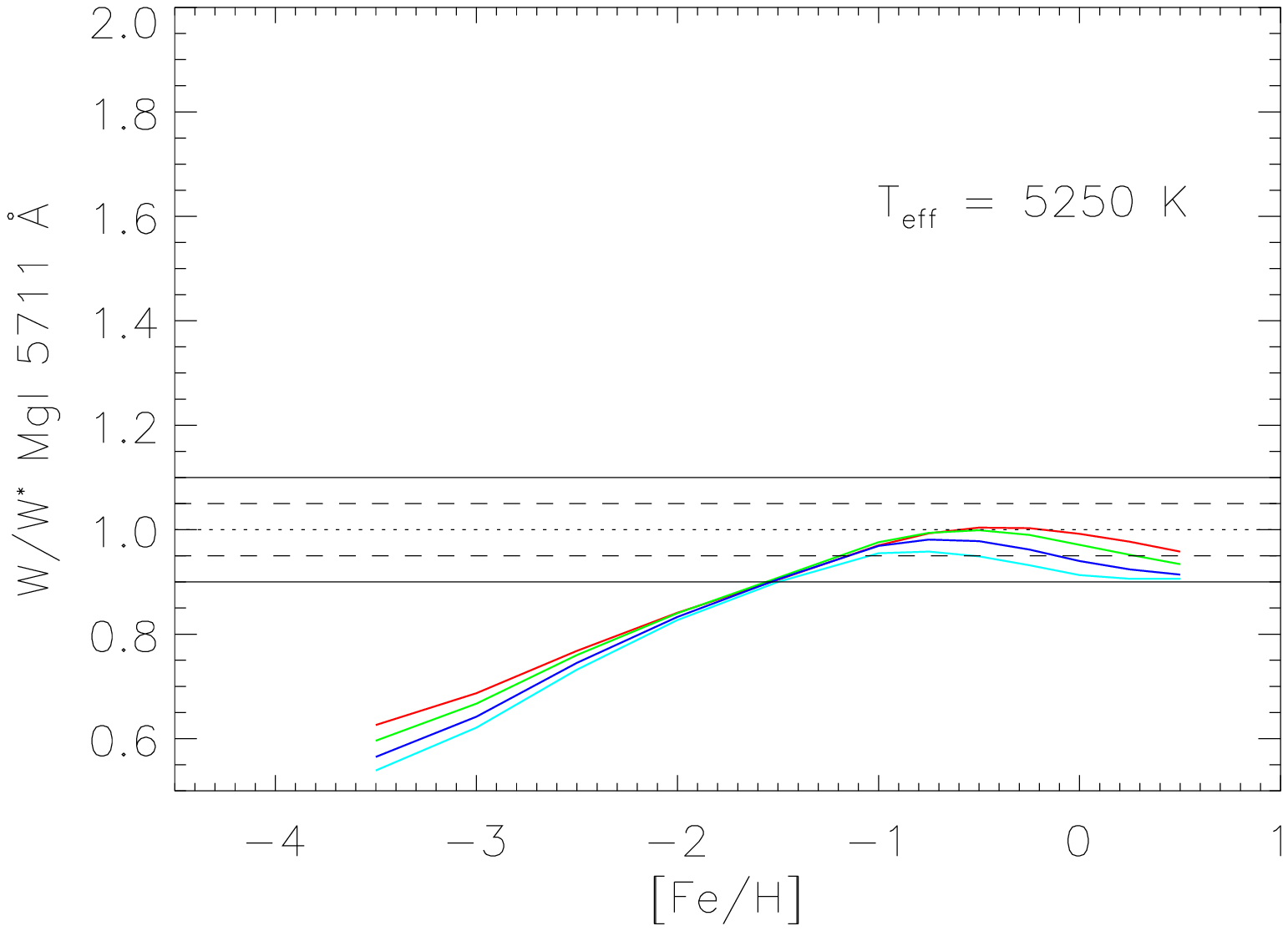}
}
\hbox{
\includegraphics[width=5.45cm]{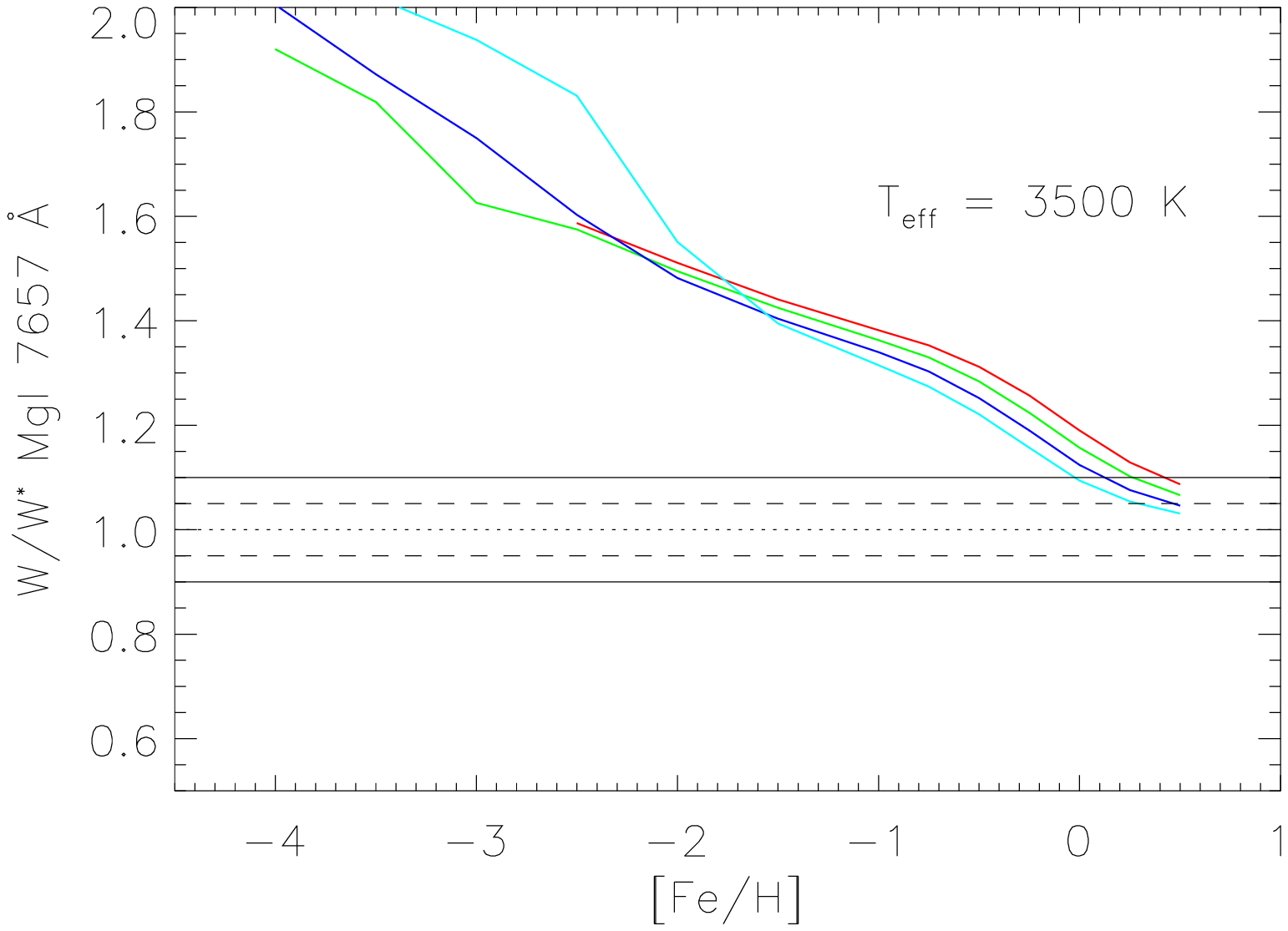}
\includegraphics[width=5.45cm]{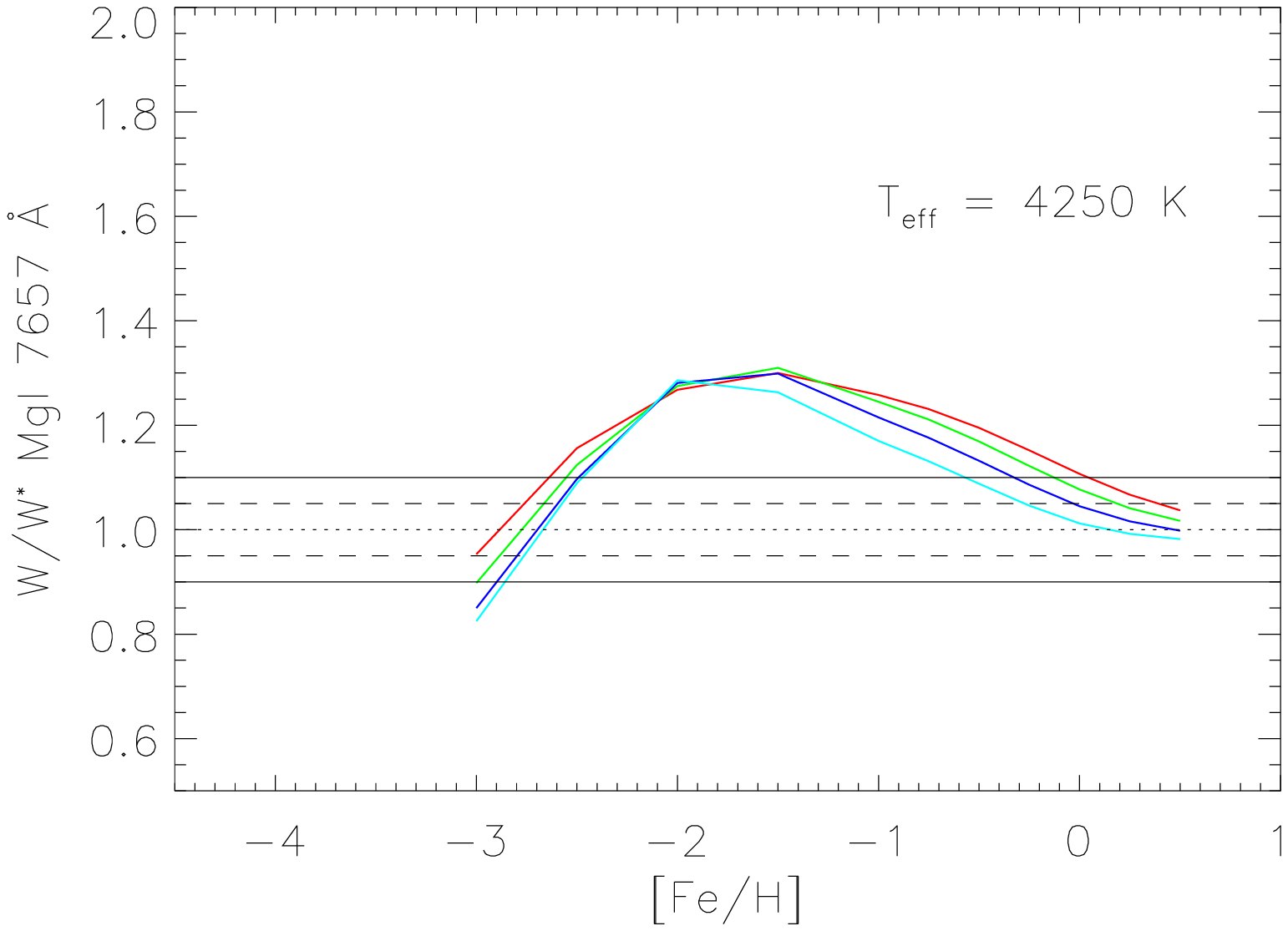}
\includegraphics[width=5.45cm]{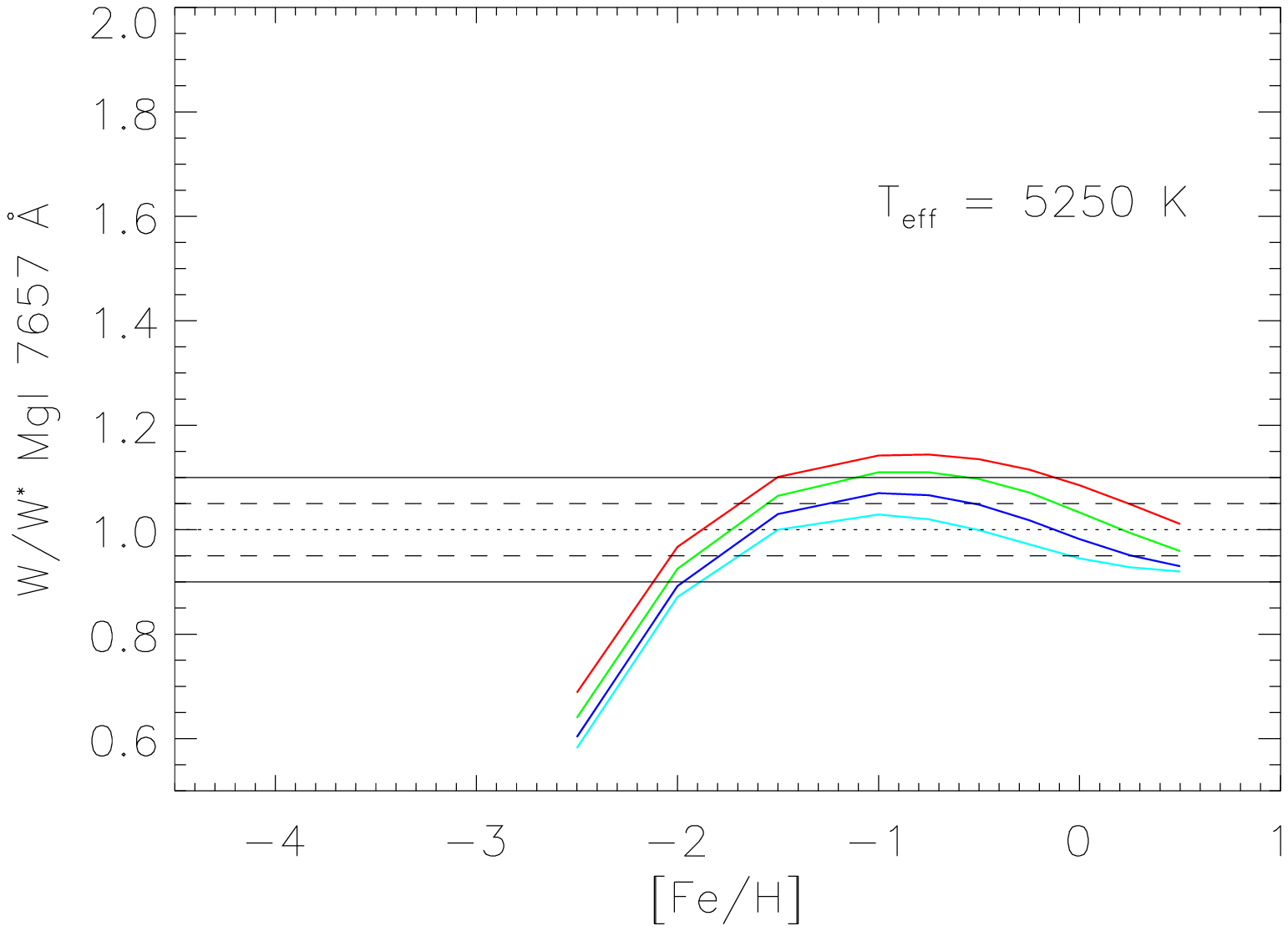}
}
\hbox{
\includegraphics[width=5.45cm]{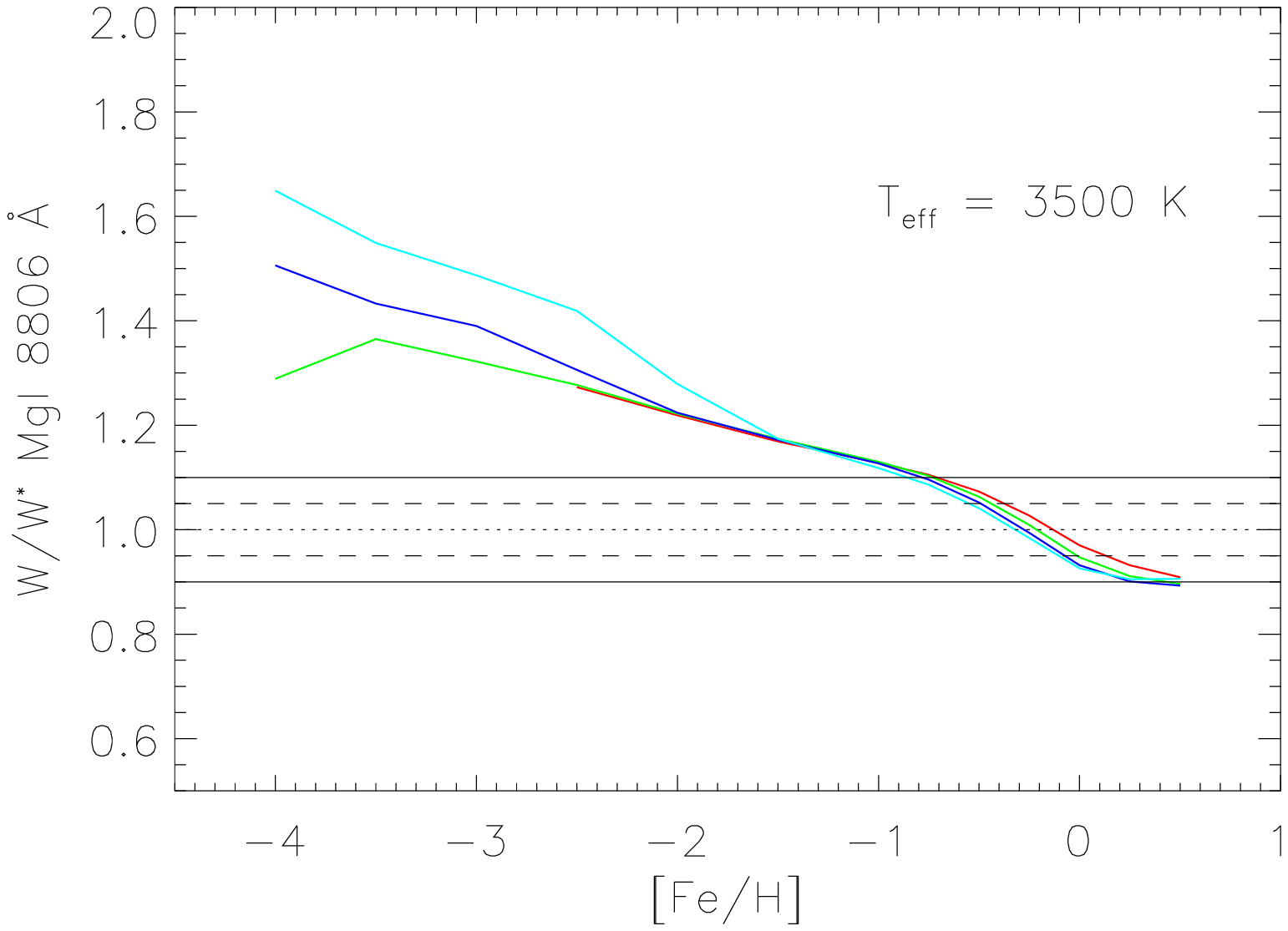}
\includegraphics[width=5.45cm]{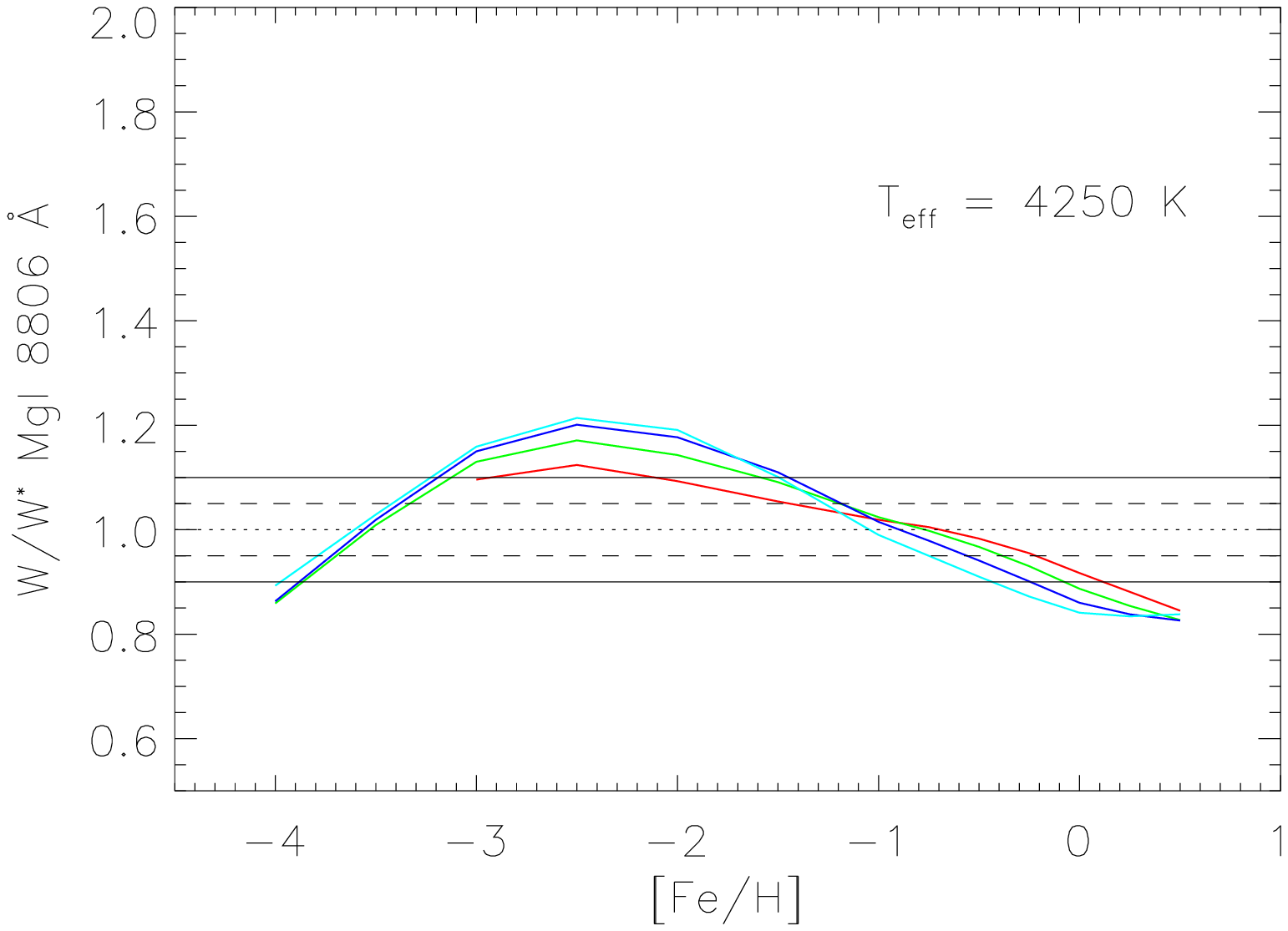}
\includegraphics[width=5.45cm]{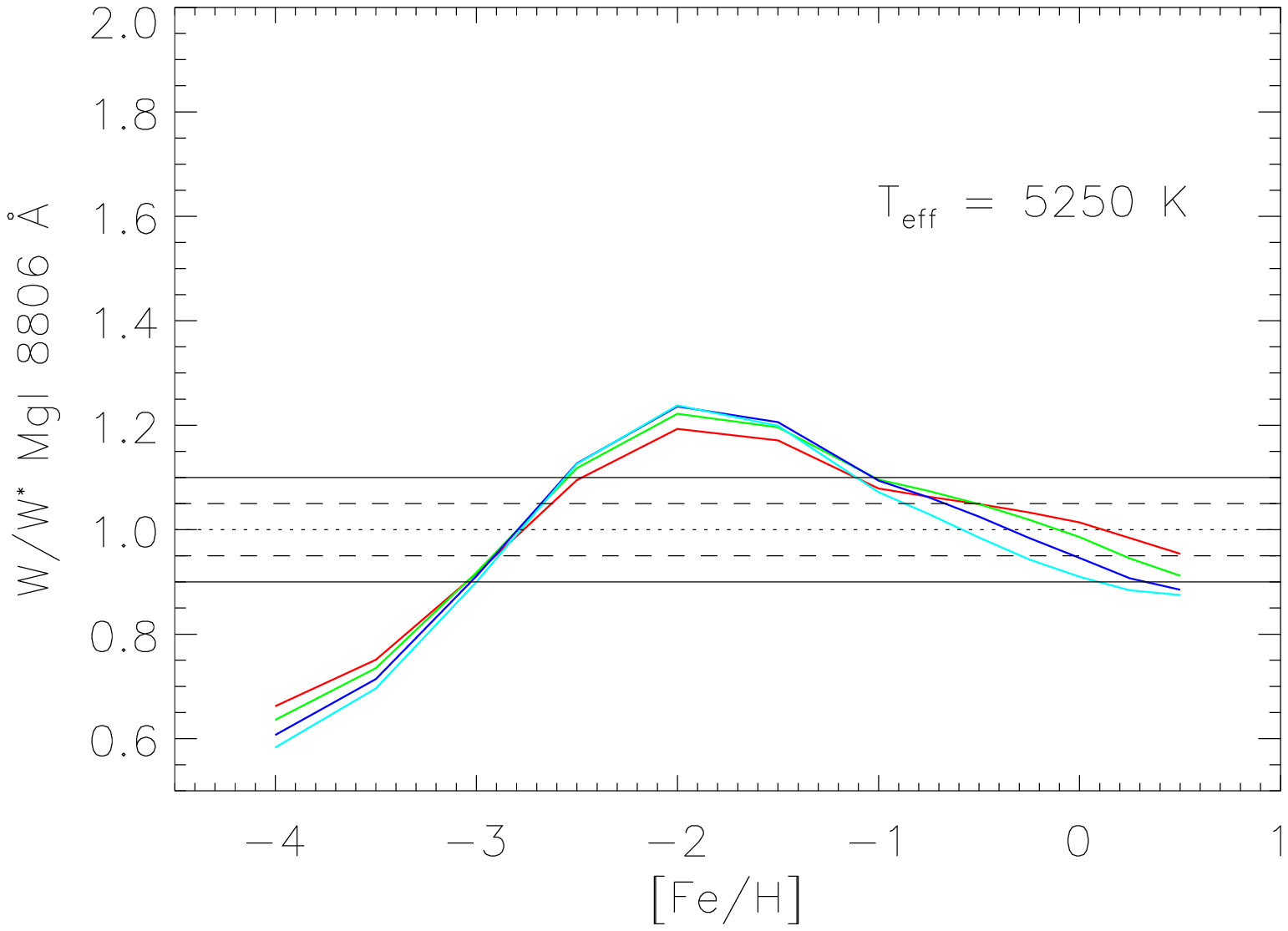}
}
\hbox{
\includegraphics[width=5.45cm]{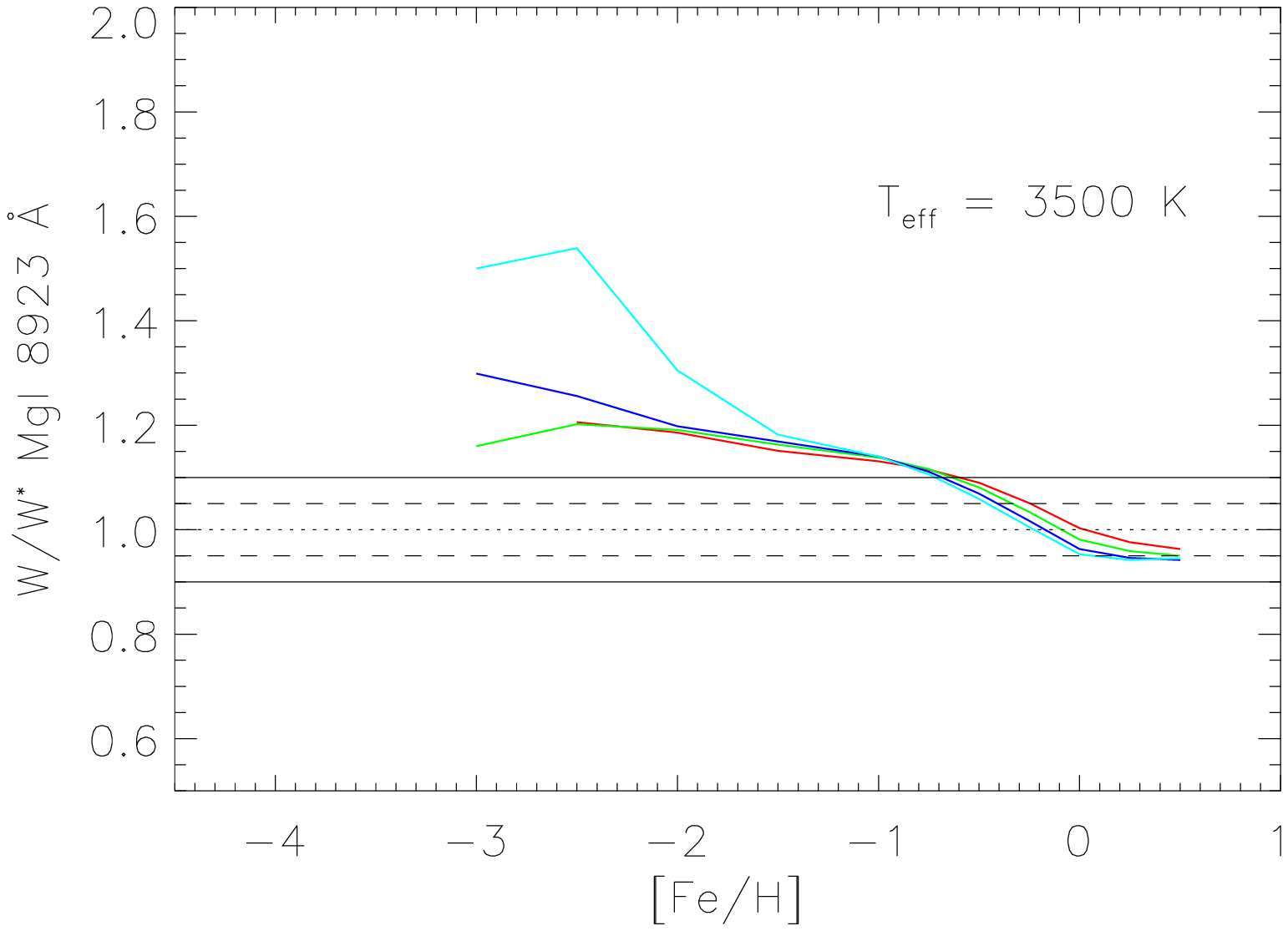}
\includegraphics[width=5.45cm]{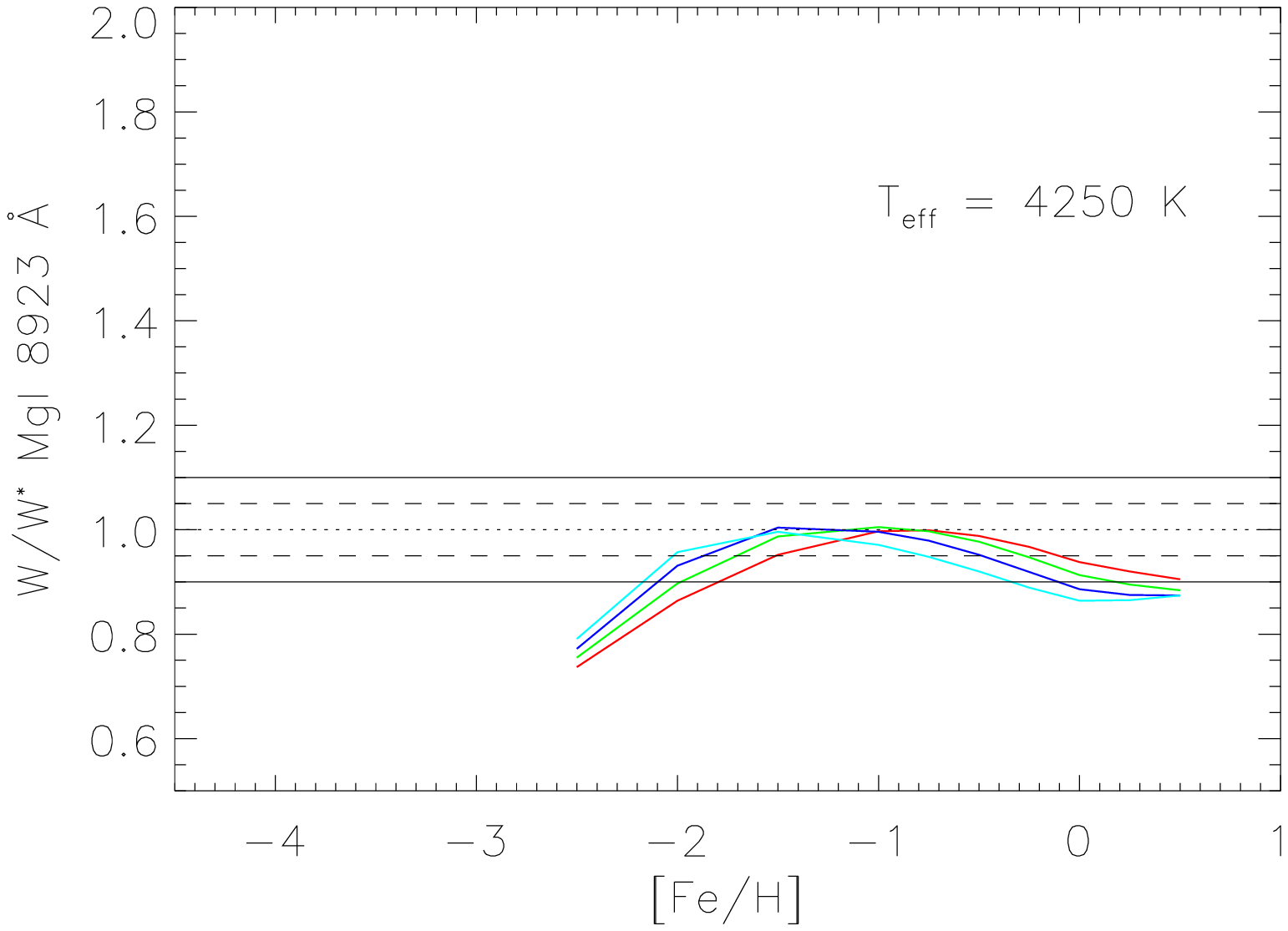}
\includegraphics[width=5.45cm]{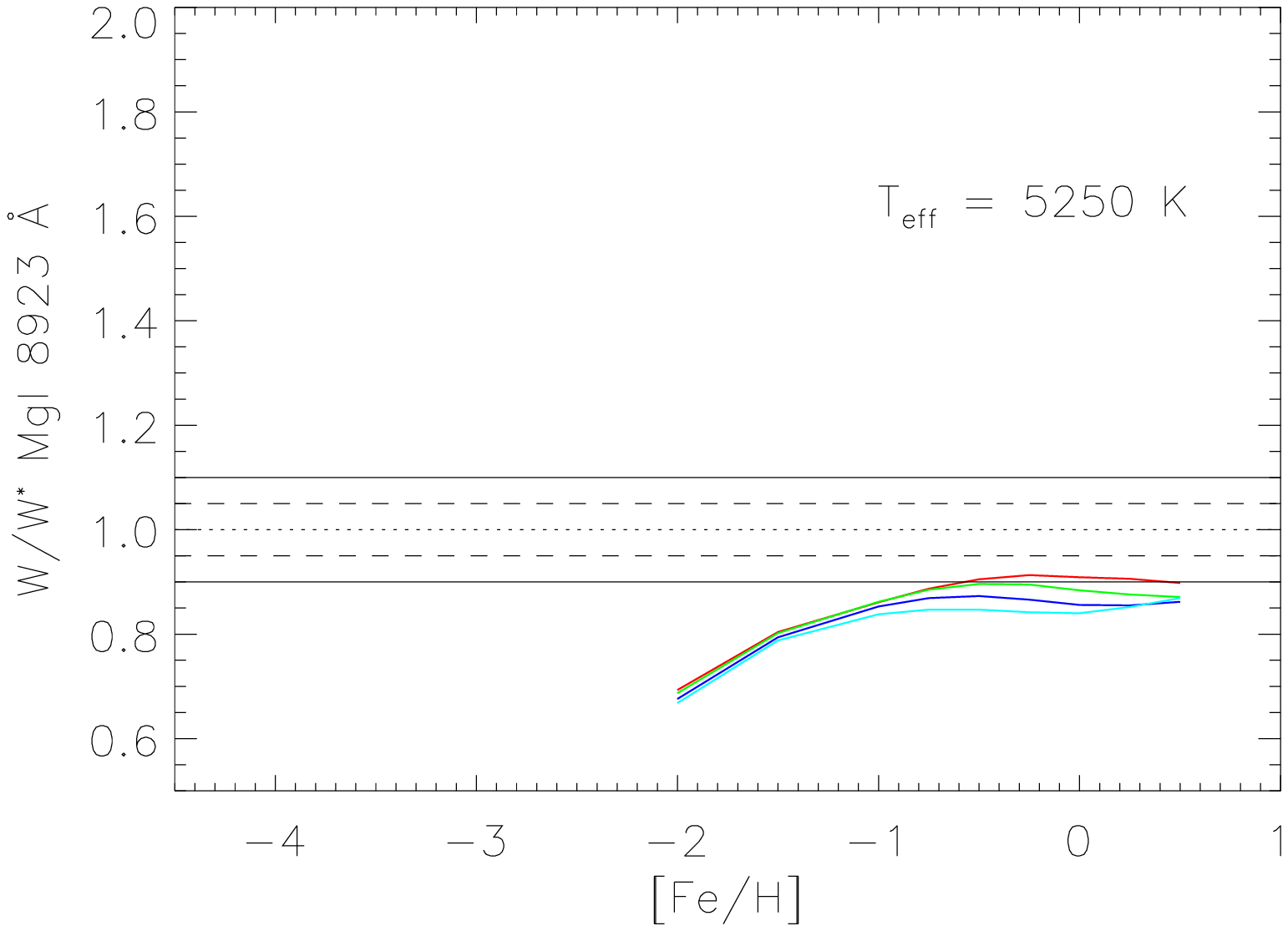}
}
\hbox{
\includegraphics[width=5.45cm]{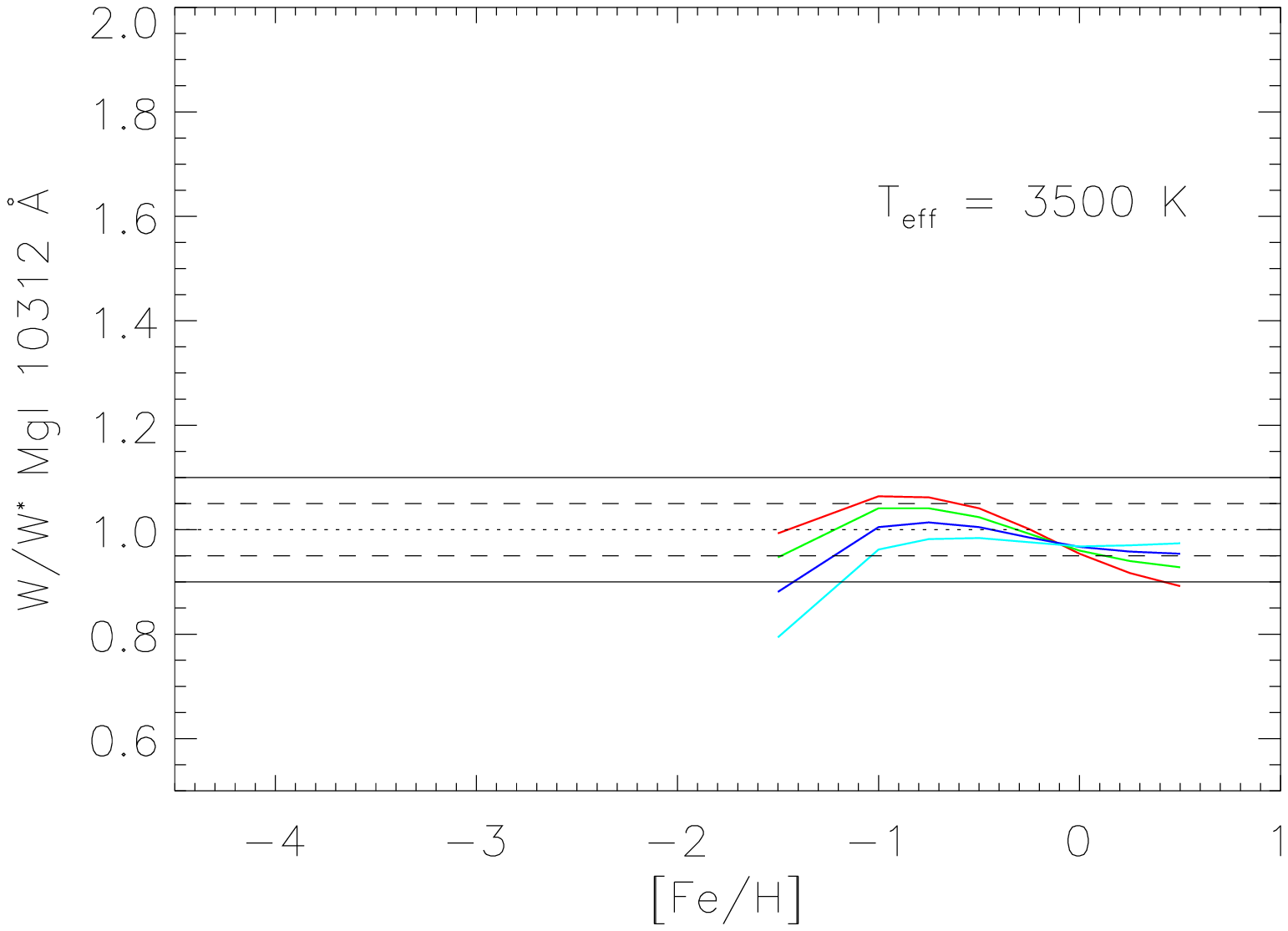}
\includegraphics[width=5.45cm]{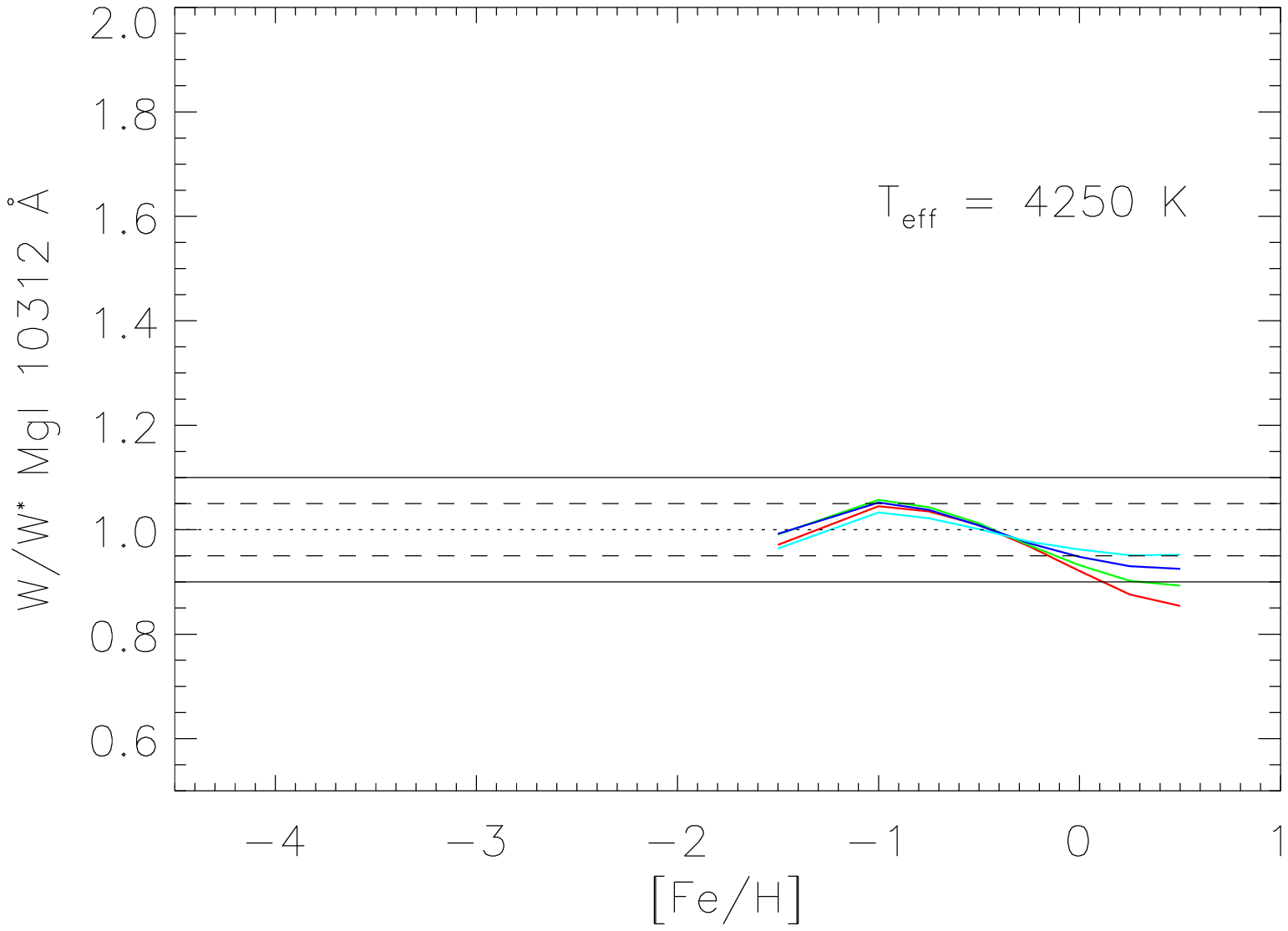}
\includegraphics[width=5.45cm]{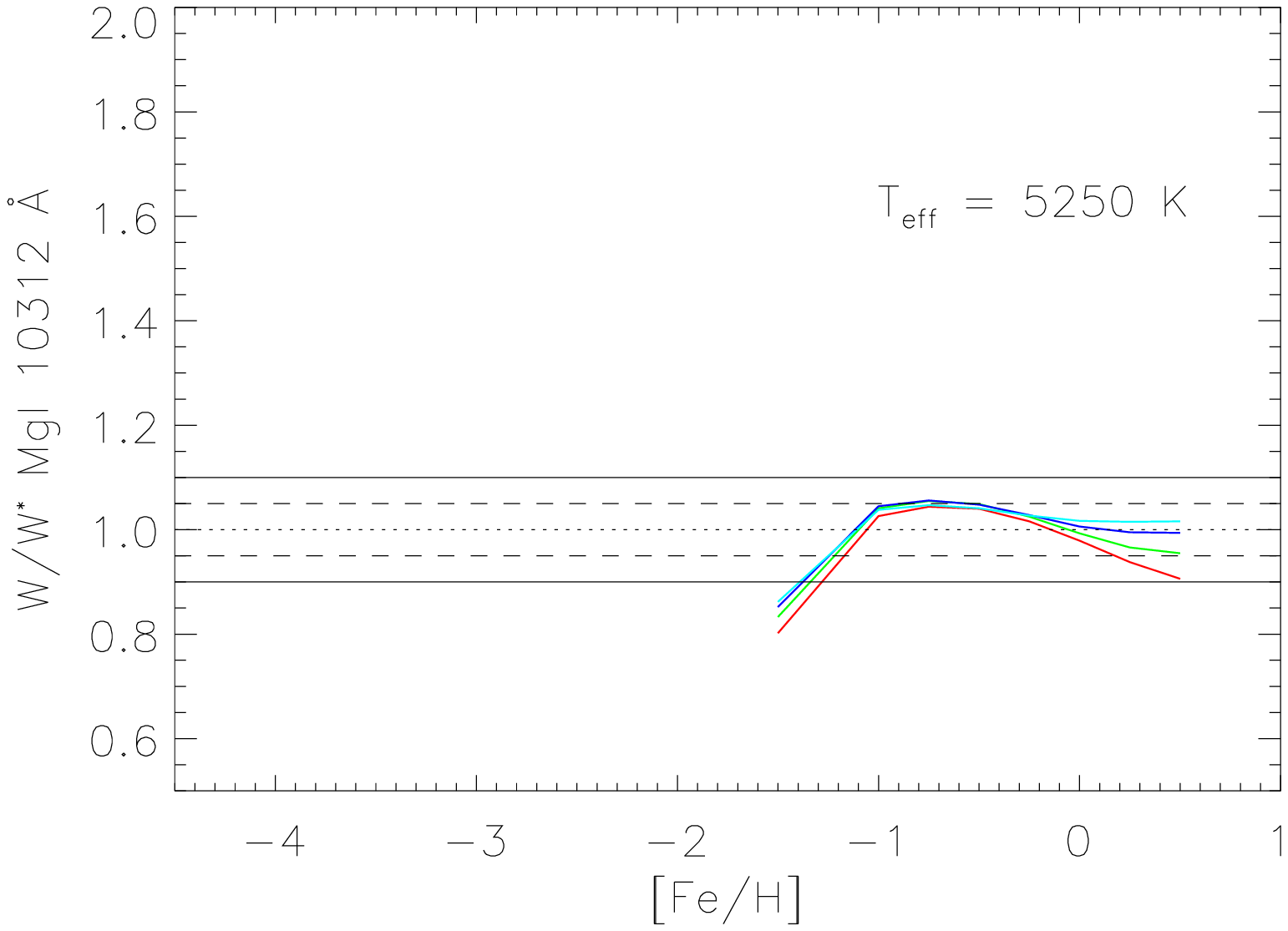}
}
\caption{$W/W^*$ for the selected \mg\ lines as function of the stellar parameters (see Appendix A for details).}
\label{MgI_lines}
\end{figure*}

\begin{figure*}
\hbox{
\includegraphics[width=5.45cm]{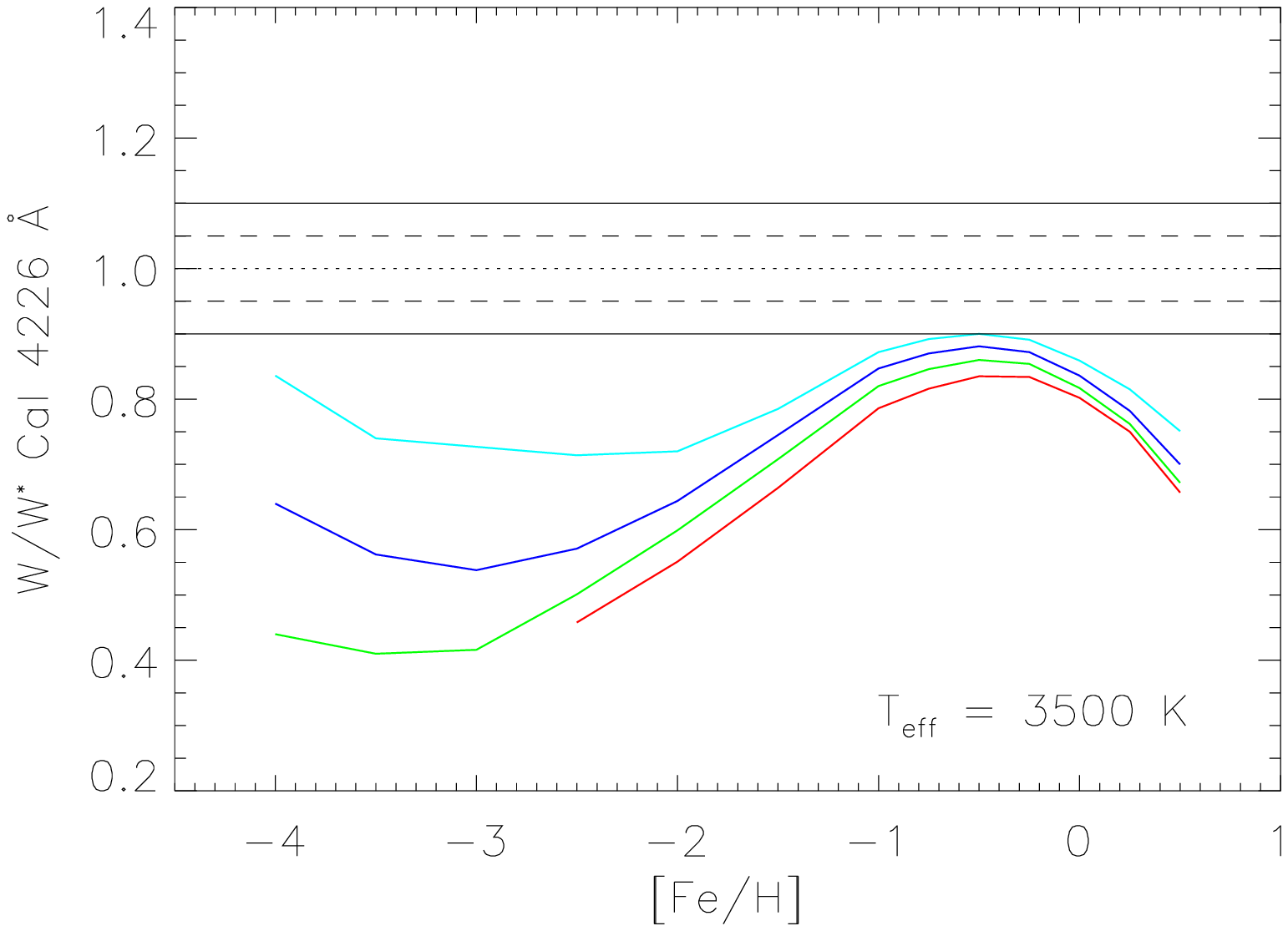}
\includegraphics[width=5.45cm]{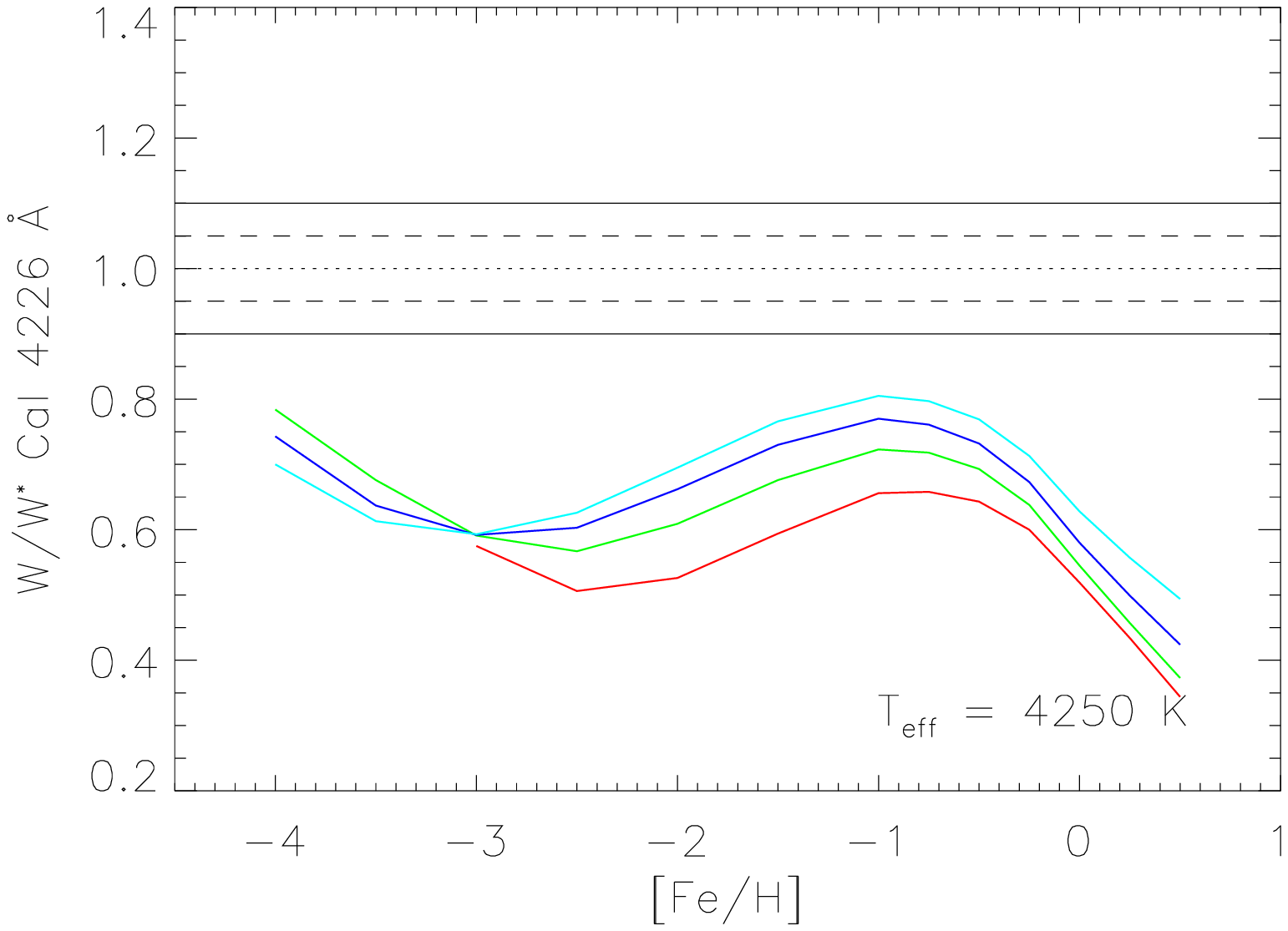}
\includegraphics[width=5.45cm]{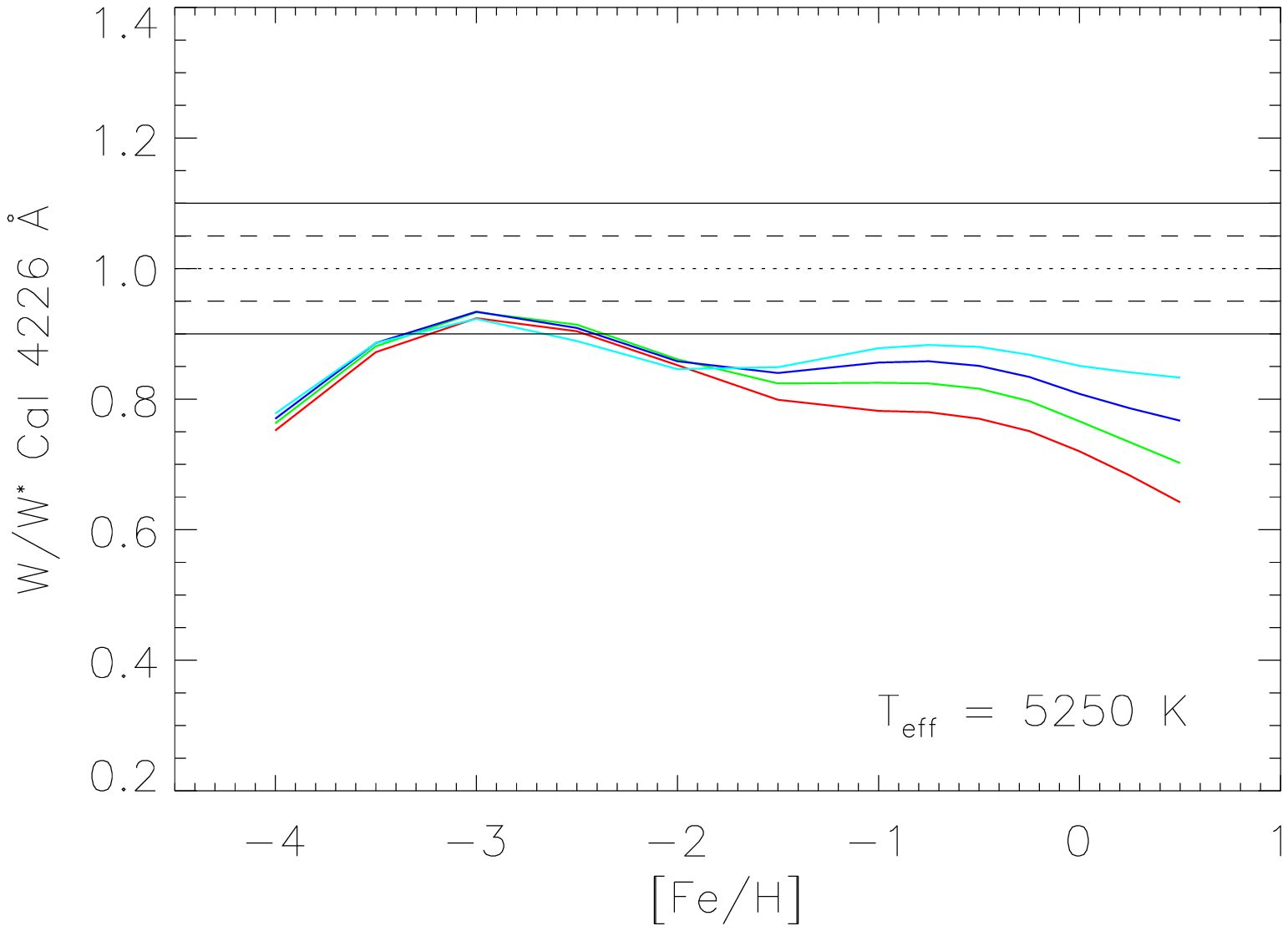}
}\hbox{
\includegraphics[width=5.45cm]{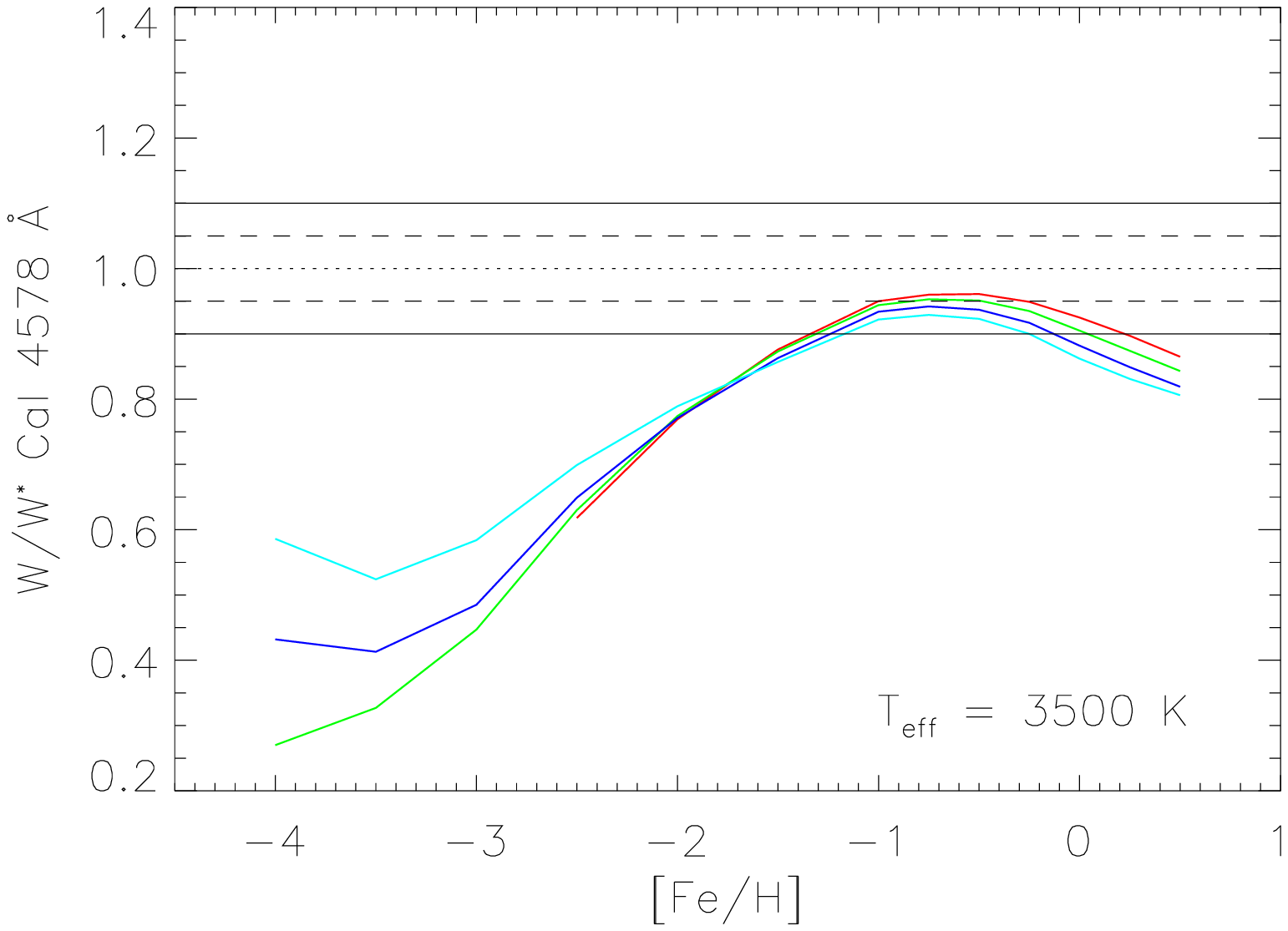}
\includegraphics[width=5.45cm]{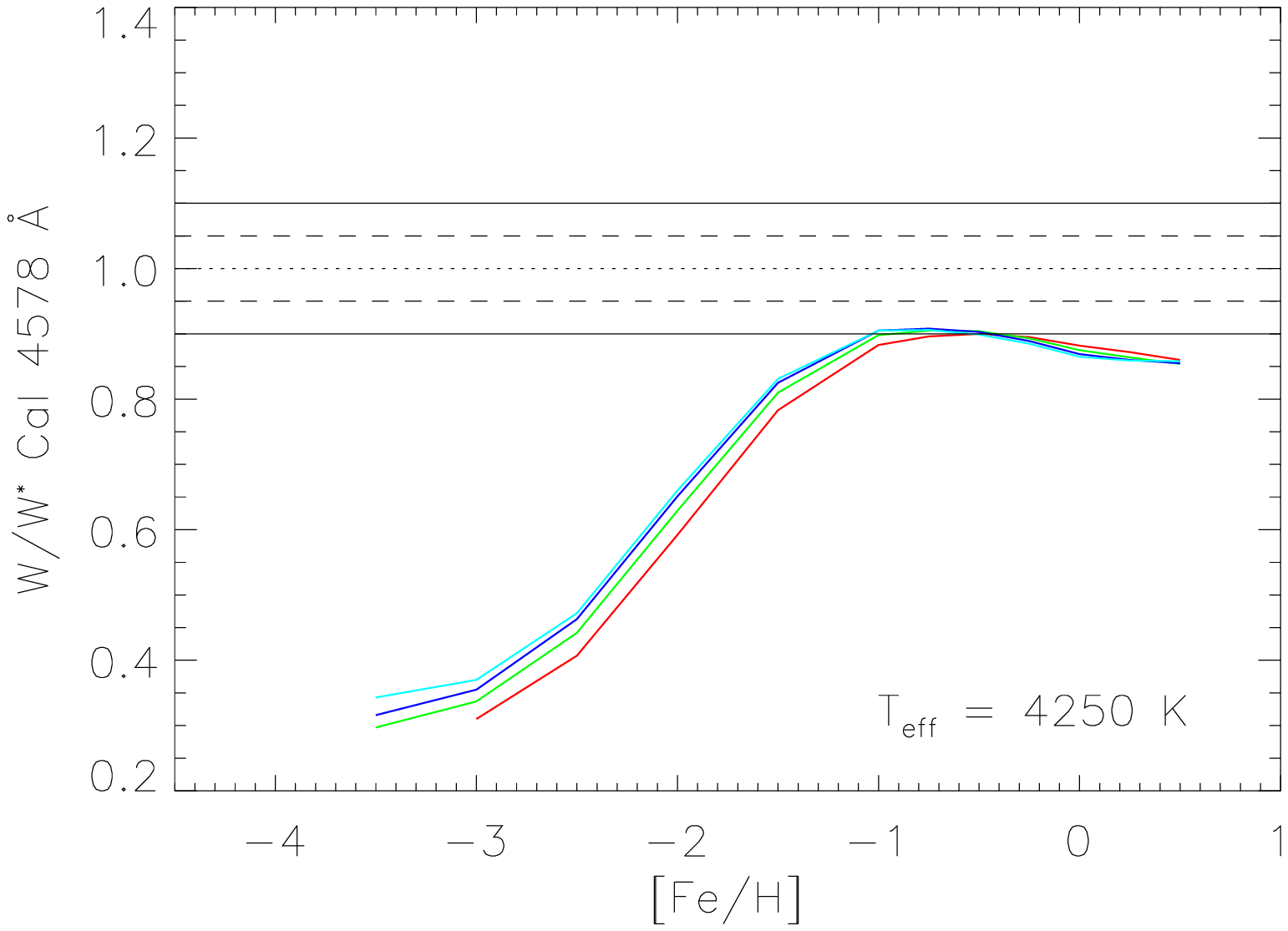}
\includegraphics[width=5.45cm]{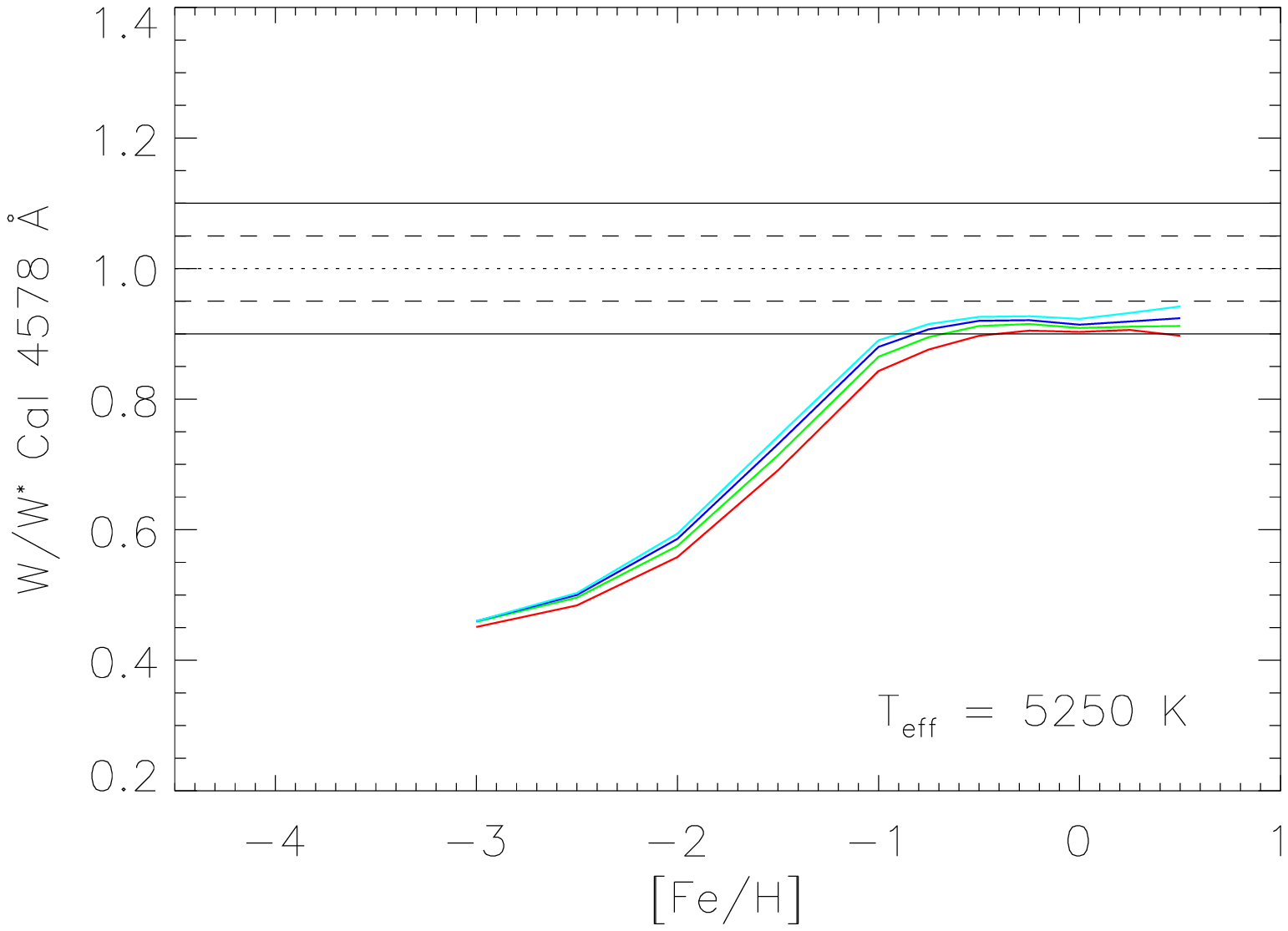}
}
\hbox{
\includegraphics[width=5.45cm]{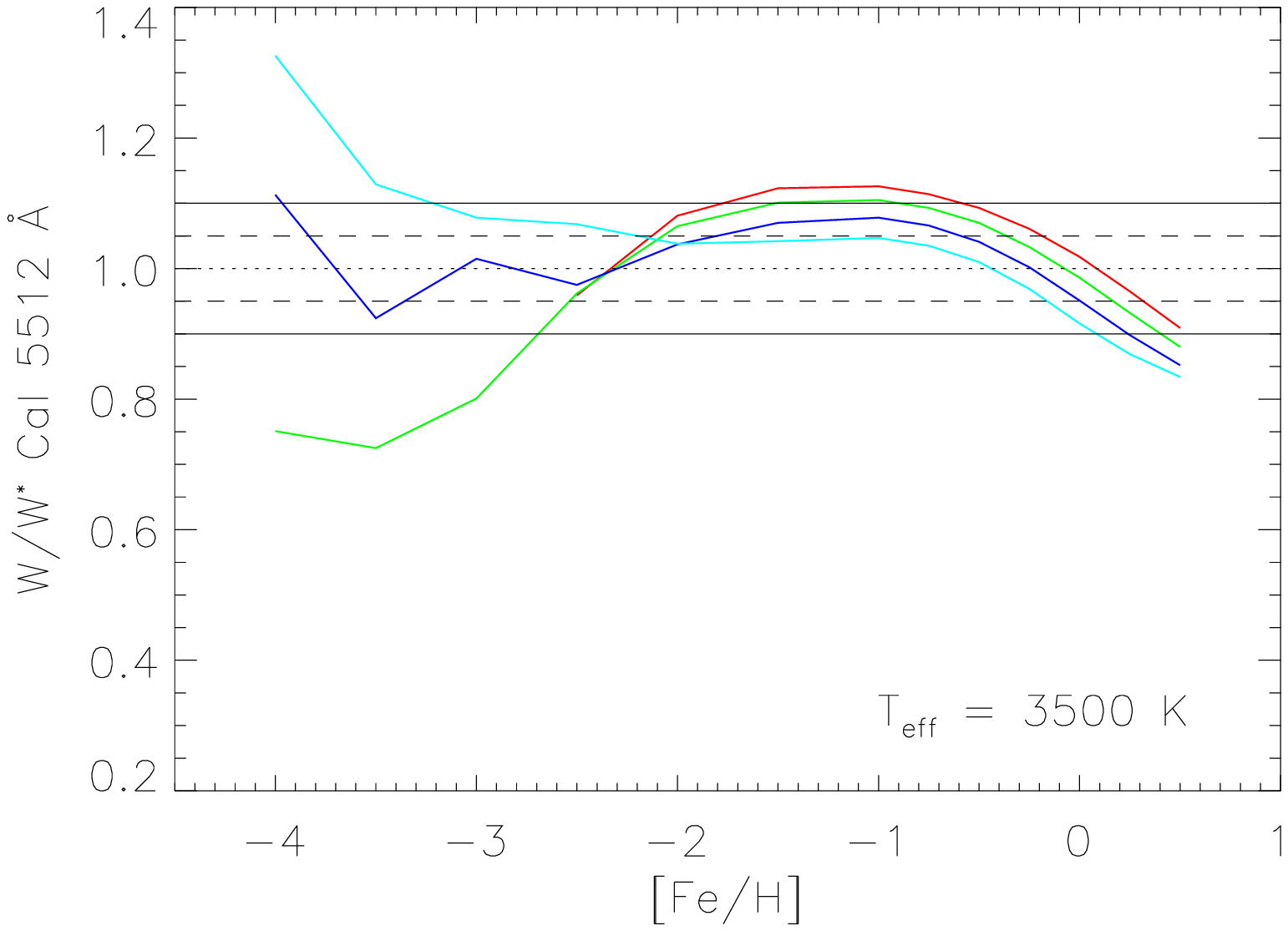}
\includegraphics[width=5.45cm]{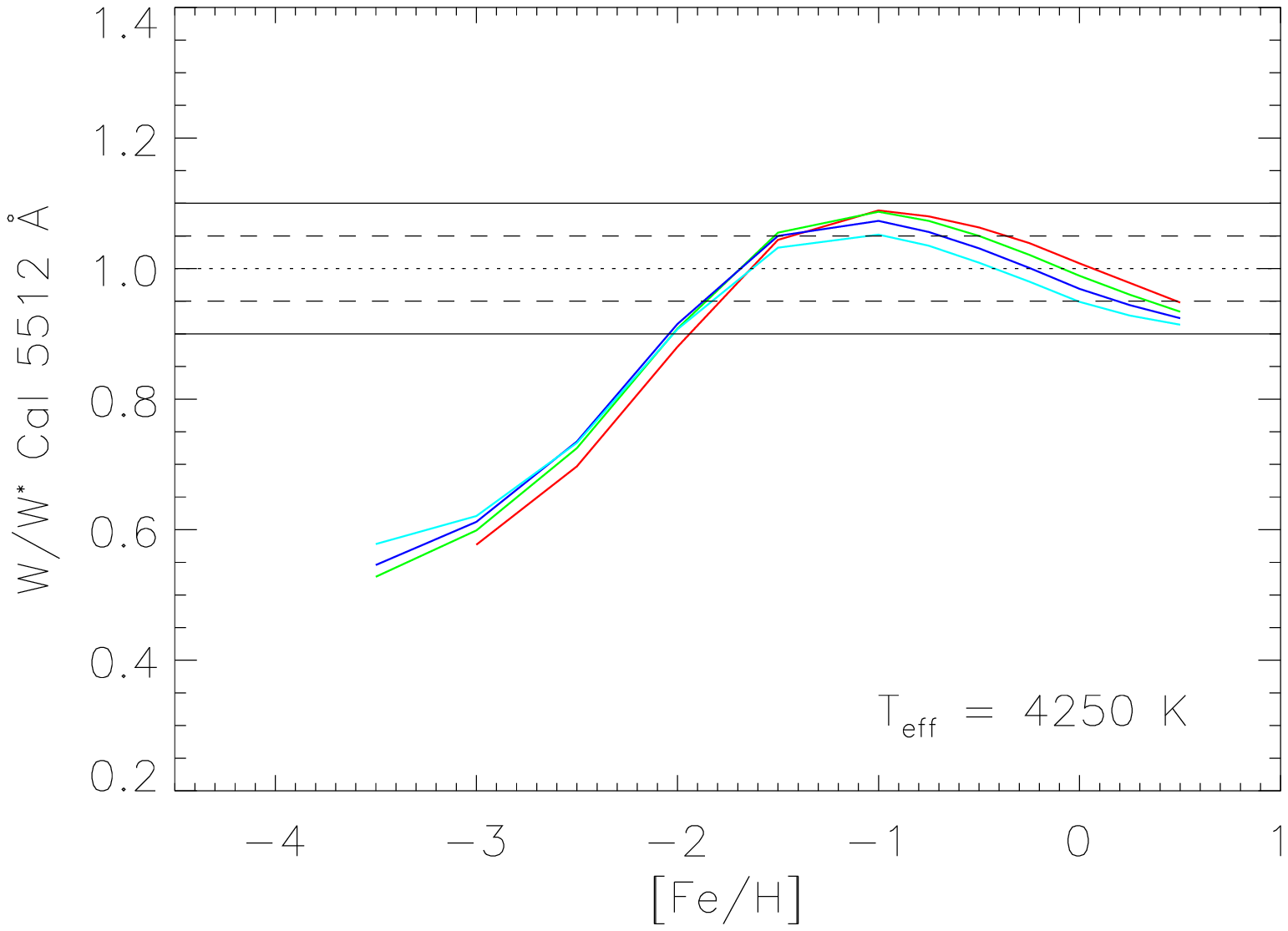}
\includegraphics[width=5.45cm]{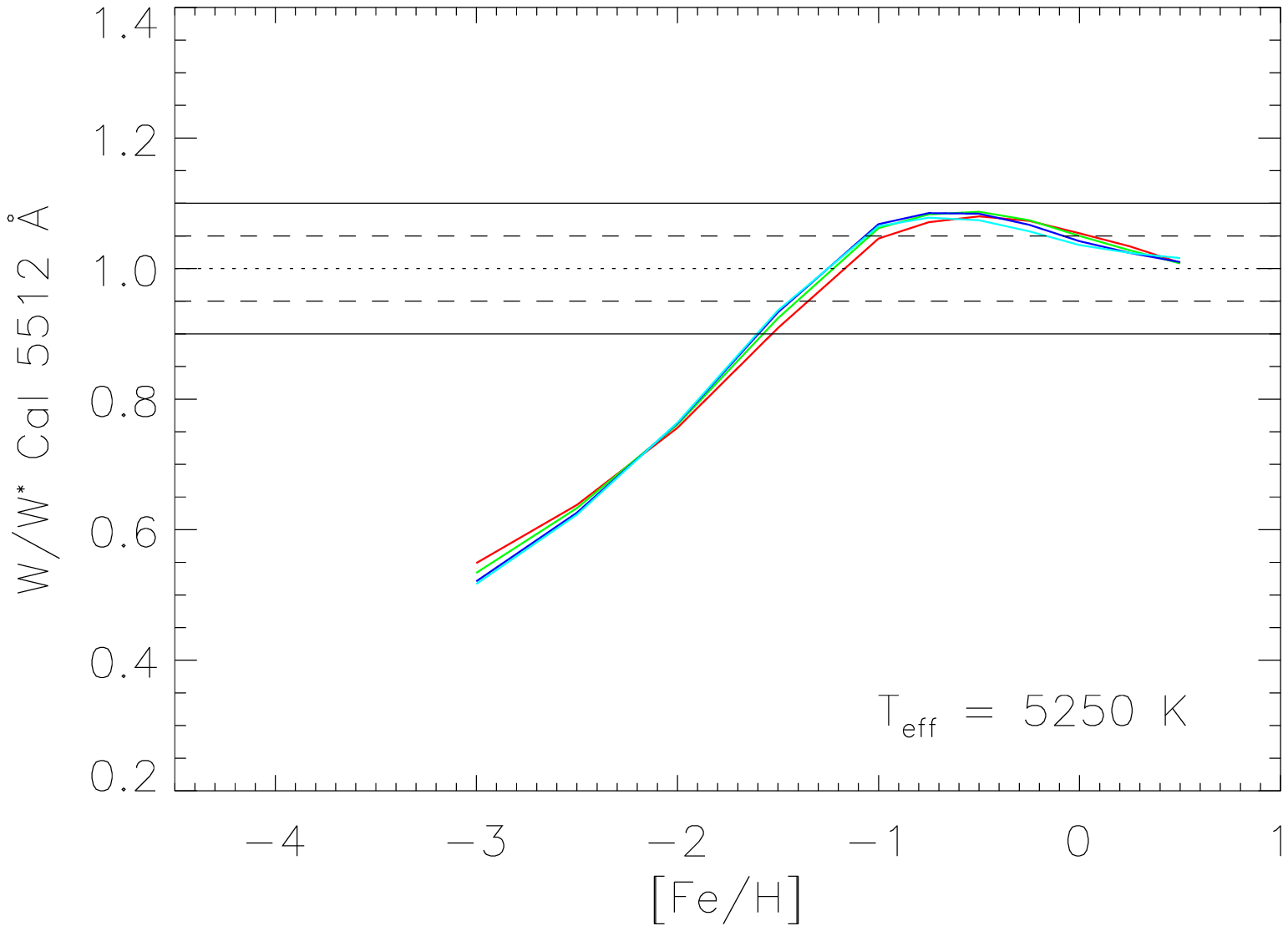}
}
\hbox{
\includegraphics[width=5.45cm]{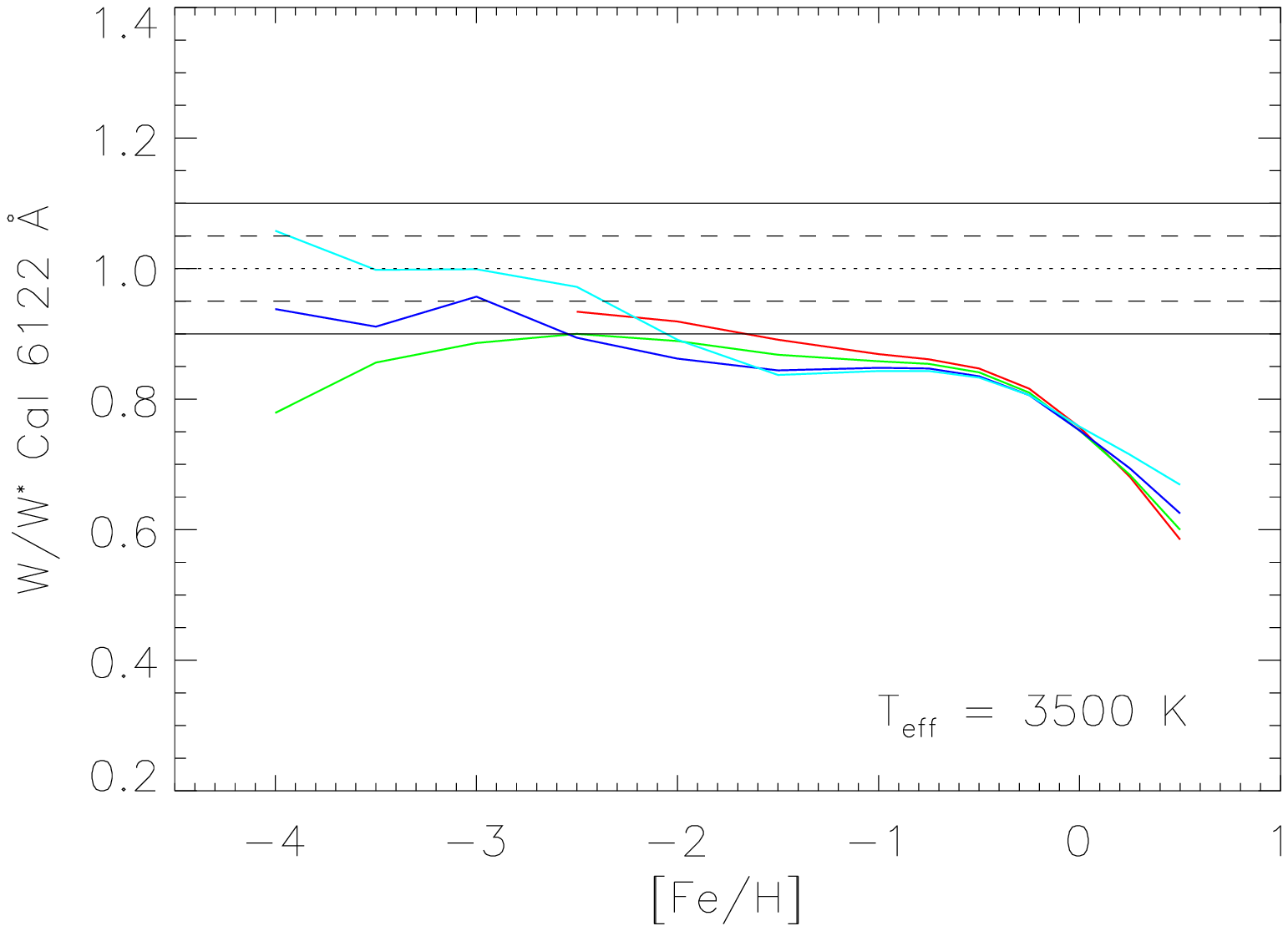}
\includegraphics[width=5.45cm]{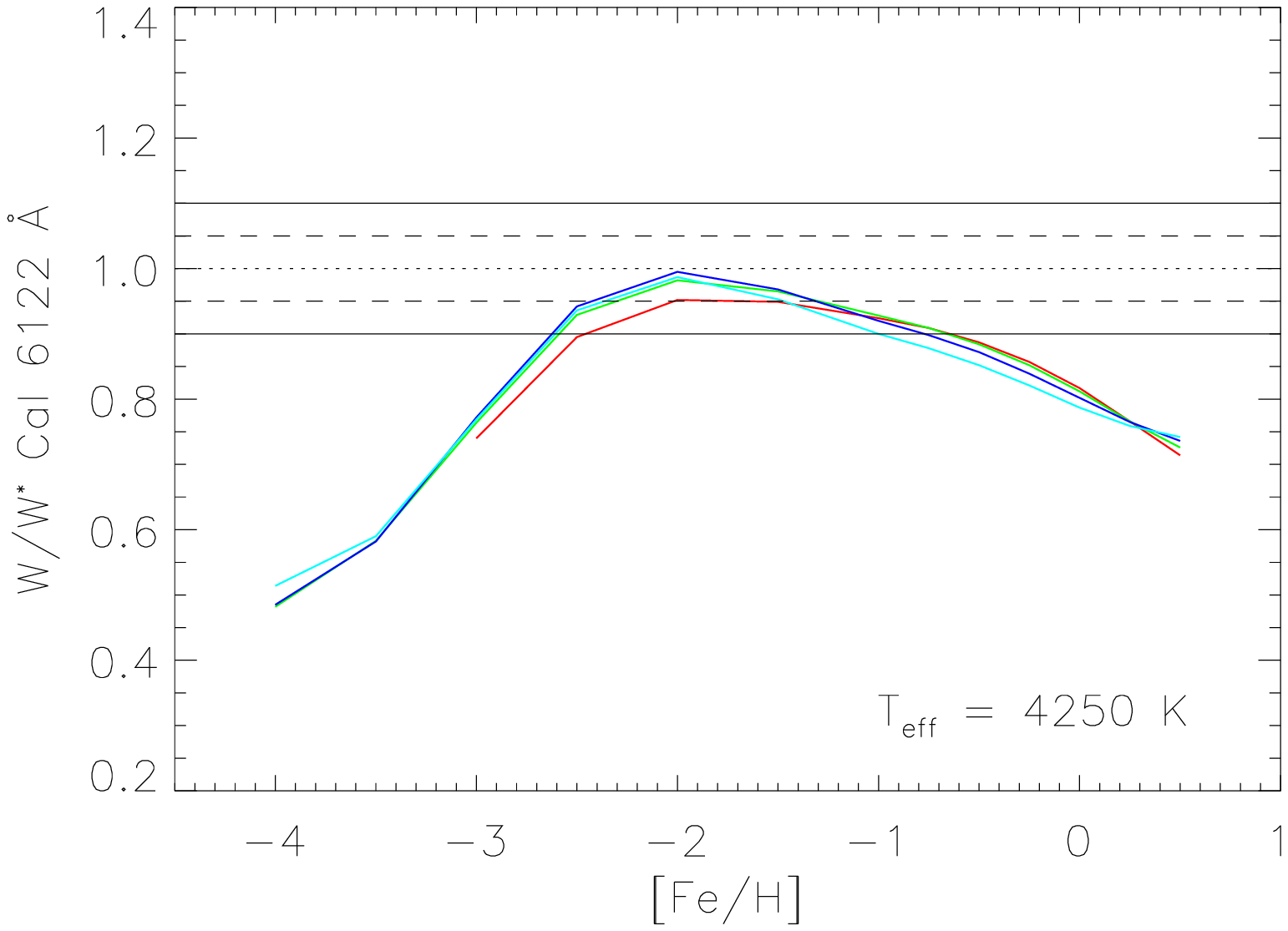}
\includegraphics[width=5.45cm]{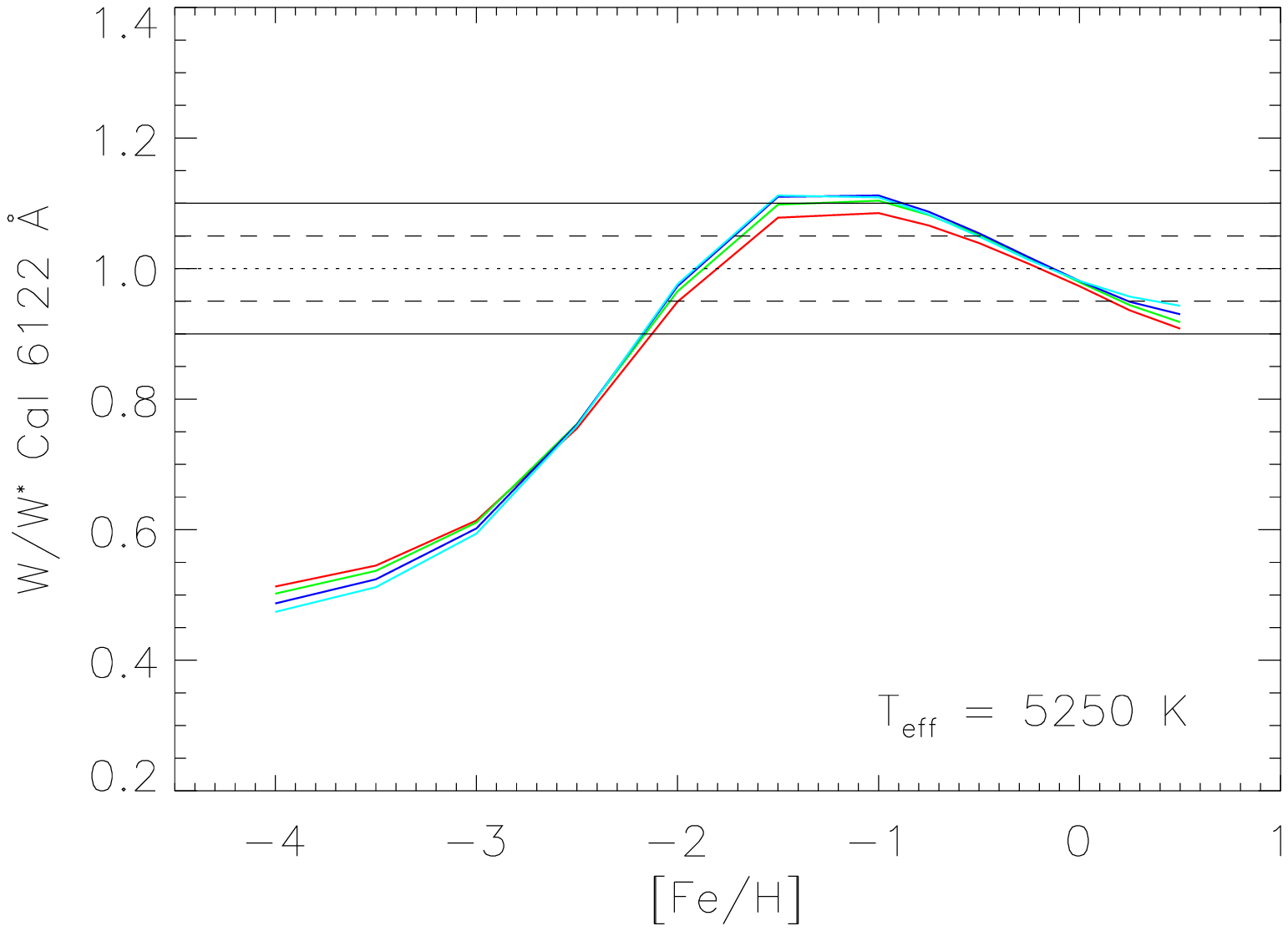}
}
\hbox{
\includegraphics[width=5.45cm]{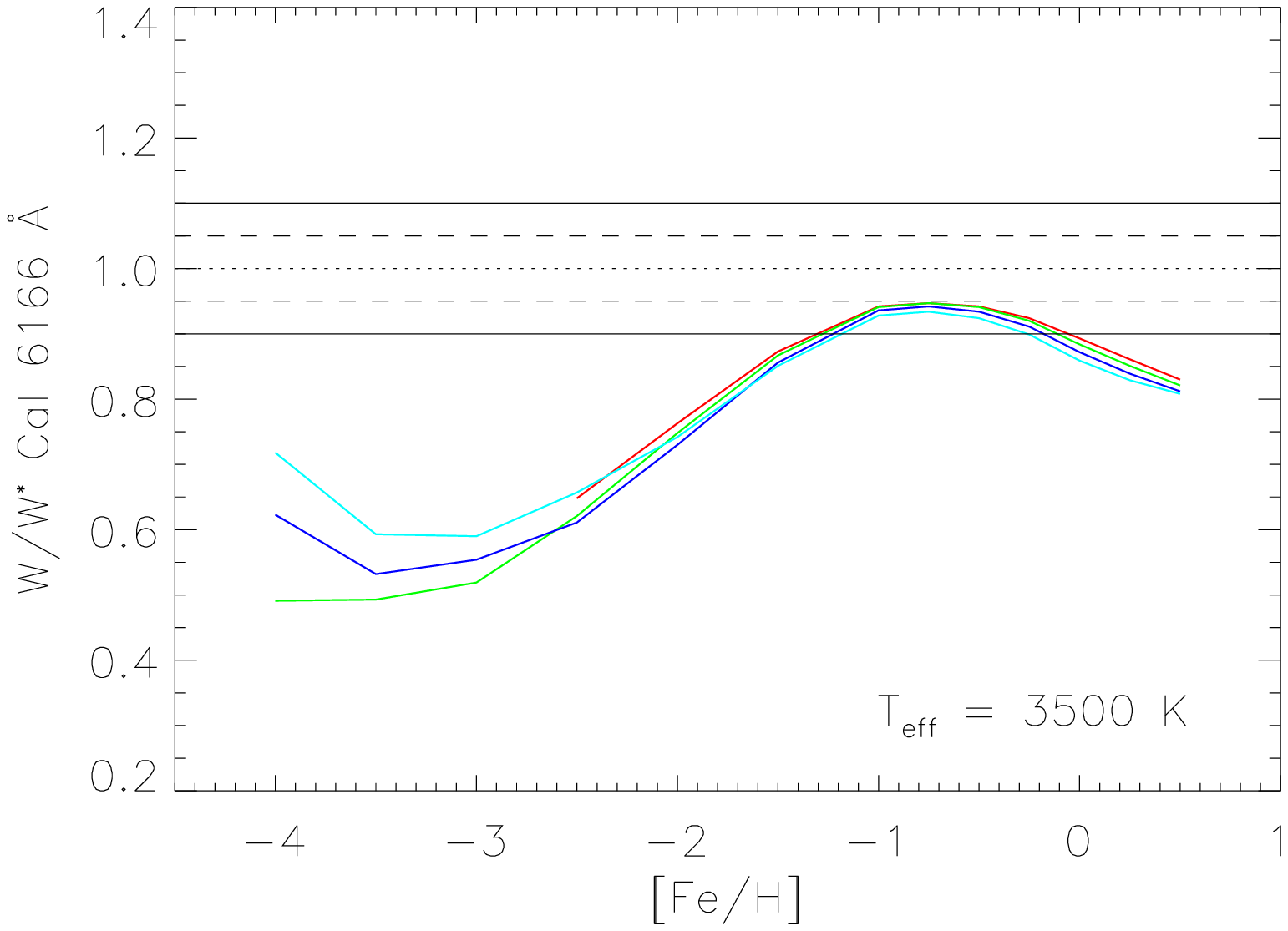}
\includegraphics[width=5.45cm]{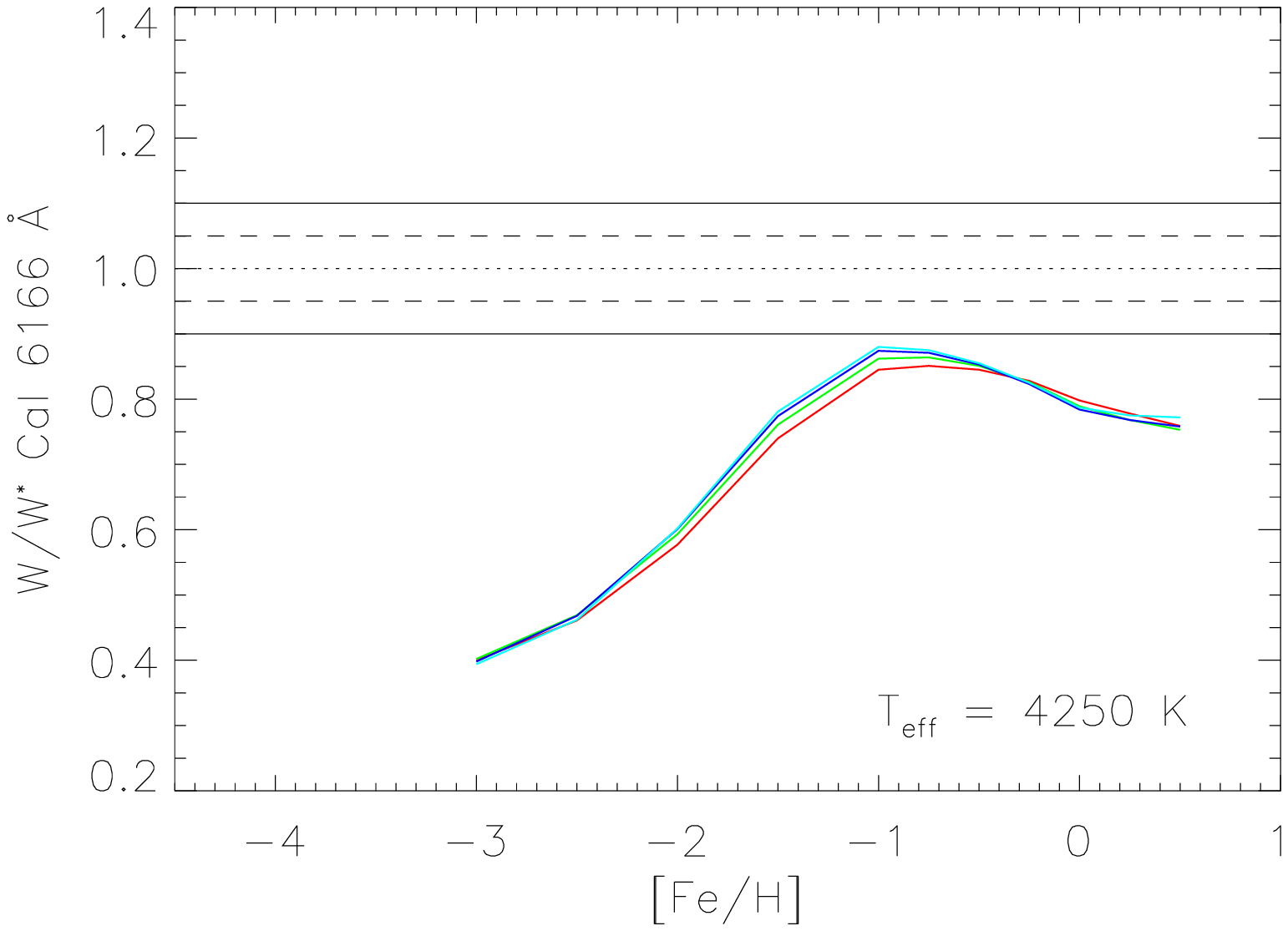}
\includegraphics[width=5.45cm]{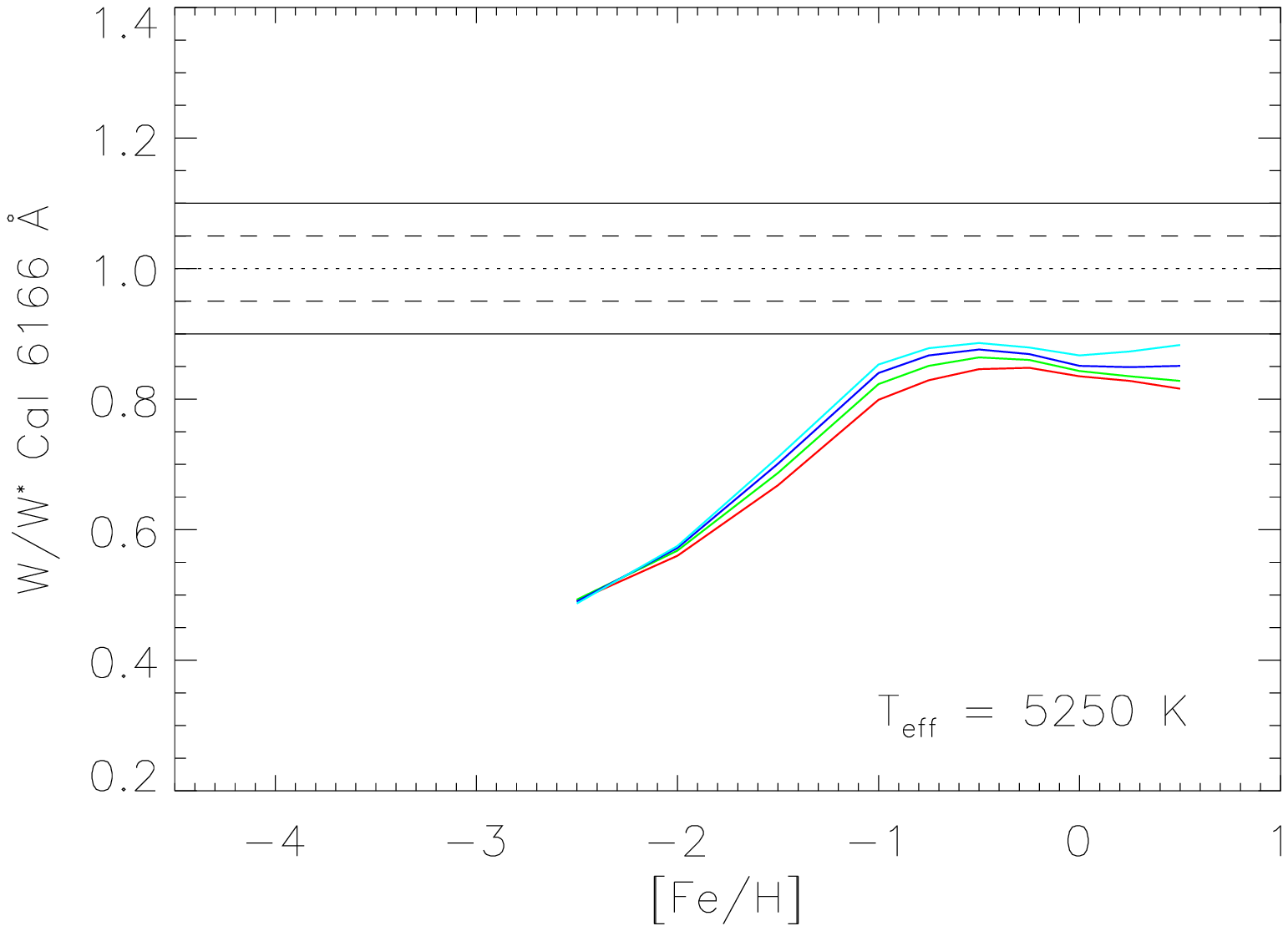}
}
\hbox{
\includegraphics[width=5.45cm]{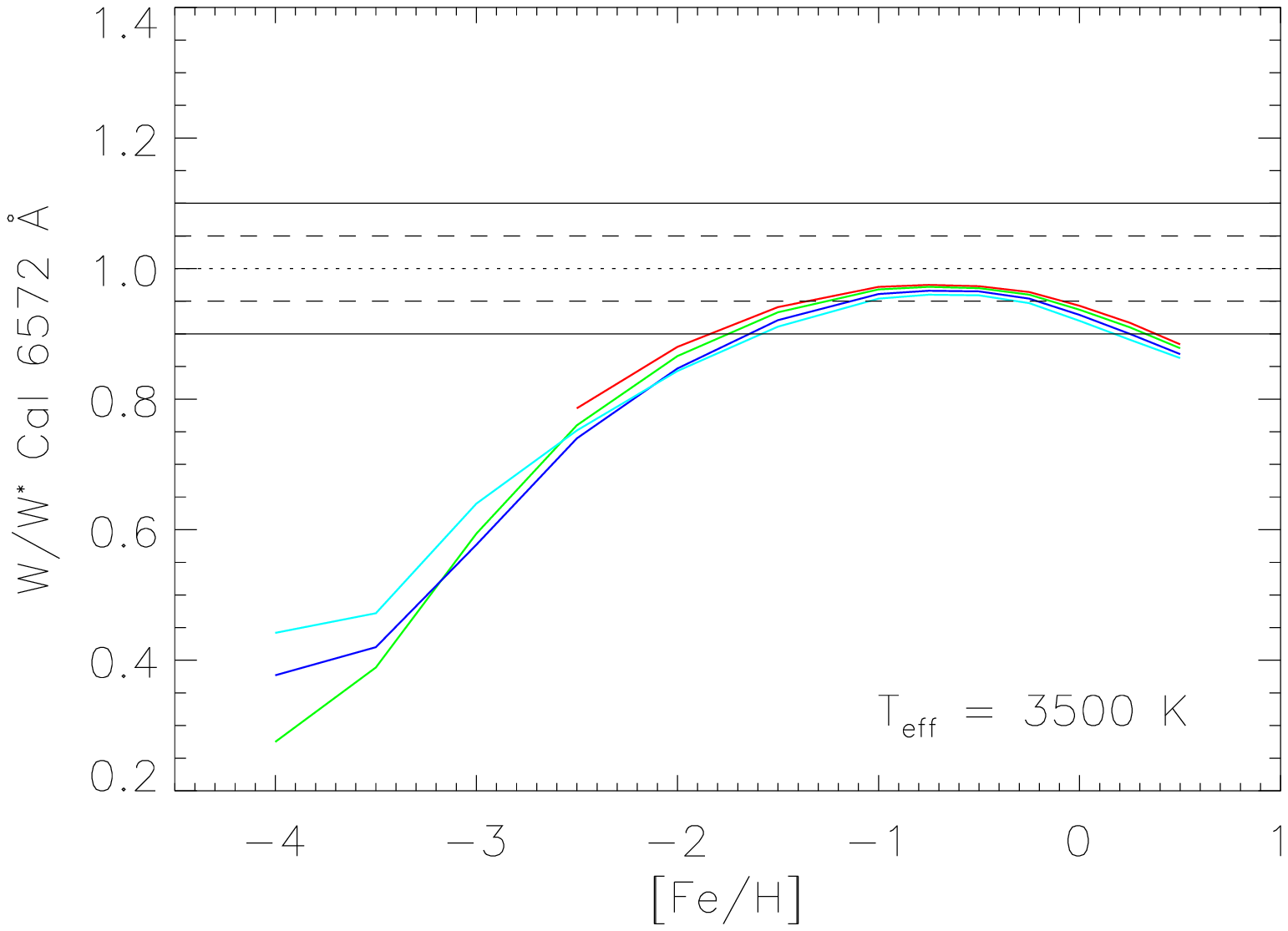}
\includegraphics[width=5.45cm]{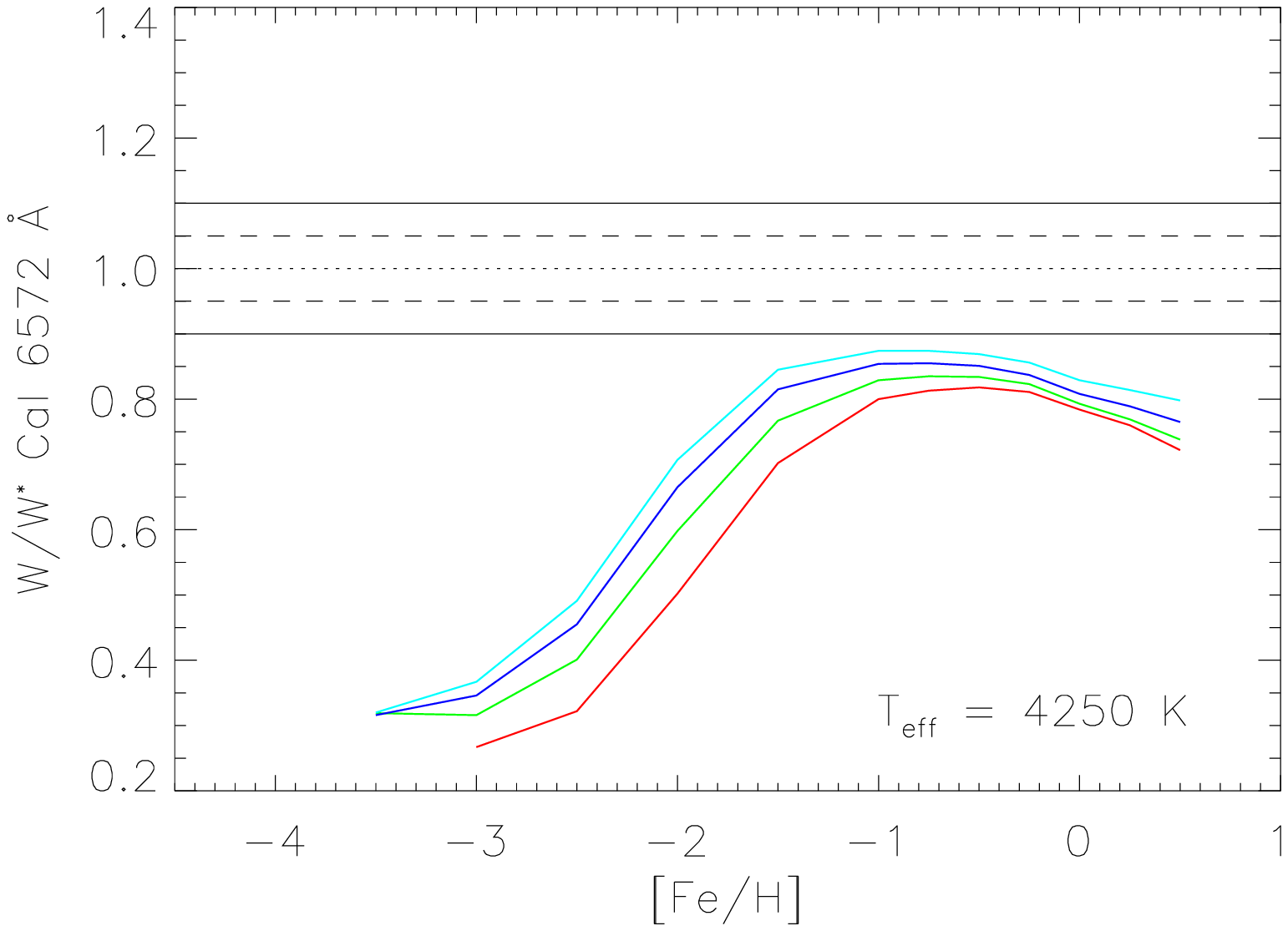}
\includegraphics[width=5.45cm]{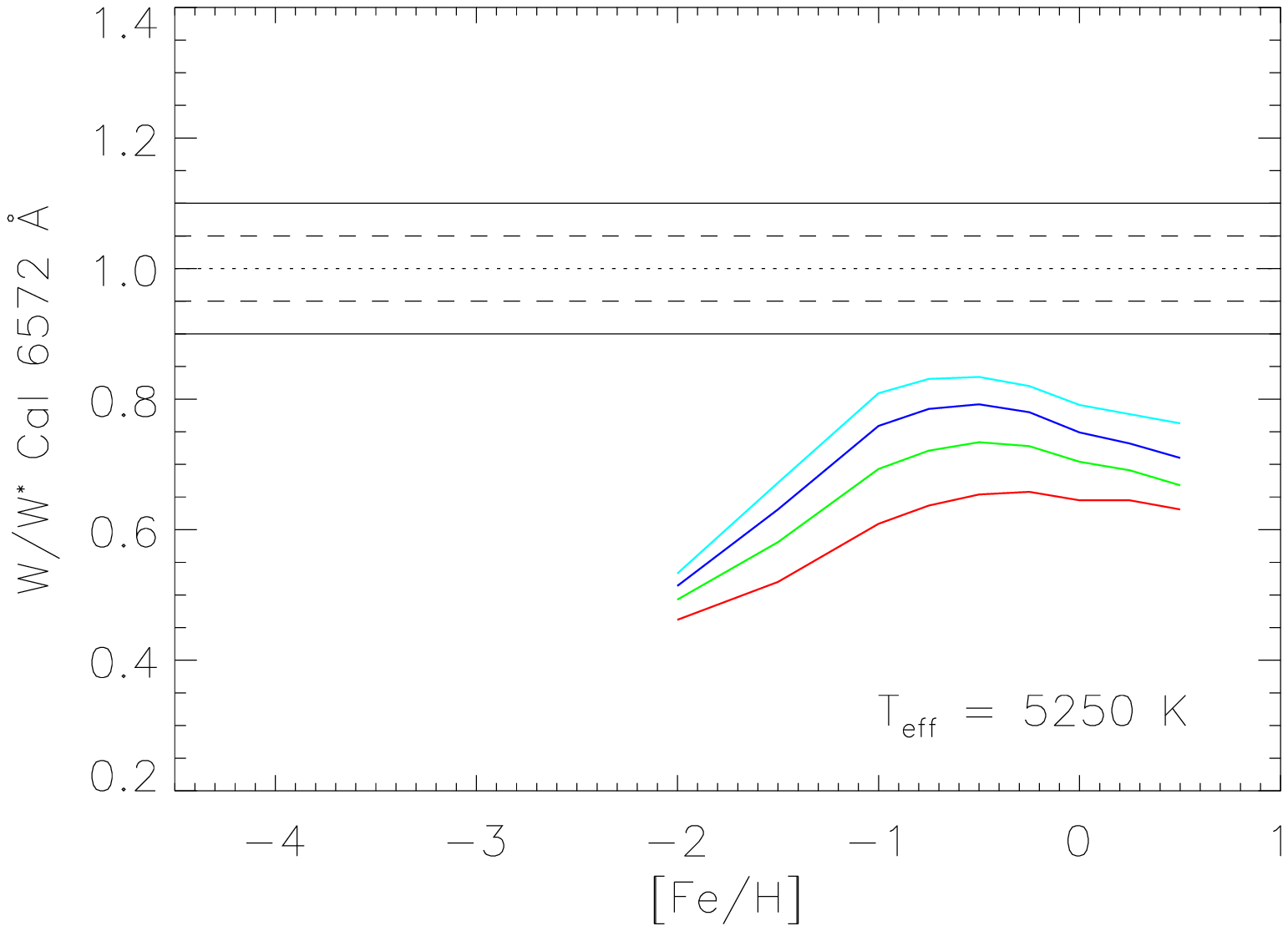}
}
\caption{$W/W^*$ for the selected \caI\ lines as function of the stellar parameters (see Appendix A for details).}
\label{CaI_lines}
\end{figure*}

\begin{figure*}
\hbox{
\includegraphics[width=5.45cm]{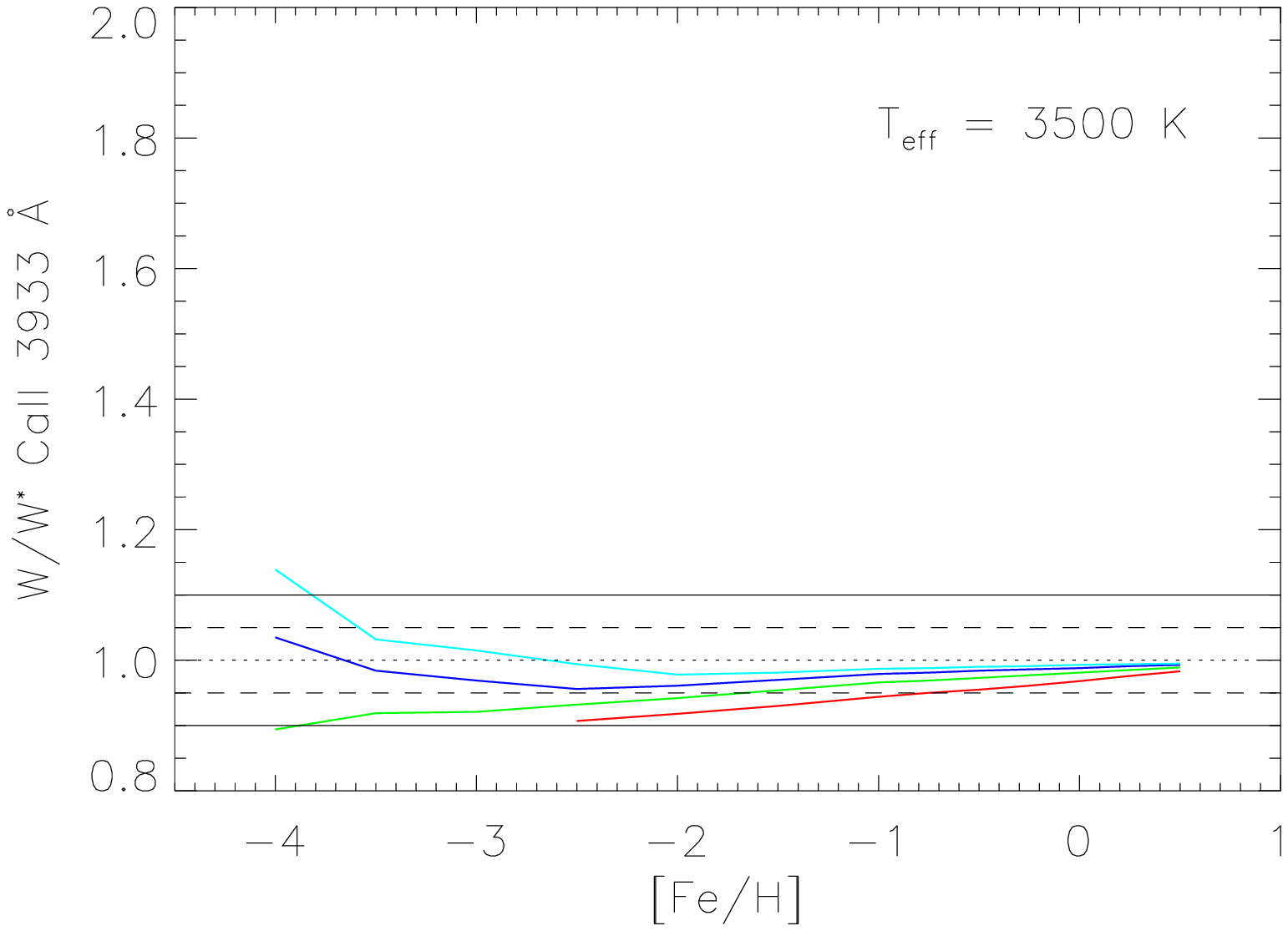}
\includegraphics[width=5.45cm]{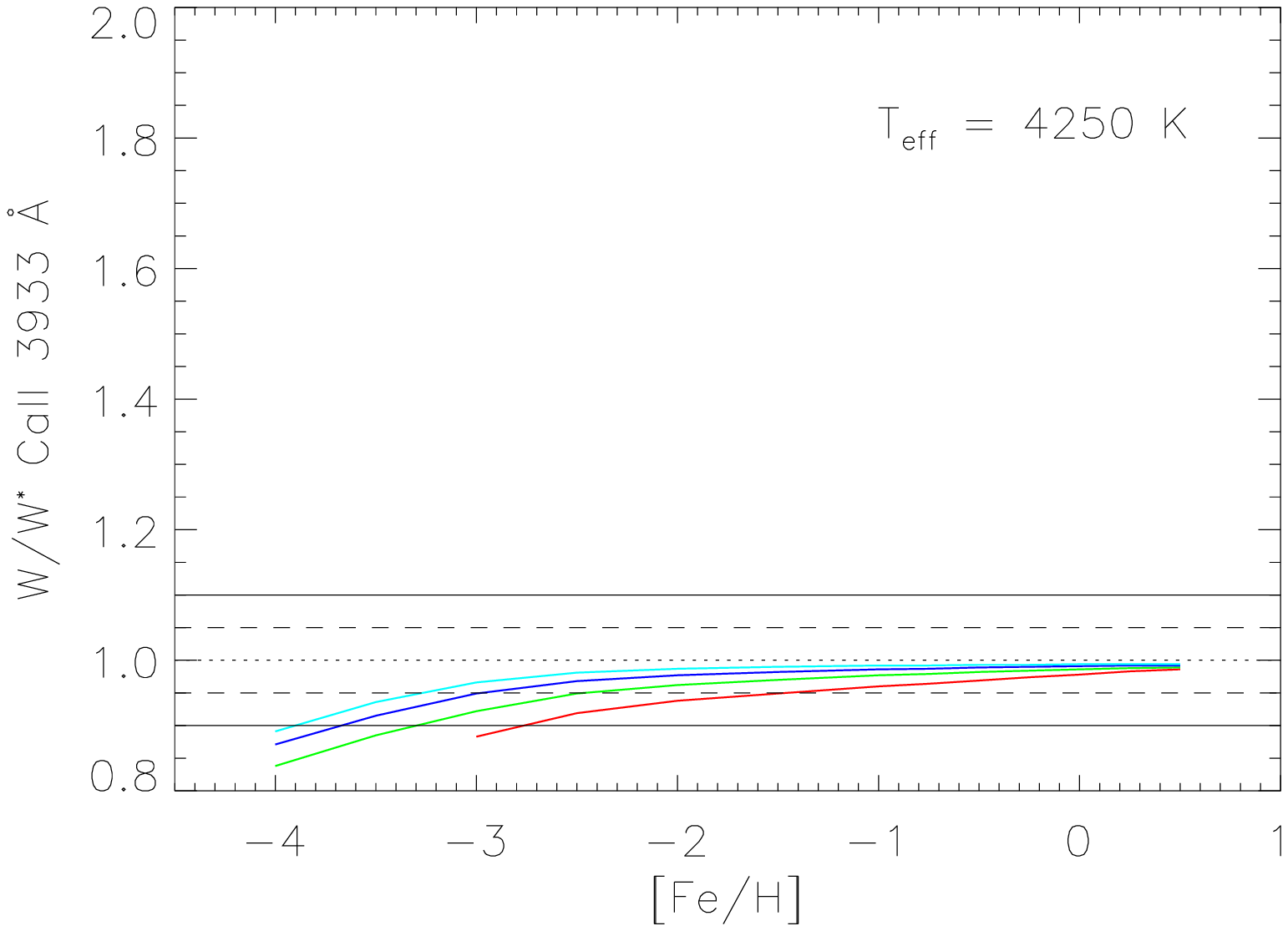}
\includegraphics[width=5.45cm]{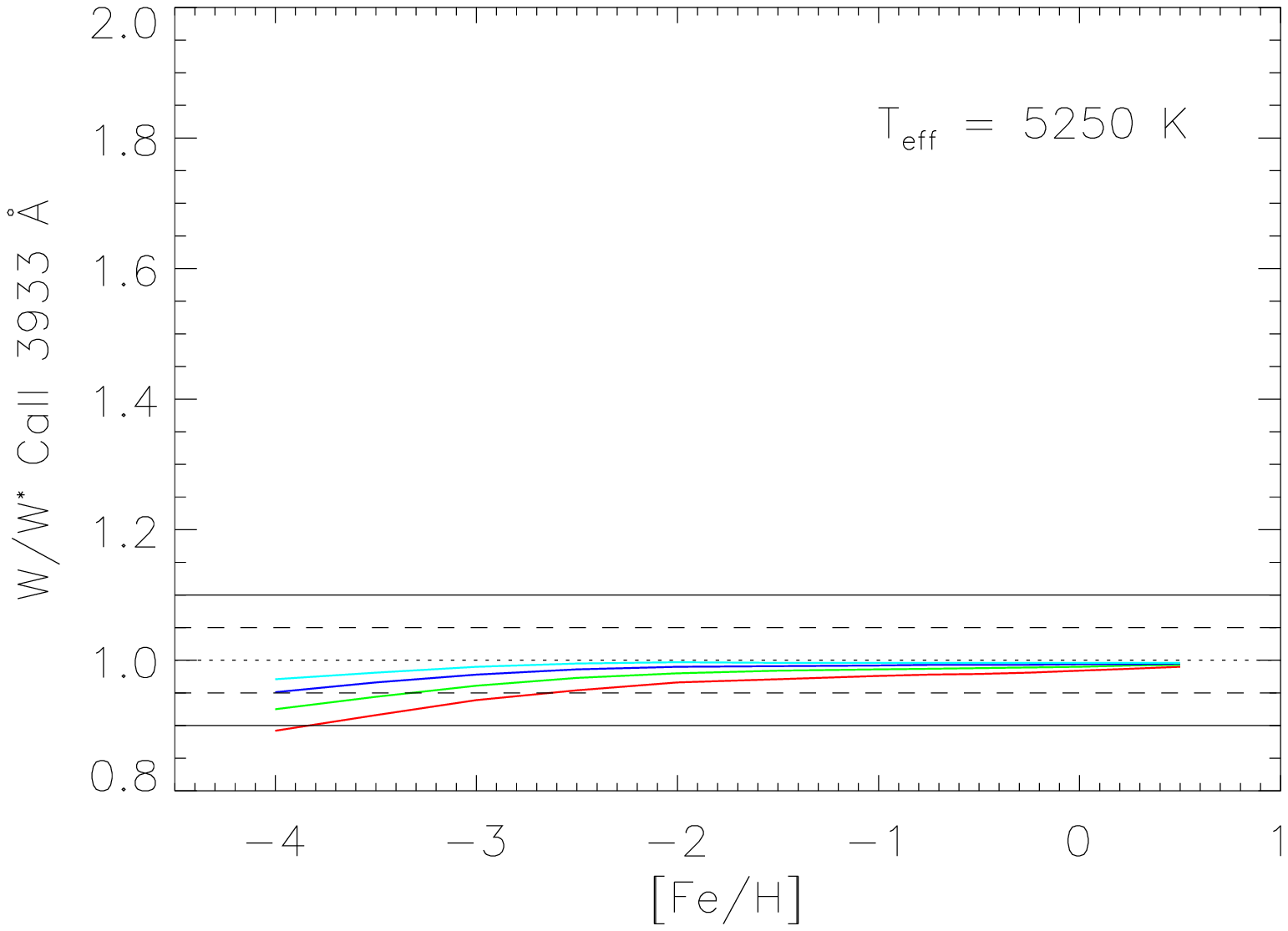}
}\hbox{
\includegraphics[width=5.45cm]{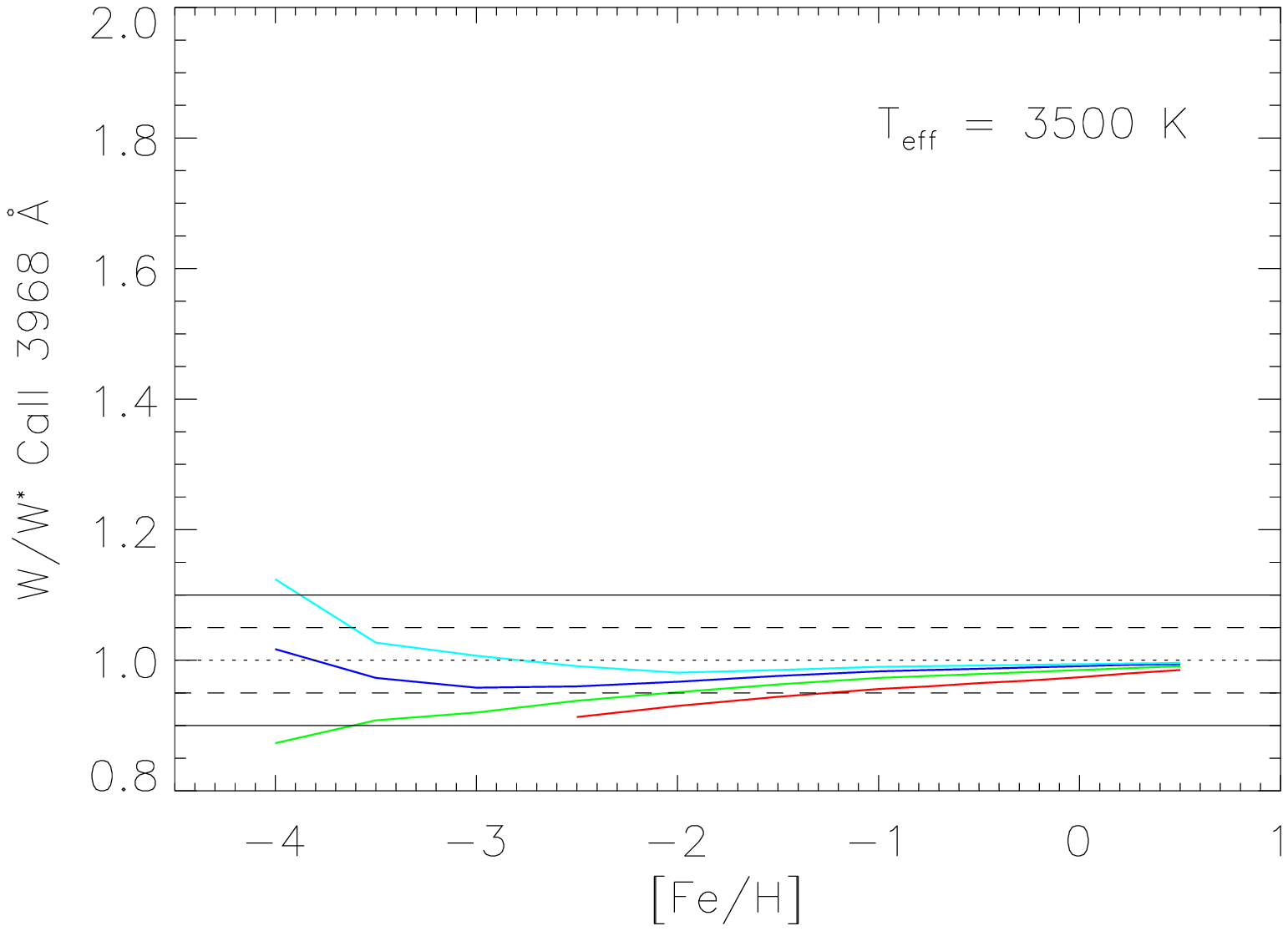}
\includegraphics[width=5.45cm]{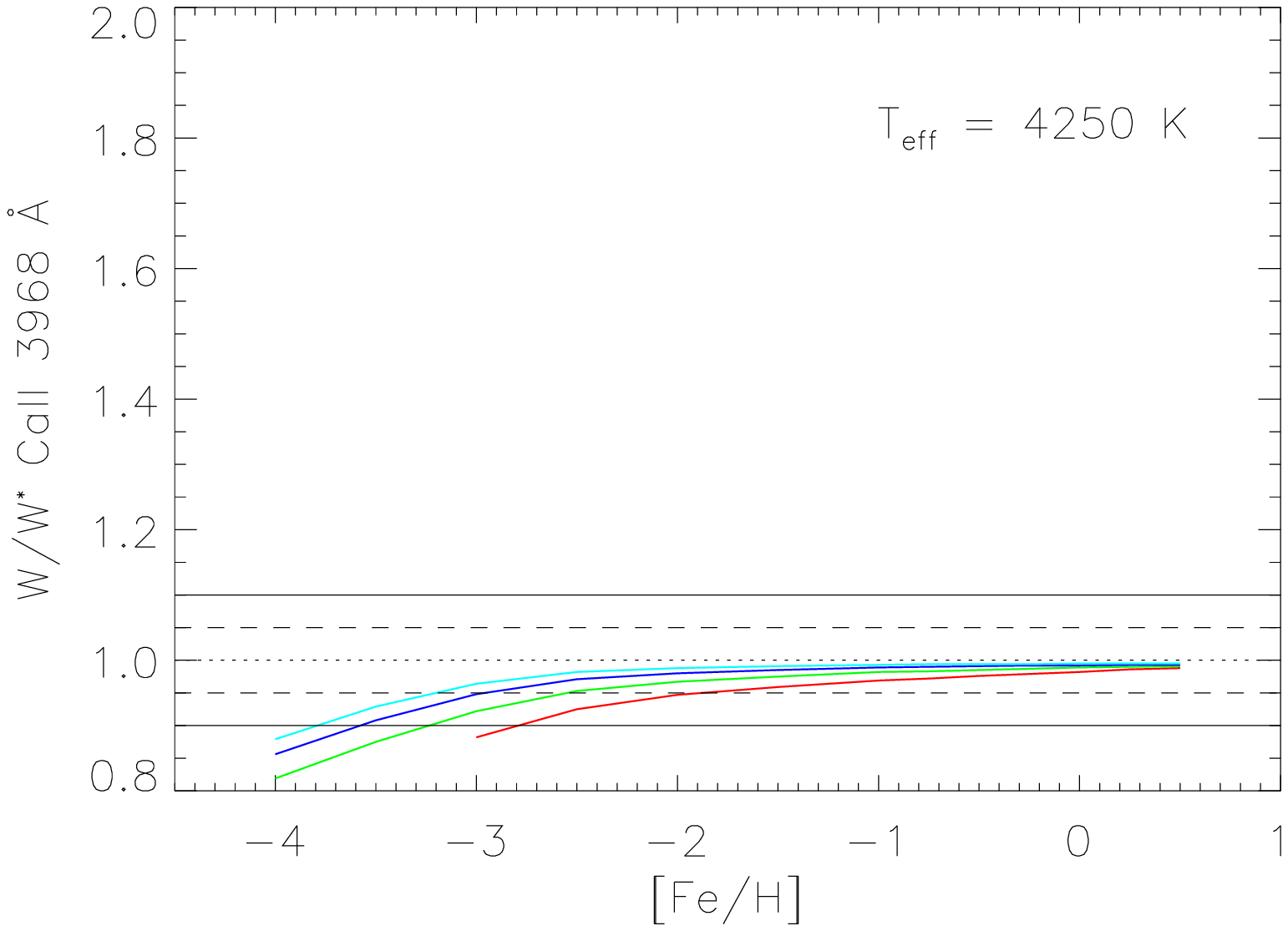}
\includegraphics[width=5.45cm]{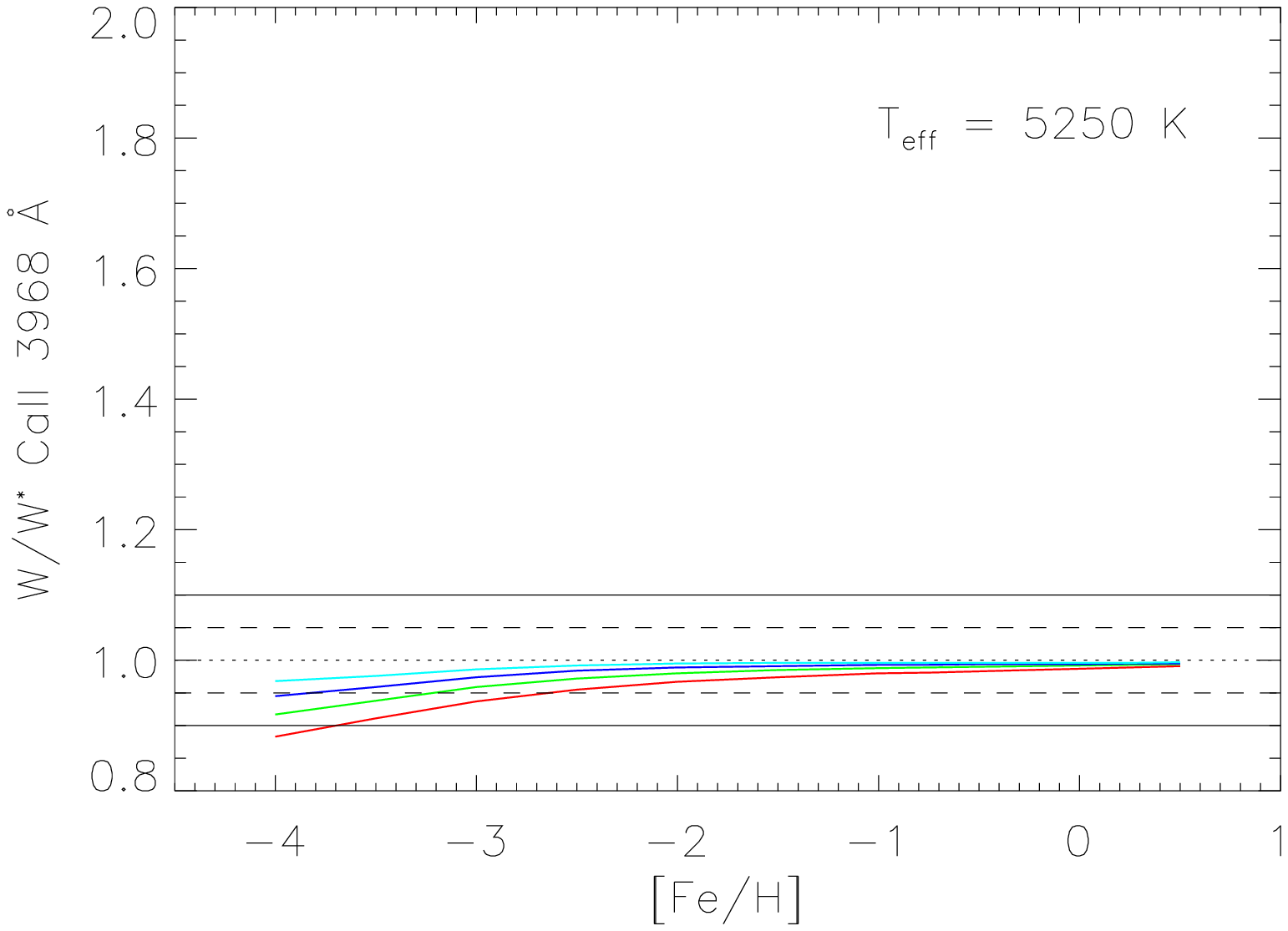}
}
\hbox{
\includegraphics[width=5.45cm]{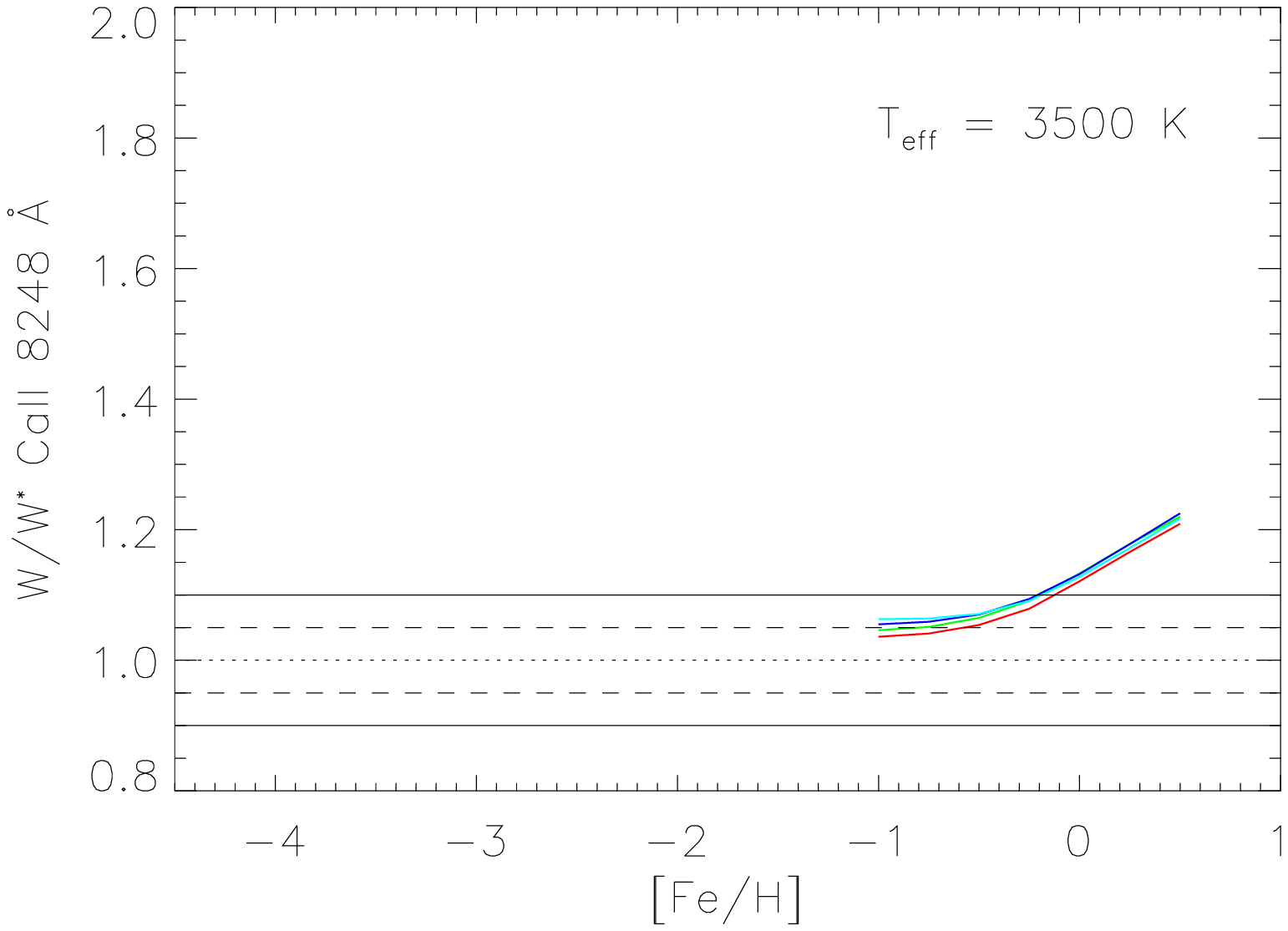}
\includegraphics[width=5.45cm]{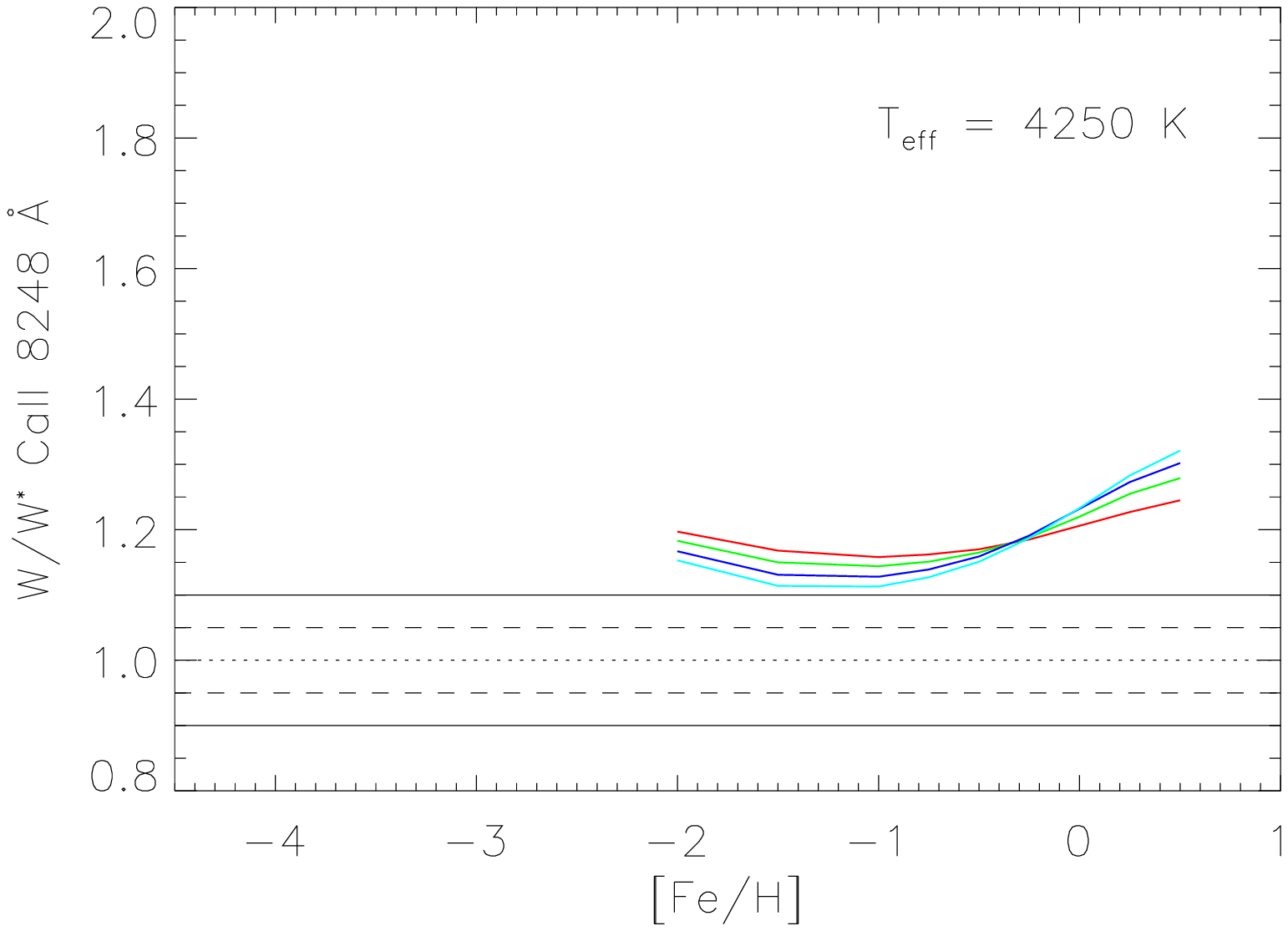}
\includegraphics[width=5.45cm]{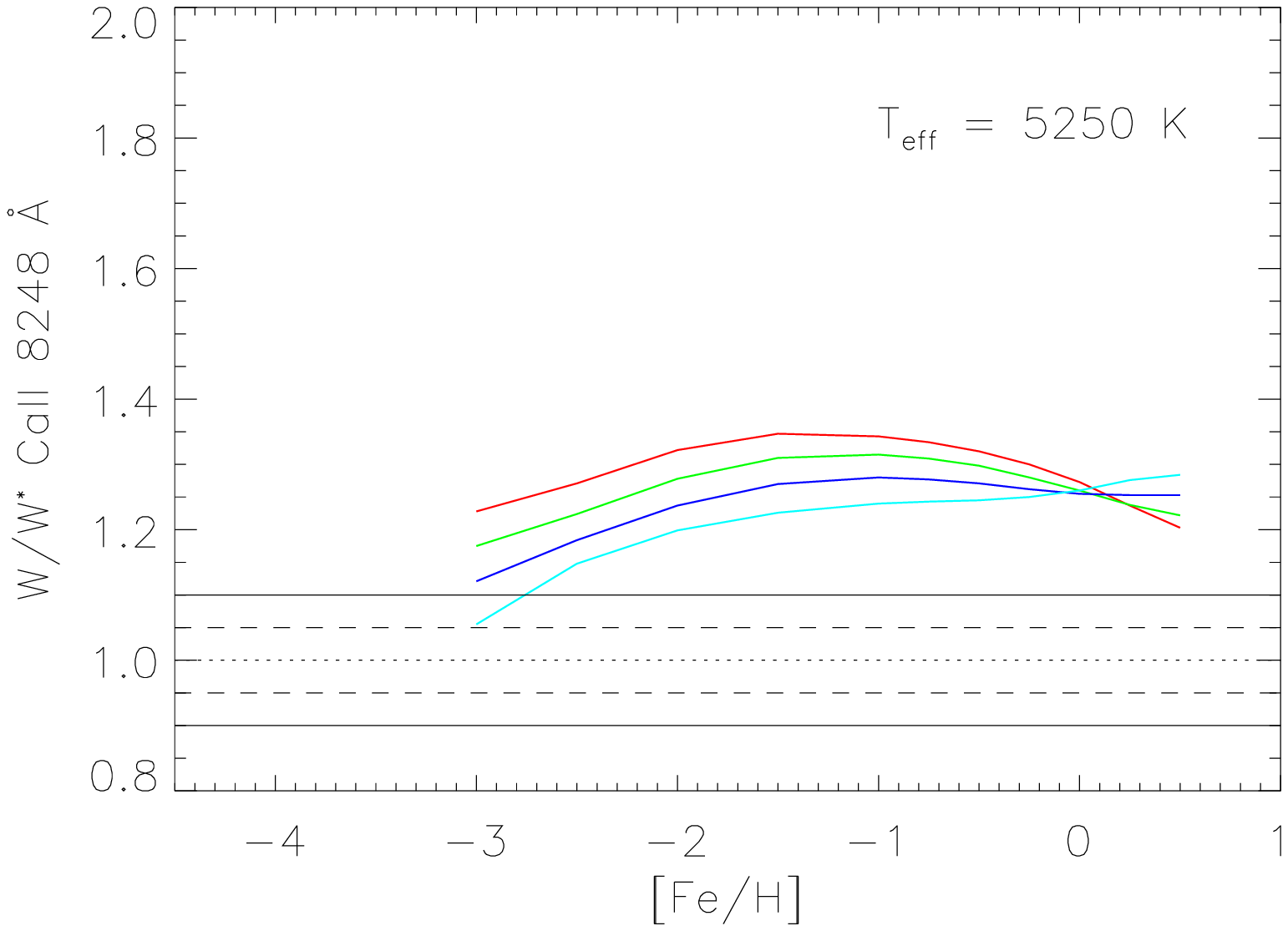}
}
\hbox{
\includegraphics[width=5.45cm]{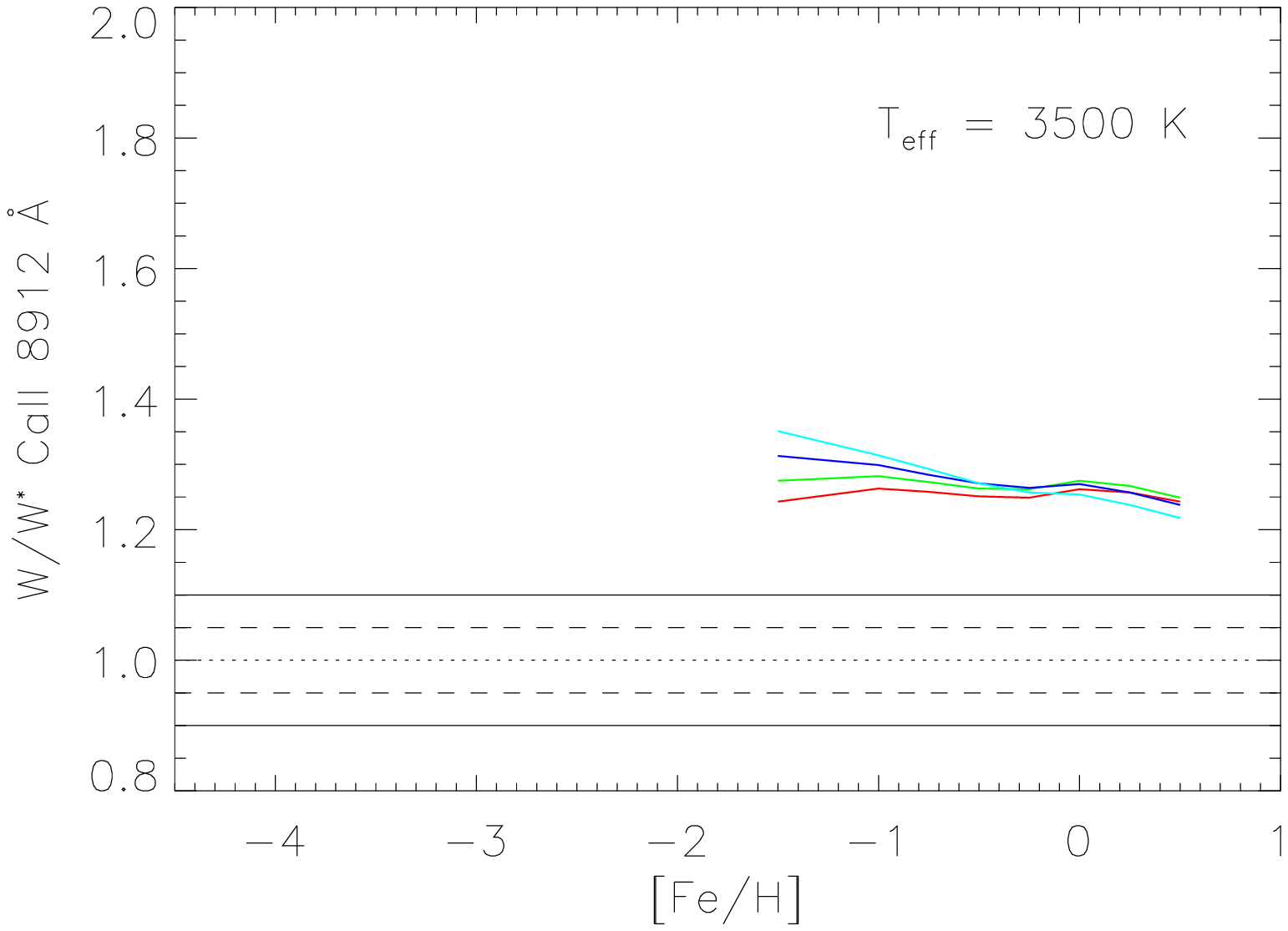}
\includegraphics[width=5.45cm]{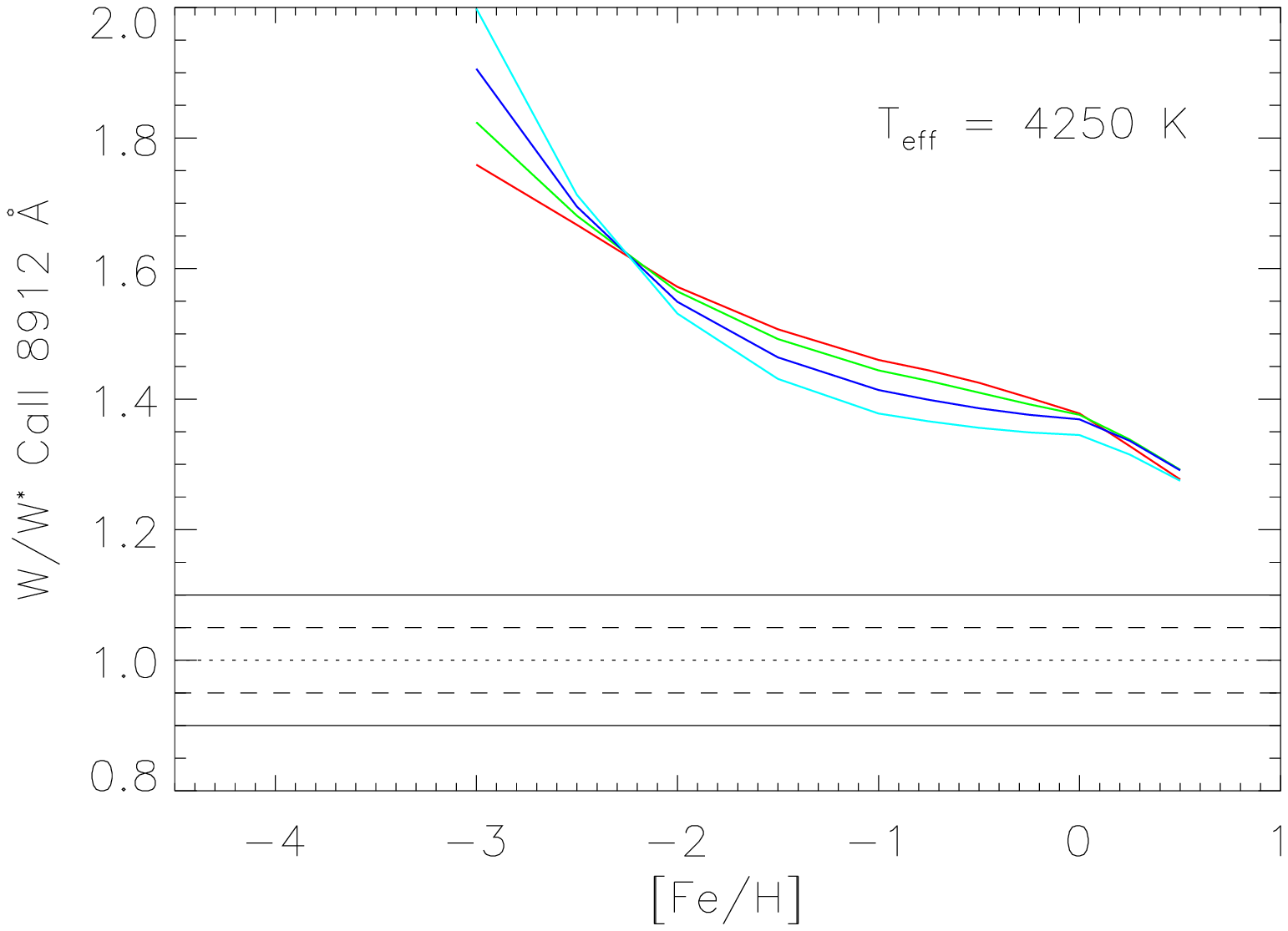}
\includegraphics[width=5.45cm]{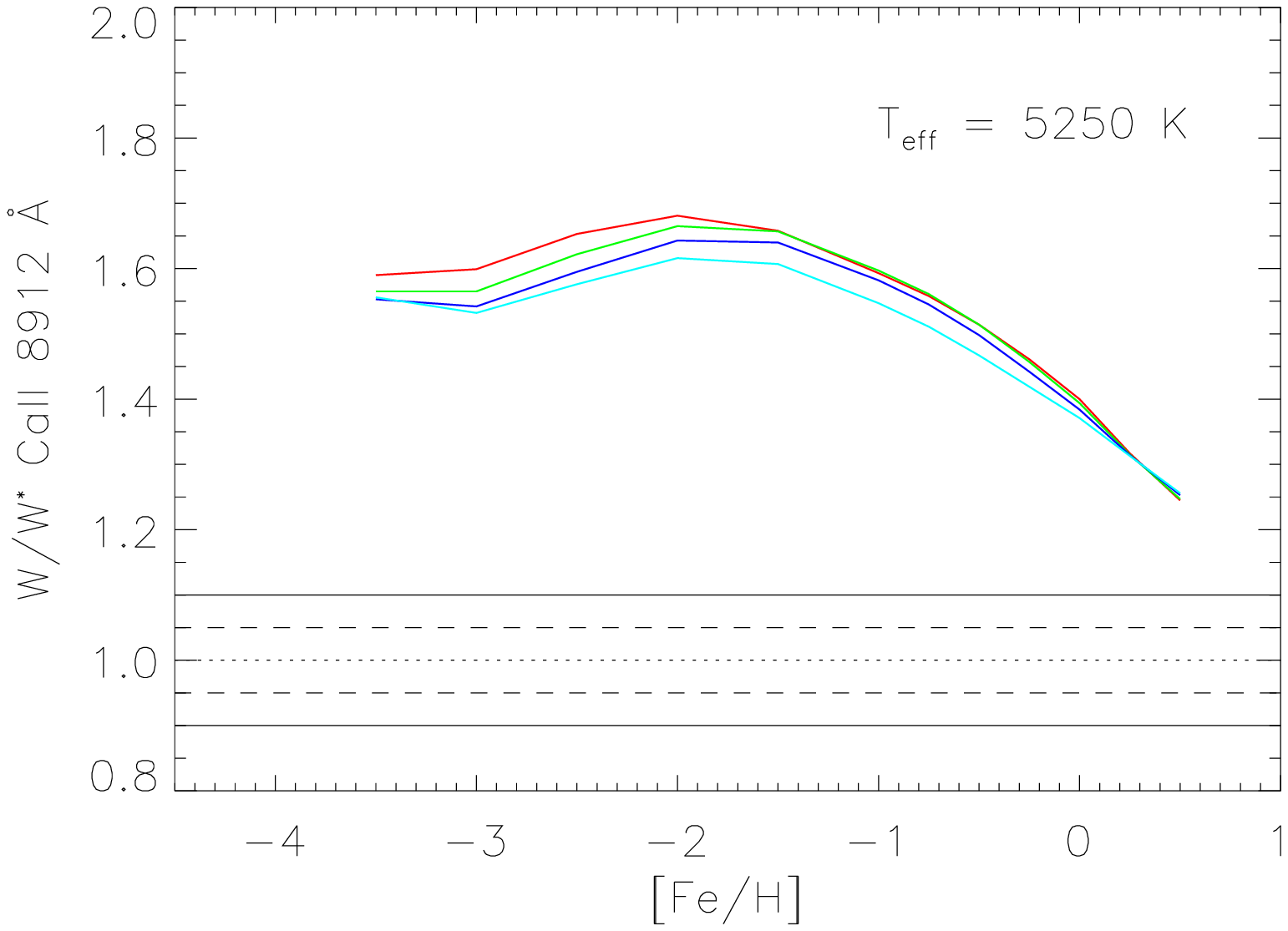}
}
\hbox{
\includegraphics[width=5.45cm]{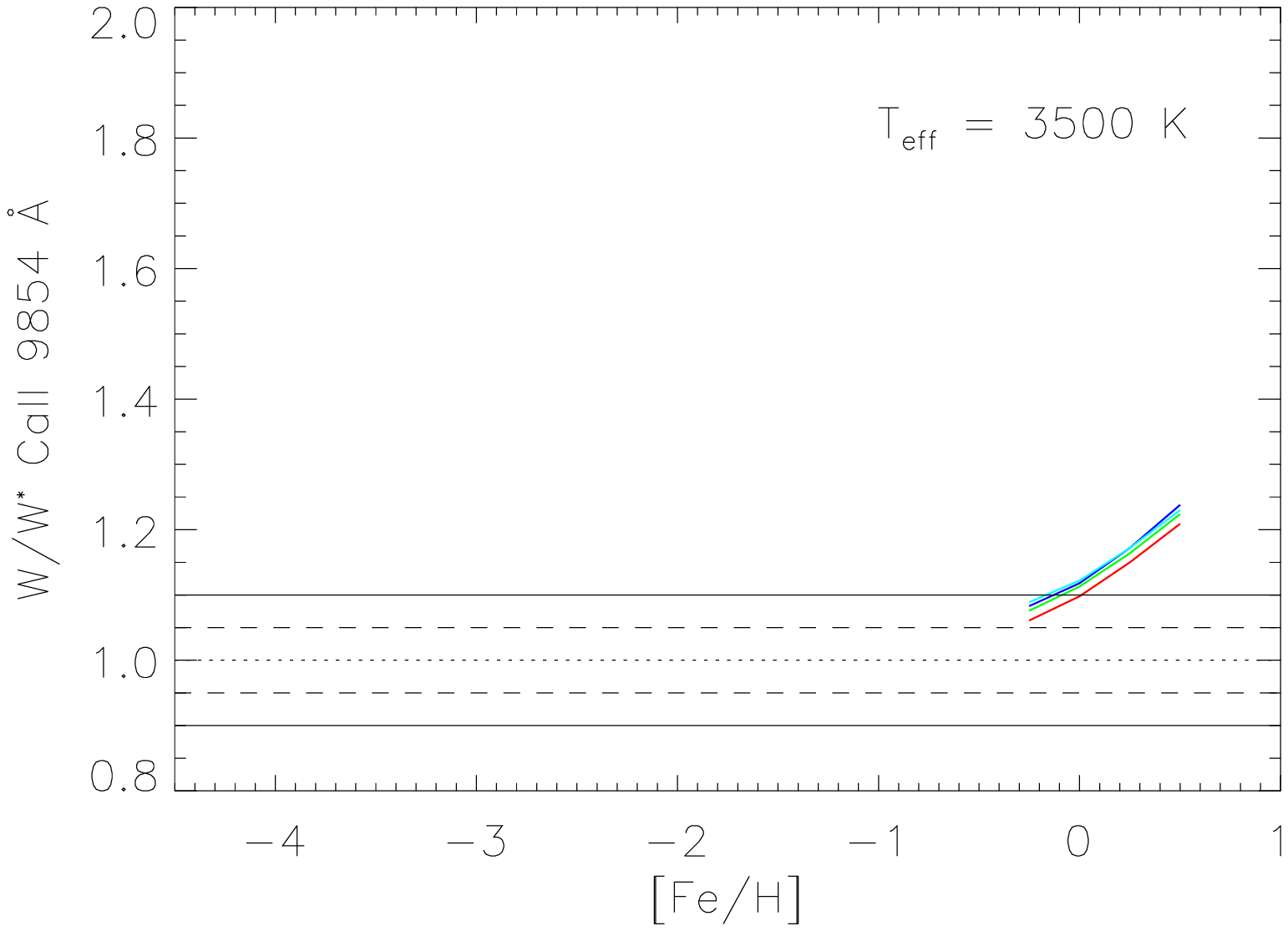}
\includegraphics[width=5.45cm]{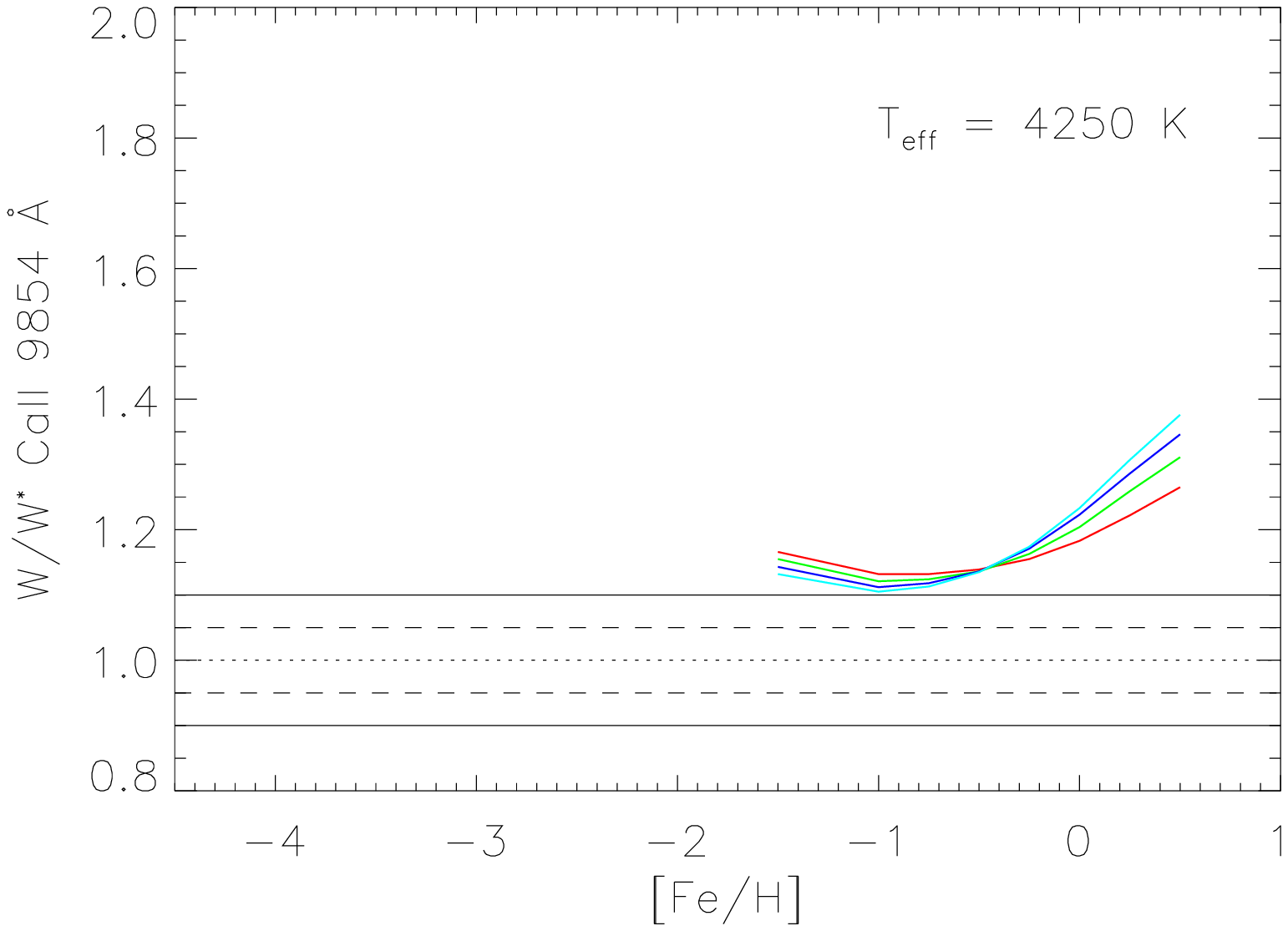}
\includegraphics[width=5.45cm]{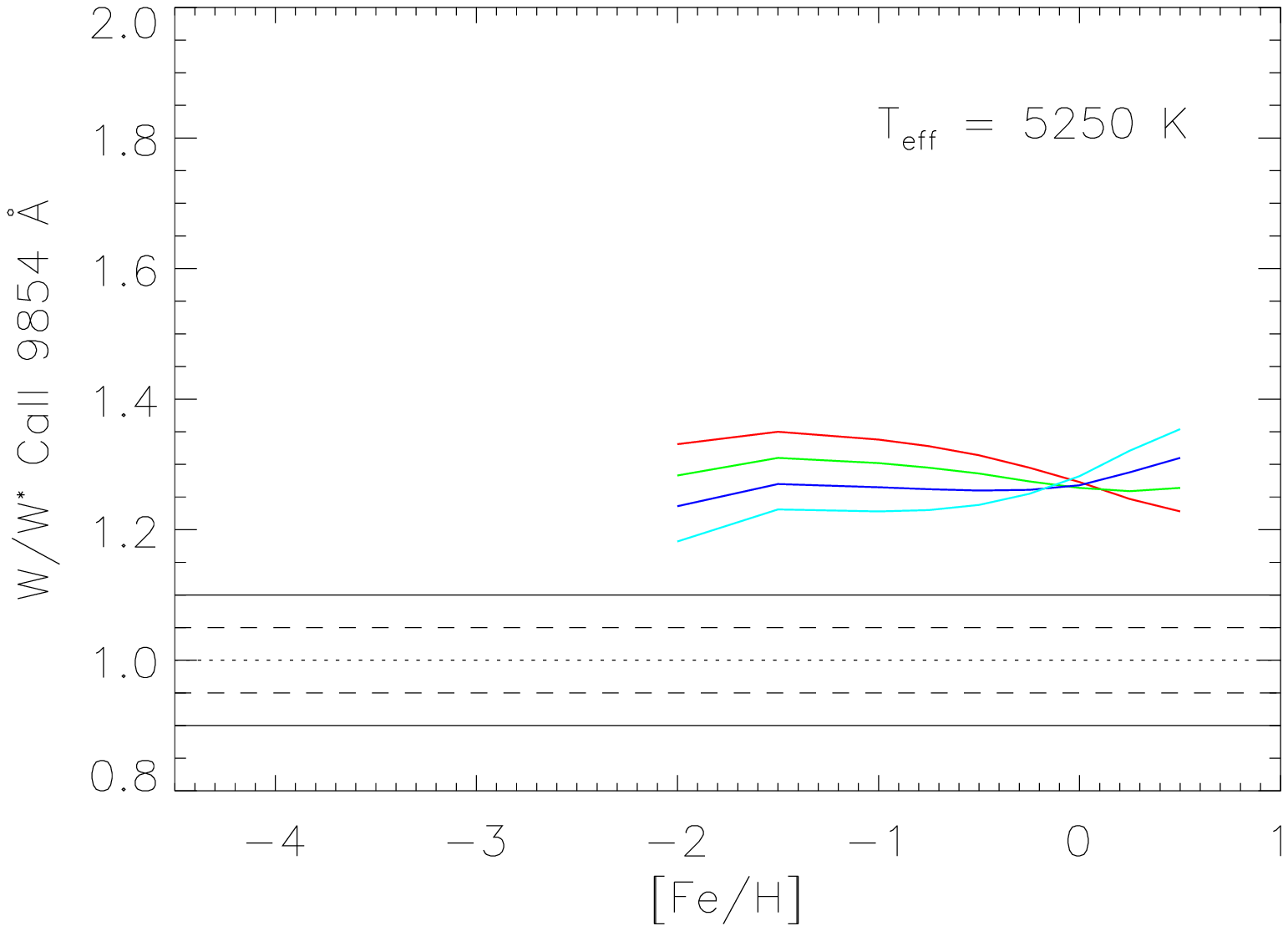}
}
\hbox{
\includegraphics[width=5.45cm]{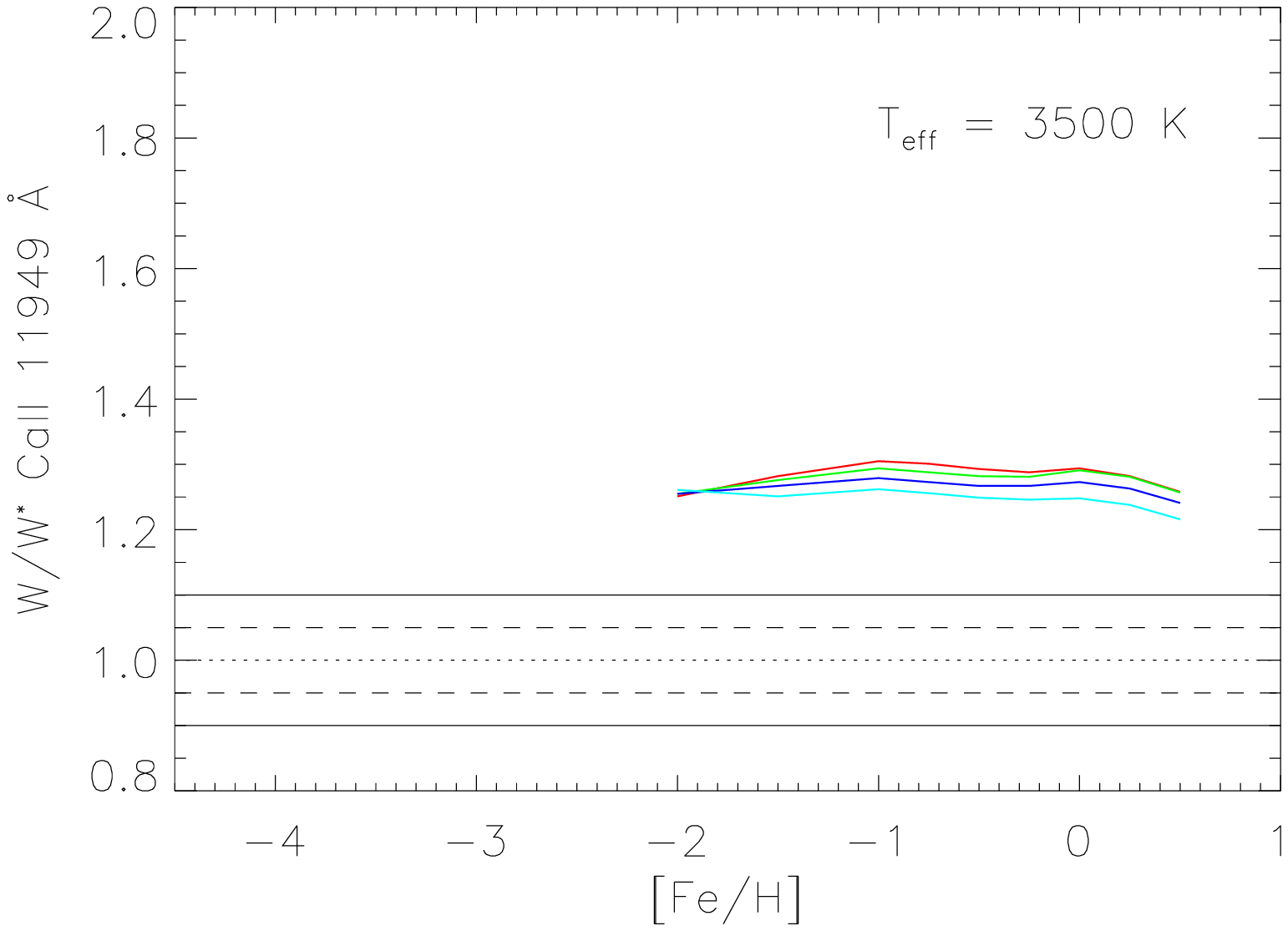}
\includegraphics[width=5.45cm]{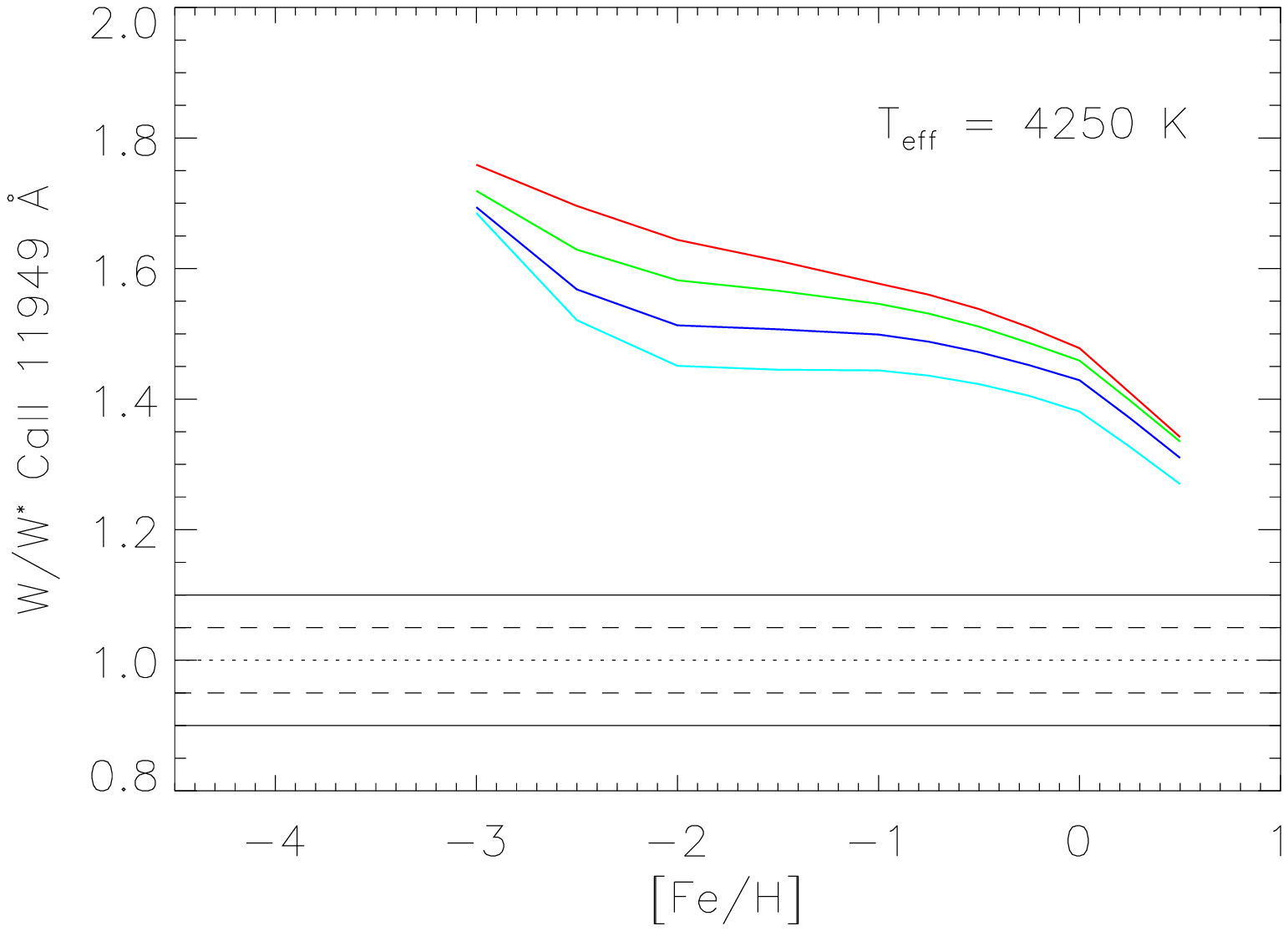}
\includegraphics[width=5.45cm]{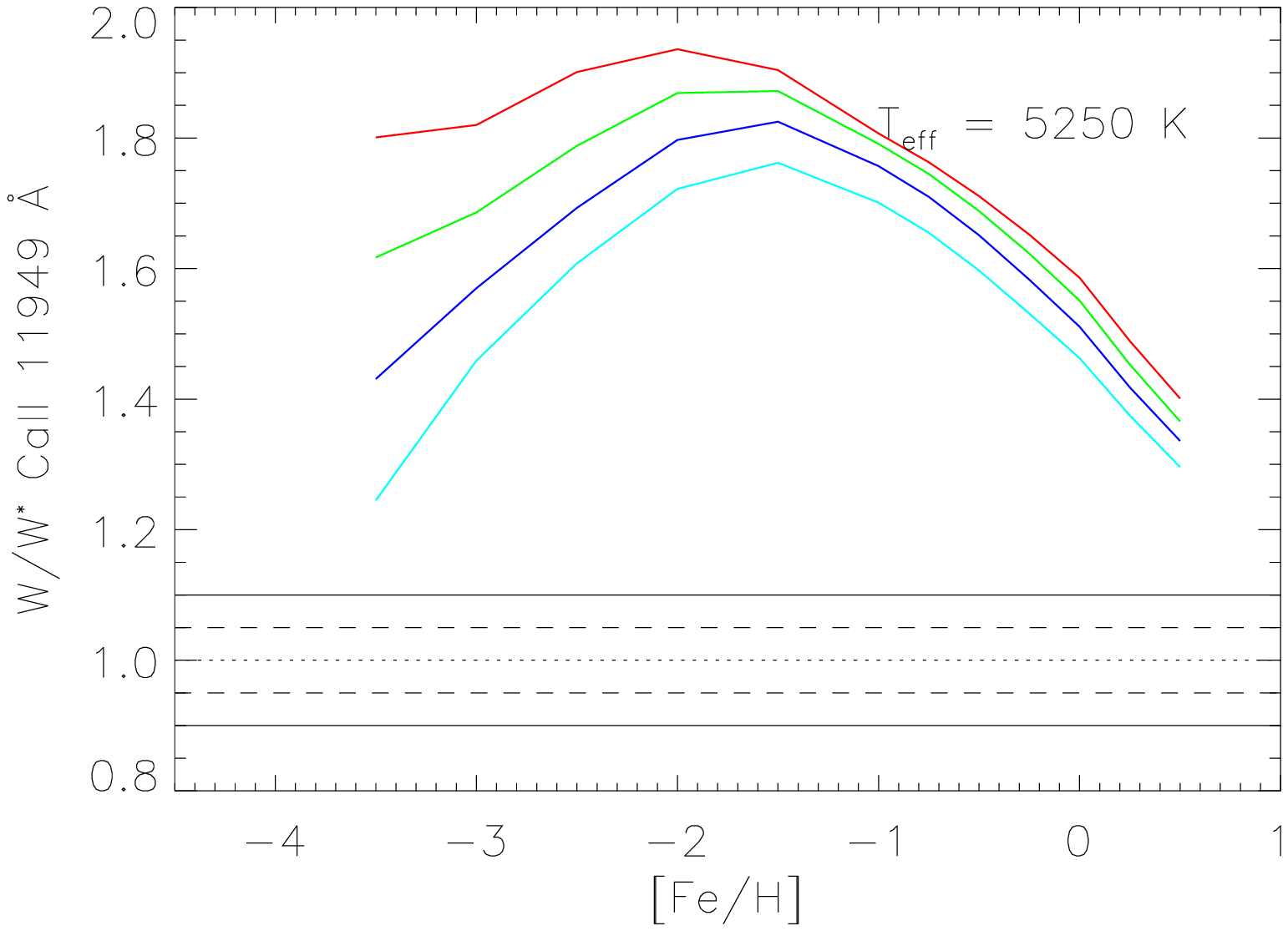}
}
\caption{$W/W^*$ for the selected \caII\ lines as function of the stellar parameters (see Appendix A for details).}
\label{CaII_lines}
\end{figure*}

We show the most representative \mg\ lines in Fig.~\ref{MgI_lines}:
\begin{itemize}
\item the 5183 \AA\ $W/W^*$ represents the \mg\ b triplet lines at $\lambda\lambda$ 5167, 5172 and 5183;
\item the 5711 \AA\ $W/W^*$ represents the similar behaviours of the $\lambda\lambda$ 4730, 5711 and 11828 lines;
\item the 7657 \AA\ $W/W^*$;
\item the 8806 \AA\ $W/W^*$ represents the similar behaviours of the $\lambda\lambda$ 4167, 4702, 5528 and 8806 lines that are issued from the $3p\ ^1P^o$;
\item the 8923 \AA\ $W/W^*$ that is the counterpart of the 7657~\AA\ in the singlet system; 
\item the 10312 \AA\ $W/W^*$ that is essentially formed in LTE.
\end{itemize}
We show the most representative \caI\ lines in Fig.~\ref{CaI_lines}:
\begin{itemize}
\item the 4226 \AA\ $W/W^*$ represents the NLTE behaviour of a resonance line;
\item the 4578 \AA\ $W/W^*$;
\item the 5512 \AA\ $W/W^*$ represents the similar behaviour of the $\lambda\lambda$ 5512, 5867 and 7326 lines;
\item the 6122 \AA\ $W/W^*$ represents the similar behaviour of the \caI\ red triplet lines at $\lambda\lambda$ 6102, 6122 and 6162;
\item the 6166 \AA\ $W/W^*$ represents the similar behaviour of the multiplet lines at $\lambda\lambda$ 6161, 6166, 6169.0 and 6169.6;
\item the 6572 \AA\ $W/W^*$ represents the NLTE behaviour of an intercombination resonance line. 
\end{itemize} 
We show all the selected lines for the \caII\ lines in Fig.~\ref{CaII_lines}.

\section{$W/W^*$ for the lines in the {\it Gaia}/RVS wavelength range}

\begin{figure*}
\hbox{
\includegraphics[width=5.45cm]{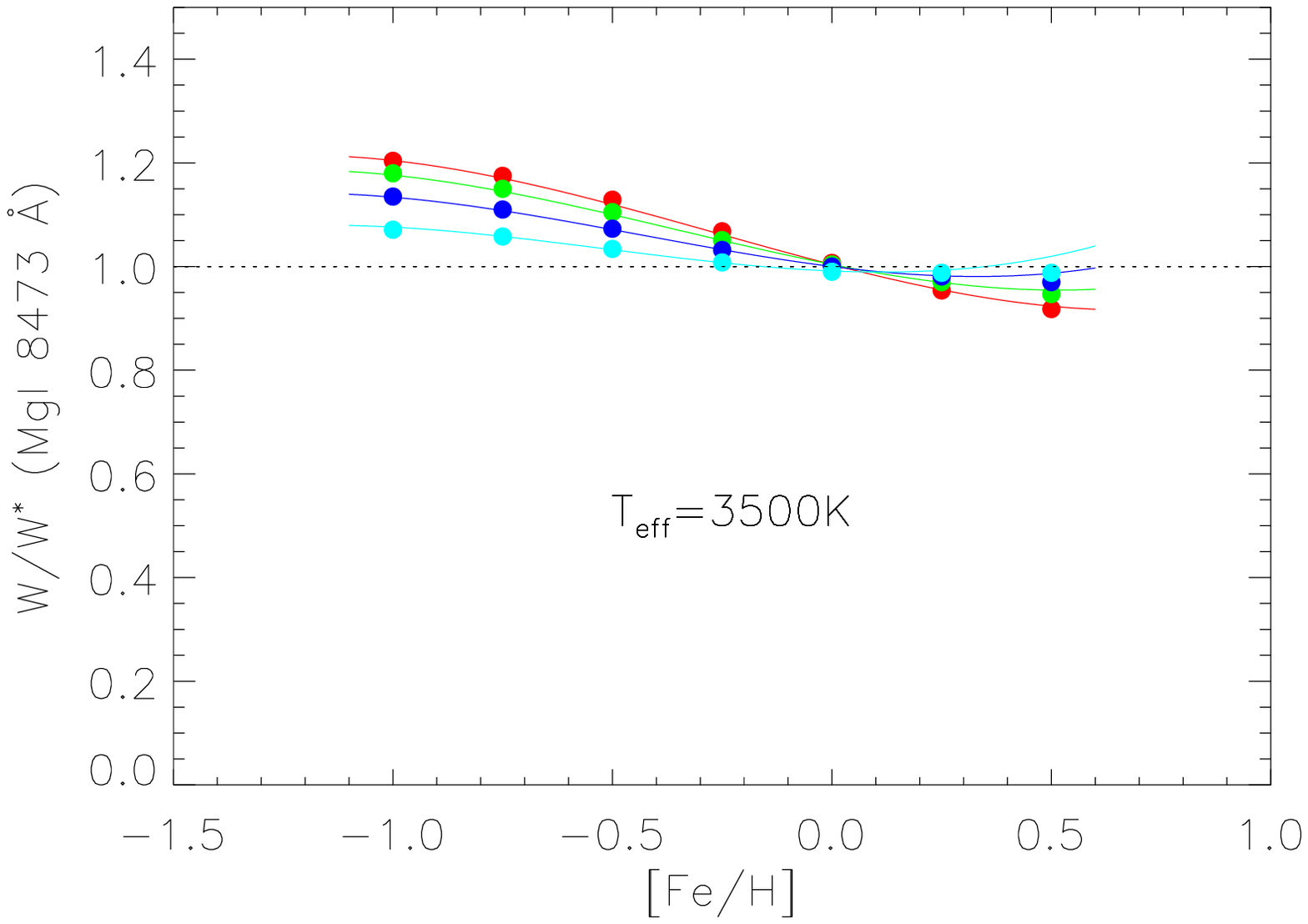}
\includegraphics[width=5.45cm]{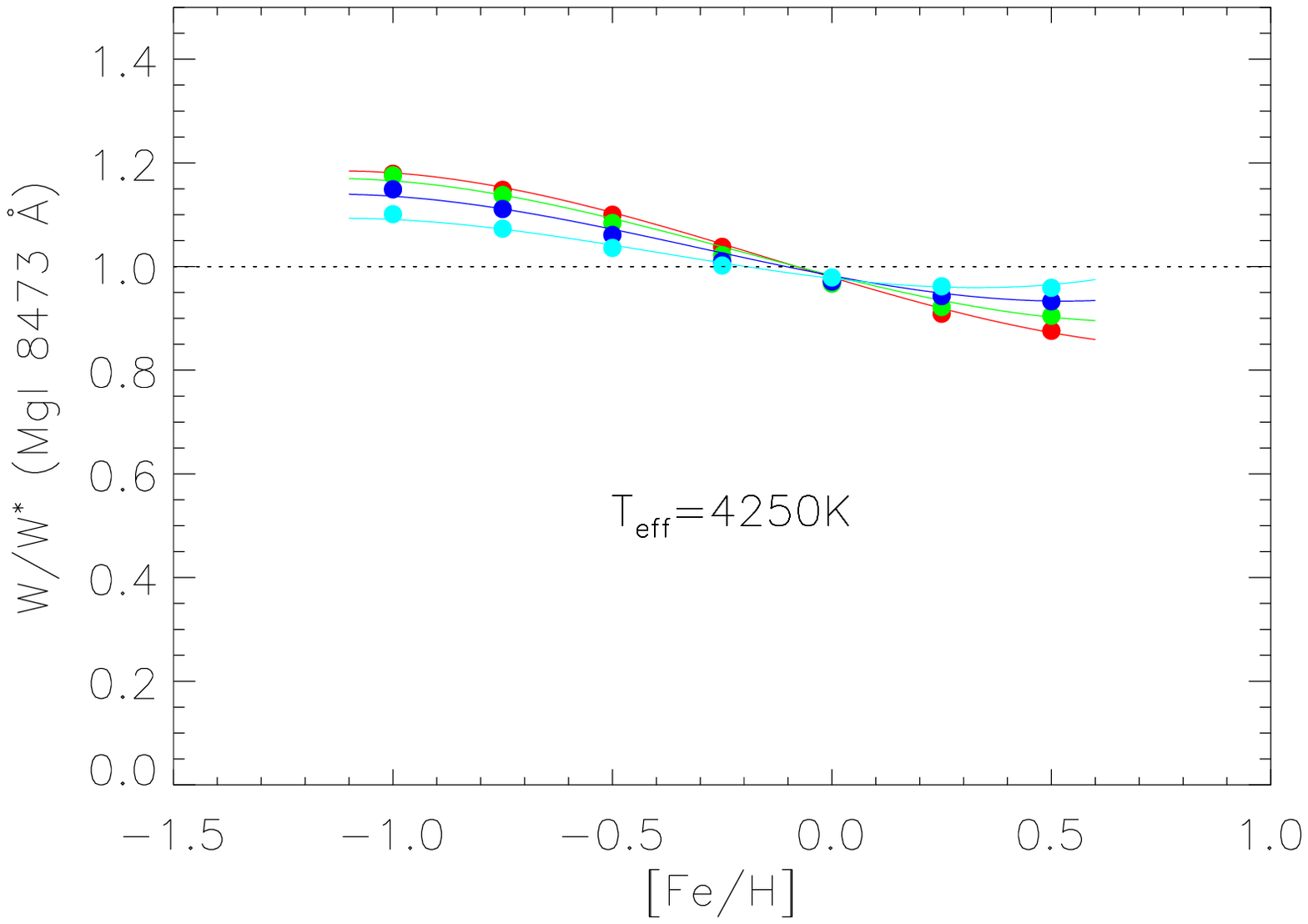}
\includegraphics[width=5.45cm]{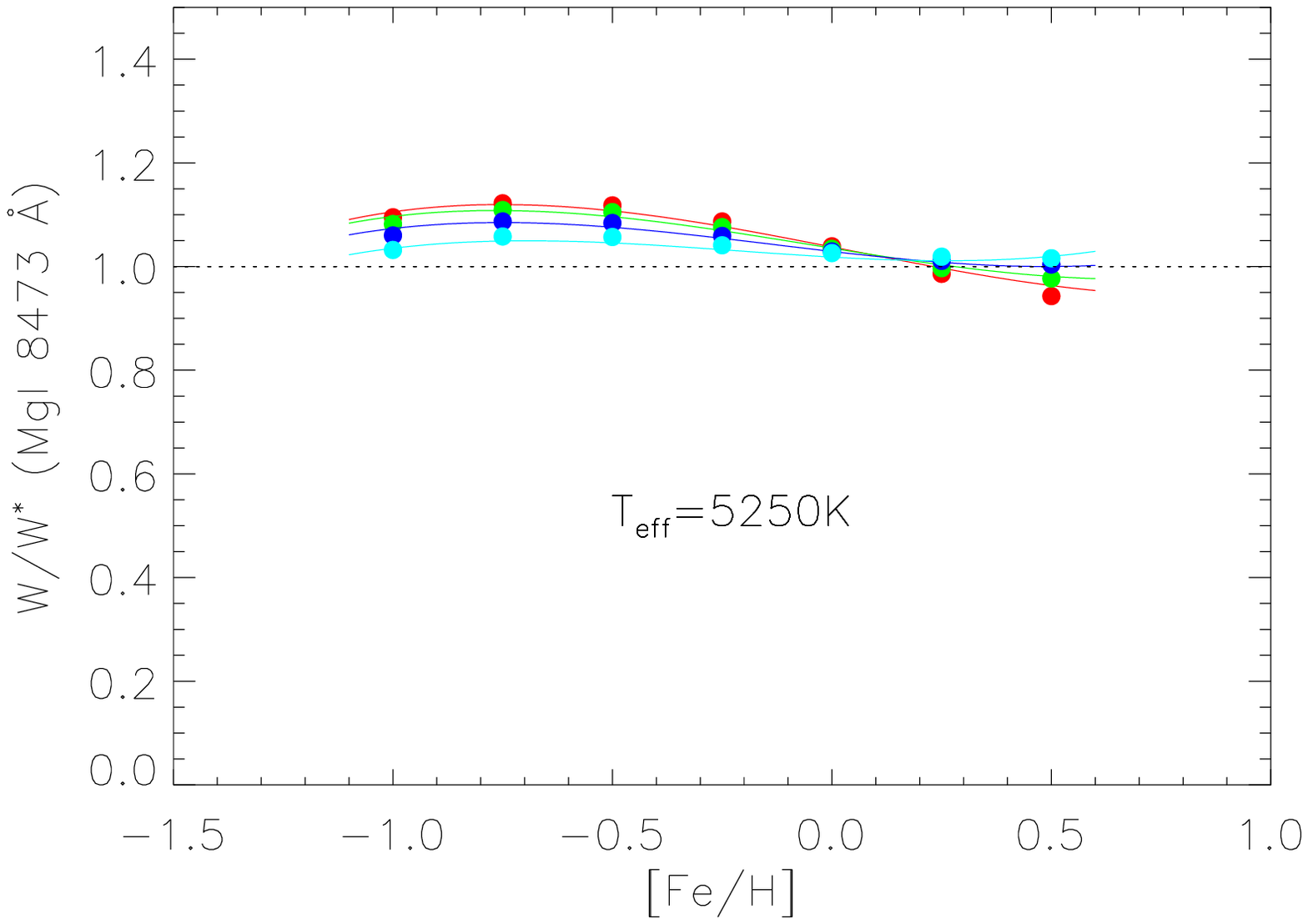}
}
\hbox{
\includegraphics[width=5.45cm]{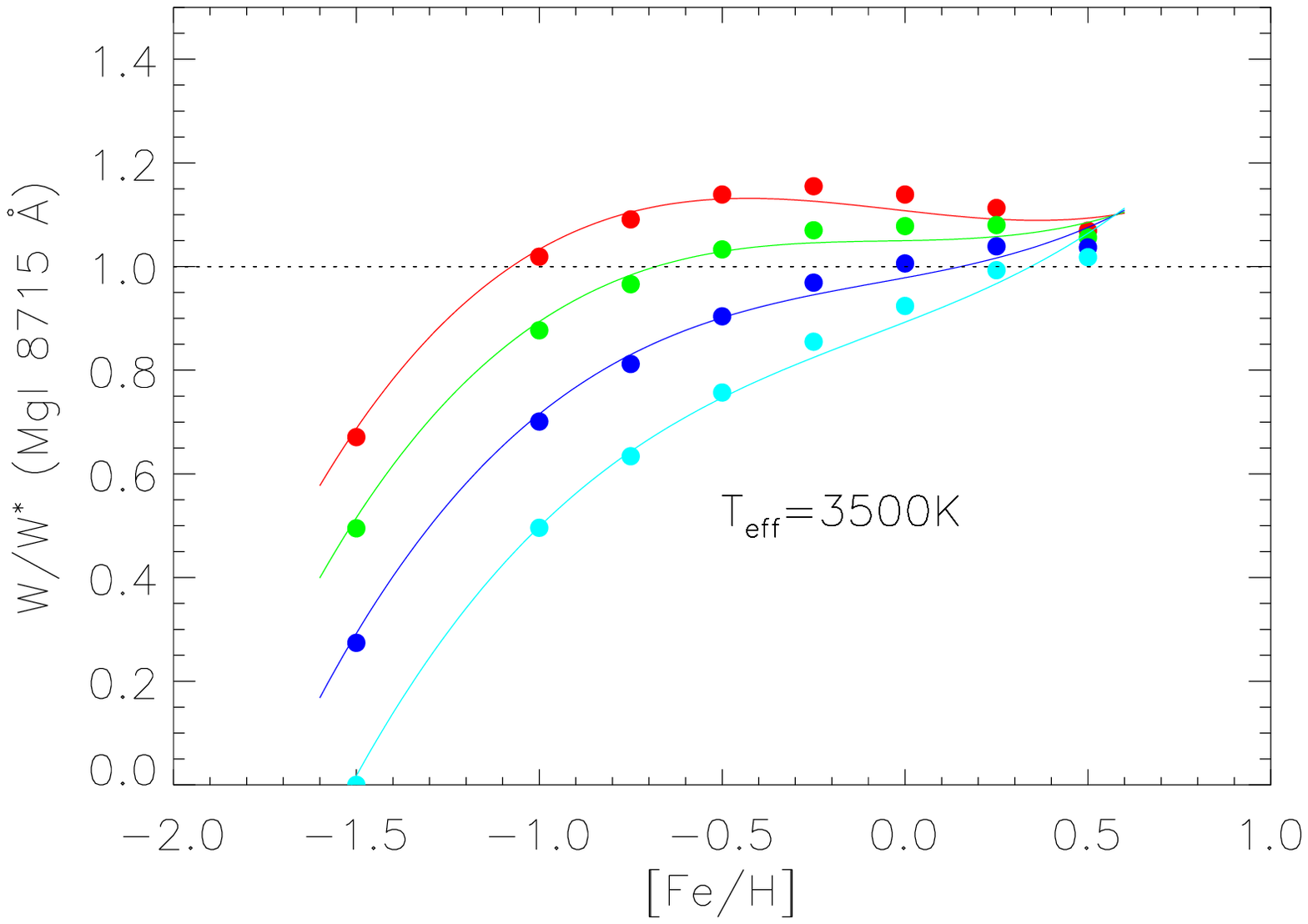}
\includegraphics[width=5.45cm]{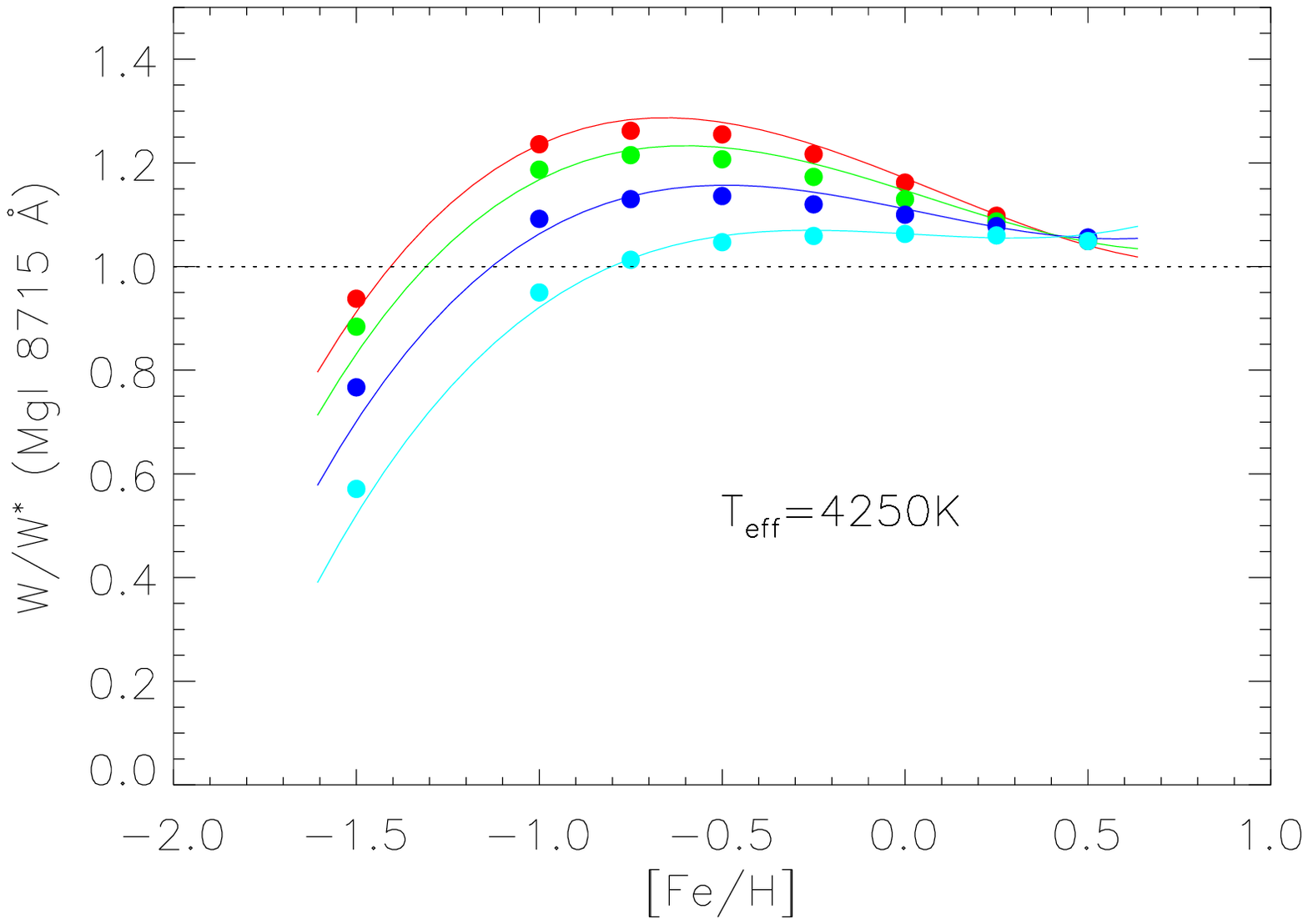}
\includegraphics[width=5.45cm]{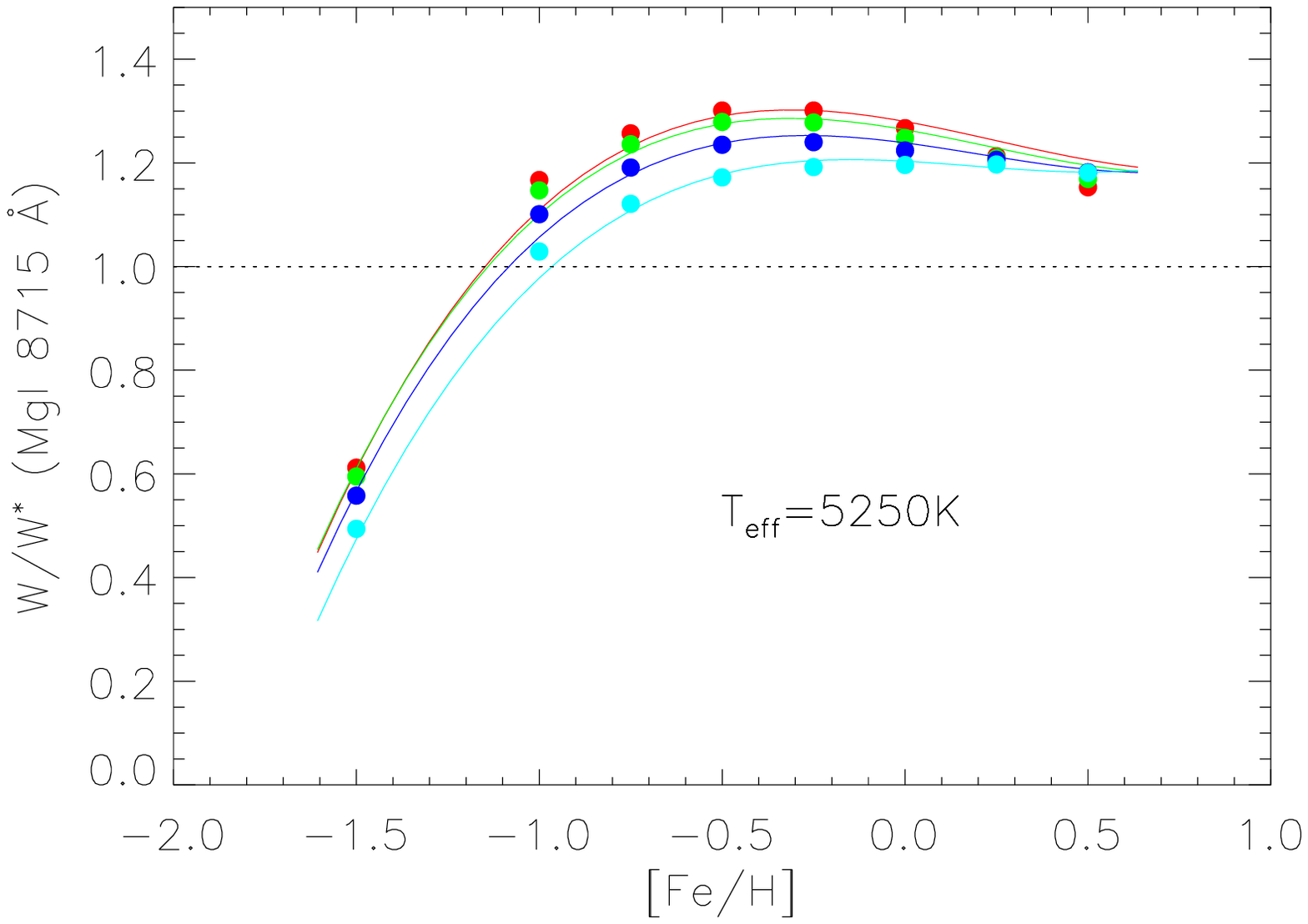}
}
\hbox{
\includegraphics[width=5.45cm]{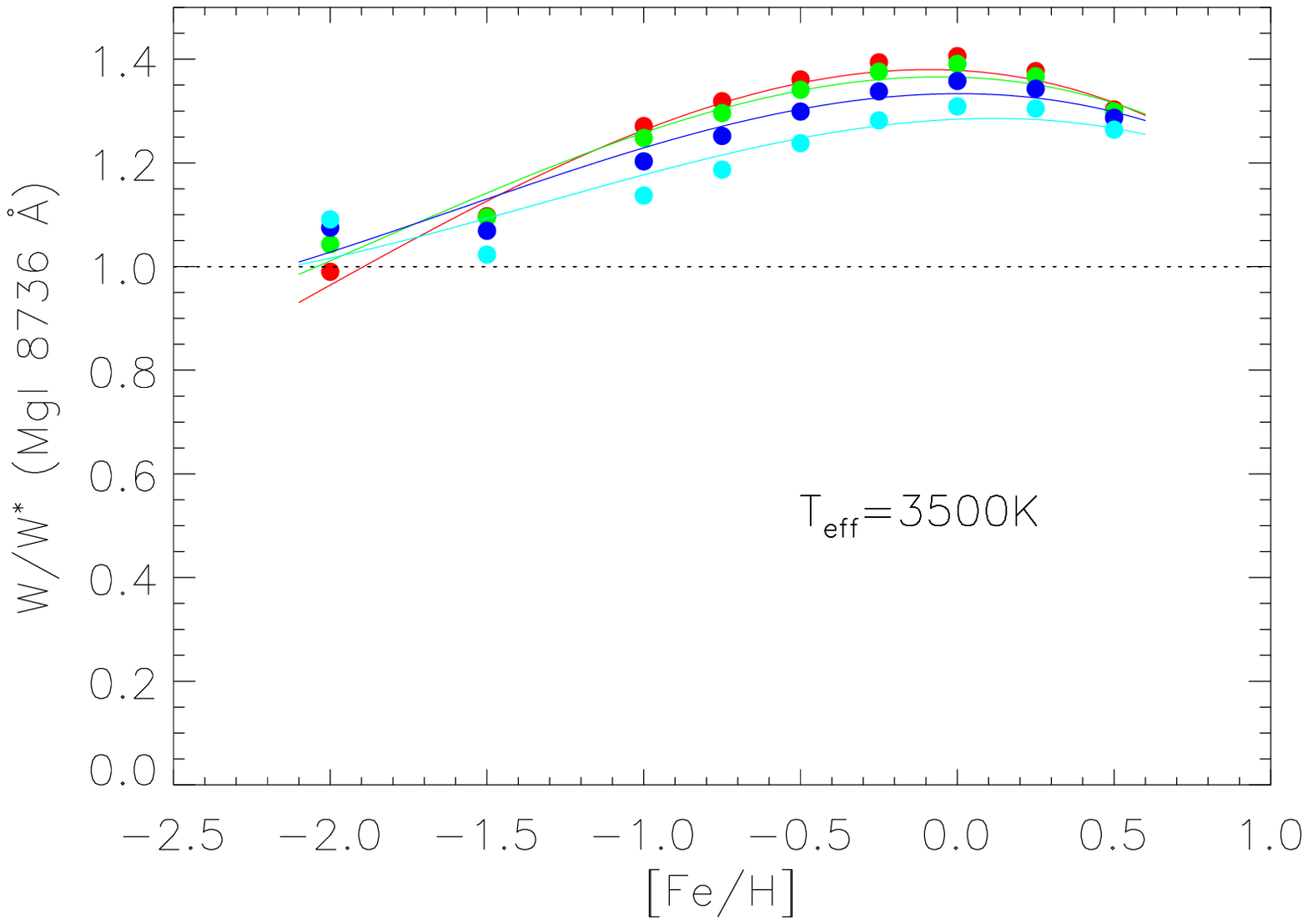}
\includegraphics[width=5.45cm]{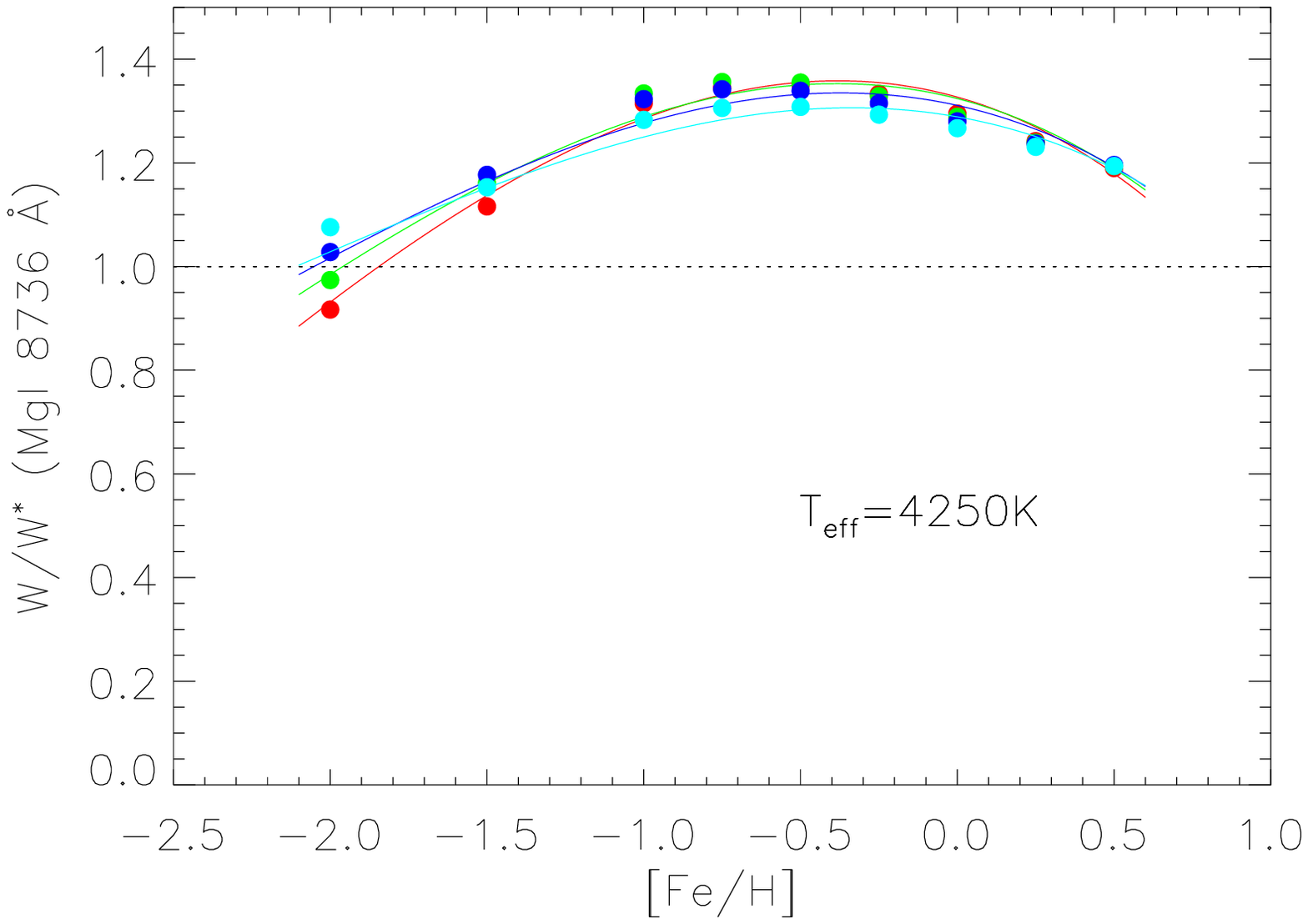}
\includegraphics[width=5.45cm]{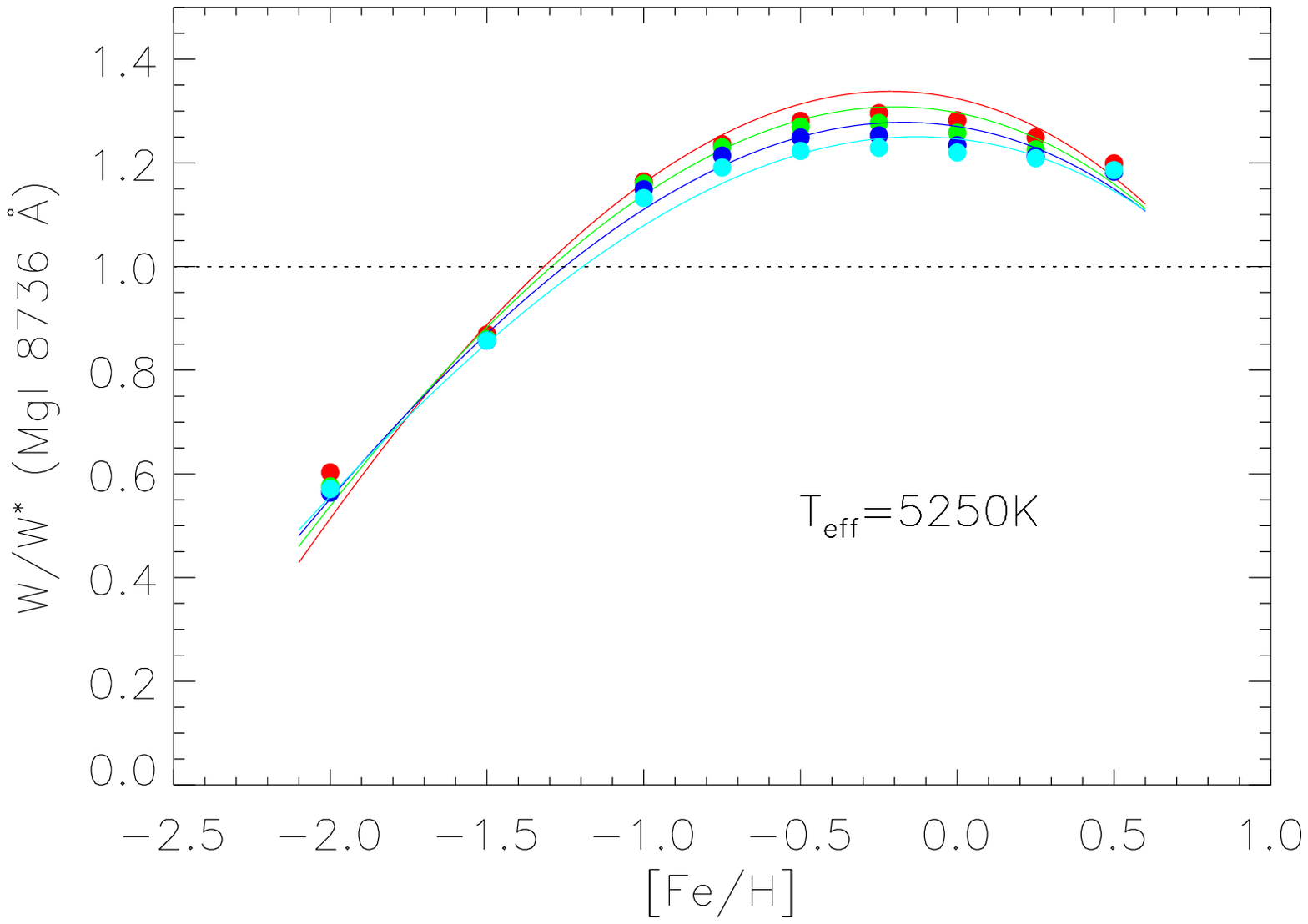}
}
\hbox{
\includegraphics[width=5.45cm]{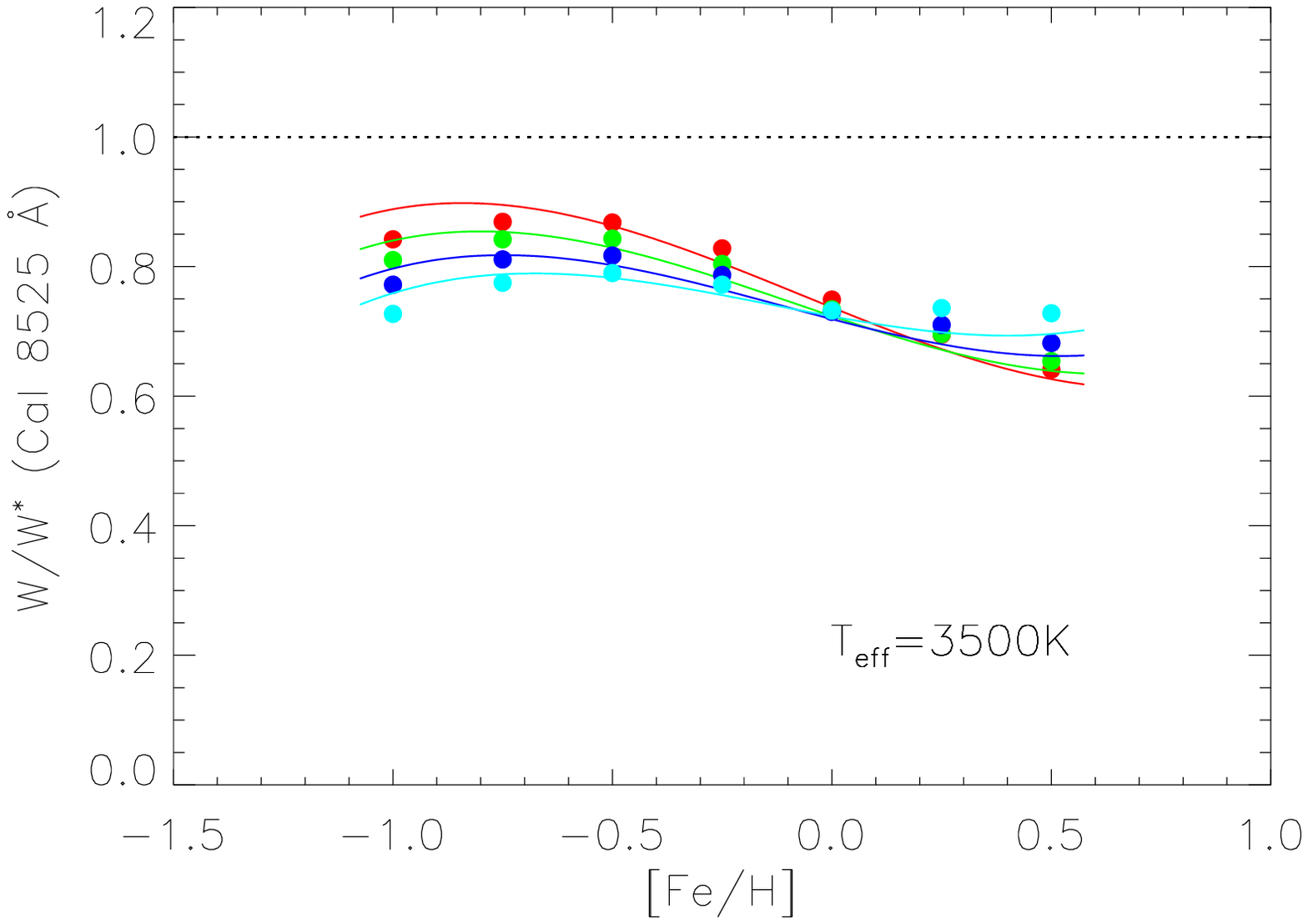}
\includegraphics[width=5.45cm]{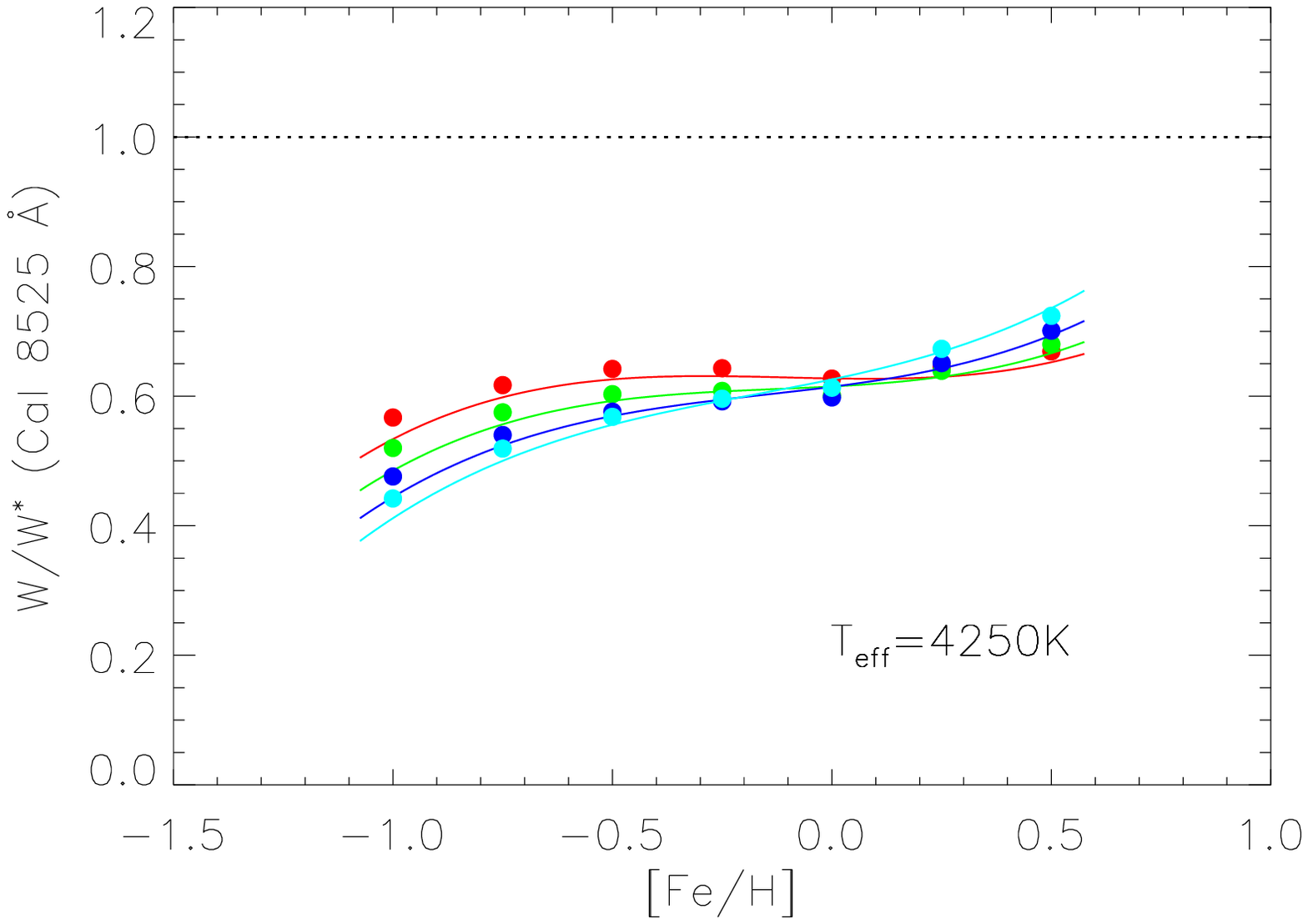}
\includegraphics[width=5.45cm]{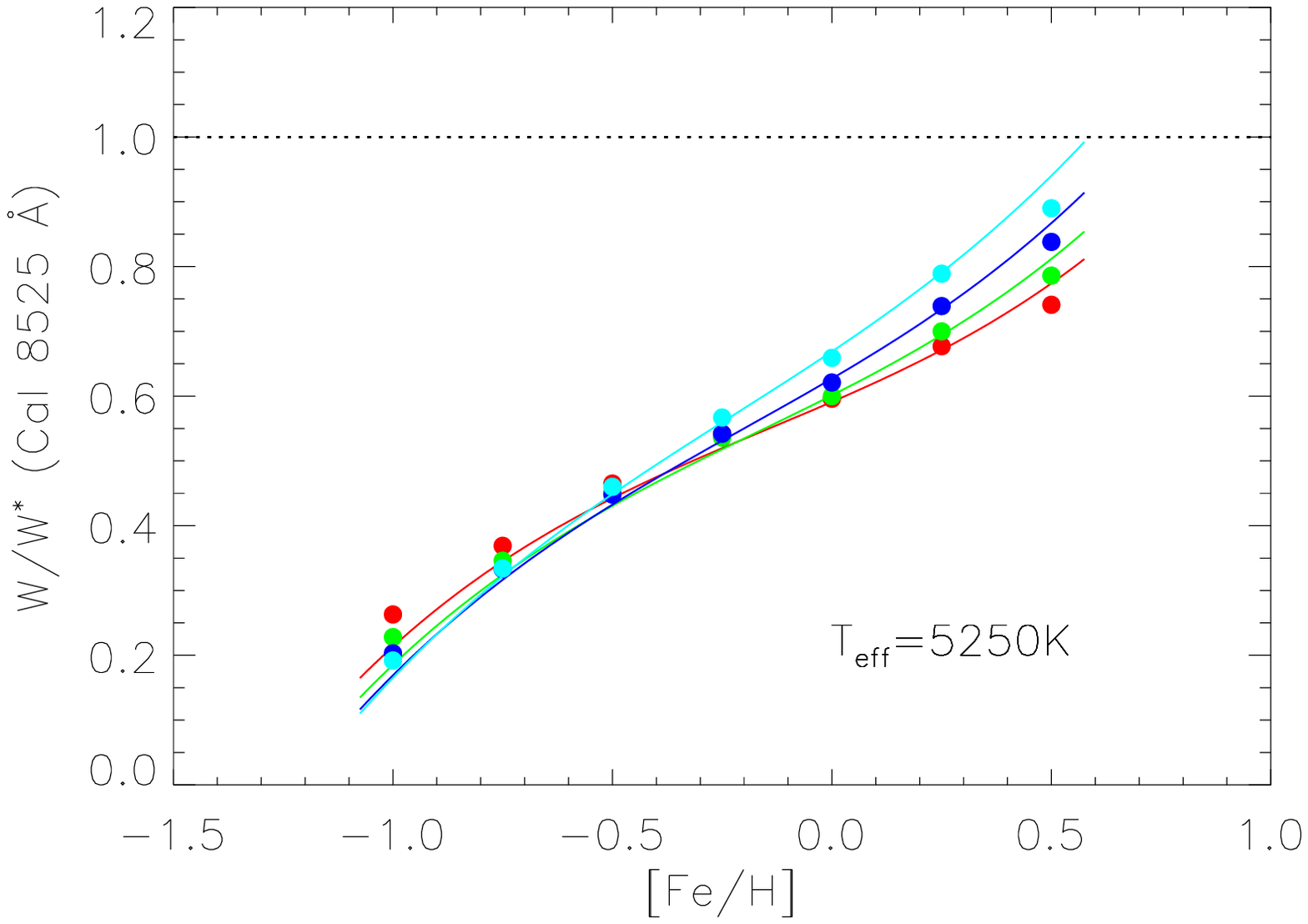}
}
\hbox{
\includegraphics[width=5.45cm]{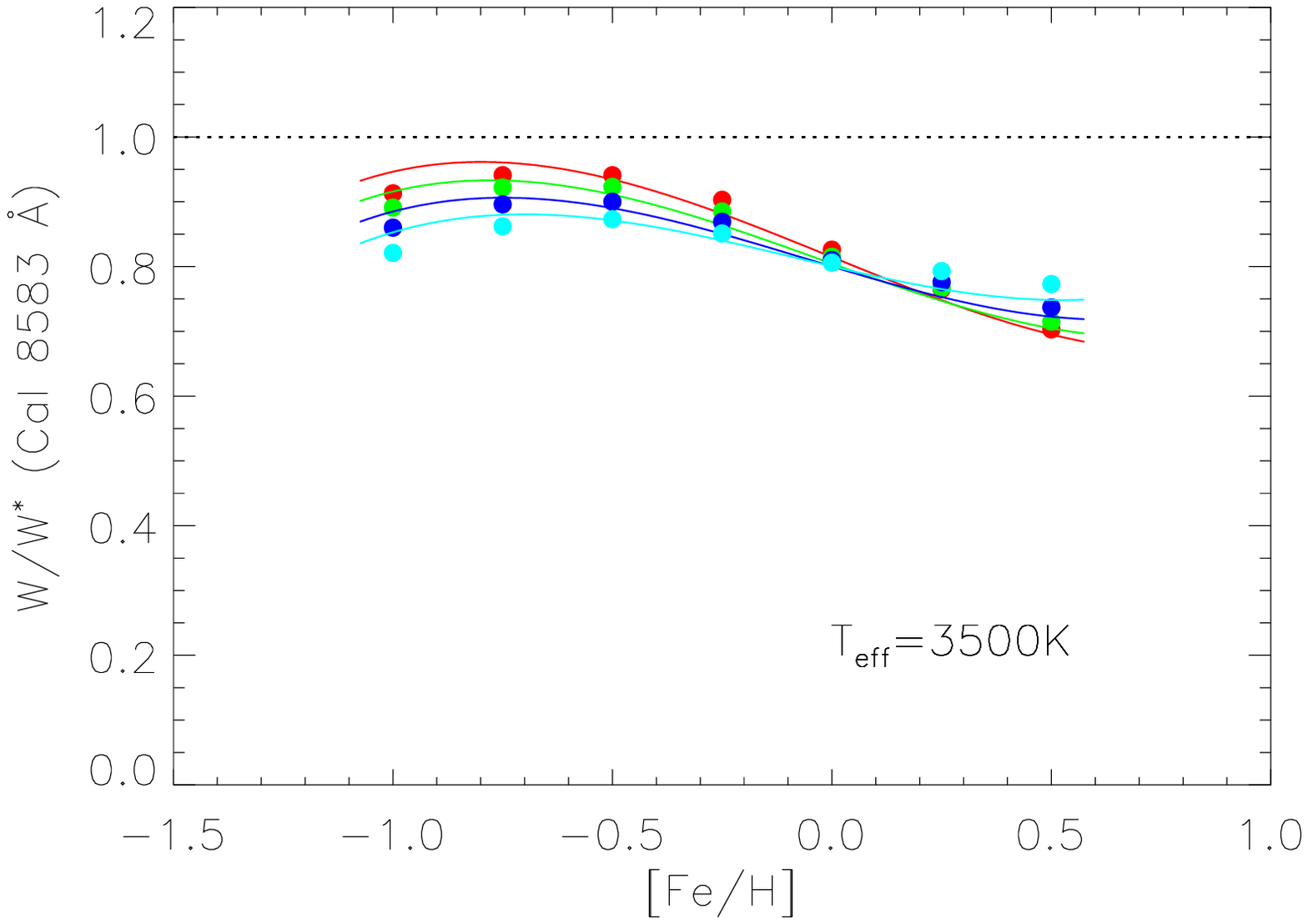}
\includegraphics[width=5.45cm]{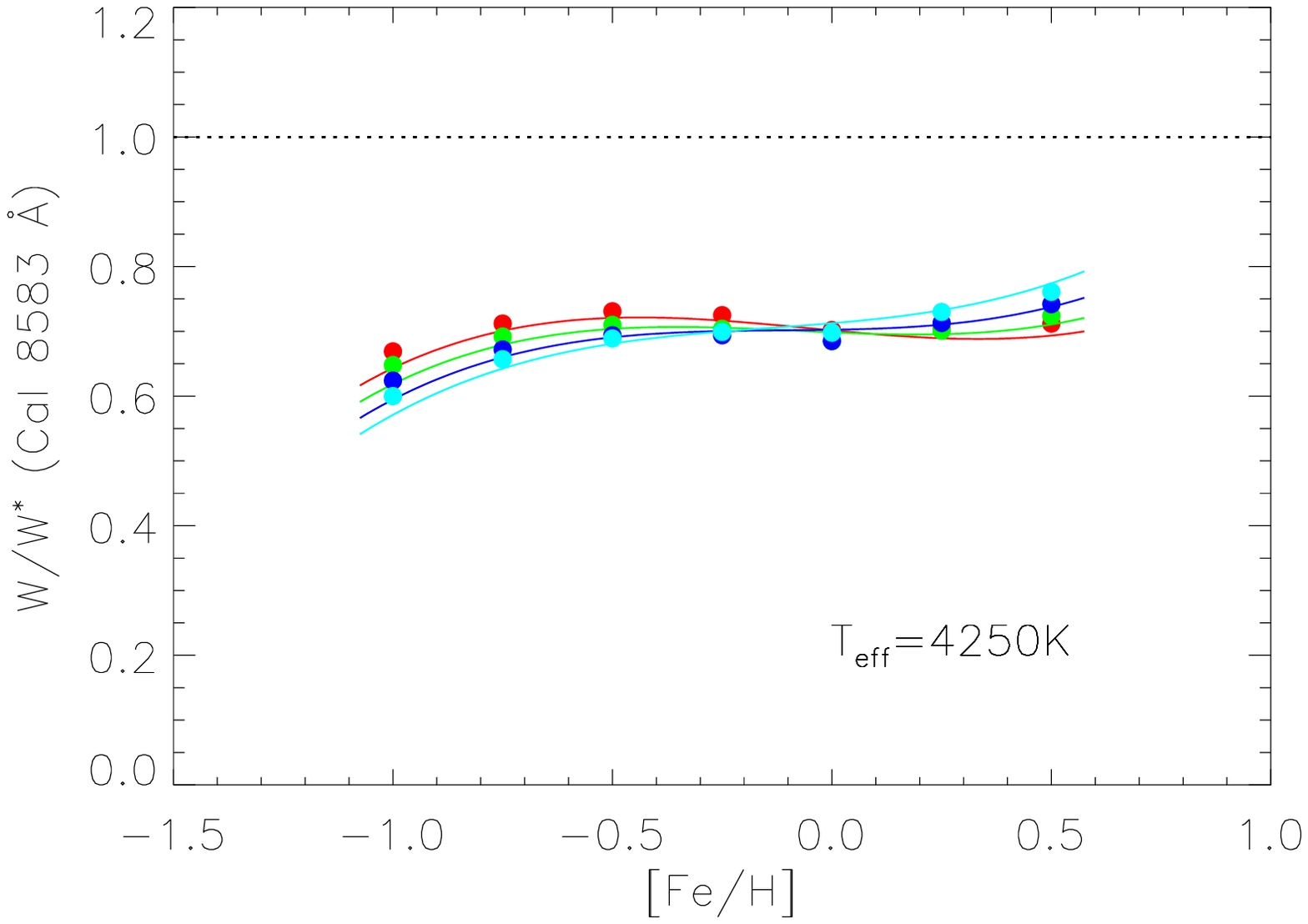}
\includegraphics[width=5.45cm]{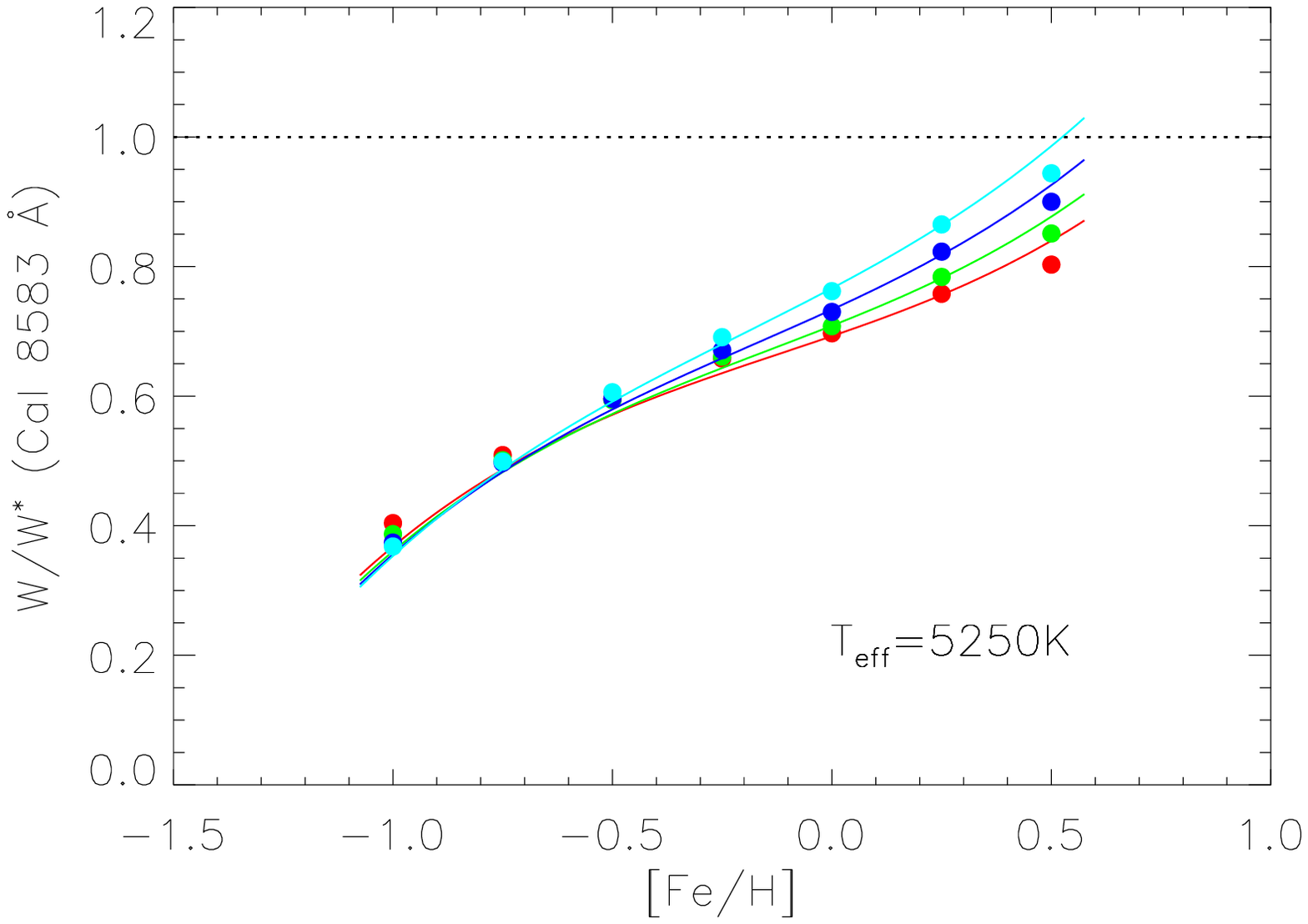}
}
\hbox{
\includegraphics[width=5.45cm]{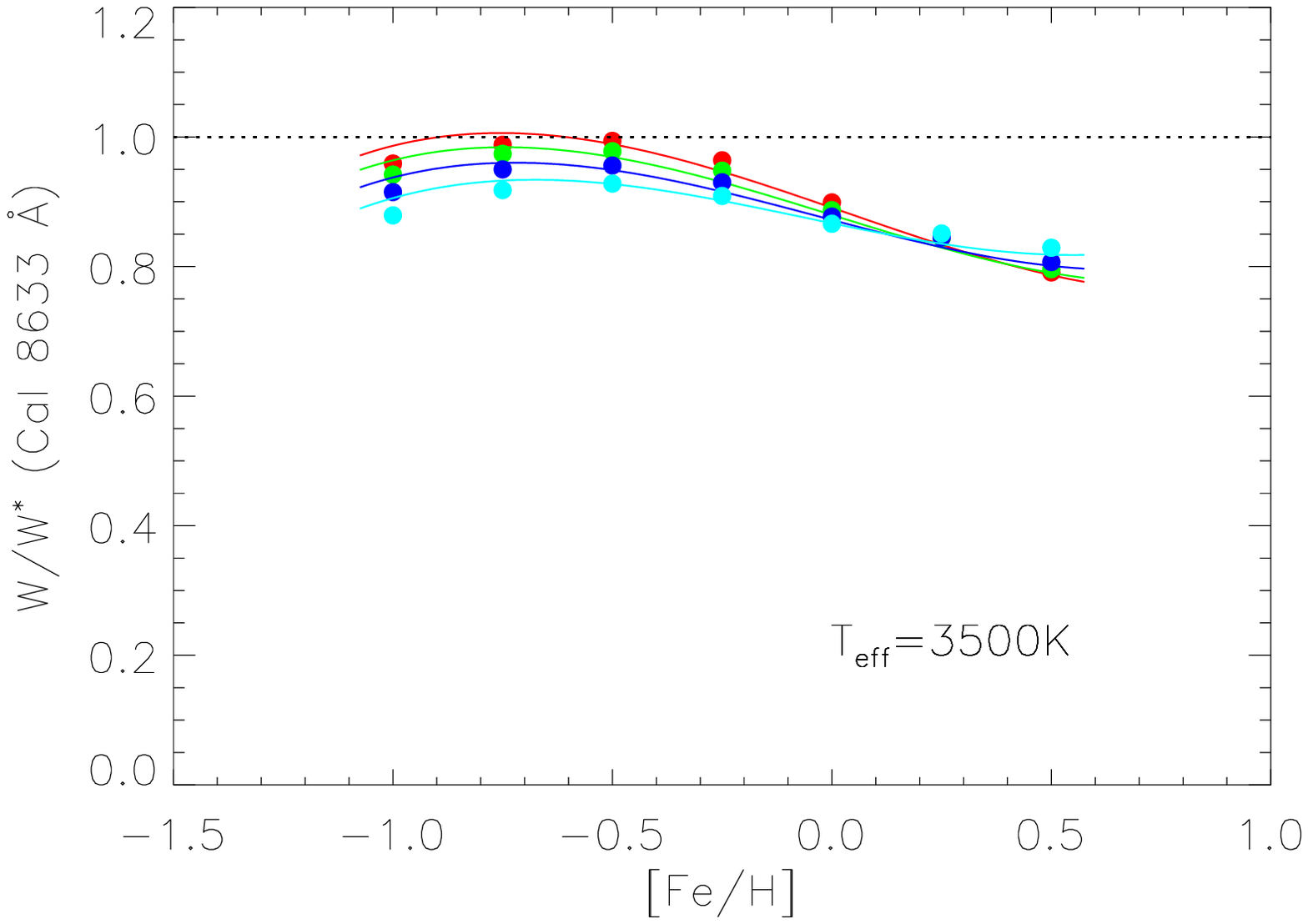}
\includegraphics[width=5.45cm]{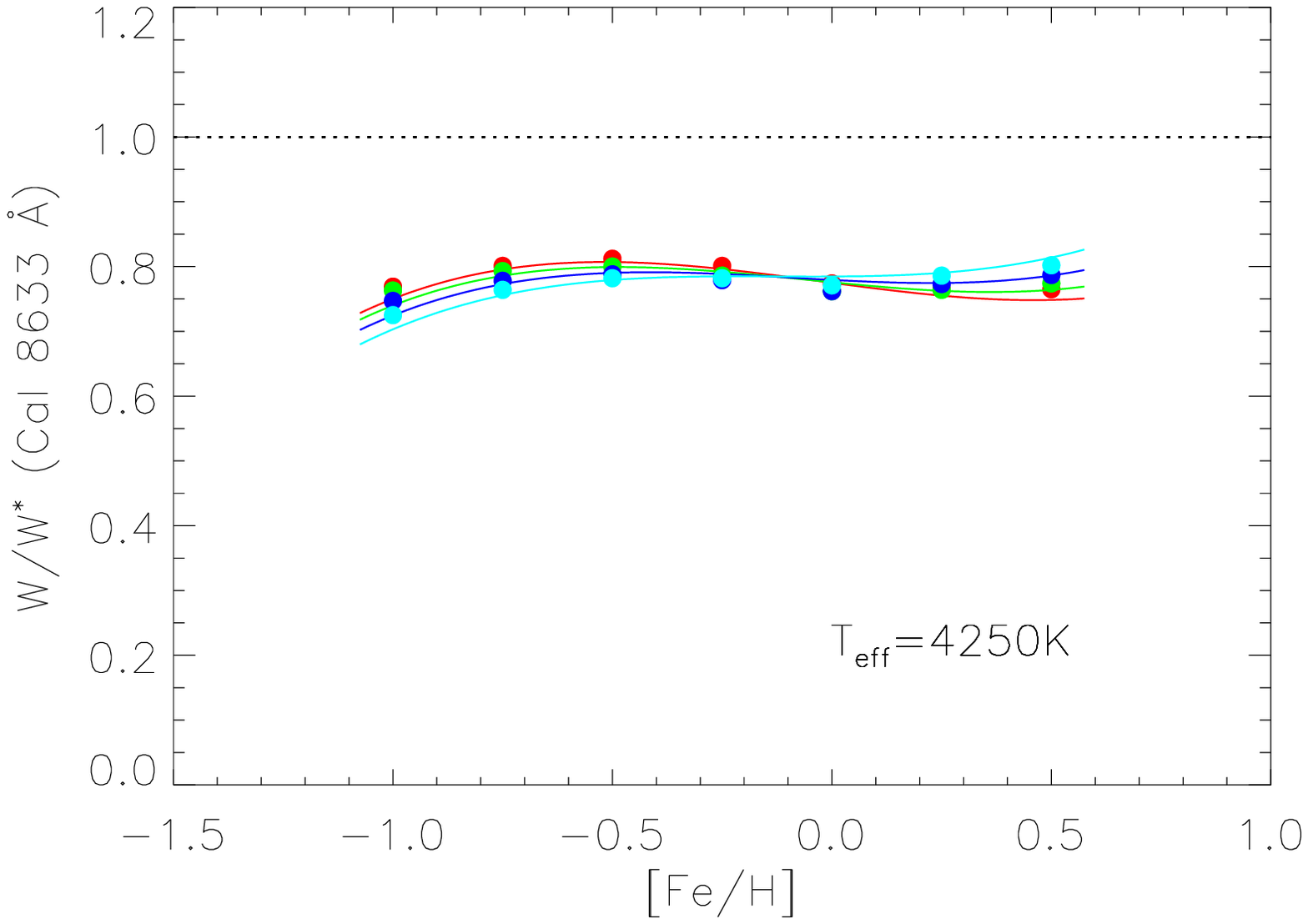}
\includegraphics[width=5.45cm]{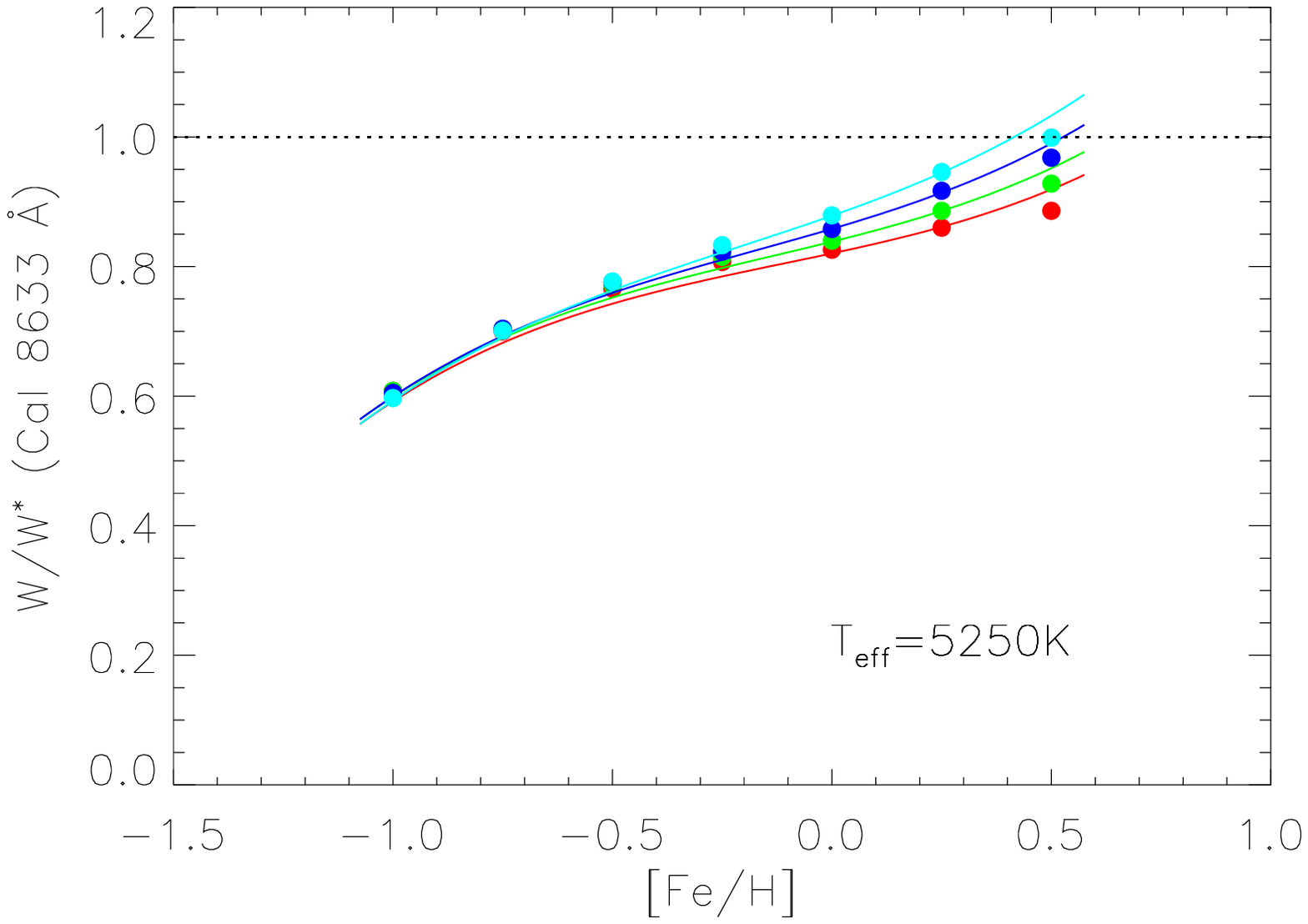}
}
\caption{$W/W^*$ for the {\it Gaia}/RVS \mg\ and \caI\ lines as function of the stellar parameters (see Appendix B for details).}
\label{Gaia_lines1}
\end{figure*}

\begin{figure*}
\hbox{
\includegraphics[width=5.45cm]{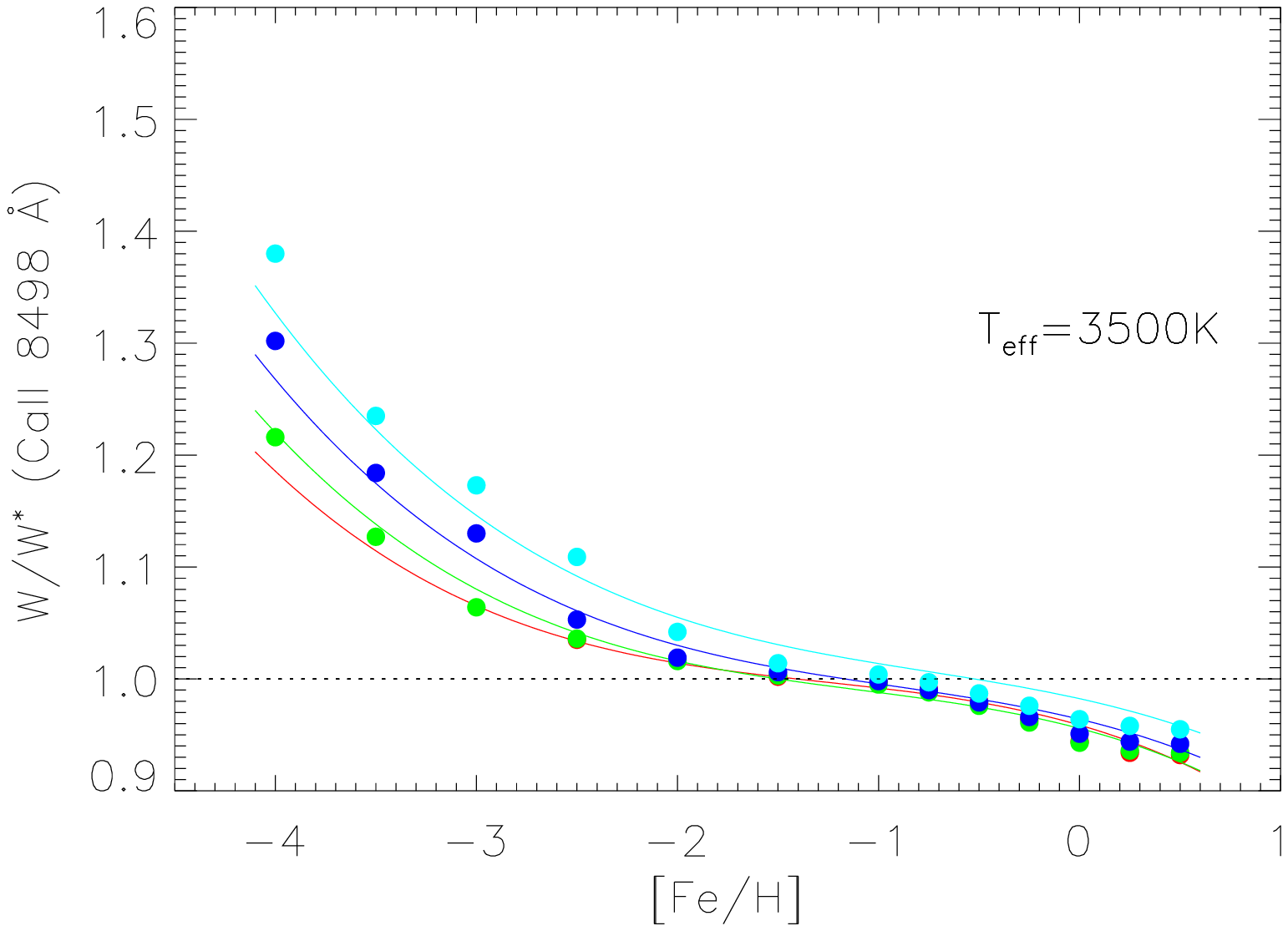}
\includegraphics[width=5.45cm]{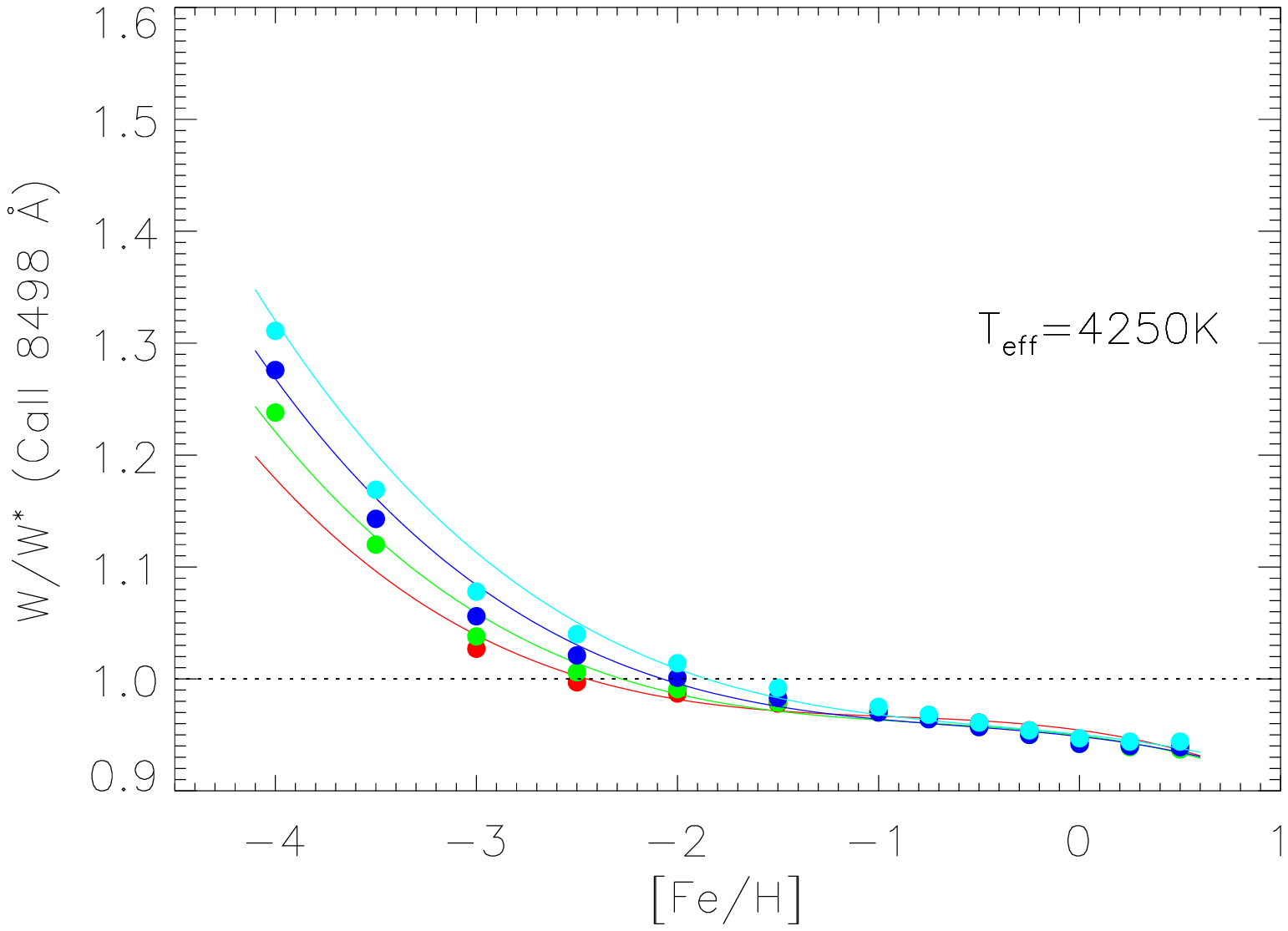}
\includegraphics[width=5.45cm]{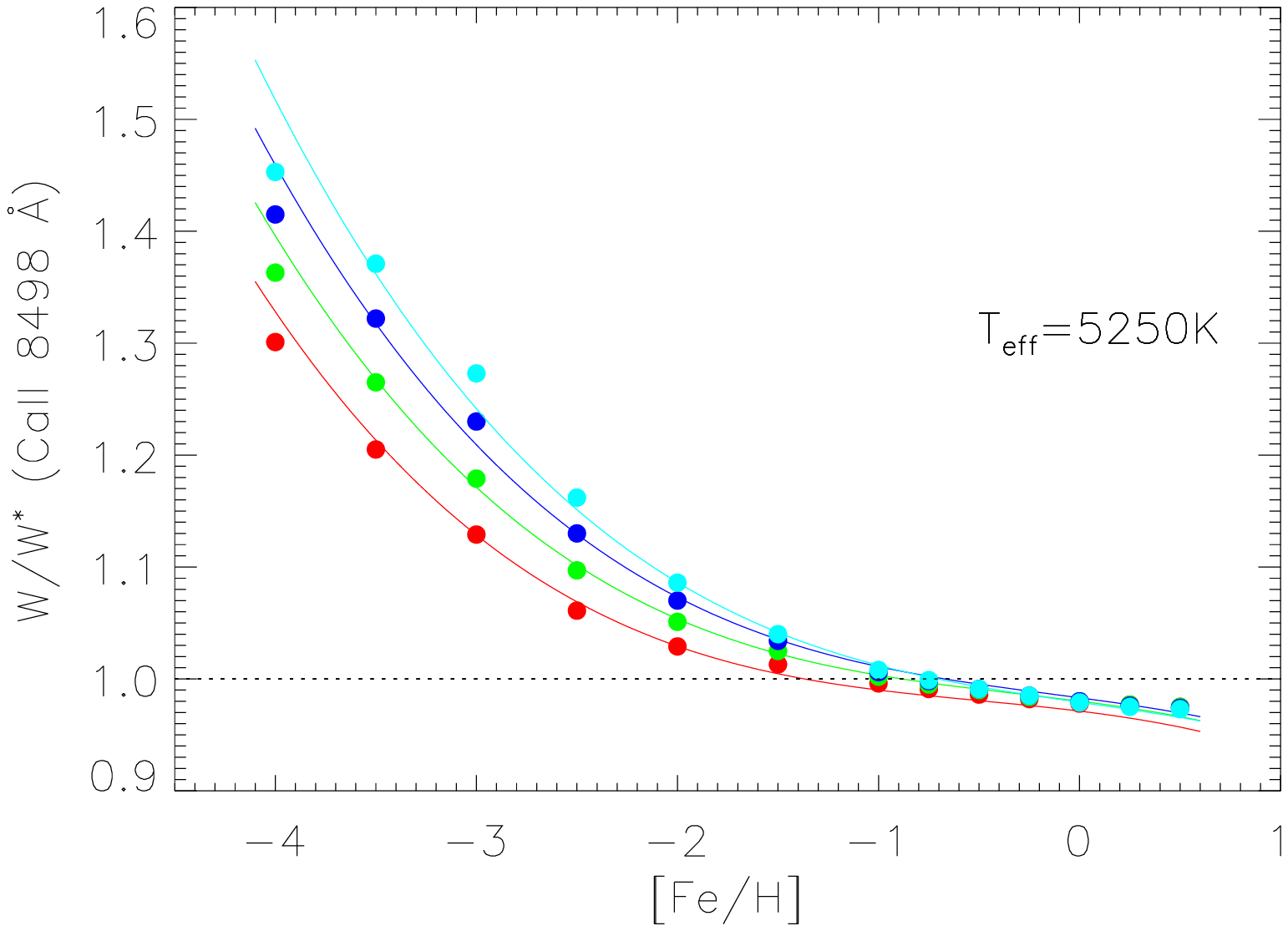}
}
\hbox{
\includegraphics[width=5.45cm]{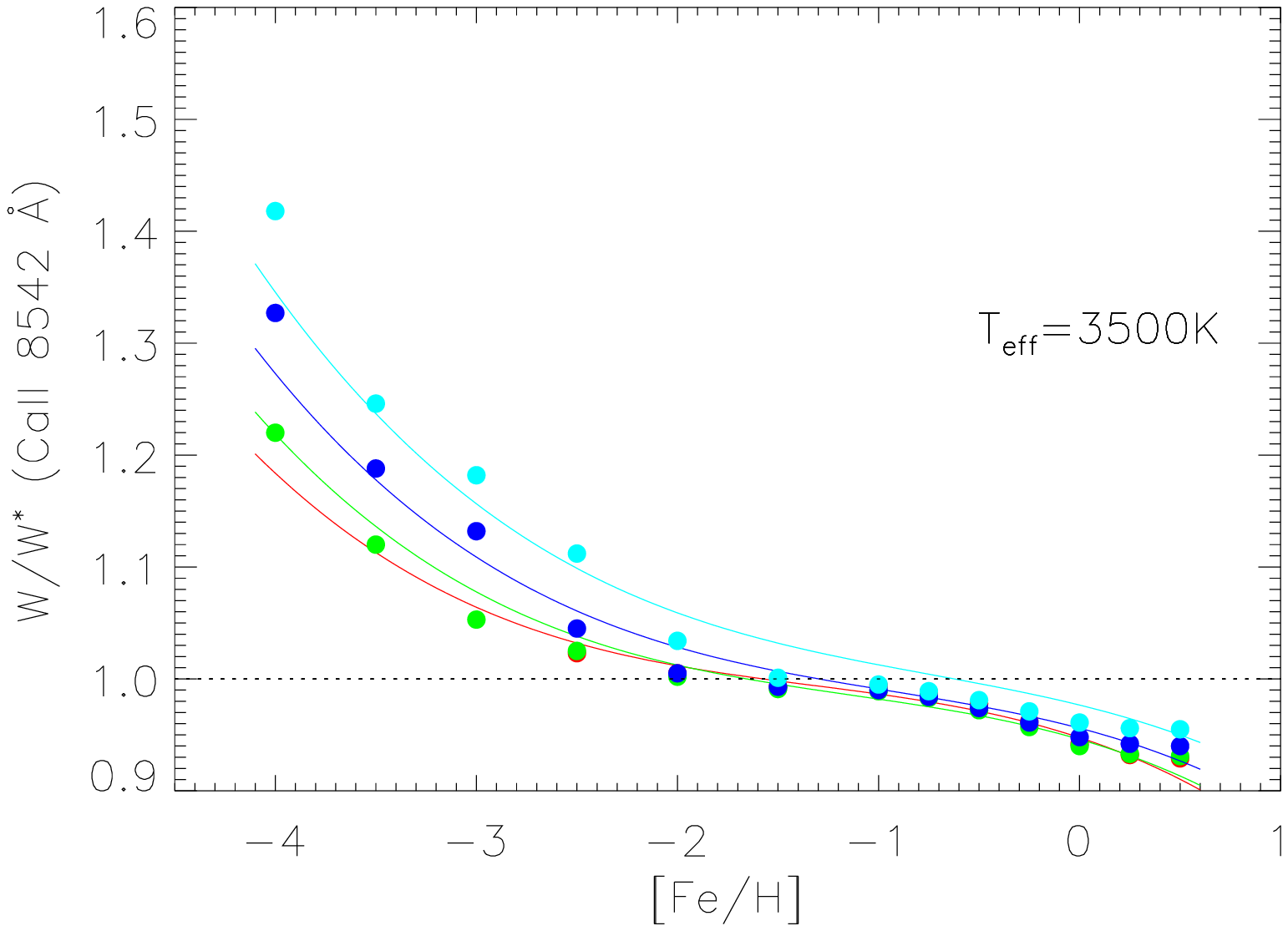}
\includegraphics[width=5.45cm]{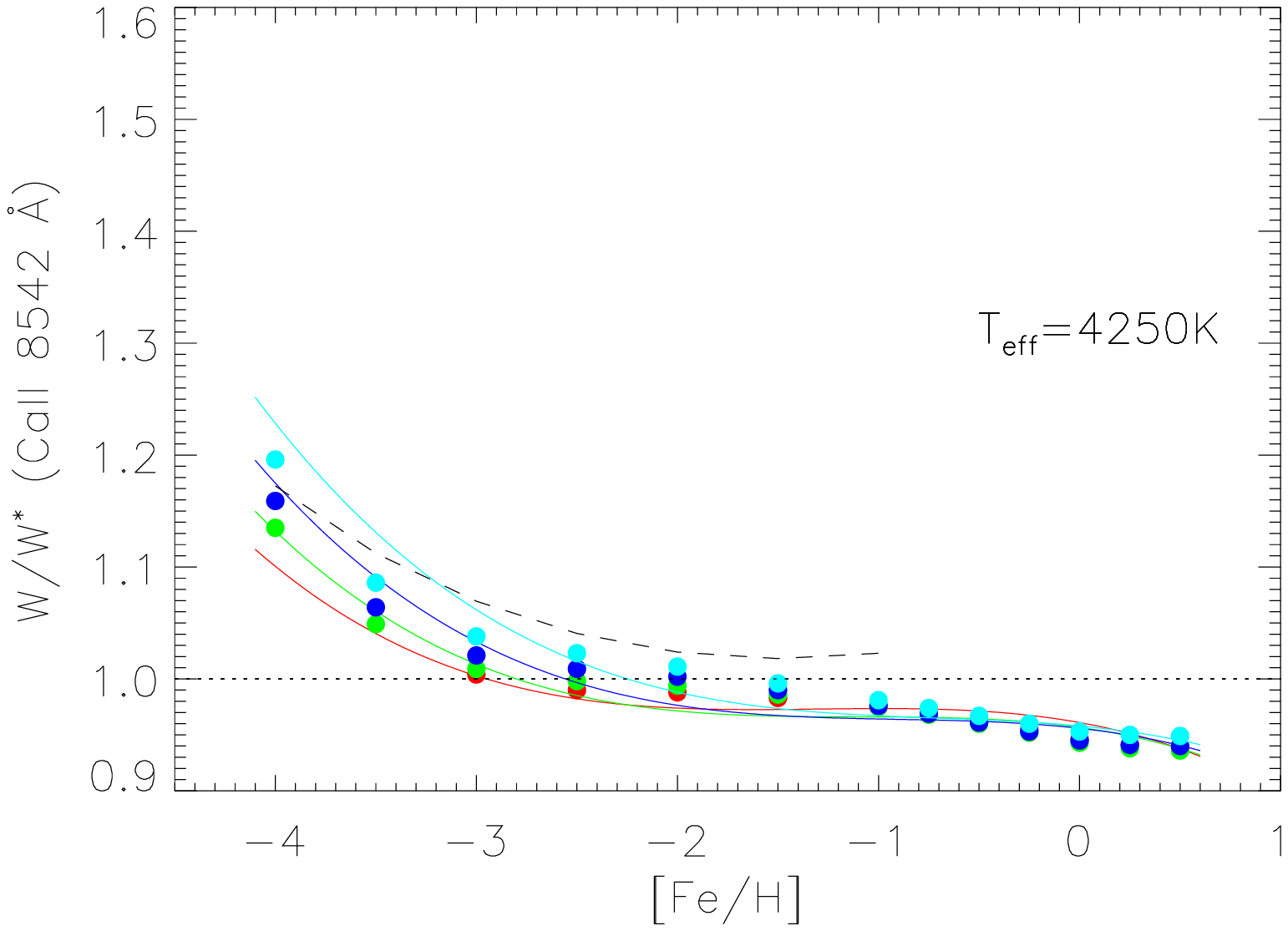}
\includegraphics[width=5.45cm]{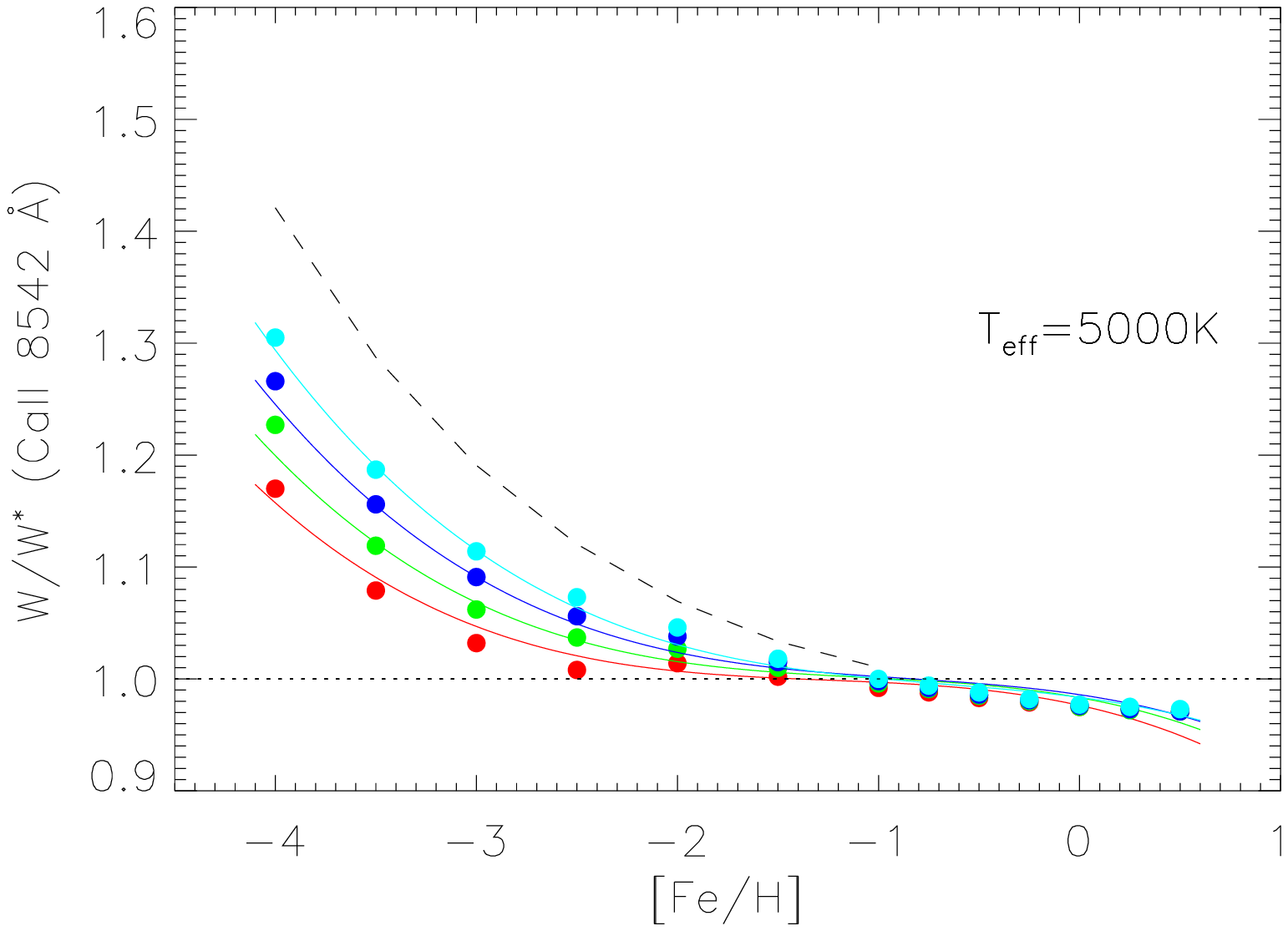}
}
\hbox{
\includegraphics[width=5.45cm]{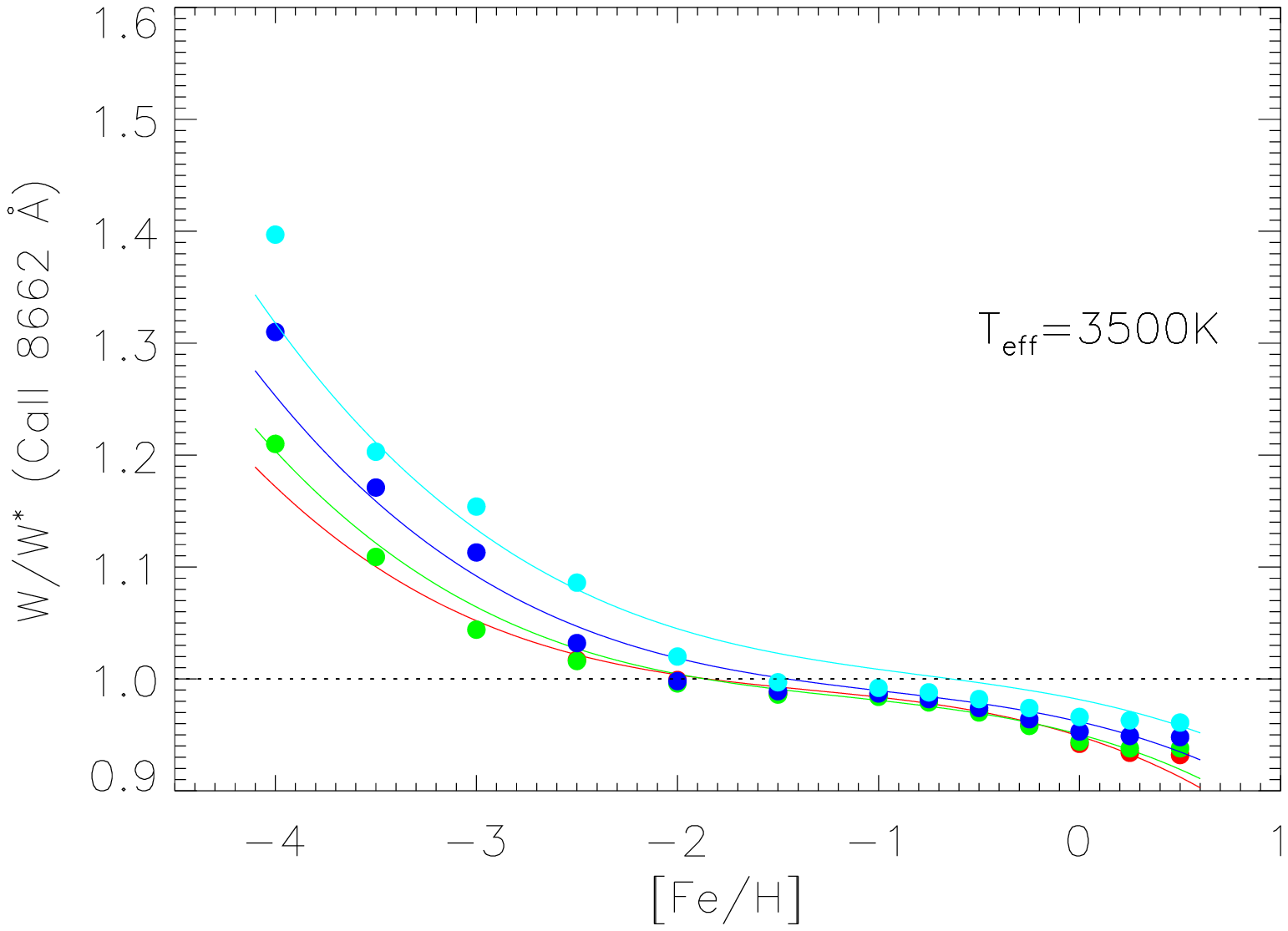}
\includegraphics[width=5.45cm]{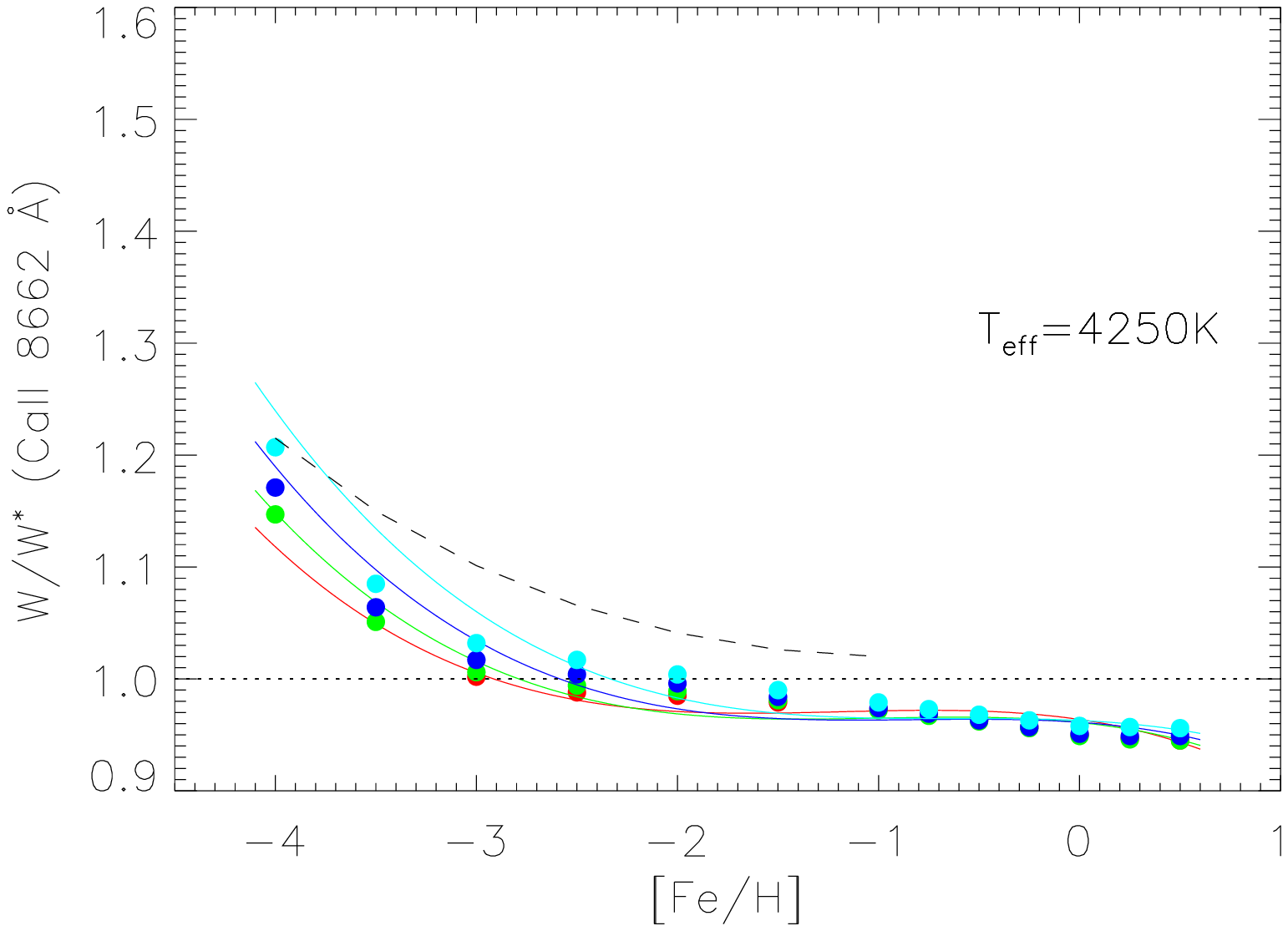}
\includegraphics[width=5.45cm]{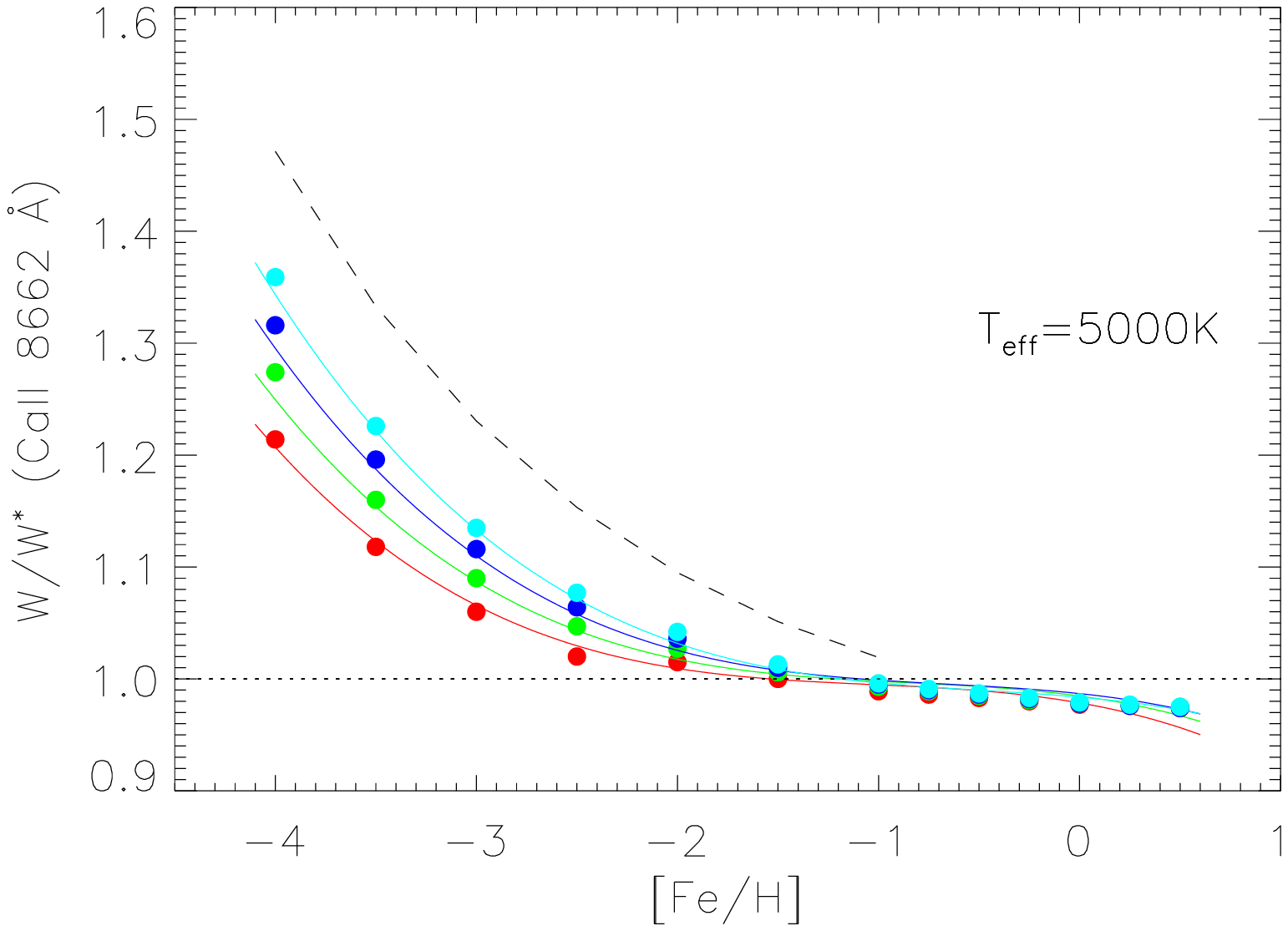}
}
\caption{$W/W^*$ for the {\it Gaia}/RVS \caII\ lines as function of the stellar parameters (see Appendix B for details).}
\label{Gaia_lines2}
\end{figure*}

We present the results of the NLTE computations and the polynomial fits as the evolution of the NLTE/LTE EW ratios $W/W^*$ as a function of the metallicity [Fe/H] for the \mg, \caI\ and \caII\ lines in the {\it Gaia}/RVS wavelength range [8470, 8740] \AA\ in Fig.~\ref{Gaia_lines1} and \ref{Gaia_lines2}. 
Each row represents an observed line, except for the quoted row at 8715 \AA\ standing for the Ritz wavelength of the \mg\ triplet at $\lambda\lambda$ 8710, 8712 and 8717.
Each panel of a row represents the $W/W^*$ for a given effective temperature (3500, 4250 and 5250~K), except for the \caII\ 8542 and 8662~\AA\ lines for which we used the following temperatures (3500, 4250 and 5000~K) in order to add polynomial fit of \citet{Starkenburg10} on the panels.
Each colour of the dots and the polynomial fits $W/W^*$ in a panel represents a surface gravity (red, green, blue and cyan for $\log g = 0.5$, 1.0, 1.5 and 2.0~dex respectively).

\bsp
\label{lastpage}
\end{document}